\documentclass[]{tADP2e}

\newcommand{\bra}{\ensuremath{\langle}}             
\newcommand{\ket}{\ensuremath{\rangle}}             
\newcommand{\tql}{\textquotedblleft}                
\newcommand{\tqr}{\textquotedblright}               
\newcommand{\re}{\text{Re}}                         
\newcommand{\ii}{\text{i}}                          
\newcommand{\ee}{\text{e}}                          
\newcommand{\HC}{\text{H.c.}}                       
\newcommand{\cSr}{\textsf{Sr}}                      
\newcommand{\cCu}{\textsf{Cu}}                      
\newcommand{\cO}{\textsf{O}}                        
\newcommand{\cBa}{\textsf{Ba}}                      
\newcommand{\cNa}{\textsf{Na}}                      
\newcommand{\cCr}{\textsf{Cr}}                      
\newcommand{\cGa}{\textsf{Ga}}                      
\newcommand{\cCo}{\textsf{Co}}                      
\newcommand{\cH}{\textsf{H}}                        
\newcommand{\cSi}{\textsf{Si}}                      
\newcommand{\cPr}{\textsf{Pr}}                      
\newcommand{\cNi}{\textsf{Ni}}                      
\newcommand{\cV}{\textsf{V}}                        
\newcommand{\cZn}{\textsf{Zn}}                      
\newcommand{\cCl}{\textsf{Cl}}                      
\newcommand{\pdagger}{\phantom{\dagger}}            

\newcommand{\bs}[1]{\boldsymbol{#1}}
\def\ea{\emph{et al.}}

\usepackage{blkarray}
\usepackage{color}
\usepackage[table]{xcolor}
\definecolor{dunkelgrau}{rgb}{0.8,0.8,0.8}
\definecolor{hellgrau}{rgb}{0.95,0.95,0.95}
\definecolor{hellgelb}{rgb}{1,1,0.5}
\definecolor{hellblau}{rgb}{0.75,1,1}
\definecolor{hellcreme}{rgb}{1,1,0.75}
\definecolor{hellrot}{hsb}{1,0.3,1}
\definecolor{hellblau2}{hsb}{0.6,0.3,1}
\usepackage{colortbl}

\begin{document}
\doi{10.1080/0001873YYxxxxxxxx}
 \issn{1460-6976}
\issnp{0001-8732}  \jvol{00} \jnum{00} \jyear{2013} \jmonth{October}

\markboth{C. Platt, W. Hanke, and R. Thomale}{Advances in Physics}

\articletype{REVIEW ARTICLE}

\title{Functional Renormalization Group for \\ multi-orbital Fermi
  Surface Instabilities}

\author{C. Platt, W. Hanke, and R. Thomale$^{\ast}$\thanks{$^\ast$Corresponding author. Email: rthomale@physik.uni-wuerzburg.de\vspace{6pt}}\\\vspace{6pt}  
    {\em{Institute for Theoretical Physics and Astrophysics,
        Julius Maximilians University of W\"urzburg, Am Hubland
        D-97074 W\"urzburg, Germany}}
    \\\vspace{6pt}\received{v2.1 released April 2008}
}

\maketitle

\begin{abstract}
Technological progress in material synthesis, as well as artificial
realization of condensed matter scenarios via ultra-cold atomic gases
in optical lattices or epitaxial growth of thin films, is opening the
gate to investigate a plethora of unprecedented strongly correlated
electron systems. In a large subclass thereof, a metallic state of
layered electrons undergoes an ordering transition below some
temperature into unconventional states of matter driven by electronic
correlations, such as magnetism, superconductivity, or other Fermi
surface instabilities. While this type of phenomena has been a
well-established direction of research in condensed matter for
decades, the variety of today's accessible scenarios pose fundamental
new challenges to describe them. A core complication is the
multi-orbital nature of the low-energy electronic structure of these
systems, such as the multi-$d$ orbital nature of electrons in iron
pnictides and transition-metal oxides in general, but also electronic
states of matter on lattices with multiple sites per unit cell such as
the honeycomb or kagome lattice. In this review, we propagate the
functional renormalization group  (FRG) as a suited approach to
investigate multi-orbital Fermi surface instabilities. The primary
goal of the review is to describe the FRG in explicit detail and
render it accessible to everyone both at a technical and intuitive
level. Summarizing recent progress in the field of multi-orbital Fermi
surface instabilities, we illustrate how the unbiased fashion by which
the FRG treats all kinds of ordering tendencies guarantees an adequate
description of electronic phase diagrams and often allows to obtain
parameter trends of sufficient accuracy to make qualitative
predictions for experiments. This review includes detailed and illustrative
illustrations of magnetism and, in particular, superconductivity for
the iron pnictides from the viewpoint of FRG. Furthermore, it
discusses candidate scenarios for topological bulk singlet
superconductivity and exotic particle-hole condensates on hexagonal
lattices such as sodium-doped cobaltates, graphene doped to van Hove
filling, and the kagome Hubbard model. In total, the FRG promises to be one of the most versatile and revealing numerical approaches to address unconventional Fermi surface instabilities in future fields of condensed matter research.


\bigskip

\end{abstract}

\section{Introduction}

Pioneered by Landau's Fermi liquid hypothesis, the quantum theory of metals can be considered one of the theoretical milestones in 
condensed matter physics~\cite{abrikosov1988fundamentals}. Interpreted as a pseudo-vacuum of correlated electrons with different 
fluctuation tendencies but no symmetry breaking, the interactions between the electrons in the gapless metallic state can seed 
various kinds of ordering transitions via the condensation of composite electronic degrees of freedom. Particle-particle condensates 
break U(1) symmetry and refer to superconductivity where the Cooper pairs represent the condensing particles, while particle-hole 
condensates can relate to various kinds of charge or spin-density wave order depending on the specific symmetry breaking in lattice 
and spin space. This itinerant perspective on ordering phenomena in electron systems, driven by electronic correlations, starts 
from an effectively non-interacting Fermi liquid point of view, on top of which electronic interactions are then 
taken into account. In many cases such as for superconductivity where, in the absence of disorder and presence of inversion 
symmetry, arbitrarily small interactions are already sufficient to drive this kind of order, many salient features can already 
be investigated from an asympotically exact perturbative approach at infinitesimal coupling~\cite{kohn-65prl524,raghu-10prb224505,PhysRevB.88.064505}.

The generic material scenario of correlated electrons is, however, located at intermediate coupling, where neither the finite 
coupling strength of interactions nor the finite bandwidth of the electronic bands is negligible. This explains why for many 
candidate materials such as cuprates, there often exists a complementary perspective from strong coupling where one starts 
from a localized Mott state of electrons or, alternatively, from weak coupling, where the non-interacting Fermi liquid state 
is the point of expansion. Choosing the latter, the crucial challenge is to appropriately include interactions. Since the 
early works of Hartree and Fock~\cite{hartree,fock}, various approximations have been formulated to treat this problem, 
which also led to a florishing use of diagrammatic methods in condensed matter physics. Until today, one central approach 
has been to incorporate an educated guess as to which kind of diagrams, or stated differently, fluctuations, are most 
important to adequately treat the given 
electronic model. This is the central idea of the random phase approximation (RPA), where a certain type of channel of 
the two-particle Green's function is specified (e.g.  the particle-particle or the direct/indirect particle-hole channel), 
which then allows to consider interactions up to infinite order within this channel~\cite{PhysRev.112.1900}. 

If, however, our purpose is to obtain a reasonable synopsis of electronic phases in a candidate material as a function 
of e.g. doping or pressure, this still may not be enough. The main reason is that condensed phases from different 
channels jointly occur in such phase diagrams, and, more importantly, that these phases exhibit crucial 
interdependencies which are indispensable for understanding the underlying microscopic nature. A paradigmatic 
example are of course the high-$T_c$ cuprates where, from many different theoretical viewpoints, the antiferromagnetic 
state, i.e. particle-hole condensate, at small doping is intimately connected to the superconducting state, i.e. 
particle-particle condensate, at larger doping~\cite{RevModPhys.78.17}.

{\it The unbiased consideration of interactions for all channels is the central advancement of the functional 
renormalization group (FRG) algorithm.}
The conceptual idea of FRG for interacting fermion systems roots back to 
Wilsonian RG~\cite{PhysRevA.8.401,polchinski-84npb269,wetterich-93plb199390}, where, however, it is 
important to note that the fermionic scenario does not have a finite number of fixed points, but a 
fixed point manifold represented by the Fermi surface. Therefore, one intends to investigate the 
scattering vertex function as a function of an RG flow parameter $\Lambda$, which in this case 
generically acts as a momentum or frequency cutoff to the Fermi surface. Note that $\Lambda$ as 
a cutoff constraint only modifies the Greens function and, thus, only the measure of the fermionic 
action. This is the core insight in the derivation of RG flow equations for many-particle vertex 
functions from this action~\cite{RevModPhys.84.299}. 
As a function of the RG parameter flow, the cutoff is lowered, and one thus effectively integrates 
out higher energy modes as one approaches the Fermi surface. In particular, most relevant for our 
subsequent investigations, the two-particle fermionic vertex function can, thus, be tracked as a 
function of $\Lambda$, allowing to investigate how the effective electronic interaction profile 
evolves as more and more modes are integrated out towards the low-energy theory. 

When a Fermi surface instabilitiy sets in, this manifests itself as a breakdown of the RG flow, as 
a certain set of scattering vertex channels flows to strong coupling, hinting the propensity of the 
system to break certain symmetries of the Fermi liquid. Comparing this approach to RPA, the FRG does 
not single out a certain parquet channel when the evolution of the two-particle scattering vertex is 
computed, which will be elaborated on in detail in this review. Even if the RPA were computed for all 
channels, i.e. an RPA series were set up for the particle-particle, as well as direct / indirect 
particle-hole channels individually, the FRG would go beyond this summation procedure because mixing 
between the channels is taken into account. 
(Stated differently, recast into diagrammatic language, the FRG allows to resum vertex corrections 
between the parquet channels at the one-loop level.) This allows for a comprehensive study of electronic 
phase diagrams, where the dominant order can switch from one channel to the other as a function of 
system parameters.

For the specific field of two-dimensional Hubbard-type models, seminal descriptions of the fermionic 
RG approach by Shankar~\cite{shankar94rmp129} and Polchinski~\cite{polchi} triggered significant 
interest in the condensed matter community concerned with Fermi surface instabilities. Note that, 
at that time, high-temperature superconductivity driven by electronic correlations had already 
emerged as one of the dominating problems in condensed matter theory. As the cuprates, however, 
at least for the most interesting low doping regime, were supposed to be located at strong coupling, 
initial theoretical attempts to address the problem of electronic pairing promoted the strong 
coupling perspective on electronically driven superconductivity established by P. W. Anderson and 
coworkers~\cite{anderson87s1197,PhysRevLett.58.2790}, while the weak-coupling perspective developed 
by Kohn and Luttinger~\cite{kohn-65prl524} had only been assumed to be suitable in certain limits 
such as the strong doping regime.

 As of today, with the advance of many new materials, this perspective has changed. In particular, 
 many families of iron pnictides as a new class of unconventional superconductors do not show any 
 Mott limit at half filling, superficially suggesting a {\it weaker coupling regime than the cuprates}~\cite{hirschfeld-11rpp124508,scalapino-12rmp1383}. 
 Even more importantly, however, it has been noted from a general perspective that the {\it Fermiology crucially affects the ordering phenomena} 
 in two-dimensional interacting electron systems, involving concepts such as distribution of density of states at the Fermi level in 
 momentum space, nesting, and Lifshitz transitions according to the emergence or disappearance of Fermi pockets. In turn, if the 
 Fermiology is the most sensitive and important aspect to describe trends of the electronic phase as a function of system 
 parameters, an itinerant electron perspective starting from a weak coupling approach appears promising.

Comprehensive FRG studies of the two-dimensional Hubbard model have been initiated by Zanchi and 
Schulz~\cite{zanchi-00prb13609}, Halboth and Metzner~\cite{halboth-00prb7364}, as well as Honerkamp, 
Salmhofer, Furukawa, and Rice~\cite{honerkamp-01prb035109} where the single-band version on the square 
lattice was studied. The approach was subsequently improved to adequately treat ferromagnetic fluctuations 
and unconventional lattices~\cite{honerkamp-01prb184516,honerkamp-08prl146404}. 
A significant enhancement of interest in the FRG approach was induced by the emergence of the iron 
pnictides. F. Wang, D.-H. Lee and collaborators provided an accurate confirmation of the extended 
$s$-wave superconducting form factor on microscopic footing~\cite{wang-09prl047005} that had previously 
been predicted by Mazin and collaborators~\cite{mazin-08prl057003}. These first steps were quickly 
followed up by more detailed FRG studies on the doping and orbital dependence of superconductivity 
in iron-based superconductors, and the generalization of the extended $s$-wave concept to extended 
$d$-wave~\cite{thomale-09prb180505,platt-09njp055058,zhai-09prb064517,thomale-11prl187003,wang-11science200}. 
For the latter, this included also concise material predictions for $d$-wave symmetry in the pnictides, 
which has found preliminary confirmation by thermal conductivity measurements~\cite{thomale-11prl117001,reid-12prl087001}. 
Furthermore, time-reversal symmetry breaking in the superconducting phase of pnictides 
has been suggested as a consequence of frustration effects in the superconducting channel~\cite{platt-12prb180502,PhysRevLett.108.247003}.

{\it The crucial progress in these FRG studies has been that the multi-orbital nature of iron pnictides has 
been taken into account, which is an indispensable feature of the generic Fermiology.} The multi-orbital 
nature substantially extends the complexity of initial conditions of the RG equations, but also the 
analysis of Fermi surface instabilities. In previous FRG works, the individual form factors have 
mostly not been extracted from the RG flow, but instead susceptibilities, which, as quantities 
integrated over the Brillouin zone, contain less specific information on the instability. It is 
the specific form, however, of the instability form factors which allows to infer many features 
of multi-orbital Fermi surface instabilities from the FRG procedure~\cite{zhai-09prb064517}. 
Important examples are form factor anisotropies or sign changes within or between Fermi pockets, 
which is a decisive property of electronic, i.e. repulsive pairing interactions, but also relevant 
for particle-hole instabilities such 
as e.g. the collinear antiferromagnetism found in the iron pnictides~\cite{ran-09prb014505}. For the 
pnictides, an alternative analytical (parquet) RG scheme has been formulated by Chubukov and 
coworkers~\cite{maiti-10prb214515} which provided, albeit approximative, independent and often 
similar insight into the competing order, as will be further discussed in Chap.~\ref{chap:pnictide}.

With this advance of the FRG approach, a plethora of correlated electron scenarios in two dimension 
has become accessible in accurate detail. This includes the multi-$d$ orbital candidate models for 
electronically driven Fermi surface instabilities such as pnictides~\cite{wang-09prl047005,thomale-11prl187003}, 
ruthenates~\cite{ruthenate}, and sodium doped cobaltates~\cite{PhysRevLett.111.097001} many of which are 
addressed in this review, but also Hubbard models on unconventional lattices such as doped graphene~\cite{PhysRevB.86.020507,PhysRevB.85.035414} 
or itinerant electron orderings on the kagome lattice~\cite{PhysRevLett.110.126405}, which will likewise be discussed in the following. 

{\it The perspective for future studies is very promising, which is what we want to convey in this review. At a technical and 
intuitive level, it aims at making the FRG approach for multi-orbital Fermi surface instabilities accessible to a broader 
community in condensed matter.} From a physics point of view, the present review strives, in particular, for providing 
further insight into a key question in the low-temperature competing phases of these fascinating multi-band materials, 
i.e. ``what is universal and what is more material-dependent?''

The review is structured as follows. In Chapter~\ref{chap:fRG}, the detailed technical description of the FRG approach 
is provided. It starts with the derivation of the flow equation of the many particle vertices, using the generating 
functional derived by the generic fermionic action. Initially following the notation of a general scheme in~\cite{RevModPhys.84.299}, 
we have paid particular attention to presenting every step of the derivation in explicit pedagogical detail, rendering it 
most accessible for an audience that intends to get started with FRG. We illustrate how the RPA diagrammatic series is a 
subset of the diagrammatic summation implied by the FRG flow equations, and in which sense the FRG allows to treat all 
parquet channels on unbiased footing. From there, we proceed by discussing the approximations imposed on the flow 
equations that render them accessible for explicit numerical solution for Hubbard-type electronic models. This 
particularly includes the neglect of single particle vertex flow and 
the projection of the two-particle vertex frequency dependence to the Fermi level. Furthermore, we discuss the 
inclusion of multi-orbital Fermiology and interactions in the FRG analysis. The mean-field analysis of the final 
RG vertex subset, that has flown to strong coupling as a function of the flow parameter $\Lambda$, is also explicated. 
This includes a detailed discussion of the role of lattice symmetries for Fermi surface instabilities, which is 
particularly relevant for subsequent discussions of electronic order on hexagonal lattices. Whoever is already 
familiar with the technicalities of FRG as well as irreducible representations of discrete lattice groups and 
their connection to electronic condensate form factors, might want to jump right to the next chapter.

Having established the technical framework of FRG and Fermi surface instabilities, the review proceeds by 
illustrating the multi-orbital FRG approach for paradigmatic models and candidate materials to which it has 
been recently applied.  {\it In Chapter~\ref{chap:pnictide}, the iron-based superconductors are discussed as 
a paradigmatic class of materials where the FRG has proven successful in describing various phenomena of Fermi 
surface instabilities, while Chapter~\ref{chap:hexa} addresses the conceptual phenomenology of Fermi surface 
instabilities on hexagonal lattices.} Both Chapters are structured such that there is a detailed introduction 
providing the reader with the fundamental phenomenology of the materials and models discussed, concluding with 
a summary and outlook section, which points out promising directions and ongoing work within these fields.   

{\it In Chapter~\ref{chap:pnictide}, the discussion of iron pnictides starts with a general characterization 
of multi-pocket pnictide Fermiology and how generic multi-orbital interaction yields unconventional Fermi 
surface instabilities for these models.} A specific analysis is provided for the 1111, 122, and 111 compounds, 
where particular emphasis is given on the universal aspects of superconductivity in these materials such as 
the role of isovalent replacement of As by P, change of Fermiology due to strong doping, interplay of 
ferromagnetic and antiferromagnetic fluctuations, and frustration phenomena of different superconducting 
ordering propensities.

{\it In Chapter~\ref{chap:hexa}, the discussion of Fermi surface instabilities on hexagonal lattices is 
split into the analysis of a triangular lattice model for the sodium doped cobaltates, a honeycomb lattice 
description of doped graphene, and the long-range Hubbard model on the kagome lattice around van Hove filling.} 
As for the pnictides in Chapter~\ref{chap:pnictide}, the synopsis of all these avenues of hexagonal systems, as 
studied by FRG, allows to precisely identify universal features related to the lattice symmetry such as the 
natural propensity to topological chiral $d$-wave superconductivity, but also important lattice-specific features 
such as the sublattice interference effect~\cite{PhysRevB.86.121105} for the kagome Hubbard model.  

{\it Several details} that are worth stating, but might be less relevant upon first 
reading, {\it have been moved to the appendices in Chapter~\ref{chap:appendix}}, which are 
accordingly referred to in the Chapters~\ref{chap:fRG},~\ref{chap:pnictide}, and~\ref{chap:hexa}. 
Finally, in Chapter~\ref{chap:conclusion}, we conclude that the multi-orbital functional renormlization 
group approach is a promising tool to address current and future questions of unconventional order in 
interacting electron systems. It particularly helps to broaden our view from the detailed investigation 
of specific materials to general trends and universal features of Fermi surface instabilities driven by electronic correlations.

 \section{Functional Renormalization Group}
 \label{chap:fRG}
 Interacting electron systems display a variety of fascinating phenomena such as superconductivity, magnetic ordering, or the
 formation of exotic quantum liquids. These phenomena usually emerge at scales far below the bare energy scale of the microscopic 
 Hamiltonian (see Fig.~\ref{fig:scales}a). In order to interpolate between these scales, it appears natural to treat 
 degrees of freedom with different energy scales successively, descending from high to low energies. Using a functional-integral formulation
 of the partition function, this idea can be implemented by integrating out high-energy modes step by step, and adjusting the action accordingly. 
 This procedure then generates a one-parameter family of actions which interpolates between the microscopic theory at high energies and an effective low-energy 
 description (see Fig.~\ref{fig:scales}b). At the same time, infrared singularities which signal an instability of the normal state are approached in a controlled way. 
 In fact, the necessary functional integration can almost never be performed in an exact way, and one has to resort to approximate treatments.
 However, for an infinitesimal mode elimination, the resulting change of the action can be expressed in a formally exact flow equation.
 In a similar way, the functional renormalization group rephrases the process of mode elimination in terms of an exact flow equation for a generating functional.
 The benefit of this flow equation lies in its transparent 
 approximation schemes as well as in its flexibility regarding the choice of flow parameters or the choice of alternative generating functionals.\par
 \textit{Starting with a general derivation of the functional flow equations at the beginning of this Chapter, we discuss the effect of
 different flow parameters as well as the implications of symmetries and possible approximation schemes. In addition, we describe the extension
 to multi-orbital or multi-sublattice systems and provide a pseudocode implementation of the functional RG method.}\par
 For first part of this chapter, we follow the notation and general derivation of~\cite{RevModPhys.84.299} and~\cite{kopietz-10book}.
 \subsection{Functional Flow Equations}\label{sec:floweq}
 In the following section, we introduce the \textit{concept of generating functionals} and \textit{the notion
 of functional flow equations}. In particular, \textit{we define the so-called effective action 
 as a generating functional of the one-particle irreducible (1PI) vertex functions} 
 (i.e. connected diagrams which cannot be separated by cutting one single propagator) 
 and \textit{derive its corresponding flow equation}. Using
 \textit{certain approximation schemes} for an efficient numerical
 treatment described thereafter, \textit{this 1PI flow equation then
 serves as computational tool throughout this review}.\par
 As a starting point, we consider an interacting
 fermion system described by the action
  \begin{equation}
 S(\overline{\psi},\psi)  = -\int_{k,k'}Q_{k,k'}\overline{\psi}_{k}\psi_{k'}
                          + \int_{k_1,k_2,k'_1,k'_2}
                          U_{k_1,k_2,k'_1,k'_2}\overline{\psi}_{k_1}\overline{\psi}_{k_2}\psi_{k'_1}\psi_{k'_2}
 \label{eq:bareaction}
 \end{equation}
 with Grassmann fields $\overline{\psi},\psi$, the inverse bare propagator
 \begin{equation}\label{eq:quadpart}
 Q_{k,k'} = \delta_{kk'}\cdot\left(G^0_k\right)^{-1} = \delta_{kk'}\cdot(ik_0 - \xi_b(\bs{k})),
  \end{equation}
 and some two-particle interaction $U$. 
  \begin{figure}[t]
 \begin{center}
 {\includegraphics[scale=0.35]{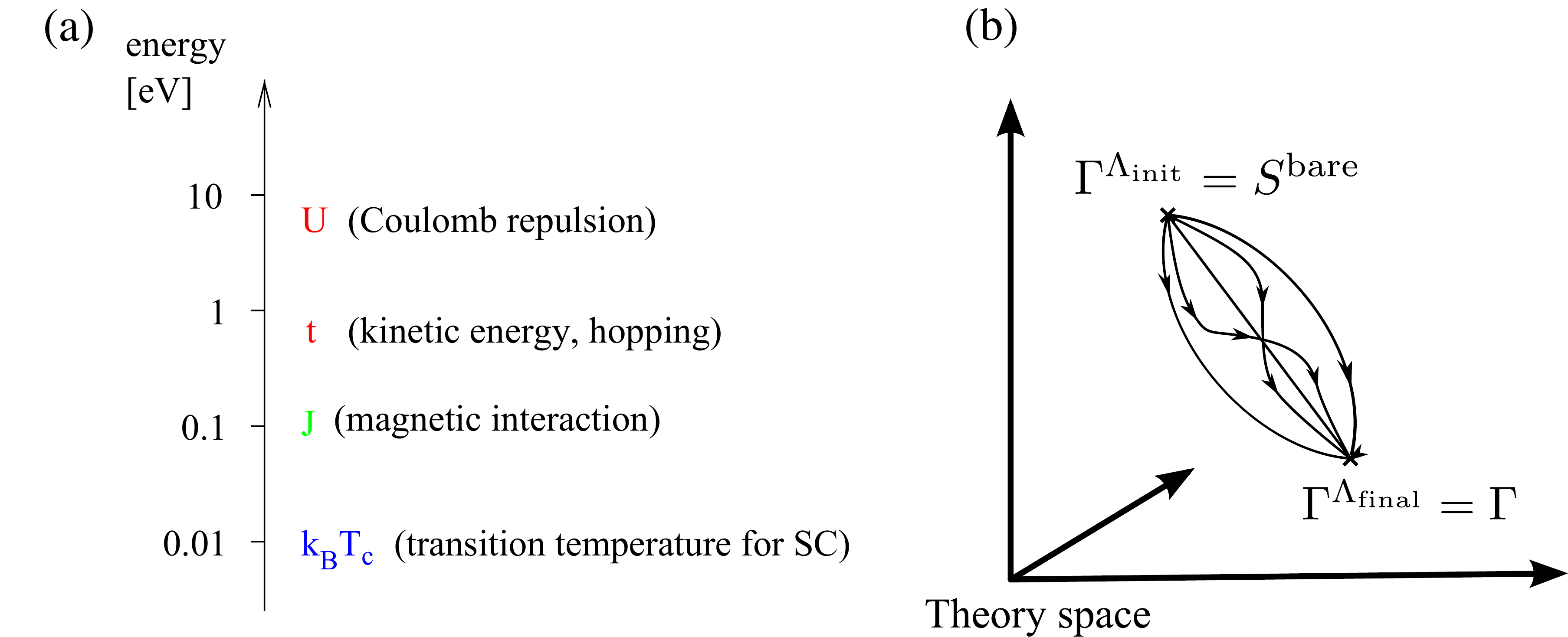}}
 \end{center}
 \caption{ (a) Important energy scales in interacting electron systems (taken from Metzner~\ea~\cite{RevModPhys.84.299}). (b) Flow of the effective action, starting at $\Gamma^{\Lambda_{init}} = S^{\text{bare}}$
          for large values of $\Lambda$ and approaching the full effective action at $\Lambda = 0$. Different trajectories correspond to distinct
          choices of the flow parameter.}
 \label{fig:scales}
 \end{figure}
 We further employ the notation of $k=(k_0,\bs{k},\sigma,b)$ including fermionic Matsubara frequencies $k_0$,
 momenta $\bs{k}$, and internal degrees of freedom such as spin projections $\sigma$ or band indices $b$. The 
 dispersion relation $\xi_b(\bs{k})$ in (\ref{eq:quadpart}) characterizes the one-particle energy as a function of momentum $\bs{k}$ and band index $b$. 
 In addition, the integral $\int_k$ contains integration and summation over each entry in $k$
 and also comprises prefactors such as temperature and volume.\par
 Based on the action $S(\overline{\psi},\psi)$ in
 (\ref{eq:bareaction}), we can infer thermodynamic quantities such as the grand-canonical
 partition function
 \begin{equation}
 Z = \int D(\overline{\psi},\psi) e^{-S(\overline{\psi},\psi)} 
 \label{eq:Z}
 \end{equation}
 or the imaginary-time ordered Green functions, i.e.
 \begin{equation}
 G_{k_1,\ldots,k_n;k'_1,\ldots,k'_n} = 
 -\frac{1}{Z} \int D(\overline{\psi},\psi) e^{-S(\overline{\psi},\psi)} \psi_{k_1}\ldots \psi_{k_n}\overline{\psi}_{k'_n}\ldots \overline{\psi}_{k'_1}
 \label{eq:G} 
 \end{equation}
 just by choosing the appropriate functional averages. It is, further, convenient to define a so-called generating functional
 \begin{equation}
 W[\eta,\overline{\eta}] = \int D(\overline{\psi},\psi) e^{-S(\overline{\psi},\psi) + (\overline{\eta},\psi) + (\overline{\psi},\eta)}
 \label{eq:W}
 \end{equation}
 with source terms
 \begin{equation}\nonumber
 (\overline{\eta},\psi) := \int_k \overline{\eta}_k\psi_k,\quad (\overline{\psi},\eta) := \int_k \overline{\psi}_k\eta_k,
 \end{equation}
 such that the functional averages in (\ref{eq:G}) can be rephrased by derivatives of $W[\eta,\overline{\eta}]$
 with respect to $\eta$ and $\overline{\eta}$. Taking the logarithm of (\ref{eq:W}), one obtains another generating functional
 \begin{equation}
 \mathcal{G}[\eta,\overline{\eta}] = -\ln\left( W[\eta,\overline{\eta}] \right)
 \label{eq:mathcalG}
 \end{equation}
 which, again by functional differentiation, provides the connected $n$-particle Green functions
  \begin{align}
 \label{eq:connectG}
 G^{c(2n)}_{k_1,\ldots,k_n;k'_1,\ldots,k'_n} &= -\left\langle \psi_{k_1}\ldots \psi_{k_n}\overline{\psi}_{k'_n}\ldots \overline{\psi}_{k'_1}\right\rangle_c\\\nonumber
                                      &=(-1)^n\left. \frac{\partial^{2n}\mathcal{G} [\eta,\overline{\eta}]}
                                                    {\partial \overline{\eta}_{k_1}\ldots \partial\overline{\eta}_{k_n}\partial \eta_{k'_n}\ldots \partial\eta_{k'_1}}\right|_{\eta=\overline{\eta} = 0}.
 \end{align}
 Note that the $(-1)^n$ factor here results from the commutation of $\partial/\partial \eta$ with the field $\overline{\psi}$ in the source term of (\ref{eq:W}).\par
 In the following, we want to derive an exact flow equation for the generating functional in (\ref{eq:mathcalG}). Here, the essential idea is to
 circumvent the integration (\ref{eq:W}), implicit in $\mathcal{G}$, by constructing a one-parameter family of generating functionals $\mathcal{G}^{\Lambda}$ that 
 interpolates between a solvable starting point at large values of
 $\Lambda$ and the full functional at $\Lambda = 0$ (Fig.~\ref{fig:scales}).  One possibility to derive this one-parameter
 dependence is by replacing $Q\rightarrow Q^{\Lambda}$ in the bare action (\ref{eq:bareaction}),
 such that 
 \begin{equation}
  Q^{\Lambda}_{k,k'} \sim
 \begin{cases}
 Q_{k,k'}, & \text{for } \Lambda\rightarrow 0\\
 \infty,   & \text{for } \Lambda\rightarrow \infty
 \end{cases}
 \label{eq:laminfty},
 \end{equation}
 which then implements the required boundary conditions for the 
 generating functional $\mathcal{G}^{\Lambda}$, i.e.
 \begin{equation}
 \mathcal{G}^{\Lambda} \sim
 \begin{cases}
 \mathcal{G}, & \text{for } \Lambda\rightarrow 0\\
 0 , & \text{for } \Lambda\rightarrow\infty
 \end{cases}.
 \label{mathcalGlimit}
 \end{equation}
 Note that the trivial case of $\mathcal{G}^{\Lambda\rightarrow \infty}=0$
 results from the infinite mass term in (\ref{eq:laminfty})
 freezing out all particle propagation. Another purpose of the replacement
 $Q\rightarrow Q^{\Lambda}$ is to regularize infrared divergencies and, thus,
 to approach a possible singularity in a controlled way. The next step is, then, to follow the 
 trajectory $\mathcal{G}^{\Lambda}$ from the known starting point 
 at large $\Lambda$ towards the full functional at $\Lambda=0$. Therefore, we determine
 the derivative $\partial_{\Lambda}\mathcal{G}^{\Lambda}$ and consider the 
 extrapolation to $\Lambda\rightarrow 0$ as an initial 
 value problem, which turns out to be the functional flow equation
 sought after. We can, thus, rephrase 
 the functional integration of Eq.~(\ref{eq:mathcalG}) into a formally exact 
 differential equation: 
 \begin{align}\nonumber
 \partial_{\Lambda}\mathcal{G}^{\Lambda}[\eta,\overline{\eta}] 
  & = -e^{\mathcal{G}^{\Lambda}[\eta,\overline{\eta}]}
         \partial_{\Lambda}e^{-\mathcal{G}^{\Lambda}[\eta,\overline{\eta}]}\\\nonumber
  & = -e^{\mathcal{G}^{\Lambda}[\eta,\overline{\eta}]}\int D(\overline{\psi},\psi)
         \left(\overline{\psi},\dot{Q}^{\Lambda}\psi\right)
      e^{-S(\overline{\psi},\psi)+(\overline{\eta},\psi) + (\overline{\psi},\eta)}\\\label{eq:mathcalGflowfirst}
  & = e^{\mathcal{G}^{\Lambda}[\eta,\overline{\eta}]}
         \left(\partial_{\eta},\dot{Q}^{\Lambda}\partial_{\overline{\eta}}\right)e^{-\mathcal{G}^{\Lambda}[\eta,\overline{\eta}]}\\\nonumber
  & = e^{\mathcal{G}^{\Lambda}[\eta,\overline{\eta}]}
         \left(\int_{k,k'}\partial_{\eta_k}\dot{Q}^{\Lambda}_{kk'}
         \partial_{\overline{\eta}_{k'}}e^{-\mathcal{G}^{\Lambda}[\eta,\overline{\eta}]}\right)\\\nonumber
  = & \int_{k,k'}\left\{\left(-\partial_{\eta_k}\mathcal{G}^{\Lambda}[\eta,\overline{\eta}]\right)
      \dot{Q}^{\Lambda}_{kk'}\left(-\partial_{\overline{\eta}_{k'}}\mathcal{G}[\eta,\overline{\eta}]\right)
   + \dot{Q}^{\Lambda}_{kk'}\left(-\partial_{\eta_{k}}\partial_{\overline{\eta}_{k'}}\mathcal{G}[\eta,\overline{\eta}]\right)\right\}\\\label{eq:mathcalGflow}
  = & \left(\left(\partial_{\eta}\mathcal{G}^{\Lambda}[\eta,\overline{\eta}]\right),
      \dot{Q}^{\Lambda}\left(\partial_{\overline{\eta}}\mathcal{G}[\eta,\overline{\eta}]\right)\right)
   + \text{tr}\left(\dot{Q}^{\Lambda}\left(\partial_{\overline{\eta}}\partial_{\eta}\mathcal{G}[\eta,\overline{\eta}]\right)\right).
 \end{align} 
 Using the Taylor expanded functional $\mathcal{G}^{\Lambda}$ in the above 
 flow equation, this provides an infinite hierarchy of 
 differential equations for the respective Taylor coefficients, the connected Green functions $G^{c(2n)}$.
 Usually, these connected Green functions contain 1PI vertex functions as subclasses of diagrams that are, in turn, 
 connected by single propagators. As discussed in \cite{morris:93ijmpa2411}, the
 isolated propagators can lead to technical problems within the flow equations,
 and also the trivial initial condition of $\mathcal{G}^{\Lambda\rightarrow\infty}=0$
 appears to be unfavorable as it absorbs any information about the original system.\par
 It is, therefore, convenient to derive a flow equation for the 1PI vertex
 generating functional, or effective action $\Gamma^{\Lambda}$, which is determined by the Legendre transformation
 of $\mathcal{G}^{\Lambda}$ 
 \begin{equation}
 \Gamma^{\Lambda}[\zeta,\overline{\zeta}] = \mathcal{G}^{\Lambda}[\eta,\overline{\eta}] 
                                          + (\overline{\eta}^{\Lambda},\zeta) 
                                          + (\overline{\zeta},\eta^{\Lambda}),
 \label{eq:gammdef}
 \end{equation}
 where the conjugate fields $\overline{\zeta}$, $\zeta$ are defined as
 \begin{equation}
 \zeta = -\frac{\partial\mathcal{G}^{\Lambda}}{\partial\overline{\eta}},\quad 
 \overline{\zeta} = \frac{\partial\mathcal{G}^{\Lambda}}{\partial \eta}.
 \label{eq:conjfield}
 \end{equation}
 Note that in (\ref{eq:gammdef}), $\overline{\eta}^{\Lambda}$ and $\eta^{\Lambda}$ are
 $\Lambda$-dependent functions of $\overline{\zeta}$ and $\zeta$ due to~(\ref{eq:conjfield}).
 The 1PI vertex functions can, then, be obtained by computing derivatives
 with respect to the conjugate fields:
 \begin{equation}
 \gamma^{\Lambda(2n)}_{k_1,\ldots,k_n;k'_1,\ldots,k'_n} 
 =\left. \frac{\partial^{2n}\Gamma^{\Lambda} [\zeta,\overline{\zeta}]}
              {\partial \overline{\zeta}_{k_1}\ldots \partial\overline{\zeta}_{k_n}\partial \zeta_{k'_n}\ldots \partial\zeta_{k'_1}}\right|_{\zeta=\overline{\zeta} = 0}.
 \label{eq:1pivert}
 \end{equation}
 We now want to derive the corresponding flow equation for the effective action $\Gamma^{\Lambda}$. Before that, we first note that (\ref{eq:gammdef})
 implies the following reciprocity relation for the second derivatives $\Gamma^{\Lambda}$ and $\mathcal{G}^{\Lambda}$:
  \begin{equation}
 \bs{\partial}^2\Gamma^{\Lambda}[\overline{\zeta},\zeta]
 = \left(\partial^2G^{\Lambda}[\overline{\eta},\eta]\right)^{-1},
 \label{eq:reciprel}
 \end{equation}
 with
 \begin{align}\label{eq:partial2gamma}
 \bs{\partial}^2\Gamma^{\Lambda}[\overline{\zeta},\zeta] &=
 \begin{pmatrix}
 \frac{\partial^{2}\Gamma^{\Lambda}}{\partial \overline{\zeta}_{k_1}\partial \zeta_{k'_1}} & 
 \frac{\partial^{2}\Gamma^{\Lambda}}{\partial \overline{\zeta}_{k_1}\partial \overline{\zeta}_{k'_1}} \\
 \frac{\partial^{2}\Gamma^{\Lambda}}{\partial \zeta_{k_1}\partial \zeta_{k'_1}} & 
 \frac{\partial^{2}\Gamma^{\Lambda}}{\partial \zeta_{k_1}\partial \overline{\zeta}_{k'_1}}
 \end{pmatrix}
 \end{align}
 and
 \begin{align}\nonumber
 \partial^2G^{\Lambda}[\overline{\eta},\eta] &=
  \begin{pmatrix}
 -\frac{\partial^{2}\mathcal{G}^{\Lambda}}{\partial \overline{\eta}_{k'_1}\partial \eta_{k_1}} & 
  \frac{\partial^{2}\mathcal{G}^{\Lambda}}{\partial \overline{\eta}_{k'_1}\partial \overline{\eta}_{k_1}} \\
  \frac{\partial^{2}\mathcal{G}^{\Lambda}}{\partial \eta_{k'_1}\partial \eta_{k_1}} &
 -\frac{\partial^{2}\mathcal{G}^{\Lambda}}{\partial \eta_{k'_1}\partial \overline{\eta}_{k_1}}
 \end{pmatrix}.
 \end{align}
 In a lowest-order Taylor expansion, this then yields 
 \begin{equation}\label{eq:gamma2}
 \gamma^{\Lambda(2)}_{k;k'} = \left(G^{\Lambda (2)}_{k;k'}\right)^{-1} = Q^{\Lambda}_{k,k'} - \Sigma^{\Lambda}_{k,k'},
 \end{equation}
 and for higher orders, all connected Green functions are obtained by summing tree-like diagrams of 1PI vertex functions
 to equal or lower order \cite{kopietz-10book}. We can now determine the flow equation~\cite{wetterich-93plb199390} for the effective action $\Gamma^{\Lambda}$ 
 by using (\ref{eq:mathcalG}) and the reciprocity relation (\ref{eq:reciprel}), i.e.
 \begin{align}\nonumber
  \partial_{\Lambda}\Gamma^{\Lambda}[\zeta,\overline{\zeta}] 
 =& \partial_{\Lambda}\mathcal{G}[\eta^{\Lambda},\overline{\eta}^{\Lambda}] 
  + (\partial_{\Lambda}\overline{\eta}^{\Lambda},\zeta) 
  + (\overline{\zeta},\partial_{\Lambda}\eta^{\Lambda})\\\nonumber
 =& \left.\partial_{\Lambda}\mathcal{G}[\eta^{\Lambda},\overline{\eta}^{\Lambda}]\right|_{\eta^{\Lambda},\overline{\eta}^{\Lambda}\text{ fixed}}\\\nonumber
 \overset{(\ref{eq:mathcalGflow})}{=}& \left(\left(\partial_{\eta}\mathcal{G}^{\Lambda}\right),
      \dot{Q}^{\Lambda}\left(\partial_{\overline{\eta}}\mathcal{G}\right)\right)
   + \text{tr}\left(\dot{Q}^{\Lambda}\left(\partial_{\overline{\eta}}\partial_{\eta}\mathcal{G}\right)\right)\\\label{eq:partialres}
 \overset{(\ref{eq:reciprel})}{=}& -\left(\overline{\zeta},\dot{Q}^{\Lambda}\zeta\right) 
 -
 \text{tr}\left(\dot{Q}^{\Lambda}\left(\left(\bs{\partial}^2\Gamma^{\Lambda}[\zeta,\overline{\zeta}]\right)^{-1}\right)_{11}\right)
 \\\label{eq:finaleq}
 =& -\left(\overline{\zeta},\dot{Q}^{\Lambda}\zeta\right) 
 - \frac{1}{2}\text{tr}\left(\dot{\bs{Q}}^{\Lambda}\left(\bs{\partial}^2\Gamma^{\Lambda}[\zeta,\overline{\zeta}]\right)^{-1}\right).
 \end{align}
 Here, the notation $(\cdots)_{11}$ relates to the $(1,1)$-element of the matrix $\bs{\partial}^2\Gamma^{\Lambda}$ given in Eq.~(\ref{eq:partial2gamma}), 
 and the bold quantity $\dot{\bs{Q}}^{\Lambda}$ in (\ref{eq:finaleq}) denotes 
 \begin{equation}\nonumber
 \dot{\bs{Q}}^{\Lambda} = \text{diag}(\dot{Q}^{\Lambda},-\dot{Q}^{\Lambda T}).
 \end{equation}
 Unlike the flow equation for the generating functional $\mathcal{G}^{\Lambda}$, the one in (\ref{eq:finaleq}) for the effective action $\Gamma^{\Lambda}$
 reveals a nontrivial initial condition of $\Gamma^{\Lambda\rightarrow \infty}=S$, where $S$ is given by the bare action (\ref{eq:bareaction})
 of the underlying system. A complete solution of the differential flow equation (\ref{eq:finaleq}) then describes the evolution from the bare action 
 towards the full effective action and hence provides all 1PI vertex functions as well as the connected Green functions in a tree-like series.\par
 Unfortunately, the direct solution of the functional flow equation (\ref{eq:finaleq}) is only possible for a very small number of systems. However, in practice, 
 one is primarily interested in a few number of 1PI vertex functions. It is therefore convenient to expand both sides of \eqref{eq:finaleq} 
 in powers of the fields, and calculate the flow only for certain coefficients.
 For this purpose, we first rewrite the effective action in a series of fields
  \begin{align}\label{eq:gammaexpansion}
 \Gamma^{\Lambda}[\zeta,\overline{\zeta}] 
 & = \sum_{m=0}^{\infty}\mathcal{A}^{(2m)\Lambda}[\overline{\zeta},\zeta],
 \end{align}
 with
 \begin{equation}\nonumber
 \mathcal{A}^{(2m)\Lambda}[\overline{\zeta},\zeta]
  =\frac{(-1)^m}{(m!)^2}
 \int_{k_1,\ldots,k_m\atop k'_1,\ldots,k'_m}\gamma^{(2m)\Lambda}_{k'_1,\ldots,k'_m;k_1,\ldots,k_m}
 \overline{\zeta}_{k'_1}\ldots\overline{\zeta}_{k'_m}\zeta_{k_m}\ldots\zeta_{k_1}.
 \end{equation}
 In order to obtain a similar series expansion for the inverse $\left(\bs{\partial}^2\Gamma^{\Lambda}[\zeta,\overline{\zeta}]\right)^{-1}$
 on the right-hand side of flow equation (\ref{eq:finaleq}), we first introduce the matrix
 \begin{equation}\nonumber
 \bs{U}^{\Lambda}[\overline{\zeta},\zeta] =    \left.\bs{\partial}^2\Gamma^{\Lambda}[\zeta,\overline{\zeta}]\right|_{\zeta=\overline{\zeta}=0}
                                                     - \bs{\partial}^2\Gamma^{\Lambda}[\zeta,\overline{\zeta}],
 \end{equation}
 together with its corresponding series expansion
 \begin{equation}\label{eq:uexpansion}
 \bs{U}^{\Lambda}[\overline{\zeta},\zeta] = -\sum_{m=2}^{\infty}\bs{\partial}^2\mathcal{A}^{(2m)\Lambda}[\overline{\zeta},\zeta].
 \end{equation}
 Making use of $\left.\bs{\partial}^2\Gamma^{\Lambda}\right|_{\zeta=\overline{\zeta}=0} = \left({\bs{G}^{\Lambda}}\right)^{-1}$ from~(\ref{eq:gamma2}),
 the inverse $\left(\bs{\partial}^2\Gamma^{\Lambda}\right)^{-1}$ is then given by the following geometric series
 \begin{align}\nonumber
 \left(\bs{\partial}^2\Gamma^{\Lambda}[\zeta,\overline{\zeta}]\right)^{-1} 
 & = \left(\left({\bs{G}^{\Lambda}}\right)^{-1} - \bs{U}^{\Lambda}[\overline{\zeta},\zeta]\right)^{-1}\\\label{eq:inversegamma}
 & = \left(\bs{1} - \bs{G}^{\Lambda}\bs{U}^{\Lambda}[\overline{\zeta},\zeta]\right)^{-1}\bs{G}^{\Lambda}\\\nonumber
 & = \left(\bs{1} + \bs{G}^{\Lambda}\bs{U}^{\Lambda}[\overline{\zeta},\zeta] 
    + \bs{G}^{\Lambda}\bs{U}^{\Lambda}[\overline{\zeta},\zeta]\bs{G}^{\Lambda}\bs{U}^{\Lambda}[\overline{\zeta},\zeta]+\ldots\right)\bs{G}^{\Lambda}.
 \end{align}
 We now insert the latter expression (\ref{eq:inversegamma}) into the right-hand side of the flow equation (\ref{eq:finaleq}) which in turn provides
 \begin{align}\nonumber
 \partial_{\Lambda}\Gamma^{\Lambda}[\zeta,\overline{\zeta}] 
   = &-\left(\overline{\zeta},\dot{Q}^{\Lambda}\zeta\right) 
     - \frac{1}{2}\text{tr}\left(\dot{\bs{Q}}^{\Lambda}
     \left(1+ \bs{G}^{\Lambda}\bs{U}^{\Lambda} 
          + \bs{G}^{\Lambda}\bs{U}^{\Lambda}\bs{G}^{\Lambda}\bs{U}^{\Lambda} 
          + \ldots\right)\bs{G}^{\Lambda}\right)\\\nonumber
   = &-\left(\overline{\zeta},\dot{Q}^{\Lambda}\zeta\right) - \text{tr}\left(\dot{Q}^{\Lambda}G^{\Gamma}\right)\\\label{eq:flowseries}
     &  + \frac{1}{2}\text{tr}\left(\bs{S}^{\Lambda}\left(\bs{U}^{\Lambda} 
        + \bs{U}^{\Lambda}\bs{G}^{\Lambda}\bs{U}^{\Lambda} + \ldots\right)\right).
  \end{align}
  Here, we exploited the cyclic invariance of the trace tr and defined the single-scale propagator
  \begin{equation}\nonumber
  \bs{S}^{\Lambda} = \text{diag}\left(S^{\Lambda},-S^{\Lambda T}\right) = -\bs{G}^{\Lambda}\dot{\bs{Q}}^{\Lambda}\bs{G}^{\Lambda} 
  \end{equation}
  with
  \begin{equation}\nonumber
  S^{\Lambda} = -G^{\Lambda}\dot{Q}^{\Lambda}G^{\Lambda}=\frac{d}{d\Lambda}\left.G^{\Lambda}\right|_{\Sigma^{\Lambda}\text{ fixed}}.
  \end{equation}
 The trace term in (\ref{eq:flowseries}) also reveals the one-loop structure of the formally exact flow equation.
 Inserting the two series expansions of $\Gamma^{\Lambda}$ in (\ref{eq:gammaexpansion}) and $\bs{U}$ in (\ref{eq:uexpansion}) 
 into (\ref{eq:flowseries}), we obtain a system of differential equations for the coefficients $\mathcal{A}^{(2m)\Lambda}$:
 \begin{align}\label{eq:flowgrandpot}
  \frac{d}{d\Lambda}\mathcal{A}^{(0)\Lambda} =& 
  - \text{tr}\left(\dot{Q}^{\Lambda}G^{\Gamma}\right)\\\nonumber
  \frac{d}{d\Lambda}\mathcal{A}^{(2)\Lambda} =& 
  - \frac{1}{2}\text{tr}\left(\bs{S}^{\Lambda}\bs{\partial}^2\mathcal{A}^{(4)\Lambda}\right) -\left(\overline{\zeta},\dot{Q}^{\Lambda}\zeta\right)\\\nonumber
  \frac{d}{d\Lambda}\mathcal{A}^{(4)\Lambda} =& - \frac{1}{2}\text{tr}\left(\bs{S}^{\Lambda}\bs{\partial}^2\mathcal{A}^{(6)\Lambda}\right) 
                                                + \frac{1}{2}\text{tr}\left(\bs{S}^{\Lambda}\bs{\partial}^2\mathcal{A}^{(4)\Lambda}\bs{G}^{\Lambda}\bs{\partial}^2\mathcal{A}^{(4)\Lambda}\right)\\\nonumber
 \frac{d}{d\Lambda}\mathcal{A}^{(6)\Lambda}  =& -\frac{1}{2}\text{tr}\left(\bs{S}^{\Lambda}\bs{\partial}^2\mathcal{A}^{(8)\Lambda}\right)+\ldots\quad.
   \end{align}
 The first equation corresponds to the flow of the grand canonical potential. It turns out that
 the infinite hierarchy of flow equations does not close at any finite order as the differential equation for a 
 given $\mathcal{A}^{(2m)\Lambda}$ always contains the next order term $\mathcal{A}^{(2m+2)\Lambda}$ in a tadpole-like
 diagram. 

\begin{figure}[t]
 \begin{center}
 {\includegraphics[scale=0.42]{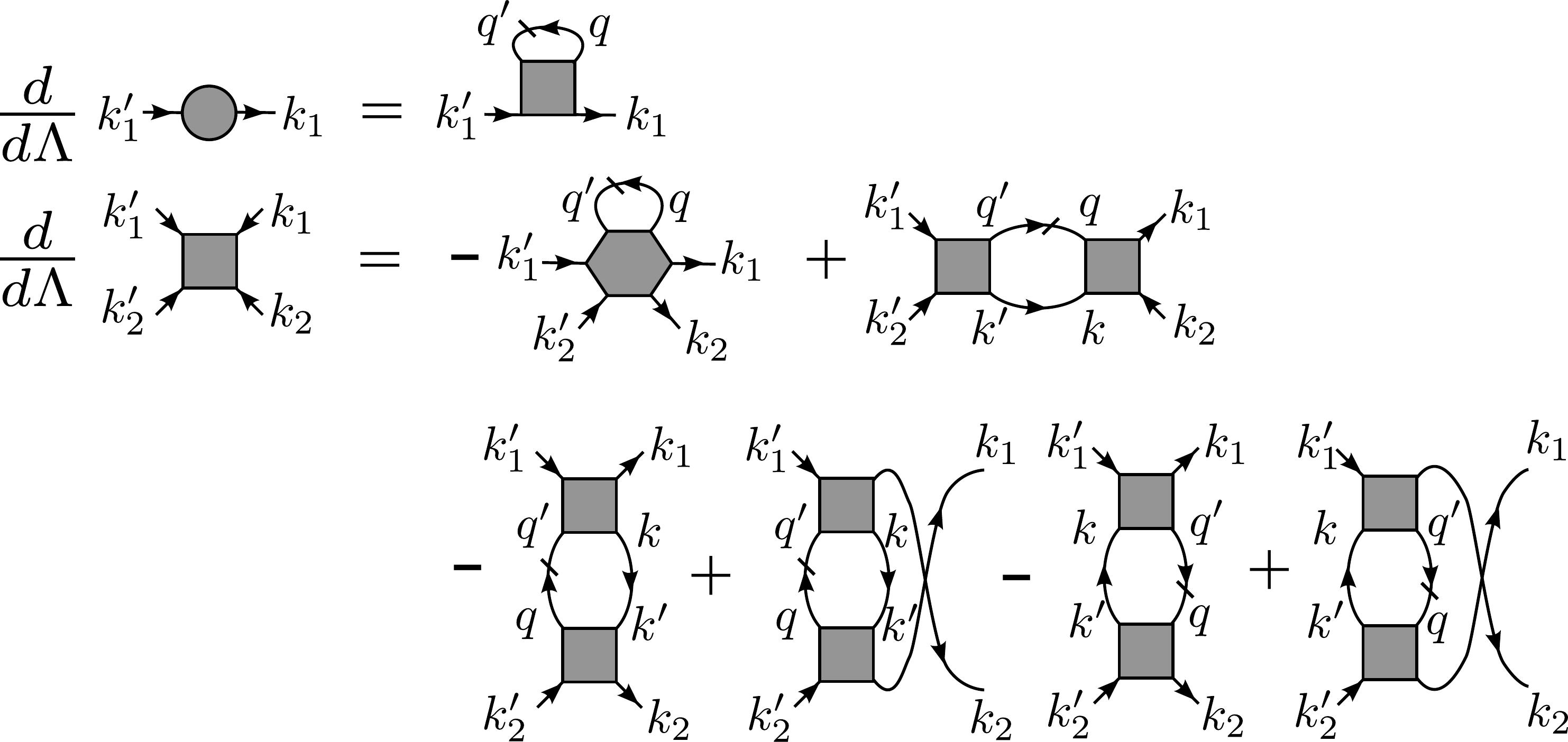}}
 \end{center}
 \caption{\label{pic:floweq} Flow equations for the first two 1PI-vertex functions in (\ref{eq:flowselfen}) and (\ref{eq:flowgamma}). 
 Slashed and full internal-lines represent the single-scale $S^{\Lambda}$
 and full propagator $G^{\Lambda}$.}
 \end{figure}

In a next step, we compare the field-independent coefficients in each of these equations and obtain
 an infinite hierarchy of flow equations for the 1PI vertex functions, of which the first two are given by:
 \begin{align}\label{eq:flowselfen}
 \frac{d}{d\Lambda}\Sigma^{\Lambda}_{k_1',k_1} 
  = &\sum_{q;q'}S^{\Lambda}_{q,q'}\gamma^{(4)\Lambda}_{k'_1,q';k_1,q}\\\label{eq:flowgamma}
 \frac{d}{d\Lambda}\gamma^{(4)\Lambda}_{k_1',k_2';k_1,k_2} 
  = &- \sum_{q,q'} S^{\Lambda}_{q,q'}\gamma^{(6)\Lambda}_{k_1',k_2',q';k_1,k_2,q}\\\nonumber
 & + \sum_{k,k'\atop q,q'}G^{\Lambda}_{k,k'}S^{\Lambda}_{q,q'}
     \times\left\{\gamma^{(4)\Lambda}_{k_1',k_2';k,q}\gamma^{(4)\Lambda}_{k',q';k_1,k_2}\right.\\\nonumber
 & - \left[\gamma^{(4)\Lambda}_{k_1',q';k_1,k}\gamma^{(4)\Lambda}_{k',k_2';q,k_2} 
   + (k\leftrightarrow q,k'\leftrightarrow q')\right]\\\nonumber
 & + \left.\left[\gamma^{(4)\Lambda}_{k_2',q';k_1,k}\gamma^{(4)\Lambda}_{k',k_1';q,k_2} 
   + (k\leftrightarrow q,k'\leftrightarrow q')\right]\right\}.
 \end{align}
 Note that in (\ref{eq:flowselfen}), we further employed 
 $\gamma^{(2)\Lambda}=Q^{\Lambda}-\Sigma^{\Lambda}$ in order to derive a flow equation for the 
 self-energy $\Sigma^{\Lambda}$. In a graphical representation, these flow equations are shown in Fig.~\ref{pic:floweq} 
 where slashed and full lines correspond to single-scale $S^{\Lambda}$ and full propagator $G^{\Lambda}$, respectively.  
 For a numerical implementation and an application to realistic material compounds, we have to justify 
 the truncation of this infinite hierarchy to a manageable closed system of flow equations. 
 In addition to that, we need to specify the flow parameter dependence in the quadratic 
 part $Q^{\Lambda}$ which was left unspecified in the definition of
 (\ref{eq:laminfty}). We will further elaborate on these aspects in
 the following.
 \subsection{Truncating the Functional Flow Equation}\label{sec:trunc}
 In order to justify the dropping of the six-point 1PI-vertex $\gamma^{(6)}$ in Fig.~\ref{pic:floweq},
 we follow the reasoning of Salmhofer~\ea~\cite{salmhofer-01prthp1}. Assuming weak to moderate two-particle interactions and $\gamma^{(6)}=0$ at bare level,
 it was shown that that for high energy scales, where $\gamma^{(4)}$ is still relatively small, the contributions of $\gamma^{(6)}$ are likewise small as they involve higher order terms
 of $\gamma^{(4)}$. At intermediate scales, a phase-space argument for sufficiently smooth and curved Fermi-surfaces then proves that the contribution of $\gamma^{(6)}$ remains small 
 even though the scale dependent $\gamma^{(4)}$ is no longer small. Only at low energy scales, where $\gamma^{(4)}$ starts to diverge, the decreasing phase-space cannot suppress
 the contribution of $\gamma^{(6)}$, and the flow has to be stopped. The remaining modes must then be treated with different approaches, for example by using mean-field analysis or
 more sophisticated methods. In general, it should be noted that the FRG
 operates most reliably for intermediate coupling. One the one hand,
 for any concise numerical implementation, it necessitates finite
 interactions to drive a flow to strong coupling at a finite number of
 flow steps. One the other hand, the neglect of higher-order vertex
 functions, which will again be essential for a concise implementation
 scheme, is unjustified in the strong coupling limit, which we will
 also be facing in the following chatpers. For further discussion we refer to the articles \cite{salmhofer-01prthp1,RevModPhys.84.299}. 
 \subsection{Flow Parameters}\label{sec:flowpars}
 In section~\ref{sec:floweq}, we have derived flow equations for the effective action and for its associated
 expansion coefficients, the 1PI vertex functions. The starting point here was the quadratic part
 of the underlying action which was then equipped with a parameter dependence such that the resulting trajectory
 of functionals interpolates between the bare theory and the full effective action. 
 In addition, we required the parameter dependence to regularize infrared
 singularities that may arise from the unbounded propagator at zero frequency and at
 Fermi surface momenta. The integration contained in the trace of the
 flow equation (\ref{eq:finaleq}) then becomes finite, which enables
 us to approach a possible singularity in a controlled way.\par
\begin{figure}[t]
 \begin{center}
 {\includegraphics[scale=0.25]{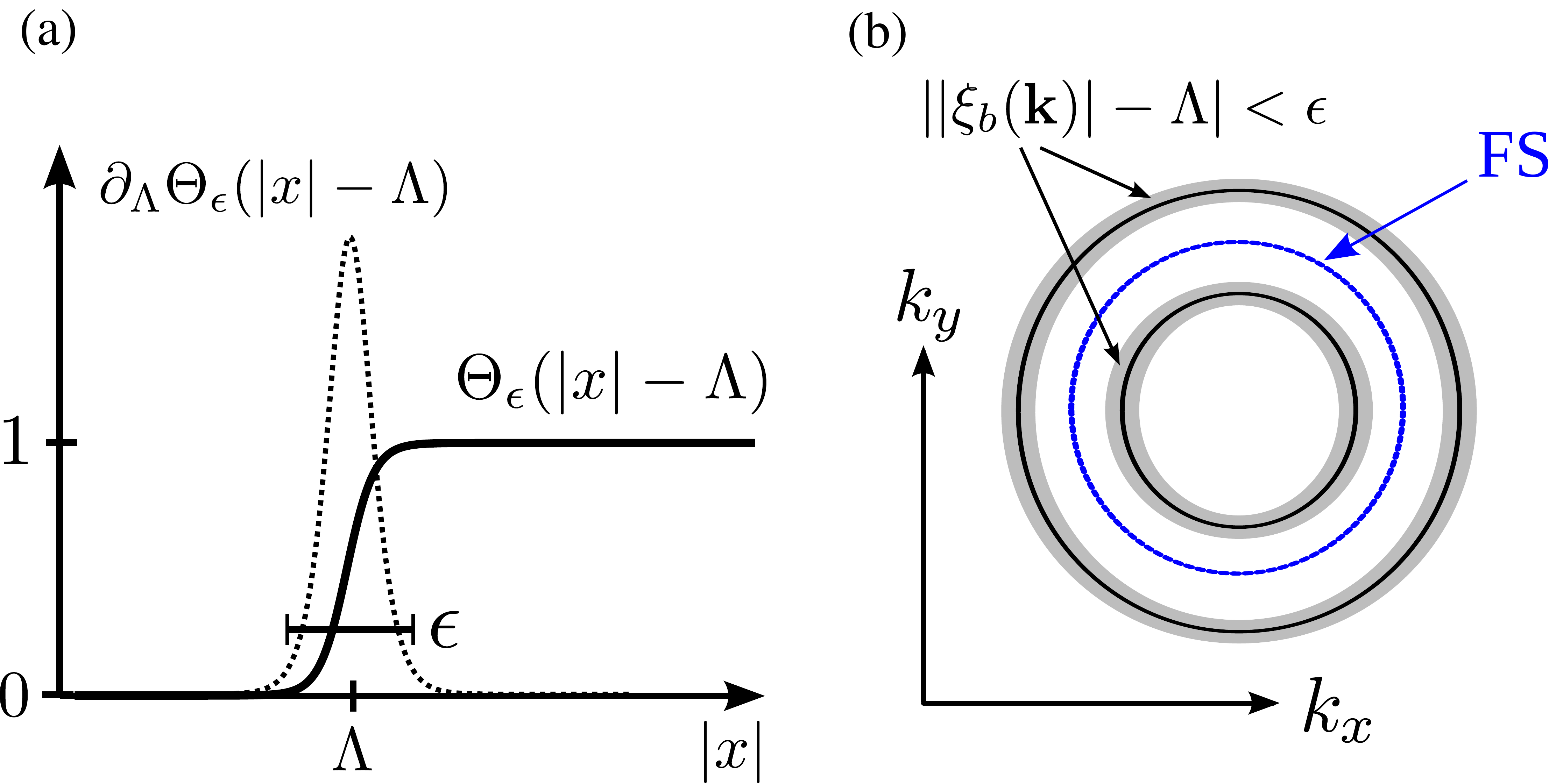}}
 \end{center}
 \caption{\label{regulator} (a) Cutoff function $\Theta_{\epsilon}$ (full line) and corresponding scale-derivative 
  $\partial_{\Lambda}\Theta_{\epsilon}$ (dashed line). (b) Momentum shells (gray) representing the finite support of the single-scale propagator $S^{\Lambda}$ at an energy
  scale $\Lambda$ away from the Fermi surface (blue).}
 \end{figure}

 One possible choice of flow parameter is implemented in the so-called cutoff scheme,
 where the inverse quadratic part reads as follows
 \begin{equation}
 \left(Q^{\Lambda}_{k,k'}\right)^{-1} = \left(G^{0\Lambda}_k\right) = \frac{\theta_{\epsilon}^{\Lambda}(k_0,\bs{k})}{ik_0 - \xi_b(\bs{k})}.
 \end{equation}
 Here, $\theta_{\epsilon}^{\Lambda}(k_0,\bs{k})$ indicates a cutoff function either in frequency 
 $\theta_{\epsilon}^{\Lambda}(k_0,\bs{k}) = \Theta_{\epsilon}(|k_0|-\Lambda)$ or in 
 momentum space $\theta_{\epsilon}^{\Lambda}(k_0,\bs{k}) = \Theta_{\epsilon}(|\xi_b(\bs{k})|-\Lambda)$
 with $\Theta_{\epsilon}$ denoting a step function of finite width $\epsilon$ as shown in Fig.~\ref{regulator}a. The full propagator
 in (\ref{eq:gamma2}) is, then, given by
 \begin{equation}\label{eq:Gscale}
 G^{\Lambda}(k_0,\bs{k}) = \frac{\theta_{\epsilon}^{\Lambda}(k_0,\bs{k})}{ik_0 - \xi_b(\bs{k}) - \theta_{\epsilon}^{\Lambda}(k_0,\bs{k})\Sigma^{\Lambda}(k_0,\bs{k})},
 \end{equation}
 and the corresponding single-scale propagator $S^{\Lambda} = -G^{\Lambda}\dot{Q}^{\Lambda}G^{\Lambda}$ reads as
 \begin{equation}\label{eq:singlescalecut}
 S^{\Lambda}(k_0,\bs{k}) = \frac{\partial_{\Lambda}\theta_{\epsilon}^{\Lambda}(k_0,\bs{k})\left[ik_0 - \xi_b(\bs{k})\right]}
                                {\left[ik_0 - \xi_b(\bs{k}) - \theta_{\epsilon}^{\Lambda}(k_0,\bs{k})\Sigma^{\Lambda}(k_0,\bs{k})\right]^2},
 \end{equation}
 with a finite support near the $\Lambda$-energy shells as depicted in Fig.~\ref{regulator}b.\par
 Using the above cutoff scheme, we obtain an implementation similar to Wilson's original idea of
 integrating out momentum modes shell by shell. For example, if we start with a momentum cutoff 
 $\Lambda$ greater than the bandwidth, all fluctuations are completely suppressed due to (\ref{eq:Gscale}),
 and the initial value of the effective action equals the bare action in Eq.~(\ref{eq:bareaction}). 
 Following the trajectory of $\Gamma^{\Lambda}$ to lower scales, 
 the calculation of $\partial_{\Lambda}\Gamma^{\Lambda}$ then comprises 
 a trace over the single scale propagator $S^{\Lambda}$ whose support is restricted to the momentum shells of Fig.~\ref{regulator}b.  
 For this reason, the calculation of $\Gamma^{\Lambda}$ with a decreasing cutoff-scale $\Lambda$ then integrates out momentum modes, shell by shell,
 and implements Wilson's picture of viewing the physical system at different length scales. 
 Although these cutoff schemes provide a very intuitive understanding of renormalization, they also involve serious drawbacks. 
 One is the violation of Ward identities at any finite cutoff value
 \cite{katanin04prb115109,enss-05phd} (see also Appendix~\ref{sec:ward}),
 as also the non-uniform treatment of particle-hole processes
 within the momentum-cutoff scheme. In order to explain the latter issue, we first write down the 
 one-loop particle-hole fluctuations arising in elementary perturbation theory,
 \begin{equation}\label{eq:chi}
 \chi^{p-h}(\bs{k},\bs{q}) = \frac{n_F(\xi_b(\bs{k})) - n_F(\xi_b(\bs{k}+\bs{q}))}{\xi_b(\bs{k}) - \xi_b(\bs{k}+\bs{q})}.
 \end{equation}
 During the flow, these contributions are taken into account successively within the trace (\ref{eq:flowseries})
 \begin{equation}\label{eq:loopaccount}
 \text{tr}\left(\bs{S}^{\Lambda}\bs{U}^{\Lambda}\bs{G}^{\Lambda}\bs{U}^{\Lambda}\right)
 \sim 
 \text{tr}\left( \chi^{p-h}(\bs{k},\bs{q})\cdot\partial_{\Lambda}(\theta_{\epsilon}^{\Lambda}(\bs{k})\theta_{\epsilon}^{\Lambda}(\bs{k+q}))\cdot\ldots\right),
 \end{equation}
 where we, for the simplicity of the argument, neglected frequency dependences and self-energy insertion.
 If we now consider particle-hole fluctuations with vanishing momentum transfer, i.e. 
 $\chi^{p-h}(\bs{k},\bs{q}\rightarrow 0)$, it turns out that the only nonzero contribution in (\ref{eq:chi})
 comes from modes $\bs{k}$ in a small energy region $(\sim T)$ around the Fermi surface. 
 However, due to the cutoff function $\theta_{\epsilon}^{\Lambda}(\bs{k})$, these modes are not taken into account until $\Lambda\approx T$.
 On the other hand, particle-hole fluctuations with large momentum transfer are already taken into account
 right from the beginning. \textit{The cutoff-scheme}, therefore, \textit{treats particle-hole fluctuations in a non-uniform way},
 and it may happen that other channels already indicate a singularity at cutoff values $\Lambda > T$, whereas 
 the small $\bs{q}$ particle-hole fluctuations have not yet contributed.\par
 In order to avoid this issue, we can exploit the flexibility in the parameter dependence
 of $\Gamma^{\Lambda}$ and \textit{regard the temperature itself as flow parameter}.
 However, we first have to shift the temperature dependences towards the quadratic part of the action (\ref{eq:bareaction}), 
 and we therefore write out all temperature prefactors
 \begin{equation}\nonumber
 S(\overline{\psi},\psi)    = -T\int_{k,k'}Q_{k,k'}\overline{\psi}_{k}\psi_{k'}
                            + T^3\int_{k_1,k_2,k'_1,k'_2}
                            U_{k_1,k_2,k'_1,k'_2}\overline{\psi}_{k_1}\overline{\psi}_{k_2}\psi_{k'_1}\psi_{k'_2}
 \end{equation}
 and rescale the field variables according to
 \begin{equation}
 \overline{\psi}_{k} = T^{-3/4}\overline{\phi}_{k},\quad \psi_{k} = T^{-3/4}\phi_{k}.
 \end{equation}
 In the rescaled action, all temperature dependences now occur only within the quadratic part, i.e.
 \begin{align}\nonumber
 S(\overline{\phi},\phi)   = -T^{-1/2}\int_{k,k'}Q_{k,k'}\overline{\phi}_{k}\phi_{k'}
                              + \int_{k_1,k_2,k'_1,k'_2}
                            U_{k_1,k_2,k'_1,k'_2}\overline{\phi}_{k_1}\overline{\phi}_{k_2}\phi_{k'_1}\phi_{k'_2}.
 \end{align}
 We can then follow the same arguments as in the cutoff case discussed
 previously, and regard the temperature $T$ itself as flow parameter.
 Here, it turns out that for temperatures on the order of the bandwidth $W$, all perturbative corrections
 to the vertex functions are suppressed with a negative power of $T$ \cite{honerkamp-01prb184516}, and 
 we obtain the required boundary condition of $\Gamma^{T\sim W} = S$. In addition to that, the temperature also 
 serves as an infrared regulator similar to the momentum or frequency cutoff, as it shifts Matsubara frequencies
 $k_0=(2n+1)\pi T$ away from zero and, therefore, cuts out the singularities of the bare propagator
 \begin{equation}
 \left(\widetilde{Q}^{T}_{k,k'}\right)^{-1} = \left(\widetilde{G}^{0T}_k\right) = \frac{T^{1/2}}{ik_0 - \xi_b(\bs{k})}.
 \end{equation}
 It is further important to note that all objects in the rescaled fields $\overline{\phi},\phi$ are highlighted with a tilde and,
 to obtain the $m$-particle Green and vertex functions back in the original fields $\overline{\psi},\psi$, we have
 to unscale these functions with a factor of $T^{-3m/2}$ and $T^{3m/2}$, respectively.\par
 The temperature flow scheme can now be implemented in a similar way as the cutoff version. We
 start again with the bare theory for temperatures of the order of the bandwidth $\Gamma^{T_{init} \sim W} = S$
 and then follow the trajectory $\Gamma^{T}$ to lower values of $T$.
 The difference between both schemes consists in the distinct scale derivatives $\partial_{\Lambda}$
 which in the temperature flow case involves the $T-$dependent Green function
 \begin{equation}
 \widetilde{G}^{T}(k_0,\bs{k}) = \frac{T^{1/2}}{ik_0 - \xi_b(\bs{k}) - T^{1/2}\widetilde{\Sigma}^{T}(k_0,\bs{k})}\nonumber
 \end{equation}
 and the single-scale propagator
 \begin{equation}\label{eq:singlescaleT}
 \widetilde{S}^{T}(k_0,\bs{k}) = -\frac{1}{2}\frac{T^{-1/2}\left[ik_0 - \xi_b(\bs{k})\right]}
                          {\left[ik_0 - \xi_b(\bs{k}) - T^{1/2}\widetilde{\Sigma}^{T}(k_0,\bs{k})\right]^2}.
 \end{equation}
 Using these propagators, the particle-hole contributions contained in the trace (\ref{eq:flowseries}) 
 are given by
 \begin{equation}
 \text{tr}\left(\bs{S}^{\Lambda}\bs{U}^{\Lambda}\bs{G}^{\Lambda}\bs{U}^{\Lambda}\right)
 \sim 
 \text{tr}\left(\partial_{T} \chi^{p-h}\cdot\ldots\right)
 \end{equation}
 and now involve the temperature derivative $\partial_{T}\chi^{p-h}$ instead 
 of a cutoff function. The broader support of $\partial_{T} \chi^{p-h}$ 
 then does not distinguish between large and small momentum transfer and, therefore, the temperature flow
 allows a uniform treatment of all particle-hole fluctuations. In addition to that, Ward identities derived in Appendix~\ref{sec:ward} 
 are also respected in the temperature flow \cite{RevModPhys.84.299},
 at least in the full hierarchy of differential equations before the
 flow is constrained to a finite subset of vertex functions. 
 For these reasons, the temperature flow is more favorable than the
 ordinary cutoff schemes
 although the intuitive picture of eliminating short-wavelength
 fluctuations and obtaining the properties on a coarse-grained scale
 is somewhat
 lost here. For completeness, while not employed in the following, another cutoff-free scheme is provided by the so-called interaction flow \cite{honerkamp-04prb235115} which multiplies the quadratic part of the action with a parameter $1/g$. By rescaling the fields, one can show that this corresponds
 to a substitution of the interacting part $U\rightarrow g^2U$ and, therefore, the flow interpolates between
 the noninteracting theory $g=0$ and the original one at $g=1$.\par
 As we have seen in this subsection, the \textit{functional 
 flow equations enable a certain flexibility in the choice of flow parameters} which \textit{can be exploited to prove 
 the robustness of the results or to justify approximations}. The most common implementations involve
 cutoff-schemes in frequency-space
 \cite{husemann-0902arxiv1651,andergassen-04prb075102,PhysRevB.83.024402},
 in momentum-space \cite{honerkamp-01prb035109}, the interaction-flow
 \cite{honerkamp-04prb235115}, and
 the beforementioned temperature-flow \cite{honerkamp-01prb184516}
 which will be employed in the following. 
 \subsection{Symmetries Properties of the Functional Flow Equations}
 The physical system to be investigated usually features certain symmetries 
 as, for example, translational or rotational invariance. According to Noether's theorem, 
 each of these continuous symmetries also entails some kind of conservation law. In the following section, we will 
 demonstrate how this \textit{symmetry statement simplifies the flow equation hierarchy and also provides useful relations
 between the different $n$-point vertex functions}.\par 
 For this purpose, we first introduce a general symmetry transformation as an invertible mapping of fields 
 \begin{equation}\label{eq:fieldmap}
 \psi_k \mapsto \int_{k'}\mathcal{M}_{k,k'}\psi_{k'},\quad
 \overline{\psi}_k \mapsto \int_{k'}\overline{\mathcal{M}}_{k,k'}\overline{\psi}_{k'}
 \end{equation}
 leaving the action $S(\overline{\psi},\psi)$ as well as the functional-integral measure $D(\overline{\psi},\psi)$ 
 invariant
  \begin{align}\label{eq:actioninvar}
 S(\overline{\mathcal{M}}\cdot\overline{\psi},\mathcal{M}\cdot\psi) &= S(\overline{\psi},\psi)\\\nonumber
 D(\overline{\mathcal{M}}\cdot\overline{\psi},\mathcal{M}\cdot\psi) &= D(\overline{\psi},\psi).
 \end{align}
 Here, we used the shortened notation $\mathcal{M}\cdot\overline{\psi}$ and $\mathcal{M}\cdot\psi$ to denote 
 the mappings of (\ref{eq:fieldmap}). Note that the very general representation of a symmetry transformation in (\ref{eq:fieldmap}) 
 comprises space-time as well as internal symmetries, and will be
 specified by various examples in the following. In order to derive
 the transformation behavior of the 1PI-vertex functions and, therefore, to simplify the flow equations, we
 first determine the transformation properties of the generating functional $\mathcal{G}[\eta,\overline{\eta}]$ for the connected Green functions.
 Using the symmetry requirements of (\ref{eq:actioninvar}), we obtain
 \begin{align}
 \mathcal{G}[\eta,\overline{\eta}] & = -\ln\int D(\overline{\psi},\psi) e^{-S(\overline{\psi},\psi) + (\overline{\eta},\psi) + (\overline{\psi},\eta)}\\\nonumber
                                   & = -\ln\int D(\overline{\mathcal{M}}\cdot\overline{\psi},\mathcal{M}\cdot\psi) e^{-S(\overline{\mathcal{M}}\cdot\overline{\psi},\mathcal{M}\cdot\psi) 
                                     + (\overline{\eta},\mathcal{M}\cdot\psi) + (\overline{\mathcal{M}}\cdot\overline{\psi},\eta)}\\\nonumber
                                   & = -\ln\int D(\overline{\psi},\psi) e^{-S(\overline{\psi},\psi) 
                                     + (\overline{\eta},\mathcal{M}\cdot\psi) + (\overline{\mathcal{M}}\cdot\overline{\psi},\eta)}\\\nonumber
                                   & = -\ln\int D(\overline{\psi},\psi) e^{-S(\overline{\psi},\psi) 
                                     + (\mathcal{M}^T\cdot\overline{\eta},\psi) + (\overline{\psi},\overline{\mathcal{M}}^T\cdot\eta)}\\\nonumber
                                   & = \mathcal{G}[\overline{\mathcal{M}}^T\cdot\eta,\mathcal{M}^T\cdot\overline{\eta}].
 \end{align}
 With this, the transformation behavior of the source fields $\zeta,\overline{\zeta}$ in the effective action $\Gamma[\zeta,\overline{\zeta}]$ of 
 (\ref{eq:gammdef}) reads as
 \begin{align}\nonumber
 \zeta(\overline{\mathcal{M}}^T\cdot\eta,\mathcal{M}^T\cdot\overline{\eta})
 &= -\frac{\partial \mathcal{G}[\overline{\mathcal{M}}^T\cdot\eta,\mathcal{M}^T\cdot\overline{\eta}]}
          {\partial\left(\mathcal{M}^T\cdot\overline{\eta}\right)}
  = -\frac{\partial \mathcal{G}[\eta,\overline{\eta}]}
          {\partial\left(\mathcal{M}^T\cdot\overline{\eta}\right)}
  =  (\mathcal{M})^{-1}\cdot\zeta(\eta,\overline{\eta})\\\nonumber
 \overline{\zeta}(\overline{\mathcal{M}}^T\cdot\eta,\mathcal{M}^T\cdot\overline{\eta})
 &=  \frac{\partial \mathcal{G}[\overline{\mathcal{M}}^T\cdot\eta,\mathcal{M}^T\cdot\overline{\eta}]}{\partial \left(\overline{\mathcal{M}}^T\cdot\eta\right)}
  =  \frac{\partial \mathcal{G}[\eta,\overline{\eta}]}{\partial \left(\overline{\mathcal{M}}^T\cdot\eta\right)}
  =   (\overline{\mathcal{M}})^{-1}\cdot\overline{\zeta}(\eta,\overline{\eta}).
 \end{align}
 Applying this result, we can easily infer the effect of (\ref{eq:fieldmap}) 
 on the effective action
 \begin{align}\label{eq:gammasym}
 \Gamma[\zeta, \overline{\zeta}]&\mapsto
 \Gamma[(\mathcal{M})^{-1}\zeta, (\overline{\mathcal{M}})^{-1}\overline{\zeta}]\\\nonumber 
 & = \mathcal{G}[ \overline{\mathcal{M}}^T\eta, \mathcal{M}^T\overline{\eta}]
   + (\mathcal{M}^T\overline{\eta},(\mathcal{M})^{-1}\zeta) + ((\overline{\mathcal{M}})^{-1}\overline{\zeta},\overline{\mathcal{M}}^T\eta)\\\nonumber
 & = \Gamma[\zeta, \overline{\zeta}],
 \end{align}
 where we made use of the following relation
 \begin{align}\nonumber
 (\mathcal{M}^T\cdot\overline{\eta},(\mathcal{M})^{-1}\cdot\zeta) &= \int_{k}\left(\mathcal{M}^T\cdot\overline{\eta}\right)_{k}\left((\mathcal{M})^{-1}\cdot\zeta)\right)_{k}\\\nonumber
                                                                  &= \int_{k,l,m}\left(\mathcal{M}^T\right)_{kl}(\mathcal{M})^{-1}_{km}\overline{\eta}_{l}\xi_{m}\\\nonumber
                                                                  &= \int_{k,l,m}\left(\mathcal{M}\right)_{lk}(\mathcal{M})^{-1}_{km}\overline{\eta}_{l}\xi_{m}
                                                                   = \left(\overline{\eta},\zeta\right).
 \end{align}
 From (\ref{eq:gammasym}) it is, then, apparent that the effective action $\Gamma$ remains invariant under all symmetry transformations 
 of the underlying physical system, i.e.
 \begin{equation}\label{eq:symmeff}
 \Gamma[\mathcal{M}\cdot\zeta,\overline{\mathcal{M}}\cdot\overline{\zeta}] = \Gamma[\zeta,\overline{\zeta}],
 \end{equation}
 and, by expanding both sides in powers of the fields, we obtain the following relation for the 1PI vertex functions
 \begin{equation}\label{eq:pisymm}
 \gamma^{(2m)}_{k'_1,\ldots,k'_m;k_1,\ldots,k_m} = \int_{q'_1,\ldots,q'_m\atop q_1,\ldots,q_m}\mathcal{M}_{k'_1q'_1}\cdots\mathcal{M}_{k'_mq'_m}\mathcal{M}_{k_1q_1}\cdots\mathcal{M}_{k_mq_m}
                                                   \cdot\gamma^{(2m)}_{q'_1,\ldots,q'_m;q_1,\ldots,q_m}.
 \end{equation}
 Since we are mainly interested in solid-state compounds with an underlying periodic crystal lattice, 
 \textit{we now assume translational symmetry under $x\mapsto x+a$}, which then translates into 
 \begin{equation}
  \psi_k \mapsto e^{i\bs{k}\cdot a}\psi_{k},\quad
 \overline{\psi}_k \mapsto e^{-i\bs{k}\cdot a}\overline{\psi}_{k}.
 \end{equation}
 Employing the symmetry relation (\ref{eq:pisymm}), we then end up with
 \begin{equation}\label{eq:transaction}
 \gamma^{(2m)}_{k'_1,\ldots,k'_m;k_1,\ldots,k_m} = e^{i(\bs{k}_1+\ldots+\bs{k}_m-\bs{k}'_{1}-\bs{k}'_m)\cdot a}\cdot\gamma^{(2m)}_{k'_1,\ldots,k'_m;k_1,\ldots,k_m}
 \end{equation}
 \textit{which in turn implies that all non-vanishing 1PI vertex functions conserve momentum up to a reciprocal lattice vector}.\par
 \textit{Besides translational symmetry, a periodic crystal lattice
 also exhibits point group symmetries} which, by definition, leave one space point fixed and therefore constitute a subgroup of the orthogonal group $O(n)$, with $n$ denoting
 the dimensionality of the lattice. If we now consider a point-group transformation $x\mapsto R x$ with $R\in O(n)$, its effect in $\bs{k}$-space is given by 
 $k\mapsto R^T k = (k_0,R^T\bs{k},\sigma)$,
 following from the duality definition of the reciprocal lattice. Accordingly, the fields transform as 
 \begin{equation}
 \psi_k \mapsto \psi_{R^Tk},\quad
 \overline{\psi}_k \mapsto \overline{\psi}_{R^Tk},
 \end{equation}
 and, due to (\ref{eq:pisymm}), the 1PI vertex functions remain invariant under all point-group symmetries:
  \begin{equation}\label{eq:ptgroupacion}
 \gamma^{(2m)}_{k'_1,\ldots,k'_m;k_1,\ldots,k_m} = \gamma^{(2m)}_{R^Tk'_1,\ldots,R^Tk'_m;R^Tk_1,\ldots,R^Tk_m}.
 \end{equation}
 \textit{Another important symmetry describes an invariance under time-reversal $t\rightarrow -t$}.
 This operation is known to be peculiar as it must have an antiunitary representation in the space of quantum states.
 Otherwise, as each symmetry has to be represented either as a
 unitary or an antiunitary operator according to Wigner's fundamental theorem,
 time-reversal would be a unitary operation which runs into fundamental problems.
 Therefore, time-reversal can be shown to act on a spin one-half state as
 \begin{equation}\label{eq:timerevop}
 \Theta = -i\tau^2K,
 \end{equation}
 where $K$ denotes complex-conjugation and $\tau^2$ labels the
 Pauli-matrix which synonymously reads $\sigma^y$ . Under the time-reversal operation (\ref{eq:timerevop}), 
 the fields then transform as 
 \begin{equation}
 \psi_{k\sigma} \mapsto \text{sign}(\sigma)\overline{\psi}_{Tk},\quad
 \overline{\psi}_{k\sigma} \mapsto \text{sign}(\sigma)\psi_{Tk}
 \end{equation}
 with $Tk=(k_0,-\bs{k},-\sigma)$ and $\text{sign}(\uparrow\downarrow) = \pm 1$ according to the matrix $-i\tau^2$. The corresponding 
 effect on the 1PI vertex function now reads as 
 \begin{equation}\label{eq:timerevaction}
 \gamma^{(2m)}_{k'_1,\ldots,k'_m;k_1,\ldots,k_m} = \text{sign}(\sigma'_1)\cdots\text{sign}(\sigma_m) \gamma^{(2m)}_{Rk_m,\ldots,Rk_1;Rk'_m,\ldots,Rk'_1},
 \end{equation}
 and the effect of other discrete symmetries like the \textit{spatial reflection} can be derived in a similar way.\par
 In addition to these discrete transformations, the underlying system also features certain \textit{continuous symmetries 
 such as the spin-rotational invariance or a global $U(1)$-phase}. The corresponding representations 
 of these continuous symmetries can be parametrized, at least locally, by some real parameters $s_1,\ldots,s_n$ such that
 \begin{equation}
 \mathcal{M}(s_1=0,\ldots,s_n=0) = 1,
 \end{equation}
 and the associated generators can be defined as
 \begin{equation}\label{eq:gendef}
 T_j = -i\left.\frac{\partial\mathcal{M}(s_1,\ldots,s_n)}{\partial s_j}\right|_{s_1=\ldots=s_n=0},
 \ \overline{T}_j = -i\left.\frac{\partial\overline{\mathcal{M}}(s_1,\ldots,s_n)}{\partial s_j}\right|_{s_1=\ldots=s_n=0}.
 \end{equation}
 Now, by using the invariance of the effective action in (\ref{eq:symmeff}), we then obtain the following relation 
 \begin{align}\nonumber
 0  = \frac{\partial}{\partial s_j} \Gamma[\zeta,\overline{\zeta}] 
   &= \left(\frac{\partial\Gamma[\mathcal{M}\cdot\zeta,\overline{\mathcal{M}}\cdot\overline{\zeta}]}{\partial(\mathcal{M}\cdot\zeta)},
      \frac{\partial(\mathcal{M}\cdot\zeta)}{\partial s_j}\right)\\\nonumber
   &+ \left(\frac{\partial\Gamma[\mathcal{M}\cdot\zeta,\overline{\mathcal{M}}\cdot\overline{\zeta}]}{\partial(\overline{\mathcal{M}}\cdot\overline{\zeta)}},
      \frac{\partial(\overline{\mathcal{M}}\cdot\overline{\zeta)}}{\partial s_j}\right)\\\label{eq:contieff}
   &= \left(\frac{\partial\Gamma[\zeta,\overline{\zeta}]}{\partial\zeta},(iT_j)\cdot\zeta\right)
    + \left(\frac{\partial\Gamma[\zeta,\overline{\zeta}]}{\partial\overline{\zeta}},(i\overline{T}_j)\cdot\overline{\zeta}\right).
 \end{align}
 In the next step, we demonstrate how this symmetry constraint restricts the form of the 1PI vertex functions.
 Therefore, we consider the case of a global-$U(1)$ phase transformation, 
 \begin{equation}
 \psi_k \mapsto e^{is}\psi_{k},\quad\overline{\psi}_k \mapsto e^{-is}\overline{\psi}_{k},
 \end{equation}
 which apparently presents a symmetry transformation of the action (\ref{eq:bareaction}). Computing the generators of this 
 symmetry group according to (\ref{eq:gendef}), one obtains
 \begin{equation}
 T = -i\left.\frac{\partial e^{is}}{\partial s}\right|_{s=0} = 1,\quad
 \overline{T} = -i\left.\frac{\partial e^{-is}}{\partial s}\right|_{s=0} = -1
 \end{equation}
 and, by using (\ref{eq:contieff}), we end up with 
 \begin{equation}\label{eq:u1state}
 \int_k\left(\zeta_k\frac{\partial}{\partial\zeta_k} - \overline{\zeta}_k\frac{\partial}{\partial\overline{\zeta}_k}\right)\Gamma[\zeta,\overline{\zeta}]=0.
 \end{equation}
 If we now expand the generating functionals in fields, the only nonzero 1PI vertex functions in (\ref{eq:u1state}) are those which correspond to monomials with an equal number
 of $\zeta-$ and $\overline{\zeta}-$fields.  Note, that we already
 assumed such a form in the expansion (\ref{eq:gammaexpansion}), which
 is now justified {\it a posteriori} with the global-$U(1)$
 phase symmetry of the bare action. As a final example, we consider the invariance under spin-rotation which, in the spin one-half representation, reads 
  \begin{equation}
 \begin{pmatrix} 
 \psi_{\widetilde{k},\uparrow} \\
 \psi_{\widetilde{k},\downarrow}
 \end{pmatrix}
 \mapsto
 e^{i\bs{\tau}\cdot\bs{s}}
 \begin{pmatrix} 
 \psi_{\widetilde{k},\uparrow} \\
 \psi_{\widetilde{k},\downarrow}
 \end{pmatrix},
 \quad
  \begin{pmatrix} 
 \overline{\psi}_{\widetilde{k},\uparrow} \\
 \overline{\psi}_{\widetilde{k},\downarrow}
 \end{pmatrix}
 \mapsto
 e^{-i\bs{\tau}^*\cdot\bs{s}}
 \begin{pmatrix} 
 \overline{\psi}_{\widetilde{k},\uparrow} \\
 \overline{\psi}_{\widetilde{k},\downarrow}
 \end{pmatrix}
 \end{equation}
 with $\tau^{1,2,3}$ denoting the usual Pauli-matrices. Together with the generators of the $SU(2)$-transformation, i.e.  
 \begin{equation}
 T_{i} = \tau^i,\quad \overline{T}_{i} = -\left(\tau^i\right)^*\quad i=1,2,3,
 \end{equation}
 the symmetry restriction on the effective action in (\ref{eq:contieff}) is given by
 \begin{equation}\label{eq:spincons}
 \int_{\widetilde{k},\sigma,\sigma'}\left(\tau^{i}_{\sigma\sigma'}\zeta_{\widetilde{k}\sigma}\frac{\partial}{\partial\zeta_{\widetilde{k}\sigma'}} 
     - \left(\tau^{i}\right)^*_{\sigma\sigma'}\overline{\zeta}_{\widetilde{k}\sigma}\frac{\partial}{\partial\overline{\zeta}_{\widetilde{k}\sigma'}}\right)
       \Gamma[\zeta,\overline{\zeta}]=0.
 \end{equation}
 Here, we explicitly spelled out the spin-projection $\sigma$ and defined the new index $\widetilde{k}$ as containing all remaining degrees of freedom
 besides $\sigma$ such that $k=(\widetilde{k},\sigma)$. Using the $i=3$ component of (\ref{eq:spincons}) as well as the $U(1)$-symmetry constraint
 of (\ref{eq:u1state}), we then obtain  
  \begin{equation}\label{eq:eqspinproj}
 \int_{\widetilde{k}}\left(\zeta_{\widetilde{k}\sigma}\frac{\partial}{\partial\zeta_{\widetilde{k}\sigma}} 
     - \overline{\zeta}_{\widetilde{k}\sigma}\frac{\partial}{\partial\overline{\zeta}_{\widetilde{k}\sigma}}\right)
       \Gamma[\zeta,\overline{\zeta}]=0,
 \end{equation}
 which restricts an expansion of $\Gamma[\zeta,\overline{\zeta}]$ to consist of monomials with an equal number of
 $\zeta_{\widetilde{k}\sigma}$- and $\overline{\zeta}_{\widetilde{k}\sigma}$-fields. Therefore, the number of particles
 with a given spin-projection is individually conserved, and we can write the fully spin-dependent 1PI vertex functions
 as 
  \begin{align}\nonumber
 \gamma^{(2m)\Lambda}_{k'_1,\ldots,k'_m;k_1,\ldots,k_m} 
 = &\gamma^{(2m)\Lambda}_{\widetilde{k}'_1\sigma'_1,\ldots,\widetilde{k}'_m\sigma'_m;\widetilde{k}_1\sigma_1,\ldots,\widetilde{k}_m\sigma_m}\\\nonumber
 = & -\sum_{p\in\pi_m}\text{sgn}(p)\cdot \widetilde{V}^{(2m)\Lambda}_{\sigma'_1,\ldots,\sigma'_m}(\widetilde{k}'_1,\ldots,\widetilde{k}'_m;\widetilde{k}_{p(1)},\ldots,\widetilde{k}_{p(m)})
   \nopagebreak[1]\\\label{eq:Uspinconserve}
   & \times\delta_{\sigma'_1\sigma_{p(1)}}\cdots\delta_{\sigma'_m\sigma_{p(m)}},
 \end{align}
 with the spin-conserving function $\widetilde{V}^{(2m)\Lambda}_{\sigma'_1,\ldots,\sigma'_m}$ depending only on the spin projection of the outgoing particles.
 Note, here, that the definition of the $\widetilde{V}^{(2m)\Lambda}$ functions is not unique and sometimes occurs with a different sign in the literature. 
 Up to now, we only implemented the $i=3$ constraint of (\ref{eq:spincons}) which is equivalent 
 to  spin-rotational invariance around the $z$-axis. Making use of the full $SU(2)$-invariance, one can even show that
 $\widetilde{V}^{(2m)}$ is independent of $\sigma'_1,\ldots,\sigma'_m$ \cite{kopietz-10book}, and we define the coupling functions
 \begin{equation}
 V^{(2m)}_{\widetilde{k}'_1,\ldots,\widetilde{k}'_m;\widetilde{k}_{1},\ldots,\widetilde{k}_{m}}
 =
 \widetilde{V}^{(2m)}_{\sigma'_1,\ldots,\sigma'_m}(\widetilde{k}'_1,\ldots,\widetilde{k}'_m;\widetilde{k}_{1},\ldots,\widetilde{k}_{m}),
 \end{equation}
 with $V^{(2m)}$ being independent of any spin-projection.
 \begin{figure}[t]
 \begin{center}
 {\includegraphics[scale=0.35]{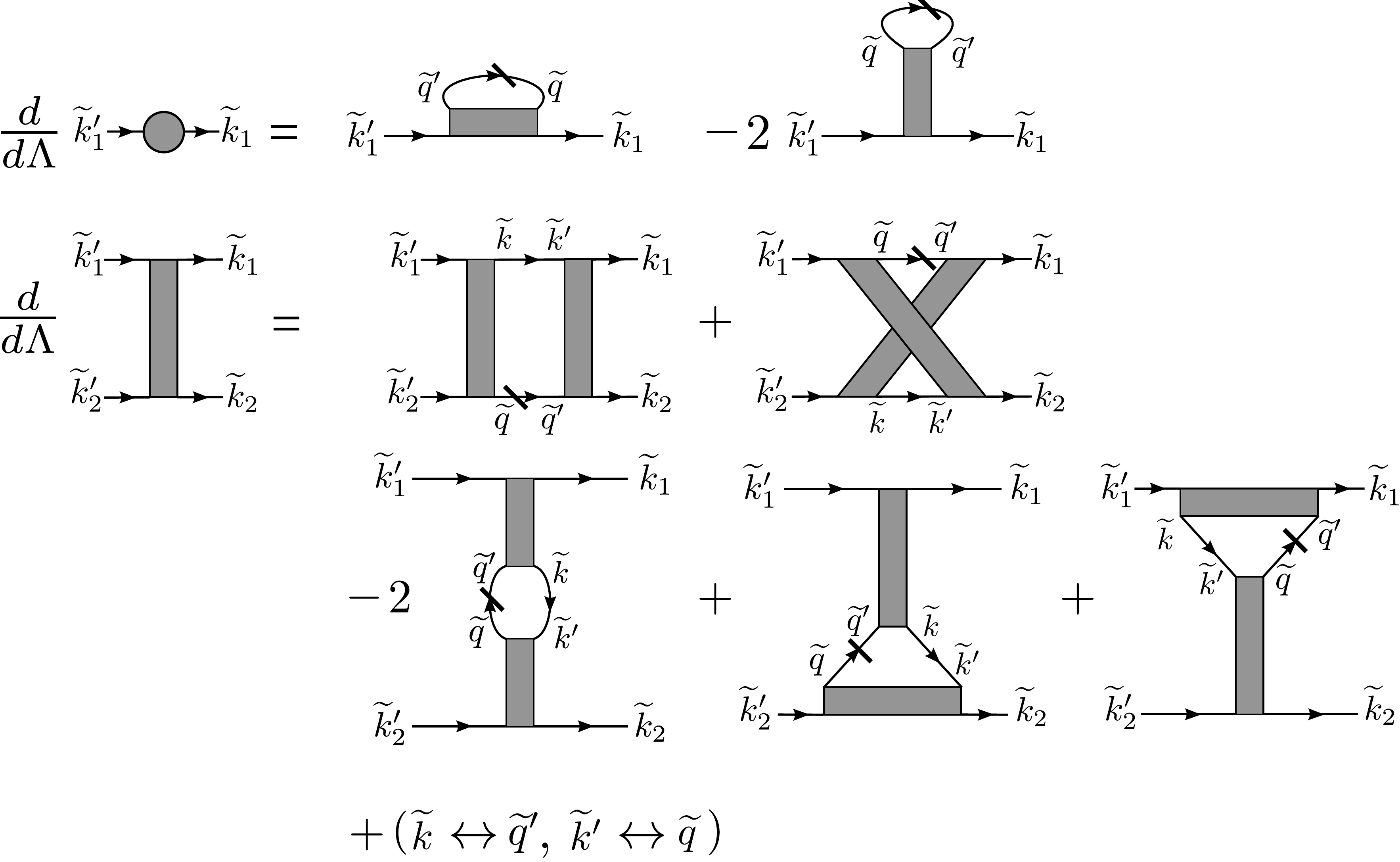}}
 \end{center}
 \caption{\label{pic:vfloweq} One-particle and two-particle vertex
   terms of the infinite hierarchy of flow equation for the coupling function $V^{(2m)}$. 
 Slashed and full internal-lines represent the single-scale $S^{\Lambda}$ and the full propagator $G^{\Lambda}$.
 In contrast to the 1PI vertex function $\gamma^{(4)}$, the coupling function $V^{(4)}$ describes the two-particle 
 scattering for a fixed spin-configuration. Here, the two incoming lines have opposite spin-projection which is conserved
 along the short edge of the rectangular box representing $V^{(4)}$.}
 \end{figure}
 Therefore, the first two vertex functions in (\ref{eq:Uspinconserve}) read 
 \begin{align}\label{eq:v2define}
 \gamma^{(2)}_{\widetilde{k}'_1\sigma'_1;\widetilde{k}_1\sigma_1}&=Q_{\widetilde{k}'_1,\widetilde{k}_{1}} - \Sigma_{\widetilde{k}'_1;\widetilde{k}_{1}}
                                                                  =V^{(2)}_{\widetilde{k}'_1;\widetilde{k}_{1}}\\\label{eq:v4define}
 \gamma^{(4)}_{\widetilde{k}'_1\sigma'_1,\widetilde{k}'_2\sigma'_2;\widetilde{k}_1\sigma_1,\widetilde{k}_2\sigma_2}
 &=-D\cdot V^{(4)}_{\widetilde{k}'_1,\widetilde{k}'_2;\widetilde{k}_{1},\widetilde{k}_{2}}
  + E\cdot V^{(4)}_{\widetilde{k}'_1,\widetilde{k}'_2;\widetilde{k}_{2},\widetilde{k}_{1}},
 \end{align}
 where we made use of (\ref{eq:gamma2}) and applied the notations 
 $D = \delta_{\sigma_1'\sigma_1}\delta_{\sigma_2'\sigma_2}$, 
 $E = \delta_{\sigma_1'\sigma_2}\delta_{\sigma_2'\sigma_1}$ in (\ref{eq:v4define}).
 In order to simplify the flow equations for the 1PI vertex functions, we insert (\ref{eq:v2define}) and (\ref{eq:v4define})
 into the truncated equations (\ref{eq:flowselfen}), (\ref{eq:flowgamma}), and, by comparing the coefficient of $D$, obtain
 the following spin-independent flow equations for the first two terms 
 \begin{align}\label{eq:flowendres1}
 \frac{d}{d\Lambda}\Sigma^{(2)\Lambda}_{\widetilde{k}'_1;\widetilde{k}_{1}}
 &= \sum_{\widetilde{q},\widetilde{q}'}S^{\Lambda}_{\widetilde{q},\widetilde{q}'}\left\{
   V^{(4)\Lambda}_{\widetilde{k}'_1,\widetilde{q}';\widetilde{q},\widetilde{k}_1}
  -2V^{(4)\Lambda}_{\widetilde{k}'_1,\widetilde{q}';\widetilde{k}_1,\widetilde{q}}\right\}\\\label{eq:flowendres2}
 \frac{d}{d\Lambda} V^{(4)\Lambda}_{\widetilde{k}'_1,\widetilde{k}'_2;\widetilde{k}_{1},\widetilde{k}_{2}} 
 &= \sum_{\widetilde{k},\widetilde{k}'\atop \widetilde{q},\widetilde{q}'}G^{\Lambda}_{\widetilde{k},\widetilde{k}'}S^{\Lambda}_{\widetilde{q},\widetilde{q}'}\left\{
    V^{(4)\Lambda}_{\widetilde{k}'_1,\widetilde{k}'_2;\widetilde{k},\widetilde{q}}
    V^{(4)\Lambda}_{\widetilde{k}',\widetilde{q}';\widetilde{k}_1,\widetilde{k}_2}\right.\\\nonumber
    V^{(4)\Lambda}_{\widetilde{k}'_2,\widetilde{q}';\widetilde{k},\widetilde{k}_1}
    V^{(4)\Lambda}_{\widetilde{k}',\widetilde{k}'_1;\widetilde{k}_2,\widetilde{q}}
 &-2V^{(4)\Lambda}_{\widetilde{k}'_1,\widetilde{q}';\widetilde{k}_1,\widetilde{k}}
    V^{(4)\Lambda}_{\widetilde{k}',\widetilde{k}'_2;\widetilde{q},\widetilde{k}_2}
   +V^{(4)\Lambda}_{\widetilde{k}'_1,\widetilde{q}';\widetilde{k}_1,\widetilde{k}}
    V^{(4)\Lambda}_{\widetilde{k}',\widetilde{k}'_2;\widetilde{k}_2,\widetilde{q}}\\\nonumber
 &\left.+ V^{(4)\Lambda}_{\widetilde{k}'_1,\widetilde{q}';\widetilde{k},\widetilde{k}_1}
    V^{(4)\Lambda}_{\widetilde{k}',\widetilde{k}'_2;\widetilde{q},\widetilde{k}_2} + (\widetilde{k}\leftrightarrow \widetilde{q},\widetilde{k}'\leftrightarrow \widetilde{q}')\right\}.
 \end{align}
 Due to this spin-independence, the complexity of the flow equations reduces by a factor of $2n$ for each of the $n$-point vertex functions.
 If the underlying system further shows a translational symmetry or an invariance under certain point-group transformations, we can apply
 (\ref{eq:transaction}) and (\ref{eq:ptgroupacion}) which further lowers the computational effort. Note that the latter symmetry relations
 of (\ref{eq:transaction}) and (\ref{eq:ptgroupacion}) were derived for the full vertex functions $\gamma^{(2m)}$ but do also hold for the 
 spin-independent coupling functions $V^{2m}$ defined in (\ref{eq:Uspinconserve}). A diagrammatic expression of the flow equations
 in (\ref{eq:flowendres1}) and (\ref{eq:flowendres2}) is depicted in Fig.~\ref{pic:vfloweq}.\par
\begin{figure}[t]
 \begin{center}
 {\includegraphics[scale=0.25]{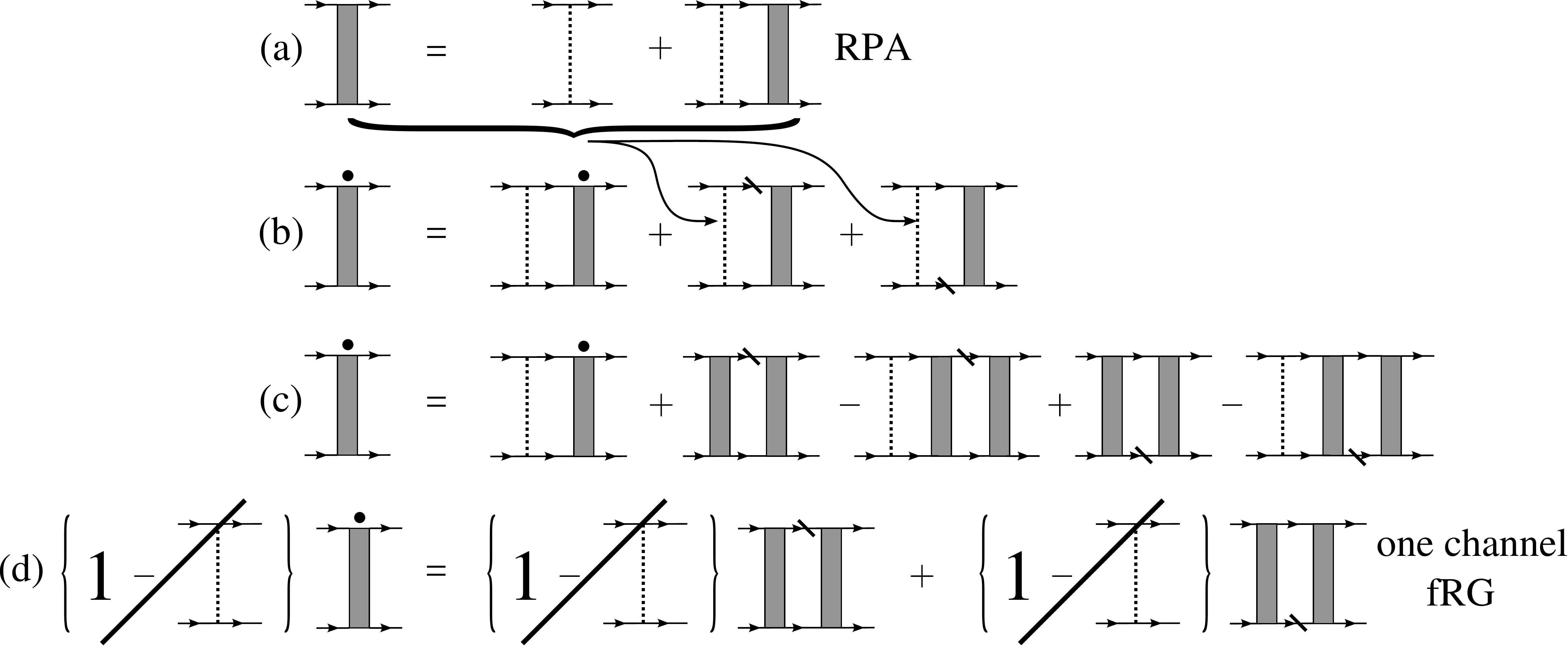}}
 \end{center}
 \caption{\label{pic:rpa} Equivalence of the ladder resummation (RPA)
   in the particle-particle channel (a) and the one-channel 
 functional RG (d). For this to hold,
 the single-scale propagators $S^{\Lambda}$ (slashed lines) have to equal the total scale derivative $dG^{\Lambda}/d\Lambda$ of the Green's function $G^{\Lambda}$.
 The dot represents the scale-derivative of the coupling
 function. Differential flow equation (b) flows towards the RPA summation (a) in the limit 
 of $\Lambda\rightarrow 0$. (c) as an intermediate step is obtained by
 inserting (a) into (b). After appropriate regrouping, (c) is
 transformed into (d), and by cancelling the
 crossed prefactor, the equivalence to one-channel functional RG
 becomes apparent.}
 \end{figure}

 
 Starting from this representation, one can easily prove pictorially that the flow equation with only one channel taken into account reproduces the ladder 
 resummations of the random phase approximation (RPA), which is depicted in Fig.~\ref{pic:rpa}. For this to hold, the single-scale 
 propagator $S^{\Lambda}$ has to be equal to the total scale derivative $dG^{\Lambda}/d\Lambda$ of the Green's function $G^{\Lambda}$.
 This is, for example, trivially fulfilled if one neglects the self-energy insertions in $G^{\Lambda}$, which then recovers the ``undressed'' RPA resummations.
 In order to obtain the ``dressed'' RPA, one has to include an additional term of the truncated $V^{(6)}$ contribution, 
 known as Katanin term~\cite{katanin04prb115109}. (See also Appendix~\ref{sec:ward} on Ward identities.)  
 
 \subsection{Implementation and Physical Interpretation}
 In the following section, we give a \textit{detailed description of the functional RG implementation for multi-orbital and multi-sublattice systems}.
 Starting from a proto-typical multi-orbital system on the square
 lattice, we first discuss the transition from orbital to band representation
 and then provide a pseudocode algorithm for the numerical solution of the RG flow equations. In addition, we present a systematic 
 channel decoupling of the two-particle interaction which allows to determine the favored type of ordering in the considered system.
 An extensive discussion of the mean-field analysis and a complete symmetry classification of particle-particle and particle-hole like order parameters on different lattice geometries
 is further provided in Appendix~\ref{sec:mftreat} and Appendix~\ref{sec:ppphcond}.
\subsubsection{Prototypical Multi-Orbital and Multi-Sublattice Models}
The non-interacting part of a multi-orbital or multi-sublattice model description is typically given by 
\begin{equation}\label{eq:freeh0}
H_{0} = \sum_{\bs{k},s}\sum_{a,b=1}c_{\bs{k}as}^{\dagger}K^{\phantom{\dagger}}_{ab}(\bs{k})c_{\bs{k}bs}^{\phantom{\dagger}},
\end{equation}
where $c_{\bs{k}as}^{\dagger}$, $c_{\bs{k}as}^{\phantom{\dagger}}$ denote the creation and annihilation operator of an electron 
with momentum $\bs{k}$ in orbital or sublattice $a$ and with spin projection $s$. In order to exemplify the numerical 
implementation of the multi-orbital (multi-sublattice) functional RG, we consider the case of iron-based superconductors (FeSCs), 
where the indices $a,b$ correspond to the five iron $d$-orbitals on the square lattice and $K^{\phantom{\dagger}}_{ab}$ is
a $(5\times 5)$-matrix determined by e.g. density functional theory (DFT). Here, the interaction part $H_{int}$ typically contains an 
intra- and inter-orbital repulsion $U_1$ and $U_2$, as well as
a Hund's rule coupling $J_{H}$ and a pair-hopping term $J_{pair}$:  
\begin{align}
H_{int}&=\sum_i \left[ U_1 \sum_{a} n_{ia\uparrow}n_{ia\downarrow} + U_2\sum_{a<b,s,s'} n_{ias}n_{ibs'} \right.\nonumber \\\nonumber
&\left.+J_{H}\sum_{a<b}\sum_{s,s'} c_{ias}^{\dagger}c_{ibs'}^{\dagger}c_{ias'}^{\phantom{\dagger}}c_{ibs}^{\phantom{\dagger}}  
 +J_{pair}\sum_{a<b}\left(c_{ia\uparrow}^{\dagger}c_{ia\downarrow}^{\dagger}c_{ib\downarrow}^{\phantom{\dagger}}c_{ib\uparrow}^{\phantom{\dagger}} + h.c.\right)\right]\hspace{-4pt}\\\label{eq:intgeneric}
&+V_1\sum_{\langle i,j\rangle}\sum_{ab,ss'}n_{ias}n_{jbs'}.\hspace{-4pt}
\end{align}
The nearest-neighbor repulsion term $V_1$ in (\ref{eq:intgeneric}) is
usually neglected in the discussion of FeSCs, but is included in our
toy model in order to demonstrate the implementation
of non-local interactions, which will become relevant in Chapter~\ref{chap:hexa}.
\begin{figure}[t]
\centering
   {\includegraphics[scale=0.3]{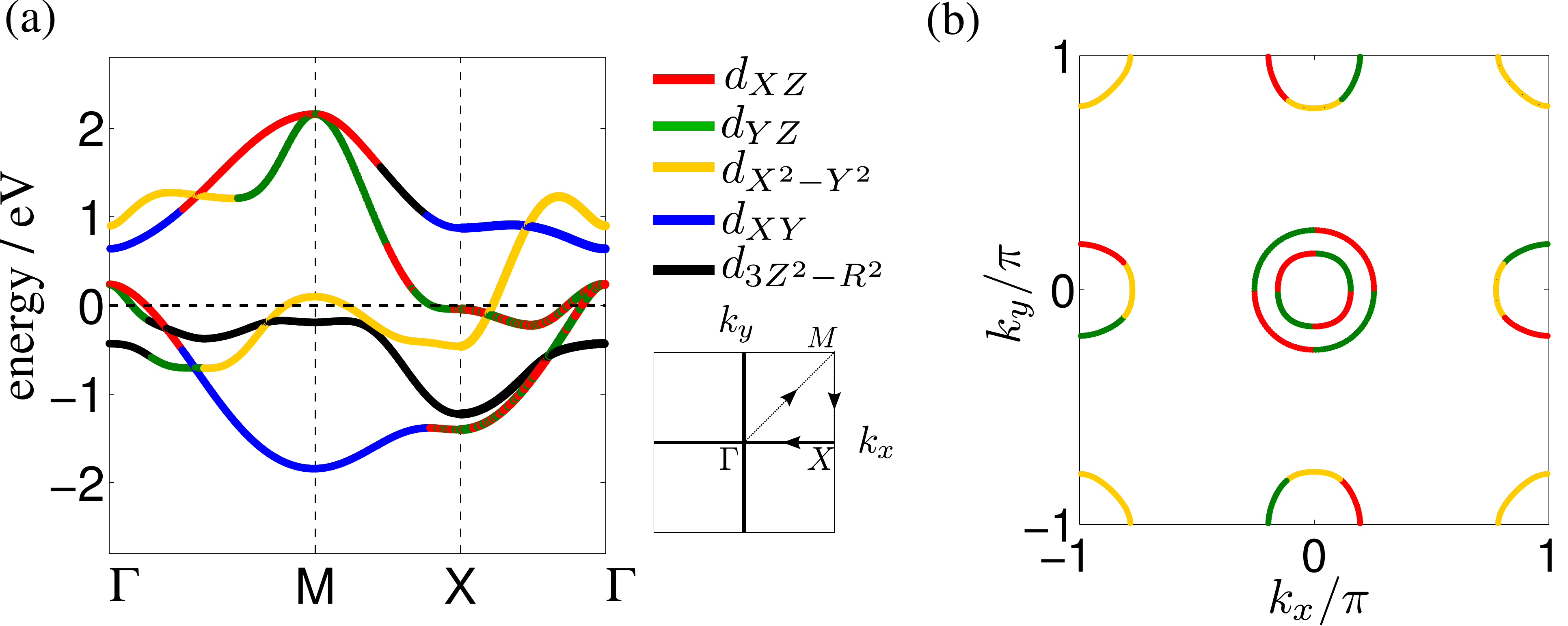}}
\caption{\label{fig:bandsandpatchv2} 
    Band structure (a) and Fermi-surface topology (b) of LaOFeAs in the unfolded (one iron per unit-cell) Brillouin zone. The colors denote the dominant orbital content.
    (Inset: Brillouin zone and orbital color-coding).} 
\end{figure}
\subsubsection{From Orbital to Band Representation}\label{subsec:bandorb}
In order to achieve an efficient implementation of the RG flow equations, it is convenient to choose an appropriate one-particle basis
in which the quadratic part $H_0$ becomes diagonal. This so-called
band basis consists of a superposition of orbital or sublattice states:
\begin{equation}\label{eq:bandbas}
|\bs{k},m,s\rangle = \sum_a u_{a,m}(\bs{k})|\bs{k},a,s\rangle,
\end{equation}
where $a$ again denotes orbital or band degrees of freedom, and $m$ labels the energy band.
The corresponding creation and annihilation operators $\gamma^{\dagger}_{\bs{k}ms}$, $\gamma^{\phantom{\dagger}}_{\bs{k}ms}$ are then defined through
$|\bs{k},m,s\rangle = \gamma^{\dagger}_{\bs{k}ms}|0\rangle$, and the Hamiltonian $H_{tot} = H_0 + H_{int}$, in this basis, reads as
\begin{align}\label{eq:hamtot}
H_{tot} & = \sum_{\bs{k},s}\sum_{m}E^{\phantom{\dagger}}_{m}(\bs{k})\gamma_{\bs{k}ms}^{\dagger}\gamma_{\bs{k}ms}^{\phantom{\dagger}} \\\nonumber
+ &\sum_{\bs{k}_1,\ldots\bs{k}_4\atop s,s'}\sum_{m_1,\ldots,m_4}V_0(\bs{k}_1m_1,\bs{k}_2m_2,\bs{k}_3m_3,\bs{k}_4m_4)
\gamma_{\bs{k}_1m_1s}^{\dagger}\gamma_{\bs{k}_2m_2s'}^{\dagger}\gamma_{\bs{k}_3m_3s}^{\phantom{\dagger}}\gamma_{\bs{k}_4m_4s'}^{\phantom{\dagger}}.
\end{align}
Here, the fourth momentum $\bs{k}_4$ is determined by momentum conservation $\bs{k_4} = \bs{k}_3 + \bs{k}_2 - \bs{k}_3$, and the eigenvalues 
$E^{\phantom{\dagger}}_{m}(\bs{k})$ provide the band structure depicted in Fig.~\ref{fig:bandsandpatchv2} for a typical FeSC model. 
It is important to note that the orbital content of the $m$-th band at momentum $\bs{k}$ is characterized by 
the matrix elements $u_{am}(\bs{k})$, whose dominant part determines the 
leading orbital content (colors in Fig.~\ref{fig:bandsandpatchv2}). In addition, the explicit form of the coupling function now reads as 
\begin{align}\nonumber
V_0(\bs{k}_1m_1,\bs{k}_2m_2,\bs{k}_3m_3,\bs{k}_4m_4) 
 = &U_1\sum_a
 u^*_{am_1}(\bs{k}_1)u^*_{am_2}(\bs{k}_2)u_{am_3}(\bs{k}_3)u_{am_4}(\bs{k}_4)
 \\ \nonumber
 + U_2&\sum_{a,b}
 u^*_{am_1}(\bs{k}_1)u^*_{bm_2}(\bs{k}_2)u_{am_3}(\bs{k}_3)u_{bm_4}(\bs{k}_4)
 \\\nonumber 
 + J_{\text{H}}\sum_{a,b}& u^*_{am_1}(\bs{k}_1)u^*_{bm_2}(\bs{k}_2)u_{bm_3}(\bs{k}_3)u_{am_4}(\bs{k}_4) \\\nonumber
 + J_{\text{pair}}\sum_{a,b} u^*_{am_1}&(\bs{k}_1)u^*_{am_2}(\bs{k}_2)u_{bm_3}(\bs{k}_3)u_{bm_4}(\bs{k}_4)\\\label{eq:bandcoupling}
 + 2V_1(\cos(\bs{k}^x_3 - \bs{k}^x_1) + \cos(\bs{k}^y_3 -
 \bs{k}^y_1))&\sum_{a,b}
 u^*_{am_1}(\bs{k}_1)u^*_{bm_2}(\bs{k}_2)u_{am_3}(\bs{k}_3)u_{bm_4}(\bs{k}_4).
\end{align}
Here, it becomes apparent that the matrix elements $u_{am}(\bs{k})$ as well as the nearest-neighbor Coulomb 
repulsion $\propto (\cos(\bs{k}^x_3 - \bs{k}^x_1) + \cos(\bs{k}^y_3 - \bs{k}^y_1))$ 
already cause a pronounced momentum dependence at the bare level (see
e.g. Fig.~\ref{fig:flowpic1}(a1) in Sect.~\ref{sec:flowtostrong}). 
The non-local interaction terms for generalized lattice geometries considered in Chap.~\ref{chap:hexa} are summarized in Appendix~\ref{sec:hexbandint}. 
As the columns of the unitary matrix $U(\bs{k}) = (u_{am}(\bs{k}))$ correspond to eigenvectors of the 
matrix $K^{\phantom{\dagger}}_{ab}(\bs{k})$ in (\ref{eq:freeh0}), the elements
$u_{am}(\bs{k})$ have a local $U(1)$-phase freedom. It is convenient here to require a smooth behavior of the matrix elements 
$u_{aw}(\bs{k})$, which otherwise would generate discontinuous sign
changes (branch cuts) in the resulting gap functions. In addition, it should be noted
that an arbitrary choice of Bloch states also leads to a nontrivial
symmetry transformation behavior of the coupling function $V_0$.
\subsubsection{Functional RG Implementation} \label{sec:frgimpl}
\begin{figure}[t]
\centering
   {\includegraphics[scale=0.3]{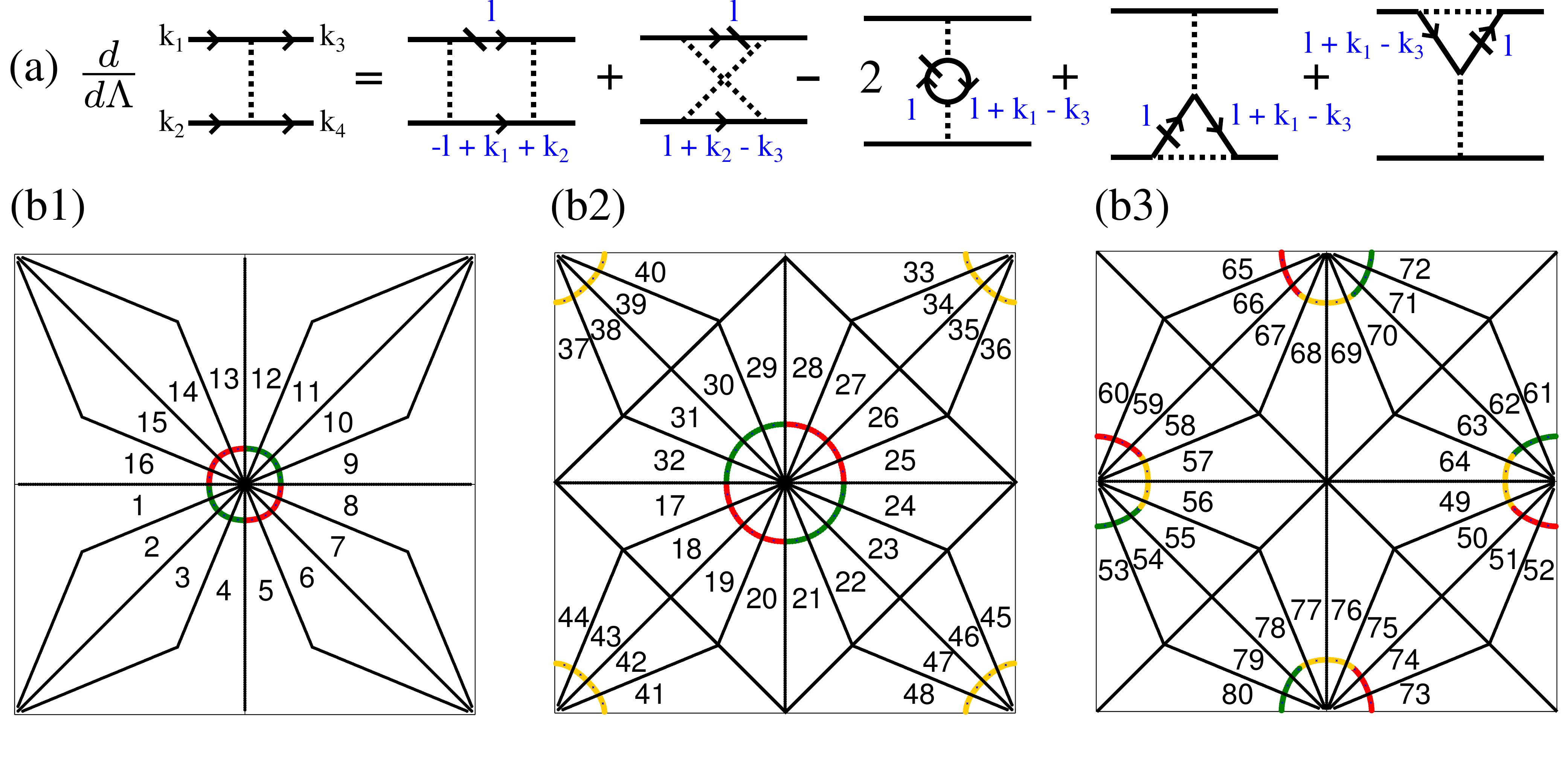}}
\caption{\label{fig:patchingscheme} 
    (a) Flow equation of the 4-point function (dashed line) according to Eq.~(\ref{eq:flowendres2}). Full internal lines denote free propagators, 
        slashed lines indicate single-scale propagators as defined in (\ref{eq:singlescaleT}) for the cutoff- and in (\ref{eq:singlescalecut})
        for the temperature-scheme. (b1-b3) Patching scheme depicted
        by the black partitioning of the Brillouin zone for each band that intersects the Fermi-level. The colors again denote the leading orbital content of the Fermi-pockets 
        with an equivalent color coding as in Fig.~\ref{fig:bandsandpatchv2}b.} 
\end{figure}

In the following section, we \textit{describe the numerical implementation of the functional RG 
as well as its application to the Hamiltonian in
(\ref{eq:hamtot})}. As discussed before, the starting point consists
of the exact hierarchy of flow equations in (\ref{eq:flowselfen}) for
the one-particle irreducible vertex functions. Restricting ourselves to the 
4-point function $V^{\Lambda}$, we then obtain the flow equation depicted in Fig.~\ref{fig:patchingscheme}a, where we 
applied the approximation of discarding the 6-point function as well as the self-energy feedback to $V^{\Lambda}$. 
Both types of approximations can be justified for sufficiently small bare interactions~\cite{RevModPhys.84.299}, 
since these two terms only generate contributions of third order in $V^{\Lambda}$. For curved and smooth Fermi-surfaces, as it is the case
in almost all FeSCs, these approximations are valid up to a scale where the 4-point vertex is very large and the flow has to be stopped.
In order to solve these differential equations numerically, we first divide up the Brillouin zone into patches for each band
that intersects the Fermi level (see
Fig.~\ref{fig:patchingscheme}b). (Note that it might occasionally also
be appropriate to include bands closely below or above the Fermi level.)
Since the leading part of the 4-point function is located at the Fermi surface and at zero frequency
\cite{shankar94rmp129,zanchi-00prb13609}, we neglect all bands that are clearly away from the Fermi-level and compute the 4-point function only at frequency zero and at 
the Fermi-surface points $\{\bs{k}^s_F\}$. For momenta $\bs{k}$ away from the Fermi-surface, we then 
approximate  
\begin{align}\nonumber
V^{\Lambda}(\bs{k}_1m_1,\bs{k}_2m_2,&\bs{k}_3m_3,\bs{k}_4m_4)
\approx \\
&
V^{\Lambda}(\pi_F(\bs{k}_1)m_1,\pi_F(\bs{k}_2)m_2,\pi_F(\bs{k}_3)m_3,\pi_F(\bs{k}_4)m_4),
\label{projectfermi}
\end{align}
where $\pi_F(\bs{k}_i)$ denotes the projection of $\bs{k}_i$ within the same patch onto the 
corresponding Fermi-surface point $\bs{k}^s_F$.
In the following, we will also employ the condensed notation of $k = (\bs{k}^s_F,m^s)$ since each $\bs{k}^s_F$ is associated to a fixed band index $m^s$ in the above patching scheme. As the 4-point function $V^{\Lambda}$ 
equals the bare interaction part (\ref{eq:bandcoupling}) at large energy scales $\Lambda$ of the order of the bandwidth $W$, i.e.
\begin{equation}\nonumber
V^{\Lambda\approx W}(k_1,k_2,k_3,k_4) = V_0(k_1,k_2,k_3,k_4), 
\end{equation}
the flow equation shown in Fig.~\ref{fig:bandsandpatchv2}a reduces to a well-defined initial value problem.
Using standard procedures for the integration of differential equations, we then track the flow of $V^{\Lambda}$ 
down to low energy scales~$\Lambda$. The following pseudocode summarizes the numerical solution of the RG flow equations,
and Fig.~\ref{fig:vertices2} in Sect.~\ref{sec:flowtostrong} illustrates the evolution of $V^{\Lambda}$.   
\begin{enumerate}\itemsep8pt
  \item[ ] \hspace{-1.3cm} \textit{\bf Functional RG Pseudocode:}
  \item[$(S0)$] \textit{Calculate bare coupling function $V_0(k_1, k_2, k_3, k_4)$}
  \item[$(S1)$] \textit{Set $V^{\Lambda_{i=0}}(k_1, k_2, k_3, k_4) = V_0(k_1, k_2, k_3, k_4)$ with $\Lambda_{i=0} = W$}
  \item[$(S2)$] \textit{Calculate right-hand side (RHS) of Fig.~\ref{fig:patchingscheme}a.
  Here, $\int_l$ denotes the summation over frequency $l_0$, momentum $\bs{l}$ and band index $m_l$}:
 \begin{align}\nonumber 
  \hspace{-3.5cm}RHS=&\int_{l} V^{\Lambda_i}(k_1, k_2, l, -l+k_1+k_2)S^{\Lambda_i}(l)G^{\Lambda_i}(-l+k_1+k_2)V^{\Lambda_i}(l,-l+k_1+k_2,k_3,k_4)\\\nonumber
  +&\int_{l} V^{\Lambda_i}(k_1, l+k_2-k_3, l, k_4)S^{\Lambda_i}(l)G^{\Lambda_i}(l+k_2-k_3)V^{\Lambda_i}(l,k_2,k_3,l+k_2-k_3)\\\nonumber
  -2&\int_{l} V^{\Lambda_i}(k_1, l, k_3, l+k_1-k_3)S^{\Lambda_i}(l)G^{\Lambda_i}(l+k_1-k_3)V^{\Lambda_i}(l+k_1-k_3,k_2,l,k_4)\\\nonumber
  +&\int_{l} V^{\Lambda_i}(k_1, l, k_3, l+k_1-k_3)S^{\Lambda_i}(l)G^{\Lambda_i}(l+k_1-k_3)V^{\Lambda_i}(k_2,l+k_1-k_3,l,k_4)\\\nonumber
  +&\int_{l} V^{\Lambda_i}(k_1, l, l+k_1-k_3,k_3)S^{\Lambda_i}(l)G^{\Lambda_i}(l+k_1-k_3)V^{\Lambda_i}(l+k_1-k_3,k_2,l,k_4)\\\nonumber
 \end{align}
 \item[$(S3)$] $dV^{\Lambda_i}(k_1, k_2, k_3, k_4) = RHS\cdot d\Lambda$
 \item[$(S4)$] $\Lambda_{i+1} = \Lambda_i - d\Lambda$
 \item[$(S5)$] $V^{\Lambda_{i+1}}(k_1, k_2, k_3, k_4) = V^{\Lambda_i}(k_1, k_2, k_3, k_4) + dV^{\Lambda_i}(k_1, k_2, k_3, k_4)$
 \item[$(S6)$] $i = i+1$
 \item[$(S7)$] \textit{GOTO (S2) UNTIL:\ \ $\max(|V^{\Lambda_{i}}(k_1, k_2, k_3, k_4)|>> W$}
\end{enumerate}
In recent works of Husemann~\ea~\cite{husemann-0902arxiv1651} and Wang~\ea~\cite{PhysRevB.85.035414}, an alternative implementation 
of the RG flow equations has been applied. According to the different transfer momenta in the one-loop diagrams of Fig.~\ref{fig:patchingscheme}a,
the 4-point function there is decomposed into superconducting~$(SC)$, magnetic~$(M)$ and forward-scattering~$(K)$ terms
\begin{align}\nonumber
V^{\Lambda} = V_0 -& \Phi^{\Lambda}_{SC}(k_1,k_3,{\color{blue}{k_1+k_2}}) + \Phi^{\Lambda}_{M}(k_1,k_2,{\color{blue}{k_3-k_1}})\\\label{eq:newpara}
+\frac{1}{2}&\Phi^{\Lambda}_{M}(k_1,k_2,{\color{blue}{k_2-k_3}}) - \frac{1}{2}\Phi^{\Lambda}_{K}(k_1,k_2,{\color{blue}{k_2-k_3}}),
\end{align}
where each of those contributions absorbs one of the transfer-momentum dependences (blue colored). This type of decomposition then 
provides a more direct parametrization of the singular momentum dependence of the right-hand side of Fig.~\ref{fig:patchingscheme}a 
and will turn out to be more efficient in numerical implementations. In the next step, the different contributions in 
(\ref{eq:newpara}) are expanded in a finite basis set of form factors $I=\{f_n\}$ for a fixed momentum transfer~$l$:  
\begin{equation}\label{eq:singvalue}
\Phi^{\Lambda}_{SC,M,K}(q,q',l) = \sum_{n,m\in I} D_{SC,M,K}^{\Lambda,mn}(l)f_m(q)f_n(q') + R_{SC,M,K}^{\Lambda}(q,q',l).
\end{equation}
Here, the expansion coefficients $D^{\Lambda,mn}_{SC,M,K}(l)$ can be considered as boson propagators, and $R_{SC,M,K}^{\Lambda}(q,q',l)$ 
accounts for the remainder part. Inserting (\ref{eq:singvalue}) and (\ref{eq:newpara}) into the flow equation (\ref{eq:flowendres2}),
one can derive a closed system of differential equations for the boson propagators $D_{SC,M,K}^{\Lambda,mn}(l)$, if one
neglects the remaining terms $R_{SC,M,K}^{\Lambda}(q,q',l)$. The study of Husemann~\ea~\cite{husemann-0902arxiv1651} demonstrated that
this alternative scheme reproduces the results of the conventional $N$-patch calculation for the square lattice Hubbard model even for a set of only
two form factors. Through the separation of leading and subleading processes, the computational effort then reduces from the solution 
of one flow equation for $V(k_1,k_2,k_3)$ to the solution of six flow equations for the boson propagators $D_{SC,M,K}^{\Lambda,mn}(l)$ 
depending only on one momentum and frequency argument. Starting with the $N$-patch scheme to identify the relevant form factors and 
to obtain a first qualitative picture that serves as a subsequent input, this singular-mode functional
RG implementation, under certain circumstances, might allow for a refined analysis of the
competing Fermi-surface instabilites. We will still constrain our
attention to the $N$-patch in this review. 
\subsubsection{Channel Decoupling and Order Parameters}\label{subsec:decoup}
In order to infer the favored type of order, we \textit{decompose the 4-point function $V^{\Lambda}$ into the following channels}:
\begin{align}\label{eq:interactionpart}
H_{int} = & \sum_{k_1,\ldots,k_4}\sum_{ss'}V^{\Lambda}(k_1,k_2,k_3,k_4)
\gamma_{k_1s}^{\dagger}\gamma_{k_2s'}^{\dagger}\gamma_{k_3s}^{\phantom{\dagger}}\gamma_{k_4s'}^{\phantom{\dagger}}\\\nonumber
 = & \sum_{k_1,k_2} W^{\Lambda,SDW}(k_1,k_2) \vec{S}^{\dagger}_{k_1,Q}\vec{S}^{\phantom{\dagger}}_{k_2,Q} 
   + \sum_{k_1,k_2} W^{\Lambda,CDW}(k_1,k_2) n^{\dagger}_{k_1,Q}n^{\phantom{\dagger}}_{k_2,Q}\\\nonumber
 + &\sum_{k_1,k_2} W^{\Lambda,FI}(k_1,k_2) \vec{S}^{\dagger}_{k_1,0}\vec{S}^{\phantom{\dagger}}_{k_2,0}
   + \sum_{k_1,k_2} W^{\Lambda,PI}(k_1,k_2) n^{\dagger}_{k_1,0}n^{\phantom{\dagger}}_{k_2,0}\\\label{eq:channeldecop}
 + &\sum_{k_1,k_2} W^{\Lambda,SCs}(k_1,k_2) p^{\dagger}_{k_1,s}p^{\phantom{\dagger}}_{k_2,s} 
   + \sum_{k_1,k_2} W^{\Lambda,SCt}(k_1,k_2) p^{\dagger}_{k_1,t}p^{\phantom{\dagger}}_{k_2,p},\\\nonumber
 + &\ldots
\end{align}
with composite operators defined by
\begin{align}\nonumber
\vec{S}^{\phantom{\dagger}}_{k,Q} = \frac{1}{2}\sum_{ss'}\gamma_{k+Qs}^{\dagger}\vec{\sigma}^{\phantom{\dagger}}_{ss'}\gamma_{ks'}^{\phantom{\dagger}},\quad
&\vec{S}^{\phantom{\dagger}}_{k,0}  = \frac{1}{2}\sum_{ss'}\gamma_{ks}^{\dagger}\vec{\sigma}^{\phantom{\dagger}}_{ss'}\gamma_{ks'}^{\phantom{\dagger}}\\\label{eq:bosonic}
n^{\phantom{\dagger}}_{k,Q} = \sum_{s}\gamma_{k+Qs}^{\dagger}\gamma_{ks}^{\phantom{\dagger}},\quad
&n^{\phantom{\dagger}}_{k,0} = \sum_{s}\gamma_{ks}^{\dagger}\gamma_{ks}^{\phantom{\dagger}}\\\nonumber
p^{\phantom{\dagger}}_{k,s} = \frac{1}{\sqrt{2}}\left(\gamma_{k\uparrow}^{\phantom{\dagger}}\gamma_{-k\downarrow}^{\phantom{\dagger}}
                                                     -\gamma_{k\downarrow}^{\phantom{\dagger}}\gamma_{-k\uparrow}^{\phantom{\dagger}}\right),\quad
&p^{\phantom{\dagger}}_{k,t} = \frac{1}{\sqrt{2}}\left(\gamma_{k\uparrow}^{\phantom{\dagger}}\gamma_{-k\downarrow}^{\phantom{\dagger}}
                                                     +\gamma_{k\downarrow}^{\phantom{\dagger}}\gamma_{-k\uparrow}^{\phantom{\dagger}}\right).  
\end{align}
Comparing the coefficients of (\ref{eq:interactionpart}) and (\ref{eq:channeldecop}), one then obtains the \textit{following relations between the 4-point function 
$V^{\Lambda}$ and the channel couplings} $W^{\Lambda,ch}$ for \textit{spin-density wave} $(ch = SDW)$, 
\textit{charge-density wave} $(ch=CDW)$, \textit{ferromagnetic} $(ch=FM)$, \textit{Pomeranchuk} $(ch=PI)$, \textit{spin-singlet} $(SCs)$ and \textit{spin-triplet pairing} $(SCt)$ 
orders:
\begin{align}\nonumber
W^{\Lambda,SDW}(k_1,k_2) &= -2V^{\Lambda}(k_1,k_2+Q,k_2,k_1+Q)\\\nonumber
W^{\Lambda,CDW}(k_1,k_2) &= -\frac{1}{2}V^{\Lambda}(k_1,k_2+Q,k_2,k_1+Q)+V^{\Lambda}(k_1+Q,k_2,k_1,k_2+Q)\\\label{eq:channelcorresp} 
W^{\Lambda,FM}(k_1,k_2) &= -2V^{\Lambda}(k_1,k_2,k_2,k_1)\\\nonumber
W^{\Lambda,PI}(k_1,k_2) &= -2V^{\Lambda}(k_1,k_2,k_2,k_1)+V^{\Lambda}(k_1,k_2,k_1,k_2)\\\nonumber
W^{\Lambda,SCs}(k_1,k_2) &= V^{\Lambda}(k_1,-k_1,k_2,-k_2) + V^{\Lambda}(-k_1,k_1,k_2,-k_2)\\\nonumber
W^{\Lambda,SCt}(k_1,k_2) &= V^{\Lambda}(k_1,-k_1,k_2,-k_2) - V^{\Lambda}(-k_1,k_1,k_2,-k_2).
\end{align}
In case of spin-rotational symmetry, the other two ($S_z = \pm 1$) channels of (SCt) are equivalent to the one with $S_z = 0$, presented above.
It is further important to note that the channel couplings $W^{\Lambda,ch}(k_1,k_2)$ can be regarded as hermitian operators due to the self-adjointness 
of $H_{int}$. For this reason, \textit{we can expand the channel-couplings $W^{\Lambda,ch}(k_1,k_2)$ 
into eigenmodes} $f^{ch}_i(k_1)$:
\begin{equation}\label{eq:eigdecomp}
W^{\Lambda,ch}(k_1,k_2) = \sum_i w^{ch}_i(\Lambda)f^{ch}_i(k_1)^*f^{ch}_i(k_2),
\end{equation}
where $f^{ch}_i(k_1)$ transforms as an irreducible representation of the symmetry group of $W^{\Lambda,ch}(k_1,k_2)$. 
For the case of zero momentum transfer ($ch=SCs, SCt, FM, PI$) this symmetry group includes
the full point group of the underlying lattice, whereas for nonzero momentum transfer ($ch=SDW$, $CDW$) it 
only includes an associated subgroup leaving the ordering vector $Q$ invariant up to a reciprocal lattice vector. 
In the case of FeSCs, the two-dimensional iron plane has a $C_{4v}$ lattice symmetry and the eigenmodes $f^{ch}_i$ with ($ch=SCs,SCt,PI,FM$) transform
as an irreducible representation of $C_{4v}$, whereas $f^{ch}_i$ for ($ch=SDW,CDW$) with ordering momentum $Q=(0,\pi)$
transform as an irreducible representation of $C_{2v}$. A systematic way of constructing all posssible eigenmodes $f^{ch}_i$ on different lattice geometries
is presented in Appendix~\ref{sec:ppphcond}.\par 
Following the \textit{flow of the eigenvalues} $w^{ch}_i(\Lambda)$ as a function of 
energy scale $\Lambda$, the \textit{most diverging one signals an ordering tendency characterized by the associated order parameter}
\begin{equation}\label{eq:orderparameter}
O_i^{ch} = \sum_{k} f^{ch}_i(k)\langle\hat{O}^{ch}_k\rangle.
\end{equation}
Here, $\hat{O}^{ch}_k$ denotes the respective bosonic operators defined in (\ref{eq:bosonic}), and
the corresponding eigenmode $f^{ch}_i(k)$ gives detailed information on the real-space ordering pattern (see Appendix~\ref{sec:realsp}) as well as on the one-particle spectrum 
within the symmetry broken phase.  \textit{At sufficiently low energy scales of the flow, the four-point function $\gamma^{(4)}$ or, respectively, the coupling function
$V^{\Lambda}$ starts to diverge and thus signals the possible onset of spontaneous symmetry breaking}. According to the discussion in Sect.~\ref{sec:trunc}, 
the flow has to be stopped before that critical energy scale $\Lambda_c$ as the applied truncation of 
the flow-equation hierarchy is no longer justified. Although, it is in principle possible to continue the flow into the symmetry broken phase 
and to account for order-parameter fluctuations~\cite{salmhofer-04ptp943, RevModPhys.84.299}, these techniques are currently
too demanding for an investigation of complex multi-orbital systems with several competing ordering channels. 
Nevertheless, it is possible to resort to a \textit{mean-field treatment of the leading correlations}~\cite{reiss-07prb075110}, i.e.
\begin{align}\label{eq:leadingeig}
H^{\Lambda}_{eff} = H_0 &+ \sum_{k_1,k_2} W^{\Lambda,ch}(k_1,k_2)\hat{O}^{ch\dagger}_{k_1}\hat{O}^{ch\phantom{\dagger}}_{k_2}\\\nonumber
\overset{\textit{MF}}{\approx}H_0 &+ \sum_{k_1,k_2} W^{\Lambda,ch}(k_1,k_2)\hat{O}^{ch\dagger}_{k_1}\langle\hat{O}^{ch\phantom{\dagger}}_{k_2}\rangle
+\sum_{k_1,k_2} W^{\Lambda,ch}(k_1,k_2)\hat{O}^{ch\phantom{\dagger}}_{k_2}\langle\hat{O}^{ch\dagger}_{k_1}\rangle\\\label{eq:meanstep}
&-\sum_{k_1,k_2} W^{\Lambda,ch}(k_1,k_2)\langle\hat{O}^{ch\phantom{\dagger}}_{k_2}\rangle\langle\hat{O}^{ch\dagger}_{k_1}\rangle,
\end{align}
which is explicitly described in Appendix~\ref{sec:mftreat} for the superconducting channel.

\section{Superconductivity and Competing Phases in Iron-Based Compounds}\label{chap:pnictide}
\textit{The central aim of this chapter is to summarize functional renormalization group studies of low-energy quantum states of matter in Fe-based superconductors (SC), 
which arise due to the interplay of interactions and Fermi surface topology.} We concentrate on the development and application of reliable 
microscopic theories to characteristic systems, combining the FRG with electronic-structure, i.e. density functional theory (DFT) determinations. 
The FRG is shown to be of particular relevance in these systems, where competing quantum fluctuations can lead to interesting new exotic 
quantum states which have been predicted to occur in Fe-based SC.\par
It should be noted again, as was already stressed in the introduction, that this chapter on the Fe-based compounds is by no means 
aiming at providing a comprehensive review of the present state of research into these materials. This also applies to the references given, which are supposed to equip the reader with a first helpful subset of the literature available. For a more complete presentation of the iron-based superconductors, 
excellent reviews on theory are available: A common thread is discussed for the pairing interaction in a wide span 
of unconventional superconductors by Scalapino~\cite{scalapino-12rmp1383}. There it is proposed, employing the 
experimental phenomenology of materials such as the cuprates and the Fe pnictides, that spin-fluctuation mediated 
pairing is the common denominator in the pairing interaction. Other theoretical-based reviews can be found in the 
detailed work on the pnictides by Hirschfeld, Korshunov and Mazin~\cite{hirschfeld-11rpp124508} and in the article 
by Chubukov~\cite{chubukov-12arcms57} as well as the references therein. Another article discussing the 
electron-pairing mechanism of Fe-based SC as well as a variety of open issues is the recent review by Wang and Lee~\cite{wang-11science200}. 
The group around Wang and Lee have initiated the application of FRG to iron pnictides, using a related functional RG scheme to the one presented here. 
Therefore, in the following discourse, we make contact with their work at various places. 
In addition, Chubukov and co-workers base their pnictide analysis on an analytical parquet-RG scheme 
(for an early paper consider e.g.~\cite{chubukov-08prb134512}), which will be mentioned below in the context of some subsequent topics.\par
What we aim here at, is to demonstrate that the functional RG scheme for a variety of challenging questions in the pnictide research 
is a powerful method: one can analyze in an unbiased manner how the competing interactions evolve, when the system flows towards 
smaller energies on the scale of $k_BT_c$. This flow involves renormalizations of both particle-particle and particle-hole 
interaction channels and goes beyond the ``single-channel'' instability approach, i.e. the RPA. The FRG scheme is constrained to intermediate coupling, 
but aside from this a rather powerful and generally applicable method: the Hamiltonian contains the (possibly also longer-ranged) screened Coulomb interaction 
obtained, for example, from state-of-the-art DFT determinations (via maximally localized Wannier functions). 
In that sense, the FRG can be combined with a-priori determinations of the starting Hamiltonian at ``high energies'' of the order 
of the (screened) Coulomb matrices and the bandwidth.
\subsection{Overview of pnictides subtopics and Golden Thread}\label{sec:goldenthread}
The \textit{discovery of unconventional superconductivity (SC) in compounds} such as the iron pnictides has opened up a new avenue for 
studying the mechanism of high-$T_c$ SC 
in a wider class of materials other than, but also including, the cuprates. There, after more than two decades of intense research, more 
and more theoretical as well as experimental studies support a scenario where the general nature of the $d$-wave SC as well as other 
salient features are accounted for by an electronic pairing mechanism extracted from a one-orbital 
Hubbard model~\cite{zhang-88prb3759,scalapino-12rmp1383} with the addition that the 
material-dependence is embedded in the multi-orbital (e.g. 3-band) extensions~\cite{hanke-cm1007}.\par
In other \textit{SC compounds such as the pnictides}, the picture seems more 
complicated: \textit{Here, at the outset, multi-band SC appears} with gaps possibly displaying different symmetries 
such as extended (sign-reversing) $s_{\pm}$-competing with $d$-wave and with nodal or also nodeless behavior on the 
disconnected Fermi surface (FS) sheets due to 
gap anisotropy~\cite{mazin-08prl057003,kuroki-09prb224511,chubukov-08prb134512,platt-09njp055058,thomale-09prb180505,wang-09prl047005,zhai-09prb064517,hirschfeld-11rpp124508}. 
Accordingly, even the simplest multi-band Hamiltonian with only on-site interactions contains four possibly relevant terms, 
the intra-orbital and inter-orbital repulsion as well as the Hund’s-rule coupling and pair hopping. Searching for SC pairing, 
these interactions have to be augmented with the orbital dependence of the FS pockets, since the interactions become matrices 
formed by local orbitals which have a dominant orbital “weight“ at the FS pockets~\cite{platt-11prb235121}. \textit{We hence investigate in 
this chapter whether this intricate interplay of 
multi-orbital band structure, FS topology and interactions still allows for insights into a more universal than 
material-dependent understanding of pnictide SC}.\par
Sect.~\ref{sec:intro} gives a \textit{summary of the structural and normal-state (magnetic) as well as superconducting properties} of Fe-based compounds.
The next Sect.~\ref{sec:fluct} reintroduces the \textit{functional RG implementation} previously discussed in Chap.~\ref{chap:fRG} in a specified way relevant for pnictides, with some details shifted to the Appendix.\par 
Sect.~\ref{sec:whyare}, then, presents a first step \textit{towards a universal picture of SC}: There, an explanation is given for the differences (nodal versus nodeless) experimentally 
observed in the order parameters of FeP-based and FeAs-based pnictide SC.
This work not only shows that  $s_{\pm}$ is the leading FRG gap instability driven by an electron pairing mechanism, but emphasizes the decisive role 
played by the FS topology: using a 5-band (orbital) model, which closely reproduces the experimentally observed FS-structure, we find that nodal (nodeless) SC on the 
e-pockets (h-pockets are always nodeless) can naturally appear when an additional orbital-sharing hole pocket (at $(\pi,\pi)$ in the unfolded BZ) is 
absent (present), as is the case for LaOFeP (LaOFeAs), see Fig~\ref{fig1-asvp}. \textit{The mechanism we describe there has far reaching implications for the As vs. P-based 
compounds in all different families of pnictides}. It also reconciles 
recent ARPES data from the viewpoint of electronic interactions~\cite{wray-12prb144515}.\par
\textit{Sect.~\ref{sec:lifeas} presents}, on the one hand, \textit{methodological progress, i.e., to combine density-functional 
and functional RG treatments and, thus, to elevate the FRG to a certain type of ``first-principle'' method.} On the 
other hand, this section contains an application to an interesting material of the (111)-class, i.e. LiFeAs, 
displaying no magnetic, but only SC order~\cite{platt-11prb235121}. LiFeAs is a stoichiometric SC that is chemically and structurally clean 
enough to avoid significant artifacts from disorder. In addition, it has a natural cleaving mirror plane, making it well-suited 
for surface-sensitive spectroscopy. As such, the 111 family of pnictides has relevance beyond the specific class of 
compounds: it allows us to access fundamental insights to pnictide SC in experiment. The absence of AF magnetic order 
in LiFeAs has fueled the suggestion that the mechanism of SC might be related to a ferromagnetic instability 
and could even result in $p$-wave triplet pairing~\cite{brydon-11prb060501}. In contrast, 
\textit{FRG results shown in Sect.~\ref{sec:lifeas}, find the SC order to be of $s_{\pm}$-type as driven by AF fluctuations at lower energies~\cite{platt-11prb235121}.} 
This finding is supported by a growing number of experiments (see e.g. the very recent STM experiments by the 
Vancouver group in Ref.~\cite{chi-12prl087002}). The finding demonstrates, in particular, the \textit{importance of 
renormalization}: As the system flows to low energies  within the RG scheme, AF 
fluctuations eventually exceed the ferromagnetic contribution. \par
\textit{ Sect.~\ref{sec:exoticdwave} elaborates further on the decisive role of the pnictide Fermiology. 
There, we present arguments why an exotic $d$-wave SC-state can exist in the strongly h-doped 122 compound K$_x$Ba$_{1-x}$Fe$_2$As$_2$}. This $d$-wave 
scenario, which has a strong resemblance, but is still different from the cuprate $d$-wave SC, appears when contingent $(\pi,0)/(0,\pi)$ e-pockets are 
absent due to the strong h-doping in the K-doped compound. 
The substantiated hope for a tunability of pnictides into different SC order via experimentally tunable parameters has become an intense field of study: while the \textit{K$_x$Ba$_{1-x}$Fe$_2$As$_2$ 122 compounds develop an $s_{\pm}$-order parameter 
in the moderately doped regime, one might tune them into $d$-wave at stronger doping}. (Other interesting exotic changes of the pairing state have likewise been suggested~\cite{PhysRevB.85.014511,PhysRevB.84.144514}.) On the basis of FRG results, \textit{we argue in Sect.~\ref{sec:exoticdwave} that the multi-pocket FS situation of the Fe-based SC promotes 
in general pairing states with competing $s_{\pm}$-wave and $d$-wave symmetry.}
 
\textit{Finally, we summarize the Fe-based results by elaborating on the general frustration principle on competing superconducting orders in  Sect.~\ref{sec:optprinciple}, where we promote our results on competing 
SC order parameter symmetries in the pnictides to a general optimization principle for multi-band SC.} Using general 
arguments to describe the relation between the SC pair wave function and the repulsive part of the e-e interaction, we reconcile the SC gap in materials with similar prototypical pnictide setups. The main point of this perspective, which again aims 
at finding out what is ``universal'' in these and other multi-band SC, is to show that \textit{the SC state, its gap, and, 
in particular, its anisotropy in momentum space is determined by an optimization principle:} it determines and optimizes 
the interplay between the attractive interaction in the SC channel and Coulomb repulsion. This optimization is unavoidable 
in a multi-band SC situation: for the pnictides, it appears because of a \textit{frustration of the $s_{\pm}$-channel}, when more 
than two FS pockets are involved in setting up the pairing interaction (Fig.~\ref{fig1-asvp}).

\textit{Furthermore, we discuss the possibility of Time-Reversal-Symmetry (TRS)
broken SC state in Sect.~\ref{sec:timereversal}}.
The frustration in the superconducting channel is considered in the extreme limit, where $s$- and $d$-channels are degenerate, leading to
the possibility of a TRS-broken $(s+id)$ state. Following Ref.~\cite{platt-12prb180502}, we demonstrate 
that \textit{the system strikes a compromise between both orders in this degenerate case, i.e. a time-reversal symmetry (TRS) breaking SC state with a complex $s+id$ gap 
order parameter.} 

\subsection{Introduction}\label{sec:intro}
In 2006, while searching for transparent semiconductors, Kamihara~\ea~\cite{kamihara-06jacs10012} found superconductivity in LaFePO with a transition
temperature ($T_c$) of~$4K$. Although iron seemed incompatible with superconductivity due to its strong local magnetic moments, its discovery was
by no means exceptional as the first iron-containing superconductors were already known since the late fifties~\cite{chandrasekha-58jpcs259}. 
However, only two years later in 2008 the same group reported another iron-based compound LaFeAsO which became superconducting 
at $26K$ upon fluorine doping. At that time, this was one of the highest transition temperatures in non-copper based superconductors,
and its discovery triggered an enormous interest within the condensed-matter community.\par
The structure of LaFeAsO is characterized by stacked layers of iron-arsenic with lanthanum-oxygen planes in between. Soon afterwards, it
also turned out that many other materials based on either iron-arsenic or iron-phosphorus layers become superconducting as well. 
During the last three years, this structural variety then gave rise to many hundreds of new iron-based superconductors (FeSCs) with transition temperatures
of up to $56K$ \cite{wang-08epl67006}. For the first time, this development suggested that high-$T_c$ superconductivity might not be limited to the
cuprates, and that the FeSCs are possibly just the beginning.\par 
In addition to its remarkably high $T_c$'s, the FeSCs also reveal interesting similarities and differences 
to the cuprates. For example, both material classes feature a close proximity of antiferromagnetic and superconducting order, which could in turn 
point to a common magnetically induced pairing mechanism. On the other hand, the magnetic order is quite different in both materials. In the cuprates,
the antiferromagnetism is of N\'{e}el-type and evolves from localized charge carriers, whereas the FeSCs show a stripe-like antiferromagnetism resulting 
from an instability of the more itinerant electrons. Related to this is another important distinction between cuprates and FeSCs, which concerns the effective Coulomb repulsion and correlation effects of the low-energy electrons. Here, the correlations in the undoped cuprates appear much stronger and lead to a Mott-insulating behavior, while 
the FeSCs remain semi-metallic with only weak to moderate correlations. The weaker correlation effects in FeSCs are also consistent with the fact that the 
low-energy electrons of the cuprates reside in one single $d$-orbital, whereas the ones of the FeSCs are distributed among all five orbitals.\par
Thus, in the pnictides, the picture seems more complicated than in the $d$-wave SC cuprates. We hence investigate whether this intricate interplay of
multi-orbital band structure, FS topology and interactions still allows for insights into a more universal than material-dependent understanding of SC in these
systems: in order to balance the interplay between the attractive interaction in the SC-channel and the Coulomb repulsion, an optimization perspective is unavoidable in a more general multi-band SC because
of the frustration in e.g. the $s_{\pm}$-channel, when more than two FS-pockets participate in the pairing interaction.\par
\begin{figure}[t]
\centering
   {\includegraphics[scale=0.2]{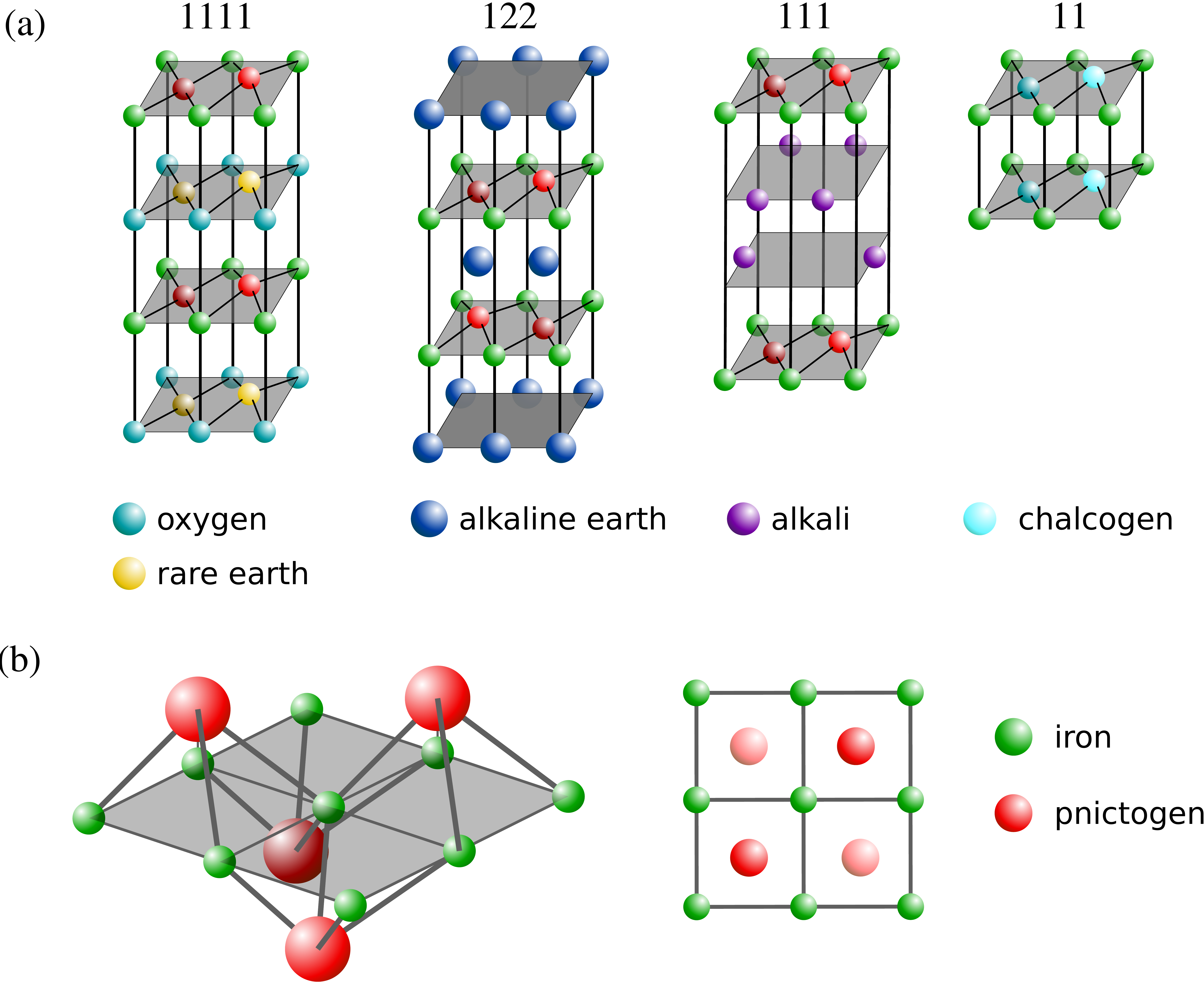}}
\caption{\label{fig:pnictides}(a) Four classes of FeSCs consisting of the iron-pnictide (1111, 122, 111)
                 and the iron-chalcogen (11) layers. The additional filling layers contain rare-earth oxygen (1111), 
                 alkaline-earth (122) and alkali (111) elements.
                 (b) Characteristic trilayer structure consisting of an 
                 iron square-lattice with pnictogen (chalcogen) atoms alternatingly protruding above and below the iron plane. 
                 } 
\end{figure}
\subsubsection{Structural and Normal-State Properties}
The class of FeSCs by now includes hundreds of different materials. It is, therefore, common to distinguish 
between the following major groups, termed after the stoichiometries of their undoped compounds:
1111 (LaFeAsO), 122 (BaFe$_2$As$_2$), 111 (LiFeAs) and 11 (FeSe). As pictured in Fig.~\ref{fig:pnictides},
the structural characteristic of all these compounds consists of an iron square-lattice in between two
checkerboard-lattices of either pnictogen (As,P) or chalcogen (Se,Te,S) atoms. 
According to these two element groups, the FeSCs are often assigned as iron-pnictides or iron-chalcogenides.\par
The additional filling-planes between the iron-pnictogen layers contain rare-earth oxygen (1111), alkaline-earth (122) and alkali (111) elements, 
or are completely absent as in the iron-chalcogenides (11). Nevertheless, superconductivity also occurs in this simplest 11 family 
with $T_c=27K$ \cite{mizuguchi-08apl152505}, suggesting that the relevant physics most likely takes places within the iron-chalcogen and iron-pnictogen layers. 
The same conclusion can also be drawn from the electronic structure. Here, it turns out that the low-energy sector~$(<2eV)$ 
is mainly governed by iron $d$-states with only small contributions from the energetically higher-lying pnictogen/chalcogen $p$-states.
Still, as opposed to the cuprates, the low-energy electrons in the FeSCs distribute among all five $d$-orbitals, which stems
from a closer iron packing in edge-sharing tetrahedrons. In addition, the 
hopping amplitude between different iron-layers turns out to be small but still leads to non-negligible effects in some of the FeSCs.
Taken together, the relevant electrons in the FeSCs are more delocalized than in the cuprates and effectively give rise to smaller electronic correlation effects.
In this context, it is also interesting to note that the calculated band structure \cite{singh-08prl237003,calder-09calder094531,kuroki-08prl087004,miyake-10jpsj044705} 
agrees well (not in all, but in many cases) with the one determined by ARPES \cite{ding-08epl47001,yi-09prb024515,wang-arXiv1201.3655} and quantum oscillation measurements 
\cite{carrington-11rpp124507,putzke-12prl047002}, which can be interpreted as another evidence for weaker correlation effects in FeSCs.\par
\begin{figure}[t]
\centering
   {\includegraphics[scale=0.4]{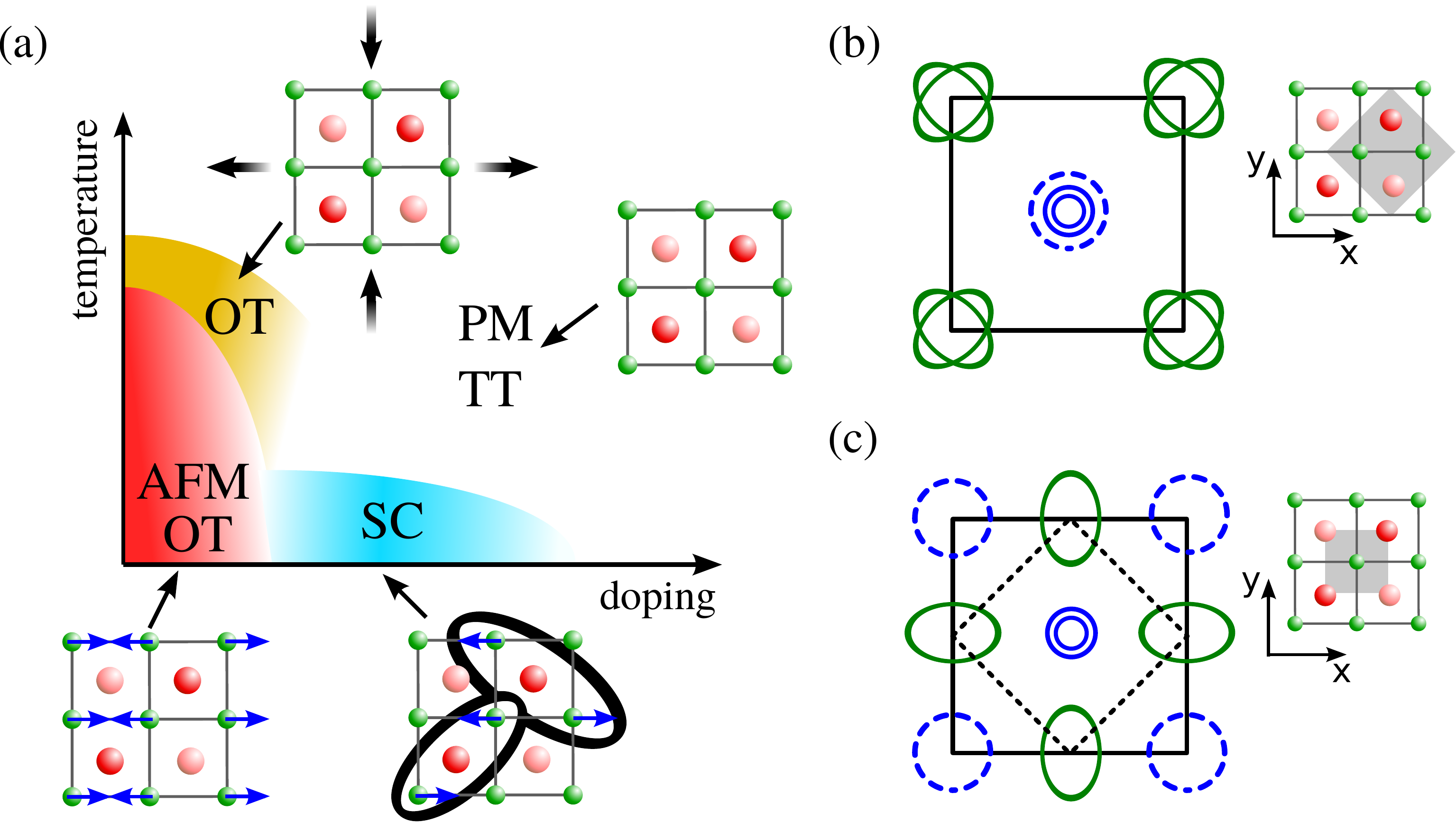}}
\caption{\label{fig:intropic}
(a) Schematic phase-diagram for FeSCs including the paramagnetic phase with a tetragonal crystal structure,
                             the structural transition to an orthorhombic crystal structure (yellow), the striped antiferromagnetic order (red) and
                             the superconducting phase (cyan). (b) Fermi-surface topology of a generic FeSCs represented in the folded 
                             Brillouin-zone (BZ), which corresponds to the two Fe-atom unit-cell shown in the inset (light-gray). (c) Fermi-surface topology
                             shown in the unfolded BZ associated with the one Fe-atom containing unit-cell depicted in the inset (light-gray).} 
\end{figure}

The band structure of the iron $d$-electrons in FeSCs then features ten bands according to the five $d$-orbitals on each of 
the two non-equivalent iron sites within the unit-cell. In Fig.~\ref{fig:intropic}a, we depicted the two-iron unit-cell as well as
the corresponding Fermi-surface with hole-like pockets (blue) around the $\Gamma$-point and electron-like pockets (green) at the $M$-point.
Here, it is important to note that the existence of the dashed pocket in Fig. ~\ref{fig:intropic}b depends in a sensitive way on material details such as the pnictogen height,
whereas the other four pockets appear quite generically in most of the FeSCs. In order to unfold the band structure to the larger Brillouin zone
of a one-iron unit cell, one exploits the glide-mirror group (translation plus $z\rightarrow -z$) under which all iron sites become equivalent \cite{andersen-11anp8,brouet-12prb075123}. 
This unfolding then provides the Fermi-surface pictured in Fig.~\ref{fig:intropic}c. Note that the difference between these two band structure representations 
becomes purely geometrical for vanishing pnictogen or chalcogen potentials. In this case, (b) is obtained from (c) by shifting all bands into the smaller (dashed)
Brillouin zone. If not stated differently, we will only use the one-iron unit cell and its associated unfolded Brillouin zone.\par
In the undoped FeSCs, the electronic filling of the iron $d$-orbitals amounts to six electrons per iron site, according to its Fe$^{2+}$ valence state.
\subsubsection{Structural Transition and Antiferromagnetic State}
Many undoped iron-pnictides develop a striped antiferromagnetic order (AFM) at about $150K$. Here, the 
magnetic moments align parallel to the iron-layer and form alternating ferromagnetic stripes along one of the two crystalline axis (Fig.~\ref{fig:intropic}).
The two possible magnetic phases are then characterized by ordering vectors $Q_1=(0,\pi)$ and $Q_2=(\pi,0)$, consistent 
with the Fermi-surface nesting of hole- and electron-like pockets. The magnetic ordering can be reconciled by
a spin-density wave (SDW) instability associated with this Fermi-surface nesting. Usually, the SDW formation then leads to a band backfolding into 
the reduced Brillouin zone and gives rise to a gap opening at the corresponding band crossings. In the iron-pnictides, however, the multi-orbital 
nature of the electronic states gives rise to symmetry protected Dirac-nodes within the SDW phase \cite{ran-09prb014505,richard-10prl137001} and 
thus implies a semimetallic behavior.\par
As apparent from the phase diagram in Fig.~\ref{fig:intropic}a, there is also a structural phase transition from tetragonal (TT) to orthorhombic (OT), 
which closely follows the magnetic transition line. Within the orthorhombic phase, the degeneracy between the 
$Q_1=(0,\pi)$ and $Q_2=(\pi,0)$ magnetic ordering is lifted, and the magnetic moments order ferromagnetically 
along the contracted and antiferromagnetically along the extended axis. 
Depending on the particular compound, the magnetic phase is either preceded by (1111) or occurs 
simultaneously (122) to the structural transition. In most cases, the simultaneous transition into 
the structurally and magnetically ordered phase turns out to be of first order, whereas the separated transition can be 
either of first or second order. On the other hand, the structural transition is completely absent in the compounds that 
show no magnetic ordering (LaFePO~\cite{mcqueen-08prb024521}, LiFeAs~\cite{wang-08ssc538}, and LiFeP~\cite{deng-09epl37004}). \par
In order to explain the interplay of structural and magnetic degrees of freedom, Fernandes~\ea~\cite{fernandes-12prb024534} proposed a nematic scenario.
Here, the magnetic ordering is preceded by an intermediate phase which first breaks the twofold degeneracy of 
$Q_1=(0,\pi)$ and $Q_2=(\pi,0)$ magnetic fluctuations ($\langle S^2_{Q_1}\rangle\neq\langle S^2_{Q_2}\rangle$)
but still respects spin-rotational symmetry ($\langle S_{Q_1}\rangle=\langle S_{Q_2}\rangle=0$). 
The breaking of rotational symmetry in this nematic phase then induces the orthorhombic
lattice distortion and also selects one of the two magnetic orderings.
Several implications of this nematic scenario have been detected experimentally, and for a detailed discussion of the rich phenomenology 
we refer to the articles \cite{fernandes-12sst0845005,fernandes-12prb024534}.\par
It is also important to note that some of the iron-chalcogenides (11) reveal a different magnetic ordering, although its electronic properties are
quite similar to the one in the iron-pnictides. The magnetic moments here are rotated by $\pi/4$ and have a doubled real-space period compared to the striped antiferromagnet. Interestingly, the ordering vector here is determined by $Q_{3} = (\pi/2,\pi/2)$   
and does not correspond to a Fermi-surface nesting \cite{bao-09prl247001,li-09prb054503}. 
A possible explanation for this striking behavior is given by 
Paul~\ea~\cite{paul-11prl0470044} who calculated the corrections due to magneto-elastic couplings and found a strong enhancement of the magnetic fluctuations at~$Q_3$.
\subsubsection{Superconducting State}
In both iron-pnictides and iron-chalcogenides, the superconductivity can be induced either by chemical 
doping or by external pressure \cite{torikachvili-08prl057006}. 
Although chemical doping is more convenient, the pressure induced method is also appealing as it 
allows to study different phases in the same sample without comparing differently fabricated crystals.   
Similar to chemical doping, the external pressure then suppresses the magnetic ordering 
and superconductivity emerges (Fig.\ref{fig:intropic}a).\par 
The vicinity of magnetic next to superconducting order appears in
most of the FeSCs and possibly suggests a common spin-fluctuation based pairing mechanism. 
\begin{figure}[t]
\centering
   {\includegraphics[scale=0.4]{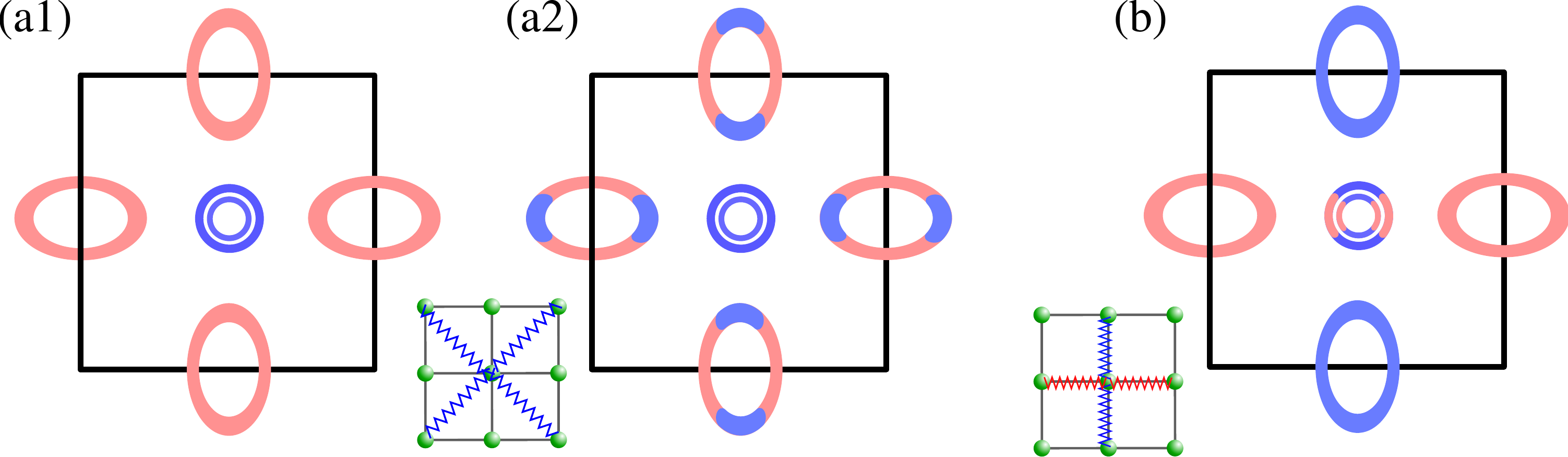}}
\caption{\label{fig:gappic} Nodeless (a1) and nodal (a2) $s_{\pm}$-gap in the unfolded (one Fe unit-cell) Brillouin zone as well as its lowest-order
                            real-space representation (inset). In (b), the $d_{x^2-y^2}$-wave gap is plotted for comparison. Red and blue springs in the insets denote
                            attractive and repulsive pair-interactions.
                 } 
\end{figure}
Here, the rigid magnetic order is destabilized and provides a polarizable background which allows to mediate pairing.\par 
Additionly, the pairing mechanism also affects the symmetry of the superconducting gap 
(pair wave function), as the spin-fluctuations (repulsive interaction) of ordering momentum~$Q$ only 
lead to pairing if the gap changes sign between $Q$-connected portions of the Fermi-surface. 
In the FeSCs, this then implies a sign-change in the superconducting gap between hole- and electron-pockets~\cite{mazin-08prl057003}. 
The corresponding pairing symmetry is commonly denoted as $s_{\pm}$-wave, since the superconducting gap reveals a sign change but breaks no additional 
point-group symmetries (see Fig.~\ref{fig:gappic}a).\par
A review discussing the pairing interaction for unconventional SC has recently been provided by D.~ J.~Scalapino~\cite{scalapino-12rmp1383}.
There, combining studies of the pairing interaction for Hubbard-type models with experimental phenomenology, the spin-fluctuation mediated pairing 
is proposed as the common link in a broad class of SC materials. Important other recent reviews on the theory of magnetically mediated pairing in 
Fe-based SC are due to Hirschfeld, Korshunov and Mazin~\cite{hirschfeld-11rpp124508} and Chubukov~\cite{chubukov-12arcms57}. Other pairing mechanisms based on the polarizability of the pnictogen or chalcogen ions \cite{berciu-09prb214507}, orbital-fluctuations \cite{kontani-10prl157001}, 
or the Hund's coupling \cite{lee-08prb144517,dai-08prl057008} were also proposed in the beginning, but mostly turned out to be inconsistent with 
certain experiments. In particular, conventional phonon-based pairing was ruled out right from the beginning, as the calculated electron-phonon coupling 
turned out to be too small \cite{boeri-08prl026403} to account for the high $T_c$'s in FeSCs. Nevertheless, it was pointed out by Yildrim \cite{yildrim-09prl037003} that phonons 
could provide an indirect contribution to superconductivity via spin-lattice coupling.\par
The pairing symmetry and pairing mechanism of FeSCs have also been the focus of numerous experimental works.
Several important indications could be obtained during the last three years (see \cite{hirschfeld-11rpp124508,RevModPhys.83.1589} for a detailed review). 
For example, nuclear magnetic resonance (NMR) measurements reported a 
vanishing Knight-shift \cite{grafe-08prl047003,matano-08epl57001} in all crystallographic directions, which points to spin-singlet 
pairing. Other experimental findings such as the spin-resonance in neutron-scattering 
\cite{lumsden-10jpcm203203,inosov-10nature178,christianson-08nature930} or 
the quasi-particle interference observed in scanning tunneling (STM) experiments \cite{hanaguri-10science474} suggest a sign-changing
pair wave function that is not inconsistent with an $s_{\pm}$-wave gap. In addition, certain Josephson interference experiments make the 
$d$-wave pairing (see Fig.~\ref{fig:gappic}b) unlikely \cite{zhang-09prl147002,hicks-09jpsj013708}, and rather support an $s_{\pm}$-wave 
scenario \cite{chen-10natphys260}. Usually, the $s_{\pm}$-wave gap was assumed to be fully gapped and a number of experiments reporting 
gap nodes seemed to be at odds with such an $s_{\pm}$-wave pairing state. However, it is now
understood that the existence of nodes in an $s_{\pm}$-wave gap (see Fig.~\ref{fig:gappic}a2) depends on details of the multi-orbital band structure 
and may vary between different FeSC compounds.
From that perspective, it was striking that angular-resolved photoemission spectroscopy (ARPES) consistently reported a full gap
\cite{yi-09prb024515,ding-08epl47001,wang-arXiv1201.3655,xia-09prl037002,kondo-08prl147003} and never revealed significant gap-anisotropies
or gap-nodes which are, one the other hand, clearly seen in the cuprate $d$-wave gap \cite{damascelli-03rmp473}. A possible explanation
for this disagreement between bulk and surface probes was provided by Kemper~\ea~\cite{kemper-10njp073030} who pointed out the existence of
an additional pocket in the surface band structure stabilizing a full gap.\par
\subsection{Electronic Correlations and the Pairing Mechanism of FeSCs}\label{sec:fluct}
In the following section, the pairing mechanism and the associated 
pairing symmetry of FeSCs is analyzed. We demonstrate how antiferromagnetic fluctuations drive
the pairing instability and also promote a number of other competing ordering tendencies.
For this purpose, we apply the functional renormalization group which, on the one hand, allows for an unbiased investigation of the competing fluctuations 
and, on the other hand, also enables the consideration of all relevant material details.
\subsubsection{Microscopic Model Description}\label{sec:micmod}
\begin{figure}[t]
\centering
   {\includegraphics[scale=0.3]{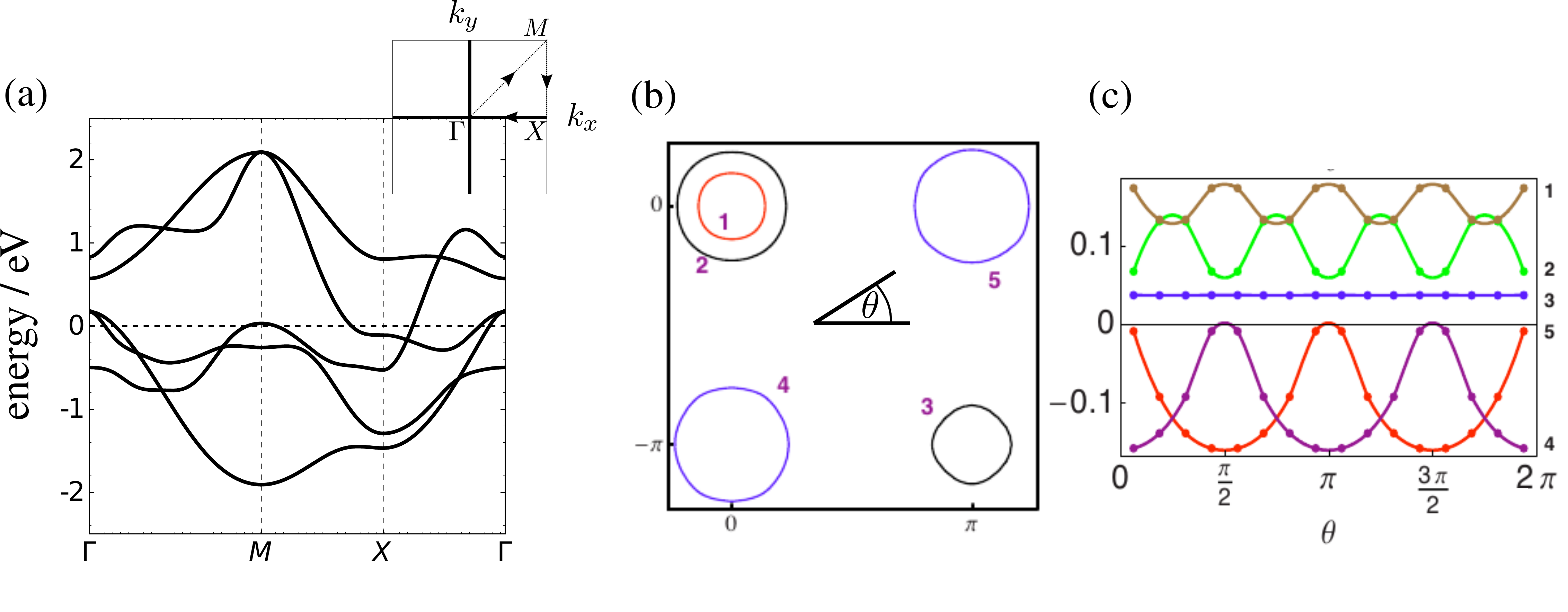}}
\caption{\label{fig:bandsandpatch} 
    Band structure (a) and Fermi-surface topology (b) of LaOFeAs in the unfolded (one iron per unit-cell) Brillouin zone. (c) Leading $s_{\pm}$-wave pairing 
    formfactor plotted along the different Fermi-surface pockets (1-5). Figures (b) and (c) were taken from one of the earliest functional RG studies on pnictides by Zhai, Wang and Lee~\cite{zhai-09prb064517}.} 
\end{figure}

In order to explore the FeSCs on a theoretical basis, it is important to start with an appropriate model description that captures all essential properties
in a type of minimal Hamiltonian. Soon after the first discovery of high-$T_c$ superconductivity in the FeSCs, a number of effective models based on 
two \cite{raghu-08prb220503} or three \cite{lee-08prb144517} of the five iron $d$-orbitals were proposed. Yet, each of these models revealed certain 
shortcomings in describing the low-energy band structure as, for example, discussed by Graser \ea~\cite{graser-09njp025016}. For this reason, we consider a five-orbital model 
suggested by Kuroki \ea~\cite{kuroki-08prl087004} which provides an almost perfect agreement to the low-energy sector of 
the band structure. 
The resulting tight-binding description, then, reads as 
\begin{equation}\label{eq:freeh0}
H_{0} = \sum_{\bs{k},s}\sum_{a,b=1}^{5}c_{\bs{k}as}^{\dagger}K^{\phantom{\dagger}}_{ab}(\bs{k})c_{\bs{k}bs}^{\phantom{\dagger}},
\end{equation}
where $c_{\bs{k}as}^{\dagger}$, $c_{\bs{k}as}^{\phantom{\dagger}}$ denote the creation and annihilation operator of an electron with momentum $\bs{k}$, 
spin projection $s$ and orbital character $a$. $K^{\phantom{\dagger}}_{ab}(\bs{k})$ stands for the (orbital) tight-binding or, more generally, for the Kohn-Sham (DFT, LDA)
matrix elements~\cite{kuroki-08prl087004}. Note that this model was constructed for the LaFeAsO$_{1-x}$F$_x$ compound
and neglects out-of-plane hopping terms along the $z$-axis. As the class of 1111 compounds generally shows a strongly two-dimensional behavior, 
this approximation is valid but may be inappropriate for other FeSCs. The corresponding band structure for the undoped case 
with $n=6.0$ electrons per Fe-site is then plotted in Fig.~\ref{fig:bandsandpatch}.
Here, the presence of a hole-like pocket at $M=(\pi,\pi)$, i.e. pocket 3 in Fig.~\ref{fig:bandsandpatch}b taken from~\cite{zhai-09prb064517}, 
strongly depends on the level of doping and on the detailed material composition, whereas the remaining four 
pockets appear quite generically in most of the FeSCs.\par 
The interaction part $H_{int}$ includes the intra- and inter-orbital repulsion $U_1$ and $U_2$, as well as
the Hund's rule coupling $J_{H}$ and the pair-hopping term $J_{pair}$:  
\begin{align}
H_{int}&=\sum_i \left[ U_1 \sum_{a} n_{ia\uparrow}n_{ia\downarrow} + U_2\sum_{a<b,s,s'} n_{ias}n_{ibs'} \right.\nonumber \\\label{eq:hintproto}
&\left.+J_{H}\sum_{a<b}\sum_{s,s'} c_{ias}^{\dagger}c_{ibs'}^{\dagger}c_{ias'}^{\phantom{\dagger}}c_{ibs}^{\phantom{\dagger}}  
 +J_{pair}\sum_{a<b}\left(c_{ia\uparrow}^{\dagger}c_{ia\downarrow}^{\dagger}c_{ib\downarrow}^{\phantom{\dagger}}c_{ib\uparrow}^{\phantom{\dagger}} + h.c.\right)\right]\hspace{-4pt}.
\end{align}
Here, $n_{ias}=c_{ias}^{\dagger}c_{ias}^{\phantom{\dagger}}$ denotes the electron density of spin~$s$ in orbital~$a$ at site~$i$. It is, further, important to note 
that the interaction parameters are actually orbital dependent (i.e. matrices) and can be computed via first-principle methods such as constrained RPA \cite{miyake-10jpsj044705}. 
The same first-principle approach then also provides parameter values for longer-ranged interaction terms. Although, both of these more precise informations can
be easily included in the functional RG, we first start out with an onsite term only 
and choose the orbital independent values of $U_1 = 4.0$, $U_2 = 2.0$, $J_{H} = J_{pair} = 0.7 eV$. (Note that the frequently applied condition of
$U_1 = U_2 + 2J_{H}$ and $J_{pair} = J_{H}$, following from rotational symmetry \cite{castellani-78prb4945}, only holds for the bare interaction values 
and not for the renormalized ones determined by constrained RPA \cite{miyake-10jpsj044705}. Here, the interaction parameters already include, for 
example, screening from high-energy bands which then may violate the above condition.)\par 
\textit{The numerical implementation of the functional RG} has already been discussed in Chap.~\ref{chap:fRG}. Here, we only shortly introduce the notations 
needed in the following discourse: For a given instability characterized by some particle-particle or particle-hole like ordering 
field $\hat{O}_{\mathbf{k}}$, the function (4PF) $V_{\Lambda}(\mathbf{k}_1, \mathbf{k}_2, \mathbf{k}_3, \mathbf{k}_4)$ 
in the particular ordering channel can be written as $\sum_{\mathbf{k},\mathbf{p}}W^{\Lambda}(\mathbf{k},\mathbf{p})\hat{O}^{\dagger}_{\mathbf{k}} \hat{O}^{\phantom{\dagger}}_{\mathbf{p}}$
(see Sect.~\ref{subsec:decoup}), where $\Lambda$ denotes the cutoff. The starting conditions are given by the bandwidth, with the bare ``high-energy'' interactions serving as the input for the 4PF. 
The diverging channels of the 4PF under the flow to the Fermi surface signal the nature of the instability, and the corresponding $\Lambda_c$ 
serves as an upper bound for $T_c$. The 4PF $V_{\Lambda}(\mathbf{k},-\mathbf{k}, \mathbf{q}, -\mathbf{q})$ in the Cooper channel is, then, 
decomposed into different eigenmode contributions (see Chap.~\ref{chap:fRG})
\begin{equation}\label{harmonicdec}
W^{\Lambda,SC}(\mathbf{k},\mathbf{p})= \sum_i w_i^{\text{SC}}(\Lambda) f_i^{\text{SC}}(\mathbf{k})^* f_i^{\text{SC}}(\mathbf{p}),
\end{equation}
where $i$ is a symmetry decomposition index, and the leading instability of that channel corresponds to an eigenvalue $w_1^{\text{SC}}(\Lambda)$ first diverging under 
the flow of $\Lambda$. $f^{\text{SC}}_i (\mathbf{k})$ is the SC form factor of pairing mode $i$ which tells us about the SC pairing symmetry and hence gap structure 
associated with it. In the functional RG, from the final Cooper channel 4PFs, this quantity is computed along the discretized Fermi surfaces (see e.g. Fig.~\ref{fig:bandsandpatch}c).
\textit{Details of the functional RG-procedure} such as the transition from orbital to band representation, the numerical implementation, the channel decoupling and the determination of the corresponding order parameters are
\textit{relegated to Chapter~\ref{chap:fRG} and according appendices.}
\subsubsection{Flow to Strong Coupling}\label{sec:flowtostrong}
\begin{figure}[t]
\centering
   {\includegraphics[scale=0.27]{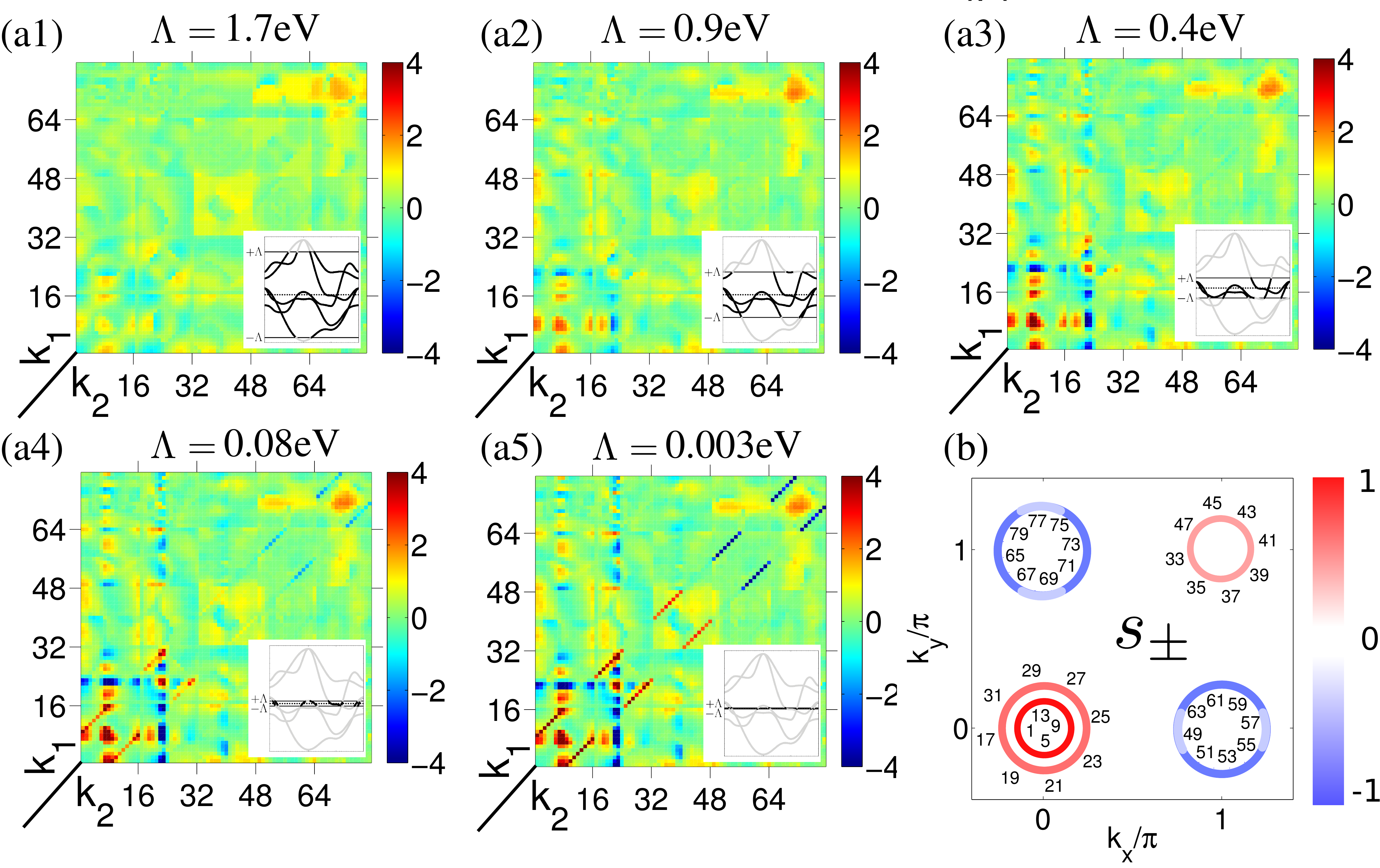}}
\caption{\label{fig:vertices2} 
    (a) $V^{\Lambda}(k_1,k_2,k_3,k_4)$ as a function of $k_1,k_2$ with $k_3$ fixed and $k_4$ determined by momentum conservation is plotted
    at five different values of $\Lambda$ (decreasing from (a1) to (a5)). The numbers 1-80 correspond to positions on the Fermi-surface as indicated in (b).
    Since $k_3$ was fixed at position 71 in all (a1)-(a5), the vertical features correspond to the SDW channel, and its positions determine the respective ordering momenta. The leading
    antiferromagnetic fluctuations then drive an $s_{\pm}$-pairing instability (diagonal features) with a sign-change between electron- and hole-like pockets. (b) Numbering
    of the Fermi-surface positions and the superconducting gap indicated by the color-scheme in units of its absolute maximum value.} 
\end{figure}

Following the flow of the full 4-point function $V^{\Lambda}$ in Fig.~\ref{fig:vertices2}a (for a typical parameter set in the Hamiltonian $H_0+H_1$), we observe 
several features which can now be related to the different channel couplings $W^{\Lambda,ch}$ derived previously (Chap.~\ref{chap:fRG}).
Therefore, we first note that Fig.~\ref{fig:vertices2}a displays $V^{\Lambda}(k_1,k_2,k_3,k_4)$ at different energy scales $\Lambda$ as a 
function of $k_1,k_2$ with $k_3$ fixed to position 71 (see Fig.~\ref{fig:vertices2}b) and $k_4$ determined by momentum conservation.
Now, using the identity (\ref{eq:channeldecop}) for the different channel couplings, one can easily verify that the \textit{vertical features} 
$(k_2 = k_3+Q)$ correspond to the \textit{spin-density wave channel}, the \textit{diagonals} $(k_2 = -k_1)$ to the pairing channel and 
the \textit{horizontals} $(k_1 = k_3+Q)$ to the \textit{charge-density wave channel}, though with a different sign from the spin-density wave case.
The positions of the vertical and horizontal features then determine the respective ordering momentum $Q$, and the 
\textit{ferromagnetic and Pomeranchuk channels} are associated with the $Q = (0,0)$ spin- and charge-density waves.
From (\ref{eq:channelcorresp}) one can further check that the singlet-pairing channel requires the same sign in the upper and lower 
diagonals of Fig.~\ref{fig:vertices2}a, whereas the triplet-pairing channel favors a relative sign change. 
In addition, the internal sign structure of each channel reflects the leading
eigenmode of (\ref{eq:eigdecomp}) and hence provides information on the gap symmetry and on the real-space ordering pattern.
\subsubsection{Antiferromagnetically Driven Pairing Mechanism}
Now that we identified different physical channels in the 4-point function $V^{\Lambda}$, we can monitor which type of long-range order is 
preferred and which type of fluctuations serves as a driving force for others. 
Starting with the initial interaction shown in Fig.~\ref{fig:vertices2}a, we again notice that the bare interaction
shows a pronounced $k$-dependence resulting from the matrix elements $u_{bm}(\bs{k})$ in (\ref{eq:bandcoupling}). This initial sign structure can then
be rediscovered in the arising SDW fluctuations and will later render the SDW-phase gapless. Upon lowering the energy scale $\Lambda$ from (a1) to
(a5) in Fig.~\ref{fig:vertices2}a, these SDW fluctuations become more and more pronounced with the two strongest features
corresponding to an ordering momentum $Q = (\pi,0)$. Of course, as we do not break lattice symmetries, the same features associated
with $Q = (0,\pi)$ are equally strong and appear if the position of the fixed $k_3$ is rotated by $\pi/2$. 
The other subdominant SDW fluctuations will have a decisive influence on the anisotropy of the superconducting gap, as will be discussed below.\par
From the flow pictures (a3) and (a4) it is, further, apparent that the SDW features trigger a \textit{repulsive interpocket pair-scattering} 
$(k,-k)\rightarrow (q,-q)$. During the flow, such pair-scatterings from the electron to the hole-pockets (interpocket) 
grow stronger (red), whereas the pair-scatterings within the electron-pockets (\textit{intrapocket}) weaken and eventually become \textit{attractive} (blue). 
At even lower scales, the attractive intrapocket and the repulsive interpocket pair-scatterings are similarly strong in absolute values. 
Therefore, in order to compensate for the repulsive interpocket pair-scattering, the associated gap function changes sign between 
the hole- and electron-pockets. This reasoning is most clearly seen from
the coupling $W^{\Lambda,SCs}$ in the singlet-pairing channel displaying the following structure 
 \vspace{-0.7cm}
\begin{equation}\nonumber
 W^{\Lambda,SCs}(k,q)\sim\begin{blockarray}{ccc}
    \textit{\Huge{\phantom{P} }} &  & \\
    \begin{block}{(rr)r}
      -1^{hh} & 1^{he} & \textit{\footnotesize{h-pocks}} \\
       1^{eh} & -1^{ee} & \textit{\footnotesize{e-pocks}} \\
    \end{block}
  \textit{\footnotesize{h-pocks}} & \textit{\footnotesize{e-pocks}} & 
  \end{blockarray},
  \vspace{-0.6cm}
\end{equation}
due to the attractive (i.e. minus sign) intrapocket and repulsive interpocket pair-scatterings. The \textit{eigenmode} associated to the leading eigenvalue of $W^{\Lambda,SCs}$ 
then equals $f^{SCs}_i(k) = (1^h\ -1^e)^T$ and therefore \textit{implies a sign-change} between hole- and electron-pockets. The corresponding gap function is also 
indicated by the colored region in Fig.~\ref{fig:vertices2}b, and the pairing symmetry is \textit{commonly termed $s_{\pm}$-wave} as it features a 
sign-change but, on the other hand, transforms trivially under all point-group actions.
Note that the above eigenvalue problem is equivalent to the solution of the linearized BCS gap-equation in Eq. (\ref{eq:lingapeq}) of Appendix~\ref{sec:mftreat}.
From these observations we, then, conclude that the $s_{\pm}$-wave pairing instability is driven by antiferromagnetic fluctuations.  
The phenomenology described above is also consistent with a two-patch analytical RG analysis presented by Chubukov~\ea~\cite{chubukov-08prb134512} where
the interpatch scattering $g_3$ grows positive and pushes the intrapatch scattering $g_4$ through zero until both diverge with different sign. Extending this previous 2-pocket
study to incorporate an angular dependence of the interactions, Maiti and Chubukov~\cite{maiti-10prb214515} obtain again similar overall results to 
the interplay of the SDW and SC channels as from the viewpoint of FRG.\par
However, note that in our functional RG analysis we take into account the full wavevector dependence around the Fermi surface, as we also incorporate the full complexity
of a realistic model description. This, in turn, allows a significantly more differentiated analysis of the competing phases in FeSCs, which will be presented in the following sections.
A related functional RG scheme which keeps the full wavevector dependence has independently been carried out for some characteristic FeSCs by the group of 
Lee and Wang~\cite{wang-09prl047005,zhai-09prb064517,wang-11epl57003}. 
\begin{figure}[t]
\centering
   {\includegraphics[scale=0.28]{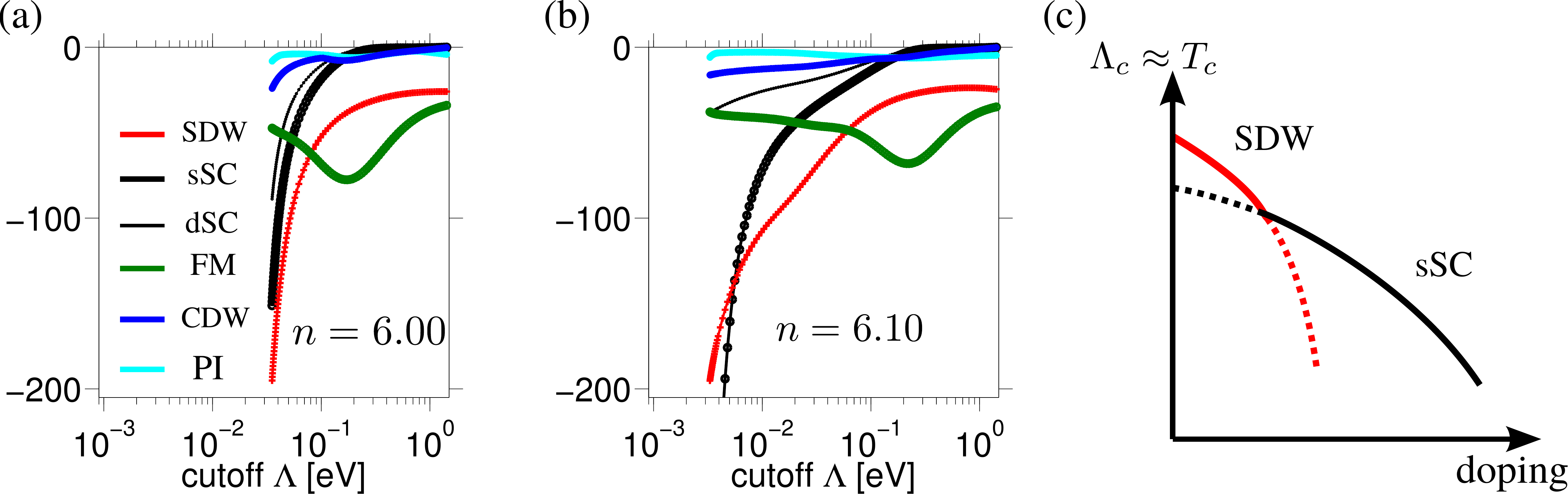}}
\caption{\label{fig:flowpic1} 
    Flow of the most leading eigenvalues $w^{ch}_i(\Lambda)$ at $n=6.00$ (a) and $n=6.10$ (b) electron filling for the spin-density wave (SDW), singlet pairing (sSC,dSC), 
    ferromagnetic (FI), Pomeranchuk (PI) and charge-density wave (CDW) channel. For the singlet pairing case, we depicted the two leading eigenvalues, of which
    the first leading one corresponds to $s_{\pm}$-wave and the second leading one to $d_{x^2-y^2}$-wave order, as will turn out from the associated eigenmodes
    in Fig.~\ref{fig:formfacs}. The charge- and spin-density wave channels correspond to an ordering vector of $(\pi,0)$ and $(0,\pi)$, which both are degenerate by symmetry.
    (c) Schematic phase diagram showing the level crossing of the two leading SDW (red) and sSC (black) eigenvalues 
    as a function of doping and critical energy scale $\Lambda_c\approx T_c$. } 
\end{figure}
\subsubsection{Channel Flow and Form Factors}
\begin{figure}[t]
\centering
   {\includegraphics[scale=0.21]{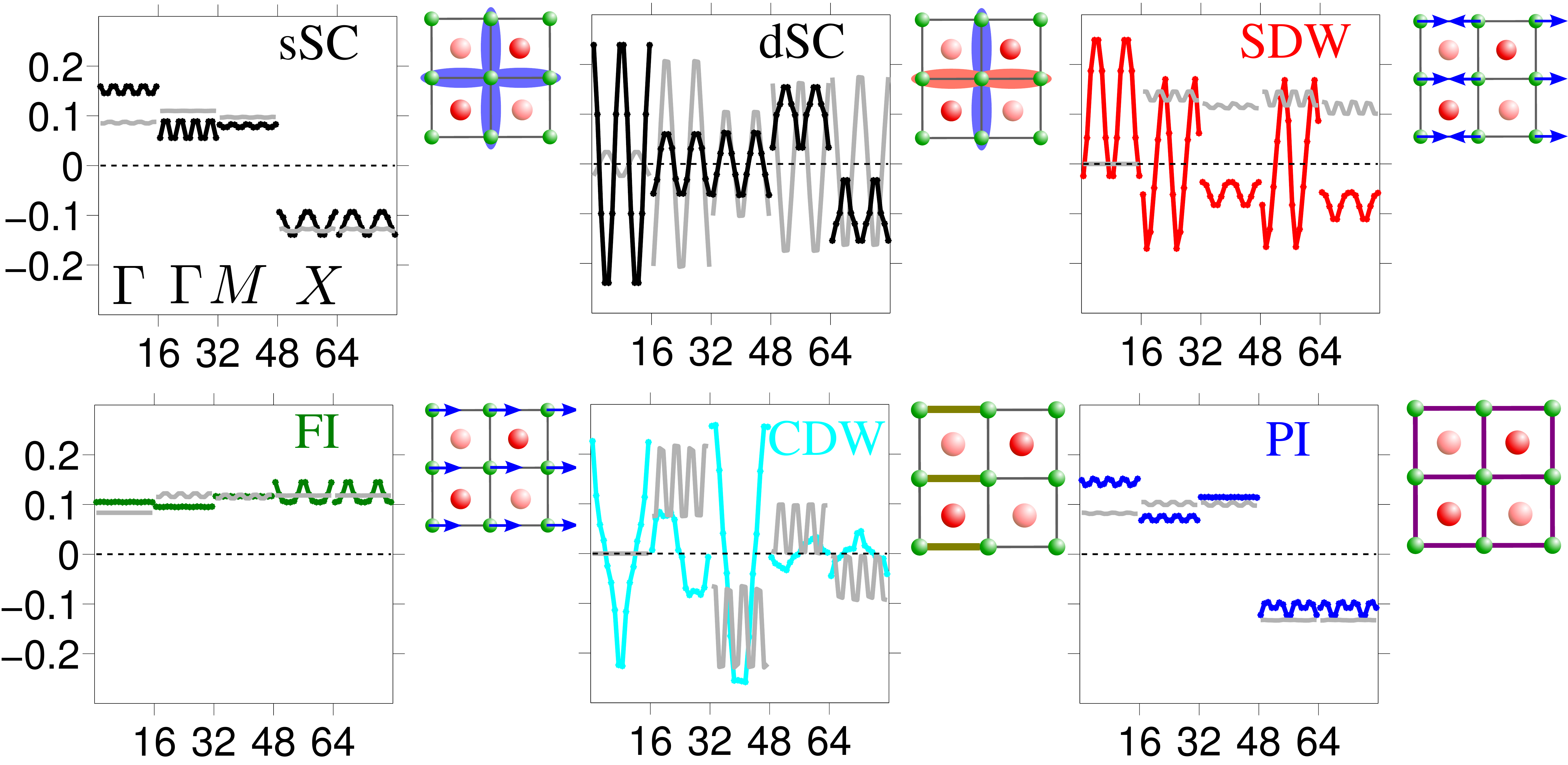}}
\caption{\label{fig:formfacs} 
    Form factors $f^{ch}(k)$ of the leading eigenvalues plotted along the different Fermi-surfaces according to the numbering of Fig.~\ref{fig:vertices2}b. The insets show the
    nearest-neighbor representation of the corresponding orders. Singlet $s_{\pm}$-wave (sSC) and $d_{x^2-y^2}$-wave (dSC) pairing orders depicted by its nearest-neighbor
    wave function, striped antiferromagnetic (SDW) and ferromagnetic (FI) order
    indicated by blue arrows, Peierls ordering ($p$-wave CDW) represented by enhanced bond hoppings (olive) and $s_{\pm}$-wave Pomeranchuk phase leading to a uniform shift of the 
    hopping parameters (purple lines). The transformation to the real-space ordering patterns in the insets is discussed in Appendix~\ref{sec:realsp}.} 
\end{figure}
In order to develop a broader understanding of the competing phases in FeSCs, we now apply the eigenmode expansion (\ref{eq:eigdecomp}) and 
study the flow of the most leading eigenvalues $w^{ch}_i(\Lambda)$. In addition, we also investigate the form factors $f^{ch}_i$ which 
tell us about the symmetry of the associated order parameter. From the eigenvalue flow in Fig.~\ref{fig:flowpic1}, it then turns out that the striped 
antiferromagnetic phase (SDW) with ordering momentum $Q=(0,\pi)$ or $Q=(\pi,0)$ is preferred at an electron 
filling of $n=6.00$ (undoped). On the other hand, singlet pairing is favored at $n=6.10$ with a critical energy scale $\Lambda_c\approx T_c$ that 
is one order of magnitude lower than in the SDW case at $n=6.00$.
A closer look at the corresponding form factors in Fig.~\ref{fig:formfacs} then reveals that 
the superconducting phase is of $s_{\pm}$-wave symmetry (sSC) with a characteristic sign-change between hole- and electron-pockets
and with a pronounced anisotropy at $X$. In the absence of the Fermi-pocket at $M$, these anisotropies can even lead to nodes (zeros) in the
superconducting gap, as will be discussed in Sec.~\ref{sec:whyare}. The form factor of the SDW phase instead shows a clear nodal behavior 
which is protected by symmetry~\cite{ran-09prb014505} due to the transformation behavior 
of the matrix-elements $u_{am}(\bs{k})$ in (\ref{eq:bandcoupling}) of Section~\ref{subsec:bandorb}. However, the real-space order in orbital-basis is still of $s$-wave type
with a striped-antiferromagnetic pattern as shown in the inset of (SDW). All these results, including the one of a nodal SDW-phase, 
are in accordance with the overall experimental picture.\par 
In addition, the FRG results also suggest a significant tendency towards $d_{x^2-y^2}$-wave pairing, as indicated by the subleading eigenvalue (dSC) in Fig.~\ref{fig:flowpic1}. 
\textit{This near degeneracy of $s_{\pm}$- and $d_{x^2-y^2}$-wave pairing channels can lead to interesting new phases, such as a time-reversal symmetry broken $(s+id)$-pairing state}
discussed in Sec.~\ref{sec:timereversal} or an extended $d_{x^2-y^2}$-wave pairing state presented in Sec.~\ref{sec:exoticdwave}.
Other ordering tendencies such as the Pomeranchuk instability (PI), spin-triplet pairing (not shown) and charge-density wave (CDW) channels remain comparably small throughout the flow.
The ferromagnetic channel (FM), though very pronounced at the beginning, decreases significantly during the flow.\par
In order to \textit{visualize the influence of the orbital make-up in the electronic states}, we artificially impose the crude approximation of neglecting all matrix elements $u_{am}(\bs{k})$
in the bare interaction (\ref{eq:bandcoupling}) and simply used $V_0(k_1,k_2,k_3,k_4)=U$ as a starting point in our functional RG implementation. 
The resulting form factors are then given by the gray lines of Fig.~\ref{fig:formfacs} and reveal a considerable deviation compared to the ones determined 
from the correct microscopic interaction. Even though the $s_{\pm}$-wave pairing symmetry is correctly reproduced, it completely fails to resolve the gap-anisotropies
and also predicts a wrong nodeless SDW gap. 
\subsection{Interplay of Competing Interactions and Fermiology: Why Are Some FeSCs Nodal while Others Are Nodeless ?}\label{sec:whyare}
Now being armed with all the details of the functional RG setup, we \textit{address} the \textit{low-energy competing fluctuations} and \textit{the emergence of the SC state in prototypical pnictide
compounds}. In particular, we analyse why some FeSCs appear 
nodal while others are nodeless, despite their similar electronic properties.  
For example, in the 1111 representative LaFeAsO, a majority of experiments point
to the existence of nodeless isotropic gaps \cite{wray-08prb184508,ding-08epl47001} on the hole-like Fermi surface (FS)
and also nodeless gaps on the electron-like FS, albeit with a larger gap anisotropy 
\cite{hashimoto-09prl207001,parish-08prb144514,matano-08epl57001,tanatar-09cm0907,checkelsky-09cm0811}.
On the other hand, in LaFePO, a clear majority of experiments support a nodal gap behavior on the electron pockets \cite{yamashita-09prb220509,hicks-09prl127003}.
This difference is even more puzzling, since both materials display similar FS pockets at the $X$- and $\Gamma$-point of the Brillouin zone, 
as pointed out in an ARPES study of Lu~\ea~\cite{lu-09pc452}. \par
\begin{figure}[t]
\centering
   {\includegraphics[scale=0.25]{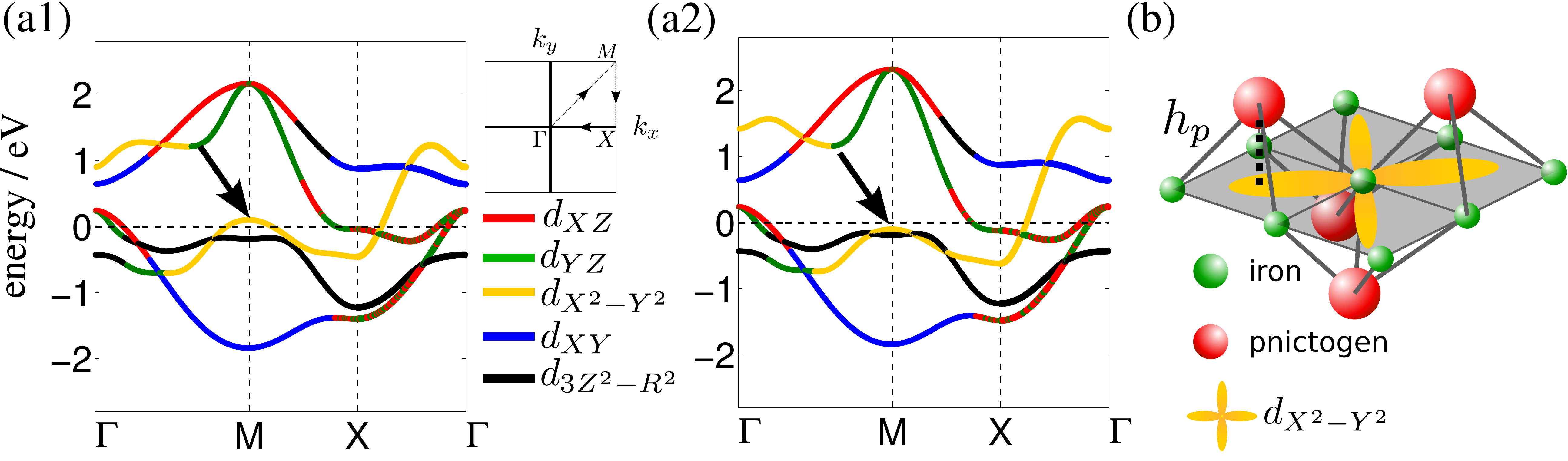}}
\caption{\label{fig:bandpnic} 
    Band structure of LaOFeAs (a1) and LaOFeP (a2) plotted in colors according to its leading orbital content (Inset: Brillouin Zone and orbital color-coding).
    The dashed horizontal lines denote the Fermi level for the respective undoped compounds.
    Here, the major difference between (a1) and (a2) is the $d_{X^2-Y^2}$-orbital dominated band, indicated by the arrow, which crosses the Fermi level in (a1), but not in (a2).
    Being still away from the Fermi level, the $d_{3Z^2-R^2}$-orbital dominated band is shifted up in
    (a1) compared to (a2). (b) Pnictogen height $h_p$ and position of the iron $d_{X^2-Y^2}$-orbital.
                 } 
\end{figure}

In what follows, the FRG analysis offers an explanation for the difference between the superconducting gaps in As- and P-based compounds. As was noted by 
Kuroki~\ea~\cite{kuroki-09prb224511}, a key difference between these two material classes consists of a modified pnictogen height, i.e. 
the distance measured from the pnictogen to the iron-plane (see Fig.~\ref{fig:bandpnic}b), which then mainly affects the spread of the iron $d_{X^2-Y^2}$-orbital. 
Therefore, the appearance of an additional hole-pocket at the $M$-point of mainly $d_{X^2-Y^2}$-orbital weight, is very sensitive to the pnictogen height and, 
as we will illustrate in the following, causes a nodal or nodeless pairing-gap.\par
As a starting point, we use a two-dimensional tight-binding model 
developed by Kuroki {\it et al.}~\cite{kuroki-08prl087004} to describe the band structure of the
1111-type iron-based superconductors $
H_{0} = \sum_{\bs{k},s}\sum_{a,b=1}^{5}c_{\bs{k}as}^{\dagger}K_{ab}(\bs{k})c_{\bs{k}as}^{\phantom{\dagger}}.$
As in Eq.~\ref{eq:freeh0} above, $c_{\bs{k}as}^{\dagger}, c_{\bs{k}as}^{\phantom{\dagger}}$ denote the electron creation and annihilation operators, 
$a,b$ represent the five iron $d$-orbitals, $s$ the spin projection and $K_{ab}(\bs{k})$ stand for the tight-binding matrix elements.\par 
While the main \textit{electronic
structure} of P-based and As-based compounds is very similar, there are
certain important differences. Figure~\ref{fig:bandpnic} shows the band
structure of LaOFeAs and LaOFeP, where the latter is obtained by
adjusting the parameters in~\cite{kuroki-08prl087004} according to the
changed pnictogen height from As to P~\cite{kuroki-09prb224511}. In
the vicinity of the Fermi surface, the most notable difference is the
presence or absence of a broad $d_{X^2-Y^2}$-orbital dominated band
at $M=(\pi,\pi)$, in agreement with ARPES data. To account for this
difference, we use a {\it $5$ pocket scenario} for the As-based
and a {\it $4$ pocket scenario} for the P-based compounds.\par
The interactions in this model are then given by Eq.~(\ref{eq:hintproto}), where intra- and inter-orbital interactions $U_1$ and $U_2$ as well as
the Hund's coupling $J_H$ and the pair-hopping term $J_{pair}$ are considered. As discussed in Sect.~\ref{sec:fluct}, we choose 
a physical interaction setting dominated by intra-orbital coupling, $U_1 > U_2 > J_{H} \sim J_{pair}$, and assume $U_1 = 3.5 eV, U_2 = 2.0
eV, J_{H}=J_{pair}=0.7 eV$~\cite{wang-09prl047005}. It is also important to note that, even 
though the interaction scales are relatively high, the bare interaction scale, taking into 
account the different orbital weights in (\ref{eq:bandcoupling}), does
not exceed $2 eV$, whereas the kinetic bandwidth amounts to $5 eV$.
From the band structure point of view, it should also be noted that the
$d_{3Z^2-R^2}$-orbital dominated band moves towards the Fermi level for the
P-based compound (Fig.~\ref{fig:bandpnic}). However, this band only plays a
marginal role since no other relevant band shares the $d_{3Z^2-R^2}$-orbital content,
and any scattering to other bands is, therefore, governed by
subleading inter-orbital interactions.\par
Using FRG, as described in the previous sections and also in Chap.~\ref{chap:fRG},
we study how the renormalized interaction described by the 4-point vertex function $V^{\Lambda}$, which links to the second quantized scattering vertex form
\begin{equation}\label{eq:reninter}
V^{\Lambda}(k_1,k_2,k_3,k_4)
\gamma_{k_1m_1s}^{\dagger}\gamma_{k_2m_2s'}^{\dagger}\gamma_{k_3m_3s}^{\phantom{\dagger}}\gamma_{k_4m_4s'}^{\phantom{\dagger}},
\end{equation}
evolves under integrating high-energy fermionic modes. Here, the flow parameter is an 
infrared cutoff $\Lambda$ approaching the Fermi surface, and \textit{we also checked the validity of our results
by implementing the temperature flow scheme described in Sec.~\ref{sec:flowpars}}. We further employed the 
condensed notation of $k_i = (\bs{k}_i,m_i)$ including momenta $\bs{k}_{i}$ and band-indices $m_i$. The label $s,s'$ in (\ref{eq:reninter}) is again used
for the spin projection. Note that in (\ref{eq:reninter}), we also applied the transformation from orbital- to band-basis $(c_{\bs{k}as}\rightarrow \gamma_{\bs{k}ms})$ 
as discussed in Sec.~\ref{subsec:bandorb}. 
The gap form factors $f_i^{SCs}(p)$ are computed along
the discretized Fermi surfaces (see Fig.~\ref{fig:gapLaFeAsO}b
and Fig.~\ref{fig:gapLaFePO}b), and the flow of the leading
eigenvalues $w_i^{SCs}(\Lambda)$ is plotted in Fig.~\ref{fig:gapLaFeAsO}e
and Fig.~\ref{fig:gapLaFePO}e.

\subsubsection{As-Based Compounds} 
For the As-based setting, we find that the
$s_{\pm}$-pairing instability, giving rise to different gap signs on electron- and hole-pockets, is the leading instability of the model at
moderate doping.  The setup resembles the situation studied
in~\cite{wang-09prl047005}.
\begin{figure}[t]
\centering
   {\includegraphics[scale=0.28]{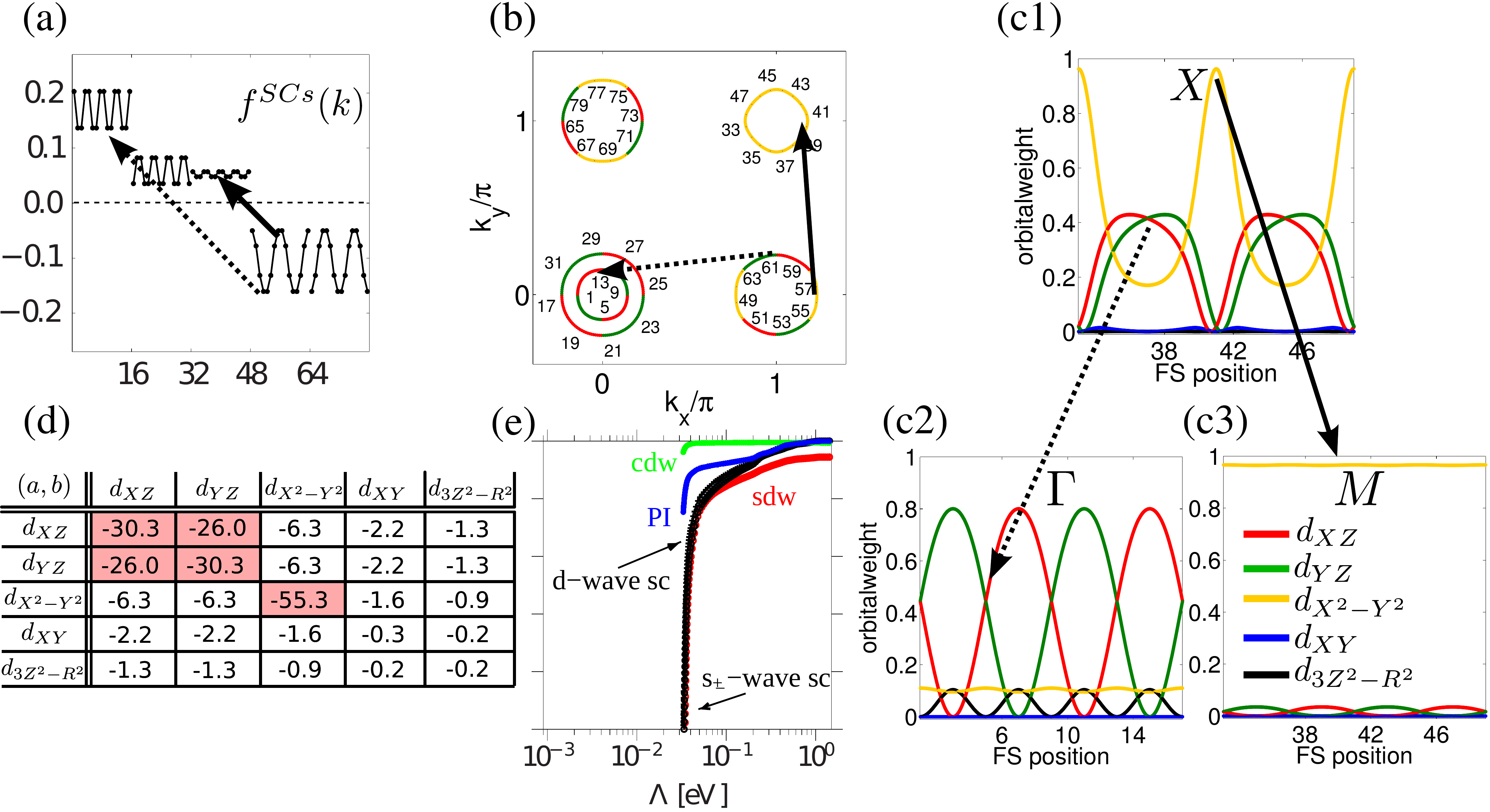}}
\caption{\label{fig:gapLaFeAsO} 
    Five pocket scenario for LaFeAsO~\cite{thomale-11prl187003}. (a) Leading pairing form factor $f^{SCs}(k)$ 
    versus patching indices (momenta) shown in (b).  The gap on the outer hole-pocket at
    $\Gamma$ is smaller than of the inner hole-pocket and of the same
    order as the $M$ pocket gap.  The gap on the electron-pockets is
    very anisotropic but nodeless and of opposite sign from hole-pocket gap.  
    (c1)-(c3) Orbital weight distribution on the different
    pockets (not shown is the outer hole-pocket at $\Gamma$, which is
    similar to (c2) shifted by 90 degrees assuring orthogonality of the
    band vectors). Dashed lines indicate most relevant scattering
    contributions for the dominating $U_1$ intra-orbital interaction.
    (d) Leading orbital pairing eigenvalues $w^{SCs}_{i,ab}(\Lambda_c)$ from Eq.~(\ref{eq:orbflow}):
    $d_{XZ,YZ}$ and $d_{X^2-Y^2}$ scattering dominates. (e)
    Flow of leading eigenvalues (charge density wave
    (CDW), Pomeranchuk instability (PI), spin-density wave SDW, and singlet superconductivity SC). The leading instability appears in the 
    $s_{\pm}$-wave pairing channel at $\Lambda_c\approx 0.03 eV$, $d$-wave pairing and SDW diverge closely (hardly distinguishable on the log scale).
                 } 
\end{figure}
We can identify the hole-pocket at the $M$-point to play a
major role in contributing to the fully gapped $s_{\pm}$-wave pairing (Fig.~\ref{fig:gapLaFeAsO}). 
In particular, we study the orbital content in detail and analyze how the pairing instability
distributes over the different orbitals (Fig.~\ref{fig:gapLaFeAsO}d). For this purpose,
consider the 4-point function in orbital space
\begin{align}\nonumber
V^{\Lambda,orb}_{c,d \rightarrow a,b}(\bs{k}_1,\bs{k}_2,\bs{k}_3,\bs{k}_4)=&\\\label{eq:orbflow} 
\sum_{n_1,\ldots,n_4 = 1}^5 V^{\Lambda}(k_1,k_2,k_3,k_4)&
u^*_{an_1}(\bs{k}_1)u^*_{bn_2}(\bs{k}_2)u_{cn_3}(\bs{k}_3)u_{dn_4}(\bs{k}_4),
\end{align}
where the $u$-coefficients relate the band basis to the orbital basis, and as such characterize the orbital components of the different bands. 
The matrix shown in Fig.~\ref{fig:gapLaFeAsO}d then gives the leading eigenvalue contributions of 
\begin{equation}\nonumber
W^{\Lambda,SCs}_{ab}(\bs{k},\bs{p})=V^{\Lambda,orb}_{a,a \rightarrow b,b}(\bs{k},-\bs{k},\bs{q},-\bs{q}) + V^{\Lambda,orb}_{a,a \rightarrow b,b}(-\bs{k},\bs{k},\bs{q},-\bs{q})
\end{equation} 
i.e. in the singlet Cooper channel of~(\ref{eq:orbflow}) where we constrain ourselves to the dominant processes of
intra-orbital pairing $(a,a)\rightarrow (b,b)$. As
above, we decompose $W^{\Lambda,SCs}_{ab}$ into different eigenmodes $\sum_i w^{SCs}_{i,ab}(\Lambda)
\tilde{f}^{SCs}_{i,ab}(\bs{k})^*\tilde{f}^{SCs}_{i,ab}(\bs{p})$,
where the leading eigenvalues at $\Lambda_c$ for different $(a,b)$ are
given in Fig.~\ref{fig:gapLaFeAsO}d and Fig.~\ref{fig:gapLaFePO}d. Intra-orbital scatterings between the $d_{XZ}$- or $d_{YZ}$-orbital
dominated parts of the electron- and hole-pockets are most important (see Fig.~\ref{fig:gapLaFeAsO}). 
They favor an $s_{\pm}$-wave pairing instability,
as was also found in \cite{wang-09prl047005}. However, the leading eigenvalue in the As scenario
comes from the diagonal part of the $d_{X^2-Y^2}$-orbital. Pointing in the direction of the $\Gamma\leftrightarrow X$ path, the electron pocket
has a high concentration of the $d_{X^2-Y^2}$-orbital. This part of the electron-pocket then
scatters strongly with the hole-pocket at the $M$-point, which is dominated by the $d_{X^2-Y^2}$-orbital band. 
The intra-orbital repulsion related to the latter scattering prefers an $s_{\pm}$-wave pairing between the hole-pocket at $M$
and the electron pockets, which reinforces the already present $s_{\pm}$-wave tendency between the $\Gamma$
hole-pockets and the $X$, ($X'$) electron-pockets. Assuming that $U_1$ is the dominant interaction, 
the three hole-pockets display a gap of identical sign: two $\Gamma$-pockets which are not nested with each other have
the same gap sign and are of different orbital content than the hole-pockets at the $M$-point. 
However, the electron pockets contain contributions 
from all three relevant $d$-orbitals. Therefore, the electron-pockets scatter strongly through $U_1$ with all three hole-pockets, which enhances the
$s_{\pm}$-wave character of the gap.\par 
So, in \textit{summary of the As scenario, the repulsive interaction  
induced by the presence of the additional $M$-pocket further increases the $s_{\pm}$-wave gap between hole- and electron-pockets}. The hole-pocket at $M$ is also responsible
for the strong SDW signal (\ref{fig:gapLaFeAsO}), as the nesting wave vector $M\leftrightarrow X$ equals the one between $\Gamma\leftrightarrow X$.
\subsubsection{P-Based Compounds}
\begin{figure}[t]
\centering
   {\includegraphics[scale=0.28]{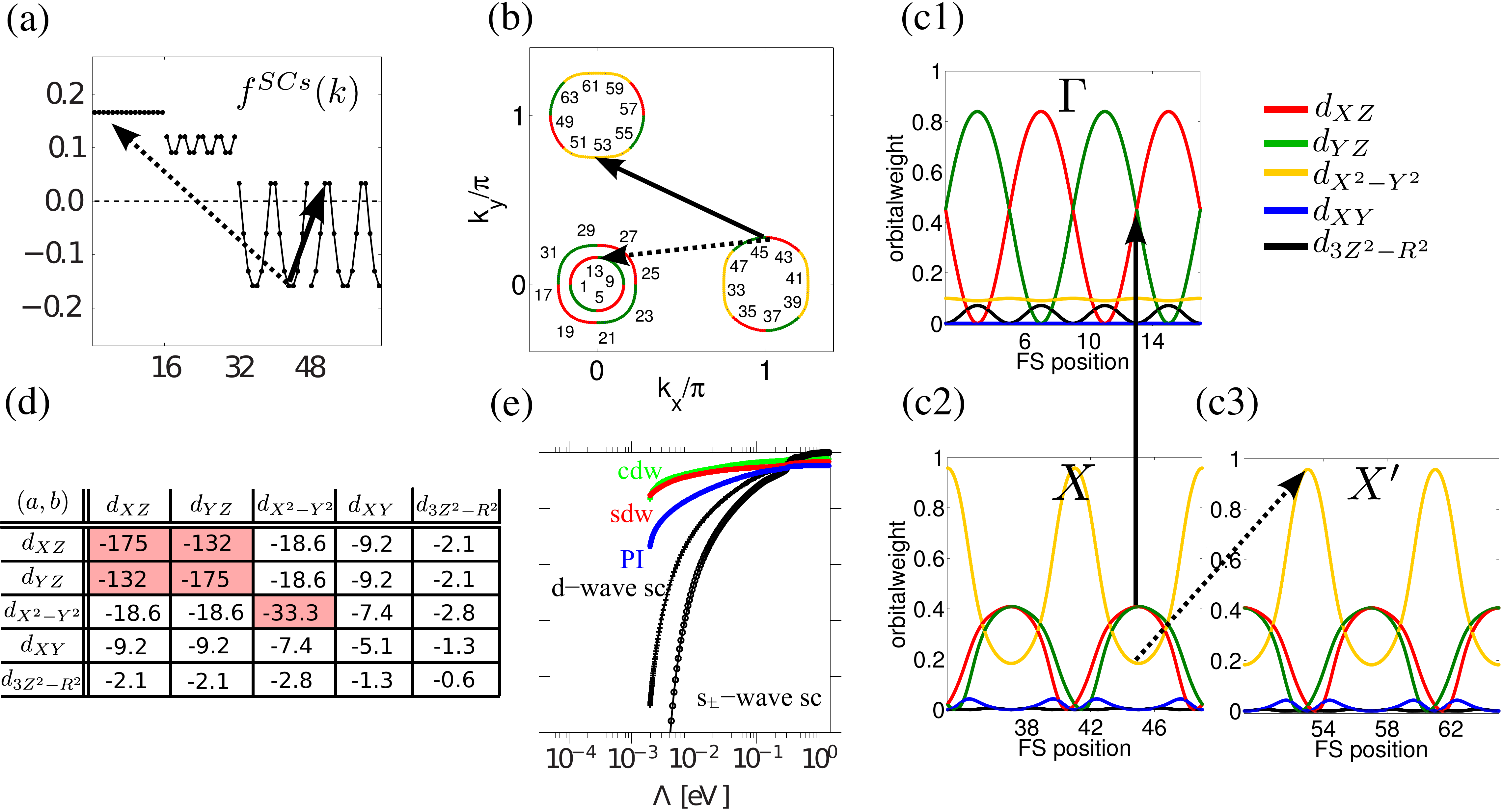}}
\caption{\label{fig:gapLaFePO} 
    Four pocket scenario for LaOFeP~\cite{thomale-11prl187003}. (a) Leading $s_{\pm}$-wave
    pairing form factor as a function of Fermi-surface position given
    in (b); the $d_{X^2-Y^2}$-dominated hole-pocket at $M$ is absent, and
    the hole-pockets at $\Gamma$ are gapped and isotropic. The electron-pockets show
    strong anisotropy, being nodal on the inner pockets tips indicated
    in Fig.~\ref{fig:gapLaFePO} by dashed arrows. (c) Relevant pair scatterings between different
    electron pockets (orbital weights of the pockets are shown in (c1)-(c3)). 
    (d) The orbital decomposition
    of the pairing instability: large $d_{XZ,YZ}$ and less relevant $d_{X^2-Y^2}$
    contribution. (e) Flow of leading instability eigenvalues (notation
    as in Fig.~\ref{fig:gapLaFeAsO}e); $\Lambda_c
    \approx 0.002 eV$ is smaller than in Fig.~\ref{fig:gapLaFeAsO}.} 
\end{figure}

In the \textit{P-based compounds}, the \textit{physical picture changes even qualitatively}. As shown
in Fig.~\ref{fig:gapLaFePO}, we find a \textit{nodal $s_{\pm}$-wave scenario} for the P-based compounds, with
lower critical divergence scale $\Lambda_c\sim T_c$ and less SDW-fluctuations. The absence of the
$M$ hole-pocket, or the presence of a less relevant $d_{3z^2-R^3}$ pocket, removes the intra-orbital scattering to the electron-pockets. This gives way to
previously subleading scattering channels such as, in particular, the pair scattering between the
$d_{X^2-Y^2}$-dominated parts of the electron pockets, but also pair scatterings from the hole pockets at 
$\Gamma$ to the electron pockets. The former acts between the $k$-points of the gap function on 
the electron pockets given by the peaks and the valleys (Fig.~\ref{fig:gapLaFePO}a) increasing the 
anisotropy and eventually giving them different signs, thus creating a nodal state. Even if the $d_{3Z^2-R^3}$-orbital dominated
band at the $M$-point (Fig.~\ref{fig:bandpnic}a2) were shifted to the Fermi level, the situation remains nearly unchanged as this pocket
does not share its orbital content with any other pocket, and hence interactions driven by $U_1$ are suppressed.\par
To further substantiate this conclusion, we perform a large sweep in parameter space to resolve the evolution of the superconducting form factor 
upon varying the interaction parameters. The corresponding results can be found in \cite{hanke-11anp638}. From the \textit{ab initio} 
data stated before, we get $U_1/U_2\approx 1.5$, $U_1/J_H = U_1/J_{pair}\approx 6.5$. Thus, the parameter regime
of the As-based and P-based compounds lies in the regime of \textit{applicability of our theory} and \textit{our findings are consistent with experiment:}\par
In the P-based compounds, we find (i) a \textit{lower divergence scale} and, hence, \textit{lower $T_c$ compared to As-based compounds}, (ii) significantly enhanced low
energy density of states in the \textit{(hence nodal) superconducting phase}, and (iii) \textit{reduced SDW-fluctuations}, which, even at pronounced nesting,
are \textit{insufficient to drive the system to a leading magnetic instability} \cite{yamashita-09prb220509,hicks-09prl127003,lu-09pc452}.
The absence of the hole pocket at the M-point also manifests itself in the orbital decomposition of the pairing instability (Fig.~\ref{fig:gapLaFePO}d);
the diagonal contribution of the $d_{X^2-Y^2}$-orbital, in comparison to the one of the $d_{XZ,YZ}$-orbital, is reduced.\par
\textit{In summary}, the \textit{broad band at the unfolded $M$-point plays the major role} in explaining the drastic change of
superconducting properties from the As-based to the P-based 1111 compounds, rendering the former nodeless and the latter nodal. The nodes
that appear in the P-based compounds are mainly driven by anisotropy of the electron pockets. Other compounds such as the 111 representatives LiFeAs 
and LiFeP display a similar phenomenology of nodal and nodeless pairing (see Sec.~\ref{sec:lifeas}), which has probably the same origin as in the 1111 compounds studied here.
\textit{In Sect.~\ref{sec:optprinciple}, a unifying picture for the findings is presented},  which addresses the unavoidable \textit{frustration due to the multi FS-pocket situation in the FeSCs.}
\subsection{Superconductivity in LiFeAs: RPA vs. FRG}\label{sec:lifeas}
Soon after the synthesis of the 1111 and 122
pnictides, LiFeAs as a representative of the 111 family, has been detected 
with a superconducting phase at $T_{\text{c}} \sim 18K$~\cite{wang-08ssc538,tapp-08prb060505}.
Unlike most other FeSCs, LiFeAs becomes superconducting without the need of doping or pressure
and shows neither a structural transition nor the usual spin-density wave order. This, together
with the report of a nonvanishing Knight-shift in some of the samples \cite{baek-arXiv11082592} and an observed fishtail effect \cite{pramanik-11prb094502} led
to an early proposal of a triplet pairing state in LiFeAs. In addition, ARPES measurements \cite{borisenko-10prl067002}
pointed out the proximity to a van-Hove singularity, which was then taken up in an RPA calculation of Brydon~\ea\ who started with a three-band ARPES fit and
predicted a ferromagnetic fluctuation induced triplet pairing state $\hat{\Delta}\sim\bs{\hat{z}}(p_x + ip_y)$, similar to the one 
proposed for strontium ruthenate~\cite{maeno-12jpsj011009}. The possible realization of such 
a chiral pairing state in LiFeAs with a relatively high $T_c$ of $18K$, opposed to strontium ruthenate with $T_c =1.5K$, is, in principle, very appealing. 
However, in the meantime, a growing number of experiments found evidence for antiferromagnetic fluctuations, as for example in NMR 
measurements \cite{jeglic-10prb140511} and neutron scattering \cite{taylor-11prb220514}. The latter measurements, 
in addition, also reported a magnetic resonance and hence provided strong indications for a usual $s_{\pm}$-wave pnictide pairing.  On the other hand, it seems commonly accepted that
the superconducting gap is nodeless as confirmed by a large body of experimental results like NMR \cite{inosov-10prl187001}, specific heat \cite{sasmal-10prb144512}, 
ARPES \cite{stockert-11prb224512} and penetration depth measurements \cite{imai-09jpsj013704,song-11epl57008}.

\begin{figure}[t]
\centering
   {\includegraphics[scale=0.40]{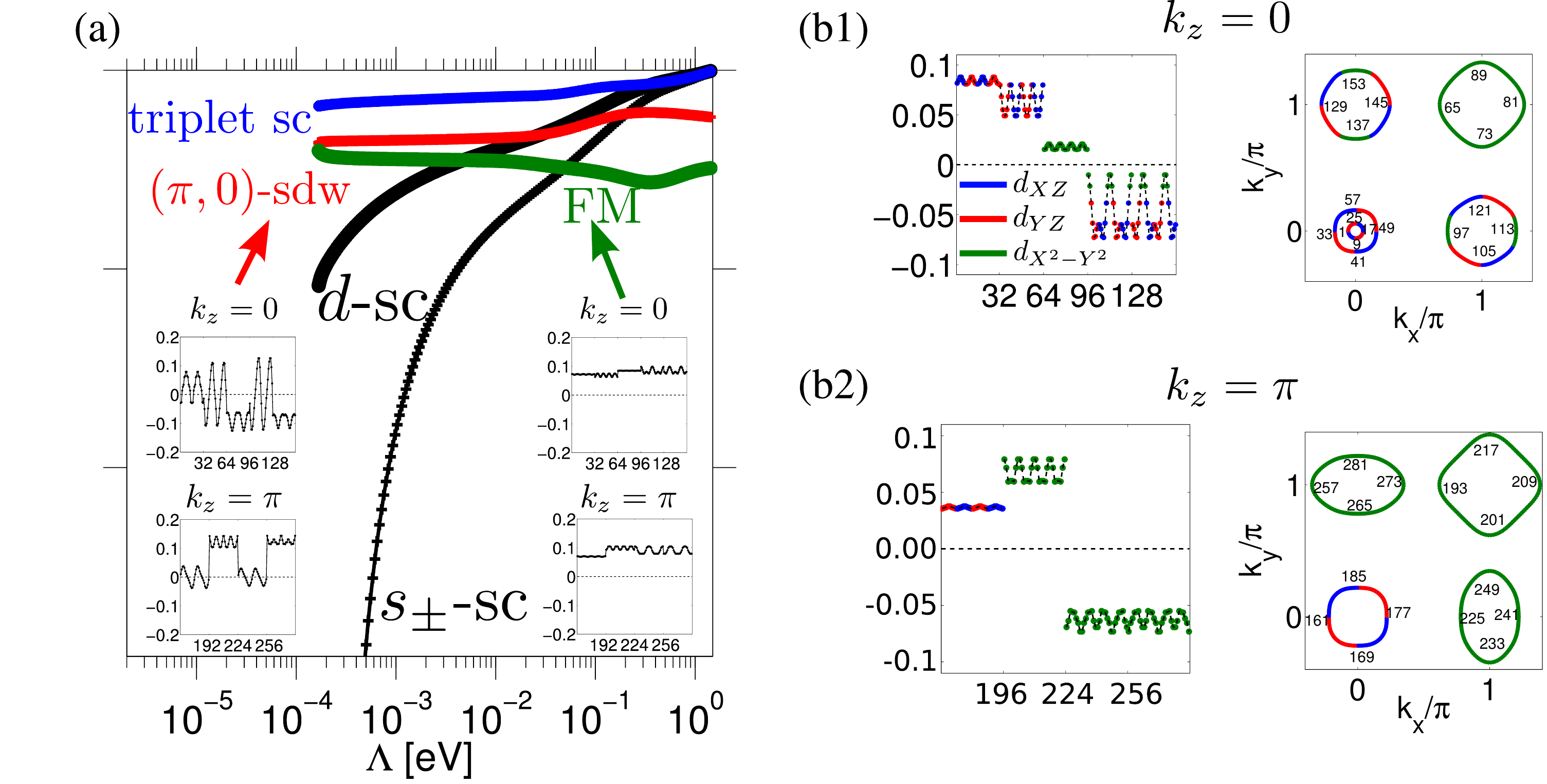}}
\caption{\label{fig3-lifeas} (a) Vertex flow from our functional renormalization group study for LiFeAs~\cite{platt-11prb235121}. 
With dominant ferromagnetic fluctuations from the outset at high energies $(\Lambda \sim 10^0)$, the RG-flow renormalization yields 
a switch to collinear AFM fluctuations as the dominant magnetic channel, which eventually drives $s_{\pm}$ in the particle-particle channel. 
Leading SC form factors and Fermi-surface discretization at $k_z=0$ (b1) and $k_z=\pi$ (b2) display a nodeless $s_{\pm}$ gap. 
The colors indicate the leading orbital weights at the corresponding FS points. } 
\end{figure}

In order to explore these different ordering tendencies, the FRG formalism allows to employ a combined approach of density functional theory (DFT) and 
FRG which connects an \textit{ab initio} description with the unbiased analysis of functional RG.
The DFT description provides a band-structure matching well the ARPES data~\cite{borisenko-10prl067002} and quantum oscillation measurements 
\cite{putzke-12prl047002} with a nontrivial $k_z$
dependence. In addition, DFT also enables to compute the orbital dependent interaction parameters via the knowledge of maximally 
localized Wannier functions~\cite{miyake-10jpsj044705}. Both of these informations are essential to explain the interesting properties of LiFeAs.
Using the combined DFT+FRG approach, we find that, despite the presence of strong ferromagnetic fluctuations at ``high energy'' (i.e. cut-off scales $\Lambda$ in Fig.~\ref{fig3-lifeas}), LiFeAs features a nodeless $s_{\pm}$-wave pairing state similar to the one in LaFeAsO at ``low energies''. Interestingly, its phosphorus based
realization LiFeP (111) exhibits nodal superconductivity, which is reminiscent of the nodeless/nodal behavior in LaFeAsO/LaFePO, discussed in
Sec.~\ref{sec:whyare}. Therefore, upon closer inspection, LiFeAs does not appear to be much different from other FeSCs.\par
Fig.~\ref{fig3-lifeas} summarizes the DFT-FRG results in terms of the RG-flow for LiFeAs. 
The SC order parameter is found to be $s_{\pm}$, driven by collinear AFM fluctuations. In the RG flow, they eventually exceed the ferromagnetic fluctuations, 
mainly stemming from the small hole pocket at the $\Gamma$-point, as the system flows to low energy. This points \textit{clearly to the importance of taking the 
competing fluctuations via an FRG-calculation into account}, when evolving to the low-energy scale of SC, $\Lambda \sim k_B T_c$. It is an explicit example where a simplified RPA approach does not yield the level of accuracy as the analogous FRG study.
\subsection{Exotic d-wave Pairing in Strongly Hole-Doped K$_x$Ba$_{1-x}$Fe$_2$As$_2$}\label{sec:exoticdwave}
In the preceding section, we found that the existence of nodes in the $s_{\pm}$-wave pairing state of FeSCs is not required by symmetry, but may
develop as a compromise between two competing fluctuation channels. Here, the node position was not fixed by symmetry and occurred somewhere
near the $\Gamma\leftrightarrow X$ axes on the electron pockets. It, therefore, came as a surprise when nodal 
superconductivity was reported in a strongly hole-doped 122 compound KFe$_2$As$_2$~\cite{dong-10prl087005,terashima-10prl259701,hashimoto-10prb014526,fukazawa-09jpsj083712,zhang-10prb012503}
where ARPES measurements~\cite{sato-09prl047002} clearly showed that the two electron pockets had nearly vanished. 
In addition, the superconducting transition temperature of $T_c = 3K$ was rather low compared to the moderately doped K$_{0.4}$Ba$_{0.6}$Fe$_2$As$_2$ 
with $T_c = 38K$~\cite{rotter-08prl107006}, where all experiments indicate a nodeless 
superconducting gap~\cite{martin-09prb020501,checkelsky-09cm0811,luo-09prb140503,rotter-08prl107006,wray-08prb184508,ding-08epl47001,zhang-10prl117003}. 
For this reason, there must be a \textit{nodal to nodeless transition or even a change of the pairing symmetry in between these doping regions}. In the following, we develop a detailed picture of how the magnetic as well as superconducting phases evolve under hole doping in
$\text{K}_{x}\text{Ba}_{1-x}\text{Fe}_2\text{As}_2$. We find that the nodal pairing phase observed for $x=1$ is of (extended) $d$-wave type.\par
\begin{figure}[t]
\centering
   {\includegraphics[scale=0.49]{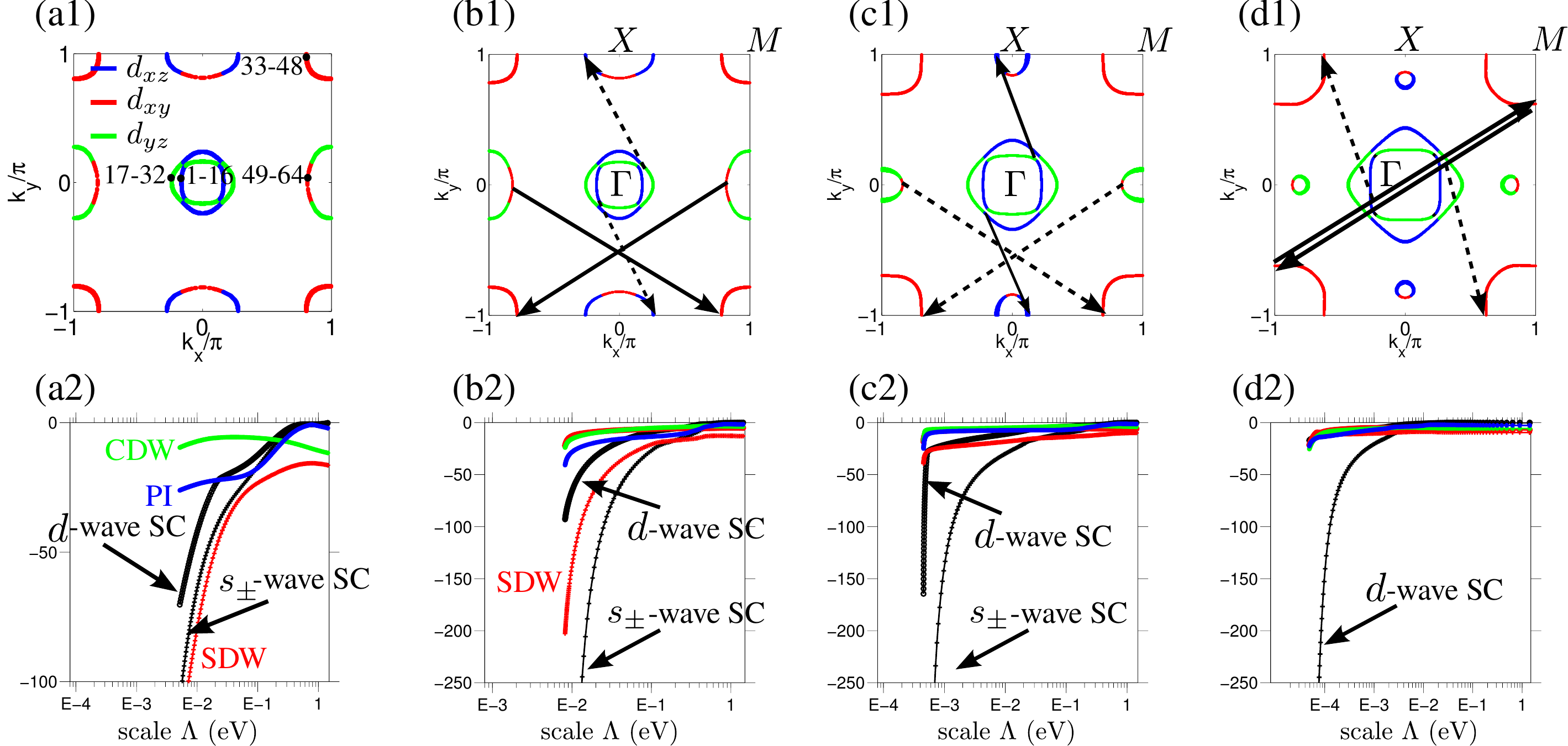}}
\caption{\label{fig:flowsKFeAs} Representative scenarios of the Fermi-surface (unfolded
    BZ) and eigenvalue flows for
    electron concentration per iron $n=6.000$ (a), $n=5.913$
    (b), $n=5.663$ (c), and $n=5.346$ (d) in K$_{x}$Ba$_{1-x}$Fe$_2$As$_2$~\cite{thomale-11prl117001}. 
    The hole doping of our model calculation in (d),
    while exceeding the experimental setup $n=5.5$, best
    matches the FS profile from ARPES~\cite{sato-09prl047002}. The dominant and subdominant
    scatterings in the Cooper channel are highlighted in (b1)-(d1) by
    full and dashed arrows. The colors and numberings along the FS label the
    dominant orbital weights (inset (a1)) and the patch indices, respectively. The leading eigenvalue flow
    of the ordering channel for different Fermi instabilities (charge-density 
    wave (CDW), Pomeranchuk (PI), spin-density wave (SDW) and
    superconductivity (SC)) are plotted in (a2)-(d2) versus the momentum
    cutoff $\Lambda$. For (a) we find SDW order as the leading Fermi instability, for 
    (b) and (c) we detect an $s_{\pm}$-pairing instability. In (c) we observe a
    transition from $s_{\pm}$-wave to $d$-wave pairing.} 
\end{figure}

We now focus on studying $\text{K}_{x}\text{Ba}_{1-x}\text{Fe}_2\text{As}_2$ starting at
the optimally doped case around $x=0.4$ and increasing the hole doping up to $x=1.0$ in
$\text{KFe}_2\text{As}_2$. An effective 5-band tight-binding
model developed by Graser {\it et al.}~\cite{graser-10prb214503} is used to
describe the band structure of the 122-type iron-based
superconductors. As seen in Fig.~\ref{fig:flowsKFeAs}, for moderate hole doping, the
conventional five pocket scenario with electron pockets at $X=(\pi, 0)$
and $M=(\pi,\pi)$ emerges. 
For larger hole doping, the electron
pockets vanish and only small disconnected lobe features are found
around $X$ (Fig.~\ref{fig:flowsKFeAs}c). The model then reduces to the
effective three-hole-pocket scenario shown in Fig.~\ref{fig:flowsKFeAs}d. Other
details, such as the irrelevance of including the 3-dim. $k_z$-dispersion can be found in Ref.~\cite{thomale-11prl117001}.
A schematic picture of the Fermi-surface topology is given in
Fig.~\ref{fig:flowsKFeAs}a, including the leading orbital character and the patch numbering. 
We use the conventional onsite orbital model for the interactions, i.e. Eq.~(\ref{eq:hintproto}), with the
corresponding parameter values chosen close to the ones obtained by constrained
RPA calculations~\cite{miyake-10jpsj044705}: $U_1 > U_2 > J_{H} \sim
J_{pair}$, and set $U_1 = 3.0 eV, U_2 = 2.0 eV,
J_{H}=J_{pair}=0.6 eV$. In Fig.~\ref{fig:flowsKFeAs},
the leading eigenvalues for different instabilities are plotted against $\Lambda$ for a variety of 
fillings between pristine filling $x=0$ from the left to strongly hole-doped to the right. 
While the leading instability is first located in the particle-hole channel exhibiting the $(\pi,0)/(0,\pi)$ SDW instability (Fig.~\ref{fig:flowsKFeAs}a), we find that, for all scenarios of hole doping decpited in Fig.~\ref{fig:flowsKFeAs}, the leading instability is in the Cooper
channel.\par 
For the moderately doped case, the electron pockets are of similar size as
the hole pockets. Figure~\ref{fig:flowsKFeAs} (b1) shows the Fermi-surface structure as
well as the dominant (full line) and subdominant scattering (dashed
arrow) processes in the Cooper channel. The two major components are
given by $\Gamma \leftrightarrow X$ as well as $M
\leftrightarrow X$ scatterings. They are particularly important for the
front tips of the electron pockets since these parts can scatter to $M$ via the
dominant $U_1$ interaction due to an identical orbital content. The spin-density wave (SDW)
fluctuations are strong, signaling the proximity
to the leading magnetic instability scenario of the undoped model
(Fig.~\ref{fig:flowsKFeAs} (b2)).\par
\begin{figure}[t]
\centering
   {\includegraphics[scale=0.31]{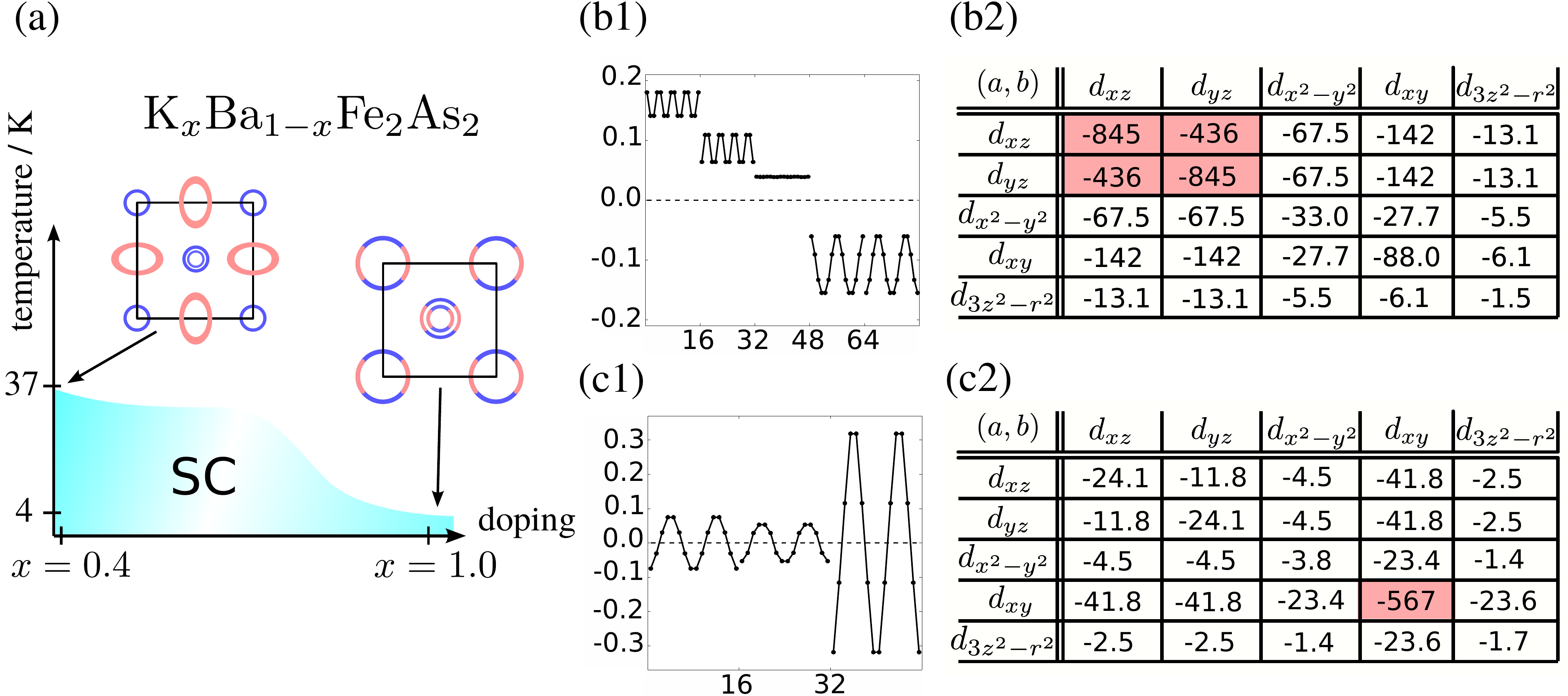}}
\caption{\label{fig:spmKFeAs} (a) Qualitative phase-diagram of K$_{x}$Ba$_{1-x}$Fe$_2$As$_2$ as a function of doping $x$ with the two determined pairing-symmetries 
    depicted in the insets. (b1,c1) Corresponding form factors~\cite{thomale-11prl117001} of the leading pairing-instability according 
    to the scenarios in Fig.~\ref{fig:flowsKFeAs}(b,c). The numbering in (b1,c1) denotes the position on the Fermi-pockets
    according to Fig.~\ref{fig:flowsKFeAs}(a1). (b2,c2) Eigenvalues of the orbital decomposition of
    the superconducting form factor in (b1,c2). Here, the ratio of the values label the
    relative importance of the orbital scattering channel $V(a,a
    \rightarrow b,b) c_{a \uparrow}c_{a \downarrow}c_{b\downarrow}^{\dagger}c_{b\uparrow}^{\dagger}$. In the strongly hole-doped case, we find an extended $d$-wave form 
    factor with nodal points located along the main diagonal of the Brillouin zone shown in the right inset of (a).} 
\end{figure}

For the intermediate doping regime, between moderate and strong hole doping,
the electron pockets are already rather small (Fig.~\ref{fig:flowsKFeAs}
(c1)). The nesting to the hole pocket is absent, and the SDW
fluctuations are strongly reduced. In addition, the SDW fluctuations
become less concentrated in the $(\pi,0)/(0,\pi)$ or $(\pi,\pi)$
channel, and spread into various incommensurate sectors~\cite{lee-11prl067003}. The $d_{xy}$-orbital weight on the
electron pocket is reduced, and the $M \leftrightarrow X$ scattering becomes
subdominant. 
The main pair scattering emerges along
$\Gamma \leftrightarrow X$. As a consequence, $s_\pm$-wave pairing is still the
leading instability, where the form factor and its decomposition into
orbital scattering contributions are shown in Fig.~\ref{fig:spmKFeAs}(c1,c2): the
largest gap is found for the inner hole pocket at $\Gamma$, followed by
the outer hole pocket and the hole pocket at $M$, where the electron pockets show
anisotropic gaps. The orbital decomposition confirms the previous 
discussion of the dominant scattering contribution, in that the largest
weight resides at intra and inter-orbital scattering of the $d_{xz}$-
and $d_{yz}$-orbital.  However, we already observe that, due to the
lack of SDW fluctuations supporting the pairing channel, the critical divergence
scale is decreased (Fig.~\ref{fig:flowsKFeAs}~(b2)-(d2)). In particular, while still subdominant, we can
already see the $d$-wave pairing evolving as the second-highest eigenvalue in the Cooper channel. 
When the electron pockets are still present, the form factor
(not shown here)  closely resembles the extended $d$-wave type involving hole pockets and
electron pockets~\cite{thomale-09prb180505}.\par
At strong hole doping, the contingent electron pockets are absent, and the hole pockets
are very large. The flow in Fig.~\ref{fig:flowsKFeAs}(d2) shows no instability up
to rather small cutoff-scales $\Lambda$, where we find a leading
instability in the Cooper channel. Its form factor and orbital
scattering decomposition is shown in Fig.~\ref{fig:spmKFeAs}(d1,d2). We observe an extended
$d$-wave instability on the three hole pockets, with nodes located
along the main diagonals in the Brillouin zones (as seen in the right inset of  Fig.~\ref{fig:spmKFeAs}a).
A harmonic analysis of the
order parameter yields a large contribution of $\cos(2k_x)-
\cos(2k_y)$ type and a subdominant $\cos (k_x) - \cos (k_y)$
component, i.e. the form factor is most accurately characterized by $(\cos k_x + \cos k_y)(\cos k_x -
\cos k_y)$. The dominant scattering is intra-pocket scattering on the large $M$ hole pocket, followed
by inter-orbital $d_{xy}$ to $d_{xz,yz}$ scattering between $M
\leftrightarrow \Gamma$. While the magnetic fluctuations are generally
weak in this regime, the dominant contribution is now given by
$(\pi,\pi)$ SDW fluctuations as opposed to $(\pi,0)/(\pi,0)$ for
smaller hole doping. For strong
hole doping, the hole pocket at $M$ is large enough to induce
higher harmonic $d$-wave pairing through intra-pocket scattering between the $d_{xy}$-orbitals 
as confirmed by the large value of $d_{xy}-d_{xy}$ pairing (Fig.~\ref{fig:spmKFeAs}(c2)).\par 
Via scattering to the other pockets, the superconductivity
is likewise induced there, however, with smaller amplitude
than for the $M$-pocket Fig.~\ref{fig:spmKFeAs}(c1). 
As opposed to conventional first harmonic $d$-wave pairing, there is no sign change
between the extended $d$-wave form factor on the $M$-pocket and the
$\Gamma$-pocket according to $\cos(2k_x)-
\cos(2k_y)$ in Fig.~\ref{fig:spmKFeAs}. This picture of a
\textit{$k$-space proximity effect from the $M$-pocket to the $\Gamma$-pockets} is substantiated by our checks with
calculations involving the $M$-pocket only, where we see a similar
evolution of a pairing instability (the divergence is lower, as the inter-orbital scatterings in the 3-pocket scenario
help to renormalize the repulsive Coulomb interactions).
This matches the orbital decomposition of the superconducting form factor in Fig.~\ref{fig:spmKFeAs}(c2), showing
dominant intra-orbital scattering of the $d_{xy}$-orbital.\par 
As apparent from the ARPES data, the nodal character of the superconducting phase in
$\text{KFe}_2\text{As}_2$ cannot originate from possible nodes on the
electron pockets (which are gapped out at these doping levels), 
but must be due to nodes on the hole pockets. It is then clear that the
order parameter cannot be $s_{\pm}$-wave as it does not tend to allow for an anisotropy that would drive the hole pockets nodal. The $d$-wave
instability, which we find for the strongly hole-doped regime, provides
an explanation for the general experimental evidence, while the
detailed gap structure certainly deserves further investigation~\cite{furukawa-11prb024507}. Electron-phonon coupling
may change the picture quantitatively, but not qualitatively,
as the nodal features tentatively tentatively linked to the $d$-wave symmetry are
unambiguously observed in experiment. An interesting alternative proposal has been communicated by Chubukov and co-workers~\cite{PhysRevB.85.014511}. While the latter has so far not been supported by FRG studies, it will be a subject of further experimental and theoretical study to ultimately understand the change of pairing symmetry in $\text{KFe}_2\text{As}_2$.
\subsection{A Common Thread in Unconventional Pairing: An Optimization Principle in BCS Mean Field Theory}\label{sec:optprinciple}
One fascinating possibility, which multi-band SC offers is to investigate how the various competing orders depend 
on the Fermi-surface topology and interactions, and how this can lead to interesting new quantum states of matter. 
For this purpose, the BCS mean field scenario for mult-band SC can be interpreted as an \textit{optimization principle}~\cite{hanke-11anp638} which equips us with a 
more universal picture on the interplay of Fermiology and interactions.\par
\begin{figure}[t]\begin{center}
  \begin{minipage}[c]{0.9\linewidth}
    \includegraphics[width=\linewidth]{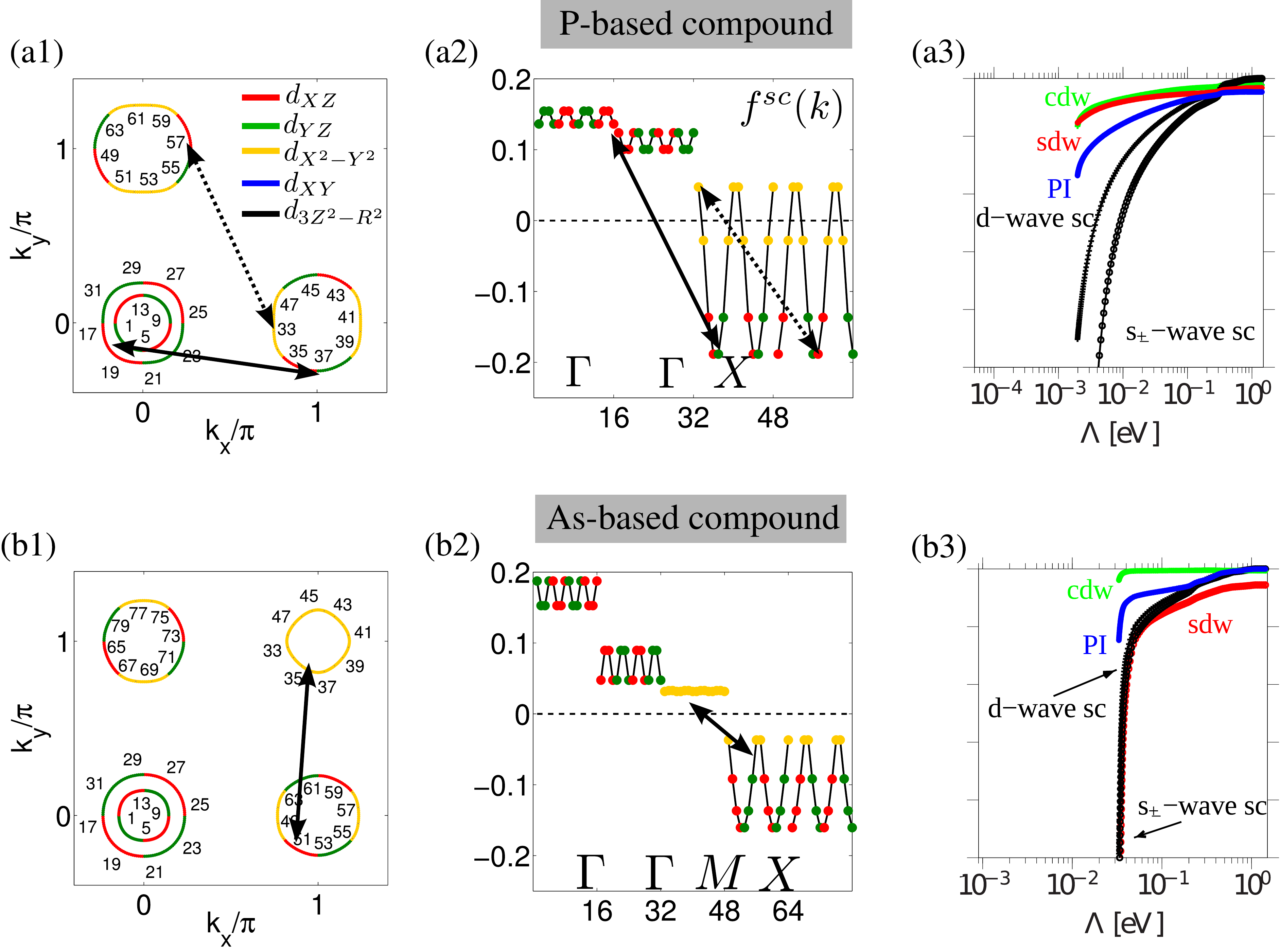}
  \end{minipage}
\end{center}
  \caption{Universal trends of FeP (a) vs. FeAs (b) based pnictide superconductors~\cite{thomale-11prl187003}, exemplified for the 1111 family. (a1) The electron-hole pocket 
  scattering along $(\pi,0)/(0,\pi)$ (full-lined arrow) is competing 
 with $(\pi/2,\pi)/(\pi,\pi/2)$ electron-electron pocket scattering (dashed-lined arrow). 
 The SC form factor (a2) yields strong, eventually nodal gap anisotropies which increases with intraorbital coupling $U_1$ (Eq.~\ref{eq:hintproto}). The channel flow (a3)
  reveals a small $T_c\approx\Lambda_c$ as well as weak spin-density wave (SDW) fluctuations. (b1) The third hole pocket at $M$, which shares orbital content with the electron 
  pockets, supports the $(\pi,0)/(0,\pi)$ scattering (full-lined arrow) and renders the SC form factor nodeless. The flow in (b3) shows a higher $T_c\approx\Lambda_c$ and also stronger SDW fluctuations.}
\label{fig1-asvp}
\end{figure}

From the BCS gap equation, one can see that a \textit{Coulomb repulsion} $W^{SC}_{pp'}$ at a finite momentum transfer \textit{can induce 
pairing only} when the wave vector of such an interaction connects regions on one FS (in the cuprate case), or regions on different FSs (in the pnictide case), 
which have \textit{opposite signs of the SC order parameter}, i.e.
\begin{equation}\label{gap}
\Delta_p = -\sum_{p'}W^{SC}_{pp'}\Delta_{p'}/2E_{p'}.
\end{equation}
This corresponds to putting the electron pairs in an \textit{anisotropic wave function such as $d$-wave} in the high-$T_c$ cuprates, or 
the \textit{sign-reversing $s$-wave ($s_{\pm}$) in the pnictides}, where in the latter case the wave vector $(\pi,0)$ in the 
unfolded Brillouin Zone connects hole (h) and electron (e) FS-pockets (i.e. $\Gamma$- and $X$-points in Fig. ~\ref{fig1-asvp}) with a 
sign-changing $s_{\pm}$-wave gap~\cite{thomale-11prl187003}. Early studies based on either RPA spin-fluctuations (SF) scenarios~\cite{mazin-08prl057003,kuroki-09prb224511} 
or analytic renormalization group (RG) studies~\cite{chubukov-08prb134512} of just one-hole and one-electron FS have reported 
a momentum-independent $s_{\pm}$-wave gap. At first glance, this similarity of the gap function obtained by so dissimilar 
approaches may appear surprising. Indeed, the repulsive part of the Coulomb interaction is treated differently in different approaches such as 
FRG and RPA, which leads to differing results for the general multi-pocket case~\cite{hanke-11anp638}.\par
The most interesting setup \textit{concerns a multi-pocket 
situation, as generally appearing in the ferro pnictides}, where more than two pockets create crucial pairing interactions (Fig.~\ref{fig1-asvp}). 
Let us look again at our \textit{prototype examples}, considered in Sec.~\ref{sec:whyare}. The DFT (LDA) band structures 
of our specific examples LaOFeAs and LaOFeP (Fig.~\ref{fig1-asvp}) are very similar: the only difference is the $d_{X^2-Y^2}$ dominated band (crossing the Fermi level at small 
h-doping in the As-, but not in the P-compound~\cite{kuroki-09prb224511}). The \textit{principal physical content of the optimization scenario} can already be observed for the 4-pocket 
Fermi surface (4pFS) and 5pFS scenarios.\par
Let us try to understand the systems starting from the unfrustrated $s_{\pm}$-limit, where the $\Gamma\leftrightarrow X$ pair scattering 
between h-pockets at $\Gamma(0,0)$ and e-pockets at $X(\pi, 0)$ is dominant. Here, a dashed-line arrow for $X\leftrightarrow X$ scattering 
(4pFS, upper FS display in Fig.~\ref{fig1-asvp}a) and a full-lined arrow (5pFS) for $X\leftrightarrow M$ in Fig.~\ref{fig1-asvp}b indicate additional 
interactions (dependent on the dominant orbital weights on the FS). The dashed interaction in Fig.~\ref{fig1-asvp}a frustrates the previous pure $s_{\pm}$-limit. 
\textit{The system then strikes a compromise by enhancing the anisotropy of the gap function} (denoted by $f^{\text{SC}}(\mathbf{k})$ in Fig.~\ref{fig1-asvp}) on 
the e-pockets at $X$ (FS positions 32 to 64 in Fig.~\ref{fig1-asvp}a), eventually reaching even a nodal situation for larger interactions. 
The dominant part of the $X\leftrightarrow X$ interaction (dashed) acts so as to push the peaks of the e-gap 
function further up (trying to achieve an $s_{\pm}$-situation for this part of the interaction), while the 
dominant part of the $(\Gamma\leftrightarrow X)$ interaction (full-lined arrow in Fig.~\ref{fig1-asvp}a) tries to push the e-gap valleys 
down (again aiming for an $s_{\pm}$-situation). Thus, a \textit{transparent understanding of the anisotropies and the nodeless versus nodal behavior emerges}: the multi-band 
SC adjusts the momentum dependence of the gap, i.e. its anisotropy, so as to \textit{minimize the effect of the Coulomb repulsion~\cite{hanke-11anp638}, which arises from frustration}.\par
In more mathematical terms, this optimization is reflected in Eq.~(\ref{eq:sc-ev}) for the dominant Cooper-channel eigenvalue $c^{\text{SC}}_1(\Lambda)$ 
taking the largest negative value:
\begin{equation}
\label{eq:sc-ev}
w^{\text{SC}}_1(\Lambda)=\langle f^{\text{SC}}({\bf k})W^{\Lambda,\text{SC}}({\bf k},-{\bf k},{\bf p})f^{\text{SC}}({\bf p})^*\rangle
\end{equation}
Here, as detailed in Eq.~\ref{harmonicdec} in Sect.~\ref{sec:micmod}, $W^{\Lambda,\text{SC}}$ denotes the
pairing function, where $\Lambda$ is the RG-flow parameter and
$f^{\text{SC}}({\bf k})$ the SC (gap) form factor associated with it. $\langle...\rangle$ stands for the
inner product and involves the ${\bf k}$- and ${\bf p}$-points on
all 4 (or 5) FS-pockets (Fig.~\ref{fig1-asvp}). We have, thus, 
\begin{equation}
\label{eq:sc-po}
w^{\text{SC}}_1(\Lambda)=\sum_{\text{FS}\ l,m}w^{\text{SC}}_{l,m}(\Lambda),
\end{equation}
and its largest negative value is determined via an optimization taking place between all
pockets $l$ and $m$. This is a frustration problem as not all
minimization conditions can be fulfilled at the same time.\par

\subsection{Pairing State with Broken Time-Reversal Symmetry}\label{sec:timereversal}
\begin{figure}[t]
\centering
   {\includegraphics[scale=0.26]{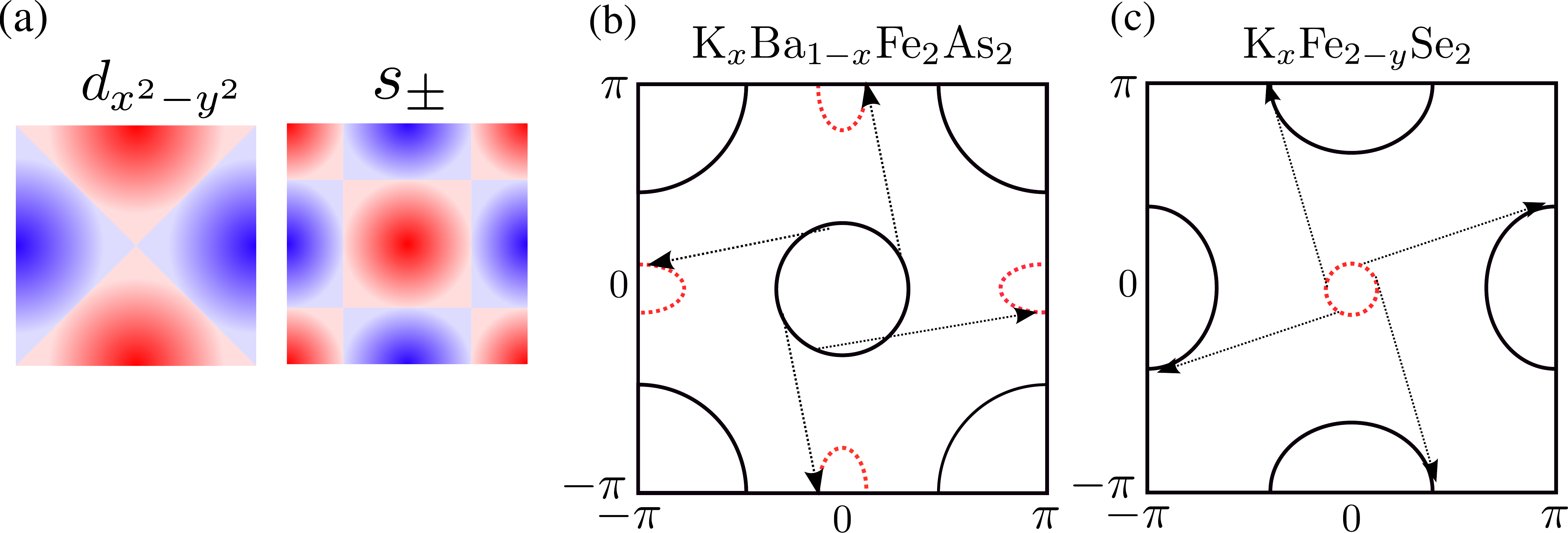}}
\caption{\label{fig:sidprop} (a) The two competing pairing symmetries in FeSCs. Frustrating the $d$-wave limits~\cite{platt-12prb180502} of
     $\text{K}_x\text{Ba}_{1-x}\text{Fe}_2\text{As}_2$ (b) and $\text{K}_x\text{Fe}_{2-y}\text{Se}_{2}$ (c). Upon doping or differently
    induced band structure effects, electron pockets appear (dashed red) in (b) and
    a hole pocket appears (dashed red) in (c), which then populate the $q \sim (\pi,0) /
    (0,\pi)$ scattering channels and enhance the $s_\pm$-wave pairing symmetry. This
    leads to frustration and provides the background for $(s+id)$-pairing.}
\end{figure}

From the results presented in the previous sections, it became apparent that the multipocket 
Fermi surfaces of FeSCs lead to a complex interplay among pairing interactions, pairing symmetries and 
Fermi-surface topology. 
Whereas a repulsive interaction between hole- and electron-pockets gives rise
to the $s_{\pm}$-wave pairing state, the interaction between the two electron-pockets 
and, as it was the case in the strongly hole-doped K$_x$Ba$_{1-x}$Fe$_2$As$_2$, the interaction within the hole-pockets
both favor $d$-wave pairing. As these two pairing symmetries cannot be satisfied simultaneously (see Fig.~\ref{fig:sidprop}a), the system may develop a
mixed $(s+id)$-pairing state, which then strikes a compromise between the two competing pairing symmetries. Of course,
this compromise is only worthwhile if the frustration between the two pairing tendencies is sufficiently strong. The resulting
$(s+id)$-pairing state then obviously breaks time-reversal symmetry (due to $\Delta_{\bs{k}}^* \neq \Delta_{\bs{k}}$) 
and shows interesting experimental signatures \cite{lee-09prl217002}.\par
Using a combined approach of functional RG and meanfield analysis, we identify the microscopic parameter regime for the $(s+id)$-state, 
which in turn provides a useful ``guiding principle'' for an experimental realization of this new pairing state.\par
In principle, there are various experimentally tunable parameters to
drive the competition between $s_{\pm}$-wave and $d$-wave pairing in FeSCs,
giving the opportunity to start from both limits. 
In $\text{K}_x\text{Ba}_{1-x}\text{Fe}_2\text{As}_2$, the Fermi surface
topology can be chosen as a paradigmatic setup for $s_\pm$-wave, consisting of hole
pockets at $\Gamma=(0,0)$ and $M=(\pi,\pi)$, as well as two electron pockets at $X=(\pi,0)/(0,\pi)$
for optimal doping $x\simeq 0.4$. Upon
increasing $x$, however, the electron pockets decrease, and have
nearly disappeared for $x=1$ as shown in Fig.~\ref{fig:sidprop}b, which has been
suggested to host a $d$-wave pairing symmetry (see Sec.~\ref{sec:exoticdwave}). In this system, it is hence plausible
that an $(s+id)$-pairing state can be realized for intermediate
values of $x$. In the chalcogenide $\text{K}_x\text{Fe}_{2-y}\text{Se}_{2}$,
the electron pockets at the $X$-points dominate, and, for a situation
seemingly inverse to $\text{K}\text{Fe}_2\text{As}_2$, a $d$-wave
pairing symmetry may likewise be expected~\cite{maier-11prb100515,PhysRevB.86.134508,wang-11epl57003}. (It
should be noted that the actual pairing symmetry in the
chalcogenides is far from settled, as a strong
coupling perspective may likewise suggest $s_\pm$-wave
pairing~\cite{fang-arXiv1105.1135}.) 
By tuning doping or other possible
parameters affecting the band structure such as pressure, one possibly induces a
pocket at the $\Gamma$-point, increasing the tendency towards $s_\pm$-wave pairing 
(see Fig.~\ref{fig:sidprop}c). In this case, one could also expect an
$(s+id)$-pairing state. By systematically tuning the Fermi-pocket topologies, one can compare the
predicted pairing symmetries with experiments, starting
from compound settings with a suspected $d$-wave symmetry (see Fig.~\ref{fig:sidprop}c).\par
In the following, we rather intend to start from an $s_{\pm}$-wave pairing
state instead, and address how we can enhance the
competitiveness of the $d$-wave symmetry to drive the system into the
$(s+id)$ regime. The reason for this is two-fold. First, the $s_{\pm}$-wave pairing
symmetry is much more generic for the different classes of FeSCs. 
Second, as we will see below, we find the most promising
setup to be located on the electron doped side of pnictides, where
high-quality samples have already been grown for different
families. We hence believe that this regime may be the experimentally
most accessible scenario at the present stage, which is why we
explicate it in detail.  In this section, we discuss the microscopic
mechanism of the $(s+id)$-pairing state by means of a functional
RG analysis of a five band model. We systematically
vary the doping level and the strength of intra-orbital interaction,
which determine the ratio between the electron-hole pocket and the
electron-electron pocket mediated pairing interactions. 
In this microscopic investigation, we
find that the $(s+id)$-pairing state can be realized in the
intermediate electron-doped regime, given that we also adjust the
pnictogen height parameter of the system appropriately.\par
We start from a representative 5-band model for the pnictides
obtained from LDA-type calculations~\cite{kuroki-08prl087004}. 
The same model has also been considered in Sec.~\ref{sec:whyare} as a starting point for explaining the
difference between the isovalent P-based and As-based 1111 compounds~\cite{thomale-11prl187003}. 
\begin{figure}[t]
\centering
   {\includegraphics[scale=0.33]{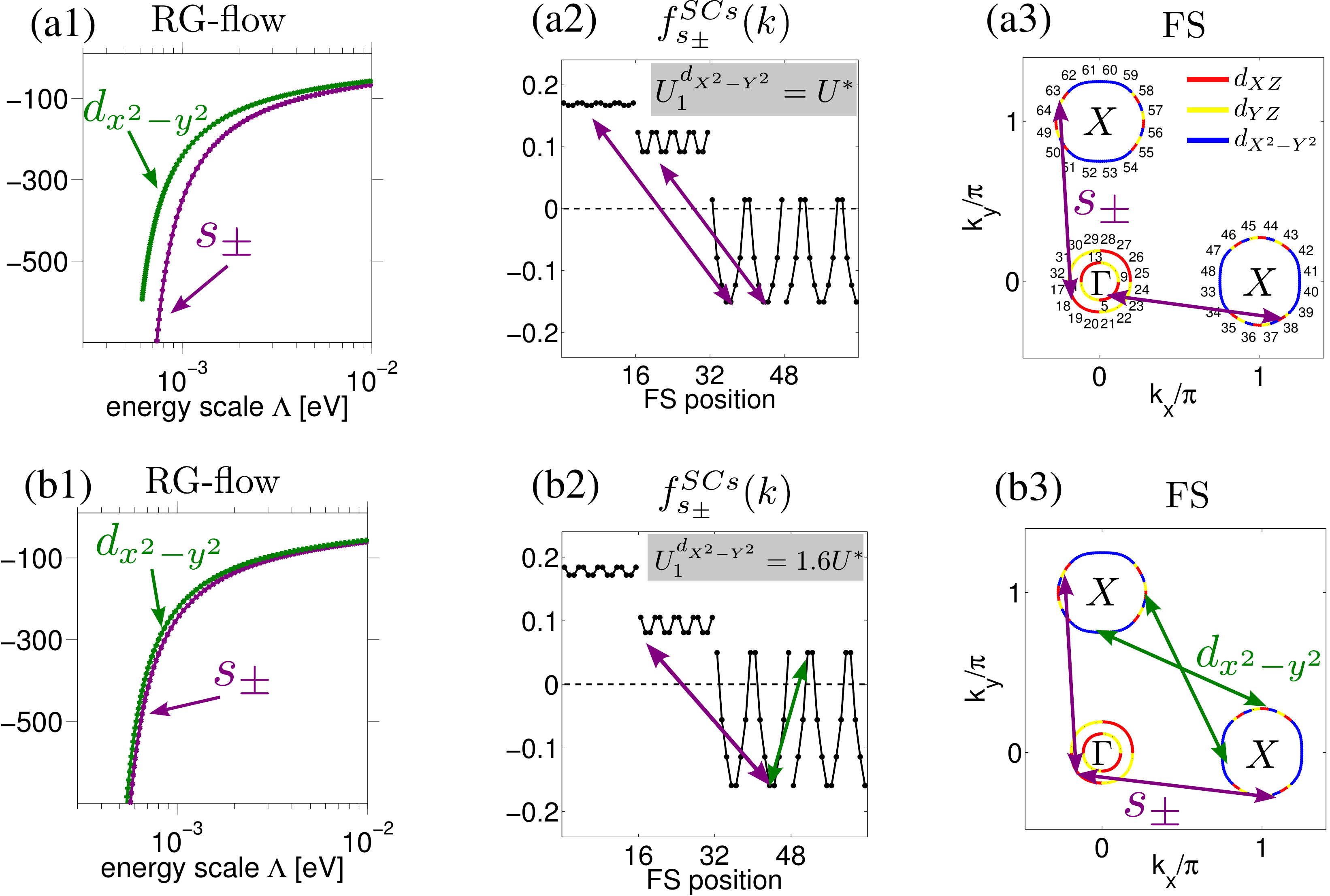}}
\caption{\label{fig:sidmech}Competing pairing orders and $s_{\pm}$-wave pairing
    form factors for $U_1(d_{X^2-Y^2})=U^*=2.5eV$ (a) and
    $U_1(d_{X^2-Y^2})=1.6U^*$ (b) at the electron doped filling of $n=6.13$~\cite{platt-12prb180502}. 
    RG channel flow (a1,b1) and $s_{\pm}$-wave gap
    form factor (a2,b2). $s_{\pm}/d$-wave
    transition from (a) to (b): increasing $U_1(d_{X^2-Y^2})$ enhances the
    gap anisotropy of the $s_{\pm}$-wave form factor on the
    electron pockets (${\bs k}$-patching: points 33-64 see (a3)) shown in
    (a2,b2) until the extended $d_{x^2-y^2}$-wave pairing becomes competitive. The
    $d$-wave form factor (not shown) does not change from (a) to (b). (a3,b3) Interactions mediated by $U_1$, inducing $s_{\pm}$-wave pairing
    tendency $(\Gamma\leftrightarrow X)$ and competing $d_{x^2-y^2}$-wave pairing symmetry due to 
    $(X\leftrightarrow X)$. (c) Variation of the
    pnictogen height $h_p$ (see Fig.~\ref{fig:bandpnic}b) mostly affects the spread of the
    $d_{X^2-Y^2}$-orbital and therefore $U_1(d_{X^2-Y^2})$, as it is oriented to the planar projection of the pnictogen.} 
\end{figure}
The situation in Fig.~\ref{fig:sidmech} is representative for moderate
electron doping and interaction scales of FeSCs, where the
$\Gamma \leftrightarrow X$ pair scattering between the hole pockets at
$\Gamma$ and the electron pockets at $X$
dominates. As discussed already in Sect.~\ref{sec:optprinciple}, 
a finite momentum transfer can induce pairing only when the wave vector of such an
interaction connects regions on one Fermi surface (such as in the cuprate case),
or regions on different Fermi surfaces (such as in the pnictide case), which have
opposite signs of the superconducting order parameter. This corresponds to putting
the electron pairs in an anisotropic wave function such as
sign-reversing $s_{\pm}$-wave in Fig.~\ref{fig:sidmech}a, where the wave
vector ($\pi,0$) in the unfolded Brillouin zone connects hole and
electron pockets with a sign-changing $s_{\pm}$-wave
gap~\cite{mazin-08prl057003,chubukov-08prb134512}.
However, in the functional RG calculation of Fig.~\ref{fig:sidmech}b with increased
$U_1$ interaction on the $d_{X^2-Y^2}$-orbital, a green arrow for
$X\leftrightarrow X$ scattering indicates additional interactions that
become similarly important as the $(\pi,0)$ channel. This increased
$U_1$ can be tuned by the pnictogen height as explained below
and frustrates the previous ``pure'' $s_{\pm}$-wave limit
($\Gamma \leftrightarrow X$). \textit{The system then strikes a
compromise~\cite{hanke-11anp638} by enhancing the anisotropy
of the $s_\pm$-wave form factor (denoted by $f^{SCs}_{s_{\pm}}(k)$ in
Fig.~\ref{fig:sidmech}) on the electron pockets at $X$}.
Throughout this variation of parameters, the sign-changing $d$-wave
form factor (not shown) remains nearly unchanged, providing nodes on the hole pockets
and gaps on the electron pockets as they do not intersect with the
nodal $d$-wave lines $k_x=\pm k_y$ in the Brillouin zone. 
\begin{figure}[t]
\centering
   {\includegraphics[scale=0.28]{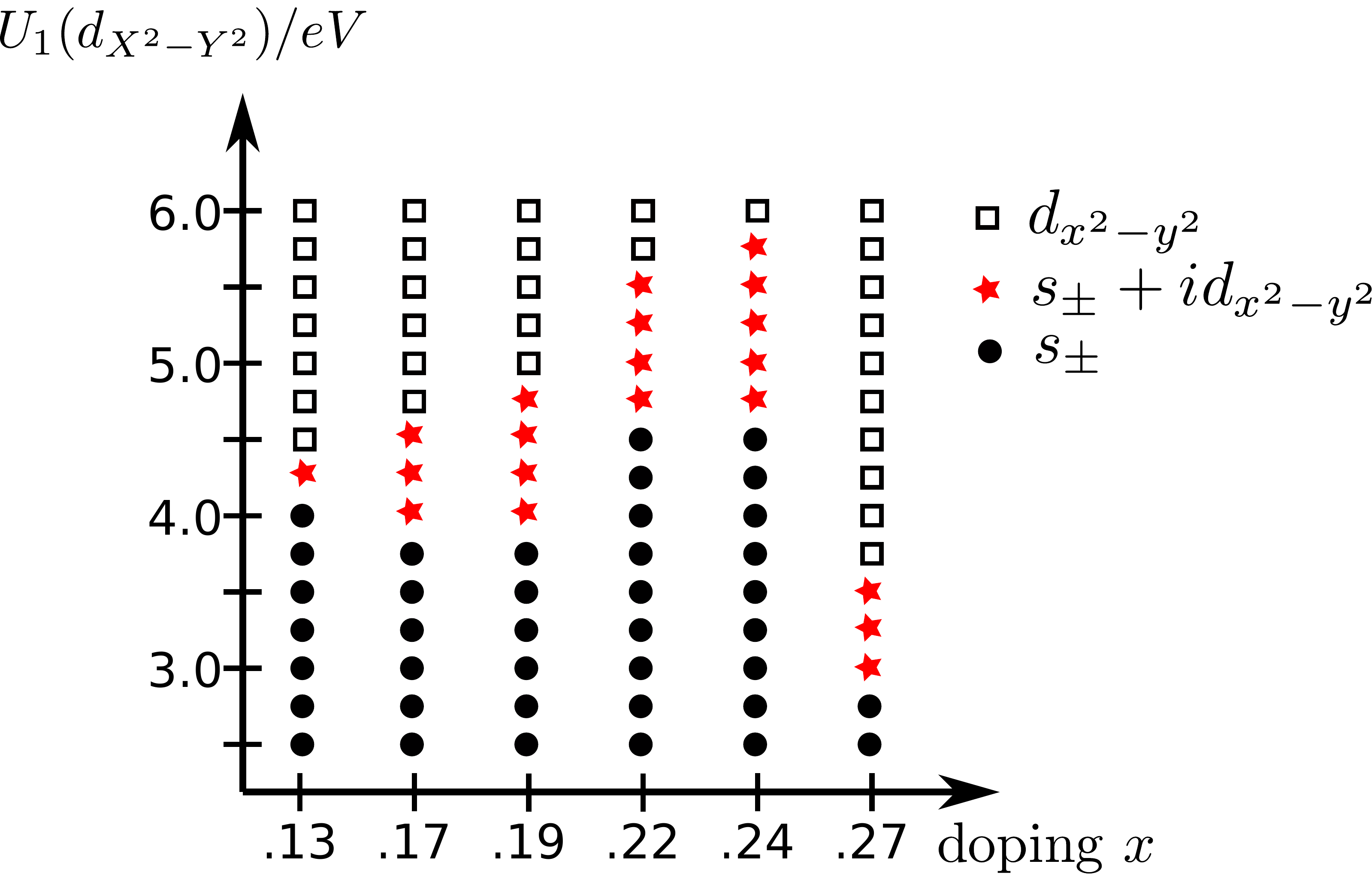}}
\caption{\label{fig:sidphase} Preferred pairing symmetry as a function
    of electron doping and intra-orbital Coulomb interaction
    $U_1(d_{X^2-Y^2})$~\cite{platt-12prb180502}. The results are
    obtained by minimizing the meanfield free energy of the effective
    theory taken from functional RG at $\Lambda\approx 0.001eV$. For $27\%$
    electron doping, the $(s+id)$-pairing state occurs at
    $U_1(d_{X^2-Y^2})=3eV$, which is comparable to the intra-orbital
    repulsion in the remaining orbitals $U_1 = 2.5eV$.} 
\end{figure}
This is because the $d$-wave-driving $X\leftrightarrow X$ scattering is
hardly affected by this change of parameters.
Instead, the $s_\pm$-wave form factor changes significantly, and adjusts the momentum dependence of the gap,
i.e. its anisotropy, so as to minimize the effect of the Coulomb
repulsion (Fig.~\ref{fig:sidmech}).\par
We now have all ingredients to \textit{tune the pairing symmetry from
$s_{\pm}$-wave to extended $d_{x^2-y^2}$-wave, and, eventually, into the
time-reversal symmetry broken $(s+id)$-phase}. In most of the FeSCs,
the tendency towards $s_{\pm}$-pairing occurs slightly more pronounced
than the competing extended $d_{x^2-y^2}$-wave pairing, and, at first glance, the
resulting frustration appears to be too small for causing
$(s+id)$-pairing.  Therefore, in order to increase frustration, we
somehow have to enhance the pair-scattering between the electron
pockets at $X$ which then promotes the subleading
$d_{x^2-y^2}$-wave channel. As shown in a-priori determinations of the
interaction parameters in~\cite{miyake-10jpsj044705}, expressed in terms of orbital matrix elements,
the pnictogen height~$h_p$ (measured from the iron-plane as pictured in Fig.~\ref{fig:bandpnic}b) has a
substantial influence on the intra-orbital interaction $U_1$ between
$d_{X^2-Y^2}$-Wannier orbitals~\cite{miyake-10jpsj044705}, which can be either
modified by isovalent doping or pressure. By increasing $h_p$, the
Wannier functions in this orbital are further localized, causing an
increase of $U_1(d_{X^2-Y^2})$. In Fig.~\ref{fig:sidmech}b, we have already
used this fact to demonstrate that, for moderate electron doping (13\%),
large values of this matrix element drive the pairing instability from
$s_{\pm}$-wave to extended $d_{x^2-y^2}$-wave symmetry. Note that in the
situation where we expect $(s+id)$-pairing to occur, both the $d$-wave and the
$s_{\pm}$-wave exhibit nodal features in the form factor.\par
For this general scenario, the predictions from FRG for time-reversal symmetry breaking is depicted in
the schematic phase diagram of Fig.~\ref{fig:sidphase}. Here, we plotted the leading $s_{\pm}$-wave, $d_{x^2-y^2}$-wave
and finally the $(s+id)$-pairing solutions as a function of $U_1(d_{X^2-Y^2})$, and electron doping.
\textit{For this purpose, we used the FRG result as a starting point for a renormalized mean-field analysis~\cite{reiss-07prb075110}}.
In this combined approach, the one-loop flow is stopped at a scale
$\Lambda$ which is small compared to the bandwidth, but still safely
above the scale $\Lambda_c$, where the 4-point function diverges. In
this range, the particular choice of the cutoff $\Lambda$ does not
significantly influence the results in Fig.~\ref{fig:sidphase}. The
renormalized coupling function
$V^{\Lambda}(k_1,k_2,k_3,k_4)$ is then taken as an
input for the mean-field treatment of the remaining modes (see Appendix~\ref{sec:mftreat}).  As shown
in Fig.~\ref{fig:sidmech}, the regime of $s_{\pm}/d$-wave pairing competition
features a single-channel pairing instability without other competing
(e.g. magnetic) instabilities and, therefore, justifies
\begin{equation}\label{meandecomp}
V^{\Lambda}(k_1, k_2, k_3, k_4) \approx W^{SCs}(k_1,k_3)\delta_{k_2,-k_1}\delta_{k_4,-k_3},
\end{equation}
with
$W^{SCs}(k_1,k_3)=V^{\Lambda}(k_1,-k_1,k_3,-k_3)$.
The effective theory for quasi-particles near the Fermi surface
$(|\xi(k)|<\Lambda)$ is modeled by the reduced Hamiltonian
\begin{equation}\label{red}
H^{\Lambda} = \sum_{ks}\xi(k)c_{ks}^{\dagger}c_{ks}^{\phantom{\dagger}} + \frac{1}{N}\sum_{k,q} W^{SCs}(k,q)c_{k\uparrow}^{\dagger}c_{-k\downarrow}^{\dagger}c_{-q\downarrow}^{\phantom{\dagger}}c_{q\uparrow}^{\phantom{\dagger}},
\end{equation}
where $\xi(k)$ is taken as the bare dispersion due to only weak
band re\-nor\-ma\-li\-za\-tion effects.  The meanfield solution of this reduced
Hamiltonian is obtained as in BCS theory, by solving the self-consistent gap-equation
and calculating the corresponding grand potential (see Eq.~(\ref{eq:omegaeval}) in Appendix~\ref{sec:mftreat}):
\begin{equation}\label{estimate}
\Omega^{stat}=-\sum_{k} \frac{|\Delta_{k}|^2 + 2\xi(k)^2}{2\sqrt{\xi(k)^2 + |\Delta_{k}|^2}} + \sum_{k}\xi(k).
\end{equation}
Within a reasonable range of parameters for the electron-doped FeSCs, we then find a regime favoring $(s+id)$-pairing due to
\begin{equation}\nonumber
\Omega^{stat}_{s+id}<\Omega^{stat}_{s_{\pm}},\Omega^{stat}_{d}.
\end{equation}
The system hence prefers to evolve into a time-reversal symmetry broken pairing
state. This is intuitive from the viewpoint of condensation energy in
the superconducting phase. While both $s_{\pm}$-wave and $d$-wave possess nodal features
individually, the combination $(s+id)$ allows to avoid the nodes
which then stabilizes the condensate.\par   
Note that the phase regime investigated by us is only a lower bound for the
possible existence of $(s+id)$-pairing which
may even be larger. This is because the functional RG setup at present only
allows us to obtain the leading pairing instability at some finite
$\Lambda_c$, while the $(s+id)$-phase may well set in below
$\Lambda_c$. This would manifest itself as a change of the superconducting phase as
a function of temperature in experiment.\par

\subsection{Summary and Outlook}
Besides its exceptionally high transition-temperatures of up to $56K$, the newly discovered class of FeSCs also attracted great interest
due to their variety of different compounds. In order to gain deeper insights into the mechanism of high-$T_c$ superconductivity,
it is therefore promising to understand the similarities and distinctions between those different compounds. For this purpose, 
we have illustrated the functional RG approach and studied the pairing symmetry as well as the underlying mechanism in different material
representatives of the FeSCs. \par
The findings presented here reveal that the pairing in FeSCs is generally driven by antiferromagnetic 
fluctuations which results at least approximate nesting features of the multi-pocket Fermi surfaces. As the mediated pairing becomes most 
effective if the superconducting gap changes sign between the nested Fermi-surface portions, the leading spin-fluctuation mode
between the hole- and electron-like pockets then gives rise to the $s_{\pm}$-wave pairing symmetry. 
However, the multi-pocket Fermi-surface as well as the multi-orbital structure of the low-lying states complicate this picture
and also lead to other spin-fluctuation modes which favor competing pairing states with, for example, $d$-wave symmetry.
Depending on this interplay of different spin-fluctuation channels, the $s_{\pm}$-wave pairing state either appears nodeless or nodal, and may even change
its symmetry to $d$-wave. At the same time, the transition temperature decreases if there are several competing pairing orders.\par
In this Chapter, we employed the functional RG to verify this picture in a microscopic model description, and we discussed 
how material specific properties such as the pnictogen height or the doping level affect the interplay of different spin-fluctuation channels. For example, we explained why the smaller 
pnictogen height in LaFePO compared to LaFeAsO causes a transition from nodeless to nodal $s_{\pm}$-wave pairing with considerably smaller $T_c$. 
The same phenomenology of a nodeless and nodal $s_{\pm}$-wave pairing also occurs in the 111 compounds, as was shown in the calculation for LiFeAs. 
We further studied the doping dependence of the pairing symmetry in Ba$_{1-x}$K$_x$Fe$_2$As$_2$ and found a transition from $s_{\pm}$-wave to $d$-wave pairing.
Finally, we also explored the conditions for a time-reversal symmetry broken $(s+id)$-pairing state by means of
a combined approach of functional RG and meanfield analysis.\par
Aiming at ``what is universal and what is more material specific'', we showed that the SC state, its gap, and in particular, its anisotropy
in momentum space (and the nodal versus nodeless behavior) is determined by an optimization principle emerging from BCS mean field theory, which optimizes the interplay between
the attractive SC channel and the Coulomb repulsion. The latter is due to the general unavoidable frustration, for example, of the $s_{\pm}$-channel.
This gives a common thread for unifying electronic structure (material) aspects of the pnictide SC.\par
Of course, there is also a variety of open questions concerning future developments in the FRG method such as the self-energy flow: In connection 
with our earlier work on the combination of the FRG with the DFT, there clearly is a need to study the impact of including the self-energy $\Sigma$ feedback and frequency-dependent interactions on 
FRG flows for our starting Hamiltonian of Eqs.~(\ref{eq:freeh0}) and~(\ref{eq:hintproto}). This particularly becomes important for multi-pocket scenarios: In the single-pocket case, the Luttinger theorem prevent the Fermi surface from significant distortions due to self energy effects unless e.g. nematic fluctuations break the given lattice symmetry. This does not hold anymore for the multi-pocket case, where a pair of pocket can in principle readily electron- and hole-dope each other without violation of any conservation law or symmetry.

Several RG studies have been performed in the past, including self-energy effects (see Ref.~\cite{giering-12prb245122} for a detailed listing). 
In the situation of an evolving FS due to self-energy flow, one faces the problem that $\bs{k}$-space cutoffs around the free FS with an 
appropriate self-energy feedback, cannot provide an adequate regularization. This poses fundamental challenges in the technical setup of FRG. 
Recently, a very promising scheme has been proposed and successfully applied by the Salmhofer group to the 2D-Hubbard model~\cite{giering-12prb245122}. 
The main idea, which facilitated the substantial progress, was the use of 
the ``channel-decomposition'' of the interaction vertex~\cite{giering-12prb245122}. It 
allows to calculate the $\omega$  and $\bs{k}$-dependence of the fermionic self-energy and the interaction vertex in the whole $\omega$-range without simplifying assumptions on its functional form. In the end, this will provide \textit{access to dynamic quantities measured in experiments}. In particular, the self-energy will provide a consistency check:
its inclusion should account for some of the known differences between low-energy ARPES data and DFT-LDA calculations.\par
Another open issue is the role of stronger correlations: In several families of recently much discussed Fe-based superconductors, such 
as the chalcogenide compounds A$_{1-x}$Fe$_{2-y}$Se$_2$ (A=K,Cs), a variety of experiments point to the emerging 
interplay of band filling and correlation effects controlling the electronic properties~\cite{werner-12nat331,liu-13jpcm125601,fang-arXiv1105.1135}. 
For example, in K$_{x}$Fe$_{2-y}$Se$_2$ systems, one finds a relatively high $T_c$ (above $30$K), a rather high N\'{e}el 
transition temperature and normal-state insulating properties at less doping, in rough analogy to the cuprates. Another prototypical compound 
of the 122 family, i.e. BaFe$_2$As$_2$, which chemically is close to FeSe, displays SC under pressure as well as h-doping and e-doping. 
There, a consensus likewise appears to emerge that stronger correlation effects (for holes than electrons) play a 
substantial role~\cite{wray-12prb144515}. In a recent analysis, P.~Werner and coworkers~\cite{werner-12nat331} found that 
this material shares properties of a strongly correlated compound with a pronounced doping dependence ranging from an incoherent 
metal (h-overdoped) over Fermi-liquid behavior (h-underdoped) to non-Fermi liquid properties near optimal doping. 
This DMFT-based work is impressive, in particular, in that it for the first time studies 
the consequences of dynamical screening of the Coulomb interaction between Fe $d$-orbital electrons. 
The scheme can be viewed as an extension of the combined DFT+DMFT method to dynamical interactions, and nurture hopes to merge them with a complementary long-wavelength view from FRG~\cite{andergas}. 
It should be mentioned, of course, that also the FRG allows for a strong-coupling analysis, so far only applied to pseudo-fermion formulations of spin systems~\cite{reuther-10prb144410,PhysRevB.83.024402,cluster}.
One future aim should be to achieve a similar broad applicability of this strong-correlation approach.
\section{Fermi surface instabilities on hexagonal lattices}
\label{chap:hexa}

It is a notable common feature of many layered compound candidates for electronically driven (high-$T_c$) superconductivity that the relevant atoms form a 
square lattice. For the cuprates, this is formed by the Cu (3$\text{d}^9$) atoms where the $d_{x^2-y^2}$ orbital largely determines the low energy degrees 
of freedom. As seen in Chap.~\ref{chap:pnictide}, most iron-based superconductors are likewise located on the square lattice where all $d$ orbitals of the 
Fe (3$\text{d}^6$) atoms become relevant. This further holds for even more exotic ruthenate compounds as the today's prime candidate for triplet superconductivity, 
where the $t_{2g}$ $d$ orbitals of the Ru atom square lattice are important~\cite{rice-95jpcm643,ruthenate}. Due to the diversity of tunable parameters and 
material classes such as found for the pnictides, the experimental evidence for many of these superconductors has already given a fairly accurate picture 
that often allows to draw connections between experimental parameter trends and 
theoretical descriptions.

The situation is different for unconventional bulk superconductors on hexagonal systems. To begin with, there are only few hexagonal materials where the 
origin of superconductivity can be unambiguously assigned to electronic interactions. Part of the reason for this is the apparently enhanced propensity 
of such hexagonal scenarios to lattice distortions, making a phonon-driven scenario of superconductivity more likely. (For example, a large class of 
organic superconductors have so far been assigned to phonon-mediated pairing. It should be noted, however, that several of them might deserve further 
investigation and that the role of electronic interaction in these materials is important as it can {\it enhance} phonon pairing~\cite{george}.) There 
are notable counterexamples where strong correlations, hexagonal symmetry, and unconventional Fermi surface instabilities come together~\cite{PhysRevLett.85.5420}. 
In particular, there are considerable indications that the Bechgaard salts constitute organic 
unconventional superconductors~\cite{PhysRevLett.95.247001,PhysRevB.88.064505}. Furthermore, SrPtAs has very recently appeared as a 
multi-layer compound where the Pt and As atoms form honeycomb rings, and initial $\mu$-SR data even shows preliminary evidence for a 
time-reversal symmetry breaking superconducting phase~\cite{PhysRevB.86.100507,srptas}. Another relevant material class on the triangular 
lattice are the water-intercalated sodium cobaltates~\cite{Takada.nature.434.53}, which are discussed in some more detail in this chapter. 
In particular, the newly emerging possibility of loading hexagonal optical lattices with fermionic isotopes~\cite{PhysRevLett.108.045305,ulf} 
of ultra-cold atomic gases might constitute another promising future playground for hexagonal Fermi surface instabilities, given that the 
limit $T<T_c, T/T_F <<1$ can eventually be reached. 

In general, however, {\it the experimental stage for unconventional hexagonal superconductors is still premature as compared to 
unconventional square lattice superconductors.} In the latter case, the diversity of different material classes, excellent crystal 
quality, and doping possibilities allow a rather complete and detailed empirical view. This similarly applies to other Fermi surface 
instabilities such as magnetism and charge density waves.

{\it Because of the comparably few prototypical examples of hexagonal scenarios, we rather preview, instead of review, fundamental 
phenomena of superconducting phases and other Fermi surface instabilities of interacting electrons on hexagonal lattices.} The 
FRG is particularly suited for this purpose, as it provides us with an approach to obtain unbiased phase diagrams of Fermi surface 
instabilties in all parquet channels, involving a rich variety of spin and charge density wave phases, triplet and singlet 
superconductivity, ferromagnetism, and Pomeranchuk instabilities on hexagonal lattices.

In Sec.~\ref{chap:cobaltates}, we investigate the multi-orbital Hubbard model on the triangular lattice. 
Motivated by the water-intercalated sodium cobaltates~\cite{Takada.nature.434.53}, we analyze a three-orbital 
model derived from CoO${}_6$ clusters~\cite{bourgeois-09prl066402} which, despite the comparably simple single-pocket 
Fermiology, exhibits a variety of different particle-particle and particle-hole instabilities. A core ingredient is 
the effect of longer-range hopping, shifting the filling of perfect nesting away from van Hove filling. The role of 
the hexagonal lattice symmetry immediately comes into play when we analzye $f$-wave and chiral $d$-wave superconductivity 
in the phase diagram. For the latter, {\it the $C_{6v}$ acts as a custodial symmetry to protect the degeneracy 
of $d_{x^2-y^2}$-wave and $d_{xy}$-wave at the instability level, which then yields chiral $d$-wave below $T_c$ 
to maximize condensation energy.} In the parameter regime relevant for the experimental superconducting cobaltate 
scenario, we explicitly illustrate how the multi-orbital character and details of the Fermi pocket contour 
affect the specific $k$-dependence of the superconducting $d$-wave form factors via a change of the particle-hole 
fluctuation profile~\cite{PhysRevLett.111.097001}.

{\it In Sec.~\ref{chap:graphene}, the discussion of hexagonal Fermi surface instabilities is transferred 
from the triangular lattice of Sec.~\ref{chap:cobaltates} to the honeycomb lattice.} Here, the multi-orbital 
nature already arises due to the two sublattices. Following proposals of unconventional superconductivity in 
graphene doped to van Hove filling~\cite{gonzalez-08prb205431,nandkishore-12natphys158,PhysRevB.86.020507}, we 
illustrate the FRG analysis of longer-range Hubbard interactions on a generalized honeycomb tight-binding model 
up to third-nearest neighbor hybridization. Several common features as compared to Sec.~\ref{chap:cobaltates} 
can be identified, such as the role of longer-range hopping in providing a distinction between van Hove filling 
and the filling of perfect nesting. In particular, we discuss the chiral $d$-wave state, which competes 
with $f$-wave further away from van Hove filling and  turns into a spin density wave state nearby van Hove 
filling beyond a certain interaction 
strength~\cite{blackschaffer-07prb13412,gonzalez-08prb205431,pathak-10prb085431,PhysRevB.86.020507,nandkishore-12prl227204,PhysRevB.85.035414}. 
We analyze how the harmonic decomposition of the superconducting form factors is directly related to the 
long-range interaction profile. Furthermore, the SDW instability is discussed in some more detail, as it 
gives a good example to illustrate that identifying the leading instability via FRG does not always yields 
a unique conclusive result. It rather triggers further investigation of possible phases following from the 
identified leading instability.

{\it Finally, Sec.~\ref{sec:theory:lattice_kagome} discusses long-range Hubbard interactions on the 
tight-binding kagome lattice~\cite{PhysRevLett.110.126405}.} As opposed to the honeycomb lattice, it 
features three sublattices, which turns out to be of fundamental importance to characterize its Fermi 
surface instabilities. While the Fermi surface topology is identical to the honeycomb model, the 
eigenstates on the kagome Fermi surface populate the three different sublattices in such a way 
that the nesting vectors that had previously been vital to characterizing Fermi surface 
instabilities in the honeycomb model are less relevant~\cite{PhysRevB.86.121105}. In turn, 
nearby and at van Hove filling, this gives rise to a remarkable competition between ferromagnetic 
fluctuations and subleading finite-$q$ particle hole fluctuations. {\it From this scenario, a series of remarkable unconventional Fermi surface instabilities emerge in the kagome Hubbard model.} This involves particle-hole condensates in charge and 
spin sector with finite relative angular momentum (charge bond order and spin bond order). 
There, the bond order forms between unequal pairs of nearest neighbor sublattice bonds, 
yielding 12 site unit cells in the ordered state. Furthermore, aside from $f$-wave superconductivity 
which is particularly promoted due to suppressed singlet-favoring SDW fluctuations, a two-fold 
degenerate $d$-wave Pomeranchuk instability emerges which opens up the possibility to break the 
lattice rotation group down to different subgroups depending on the choice of superposition.

\subsection{Triangular lattice: Multi-orbital descripton of superconductivity in $\cNa_{0.3}\cCo\cO_2$} \label{chap:cobaltates}

Electronically-driven superconductivity with large absolute $T_c$ such as found in the cuprates or the 
pnictides (Chap.~\ref{chap:pnictide}) has predominantly been constrained to layered materials which form a square lattice. 
It is apparently a non-trivial task to identify electronically-driven superconductors on the triangular lattice. 
Aside from certain heavy-fermion compounds which are beyond the range of candidate models we wish to analyze 
from the viewpoint of Fermi surface instabilities in this review, organic charge-transfer salts in principle 
provide promising arena for such phases. Therefore, it is likely that intensified research towards this 
direction will lead to further material candidates to compare against theoretical 
predictions~\cite{baskeran89prl2524,kontani,PhysRevLett.111.097001,PhysRevB.88.041103}. 
Formed by BEDT-TTF molecules, strong electronic correlations in the organic salts are indicated 
by their strongly frustrated magnetic properties, hinting at an RVB scenario suggested by 
Anderson as early as 1987~\cite{anderson87s1197}. 
{\it Drawing a more direct connection to the cuprates, however, water-intercalated sodium-doped 
cobaltates have become a further, even more prominent candidate scenario~\cite{Takada.nature.434.53}.} 
The Co atoms form a triangular lattice and take the role of the Cu atoms in the cuprates. As the most 
important difference, the whole $t_{2g}$ $d$-orbitals are crucial to adequately describe the cobaltate 
superconductors, which is also our starting point of investigation. {\it This suggests some resemblance 
to the iron pnictides where, analogously, the valence structure of the Fe atoms necessitates a multi-orbital description.} 

It is, thus, an important task to investigate the application of the multi-orbital FRG to such triangular 
lattice Hubbard models. In the following, as an exemplary scenario, we investigate a band structure model 
for the cobaltate superconductors and highlight how aspects of nesting, multi-orbital character, and 
interactions manifest themselves in the FRG analysis. Note that many phases appearing in the phase 
diagram of sodium cobaltates might be suited to be discussed from a strong-coupling perspective. For 
the superconducting phase, however, it turned out recently that an FRG perspective from intermediate 
coupling has allowed to advocate an anisotropic chiral $d$ superconductor as a candidate state of 
matter for the Na doping regime where superconductivity is found in the material. (Further connection of this proposal 
to the experimental evidence and a comprehensive discussion of alternative approaches applied to the 
cobaltates can be found in~\cite{PhysRevLett.111.097001}.) In the following, we wish to 
broadly analyze a given three-orbital model for the cobaltates regarding all phases which 
appear in the FRG analysis of it, and point out the general mechanisms of Fermiology, multi-orbital character, and interactions that give rise to such phases.

    \subsubsection{Effective Three-Band Model}
 $\cNa_x\cCo\cO_2$ (NaCoO) features alternating $\cNa$- and $\cCo\cO_2$-layers, where the latter forms a triangular lattice and the $\cNa$-layers 
 act as an electron donor. In particular, the $\cNa$-content between the $\cCo\cO_2$-layers can be suitably tuned. The doping range of $x=[0,1]$ 
 induces a diverse phase diagram, featuring superconducting and metallic phases, together with a charge-ordered insulating regimat at $x\sim 0.5$ 
 and weak-moment magnetically ordered states~\cite{foo.2004.PhysRevLett.92.247001,lang.2008.PhysRevB.78.155116,boehnke.2012.PhysRevB.85.115128}. 
 Superconductivity was detected at a doping level of $x\approx0.3$, when water is immersed to increase the inter-layer distance~\cite{Takada.nature.434.53}. 
We employ an effective model proposed by \textit{Bourgeois et al.}~\cite{bourgeois.2007.PhysRevB.75.174518, bourgeois-09prl066402} which 
takes into account all relevant orbitals within the layer, i.e. stemming from $\cCo_{3d}$ and $\cO_{2p}$. \textit{Bourgeois et al.} start 
with a $\cCo\cO_6$ cluster model and map it down to an effective three-orbital Hamiltonian. The specific model parameters were fitted to 
X-ray absorption spectroscopy (XAS) experiments, and the direct $\cCo$-$\cCo$-hopping was set to fit a FS obtained by ARPES experiments. 
While we take this specific cobaltate model as our starting point, note that its correspondence to the experimental scenario might be 
particularly constrained to the moderate Na doping regime where superconductivity occurs. We concentrate on the main conceptual 
features we can resolve in this model which, to a large extent, are generic for Hubbard models on the triangular lattice.
The resulting tight-binding model of CoO${}_6$ calculation includes three hybridized orbitals per site $(\tilde{d}_{xy},\tilde{d}_{yz},\tilde{d}_{zx})$, 
and the Hamiltonian reads
      \begin{figure}[t]
        \centering
        \includegraphics[width=0.99\linewidth]{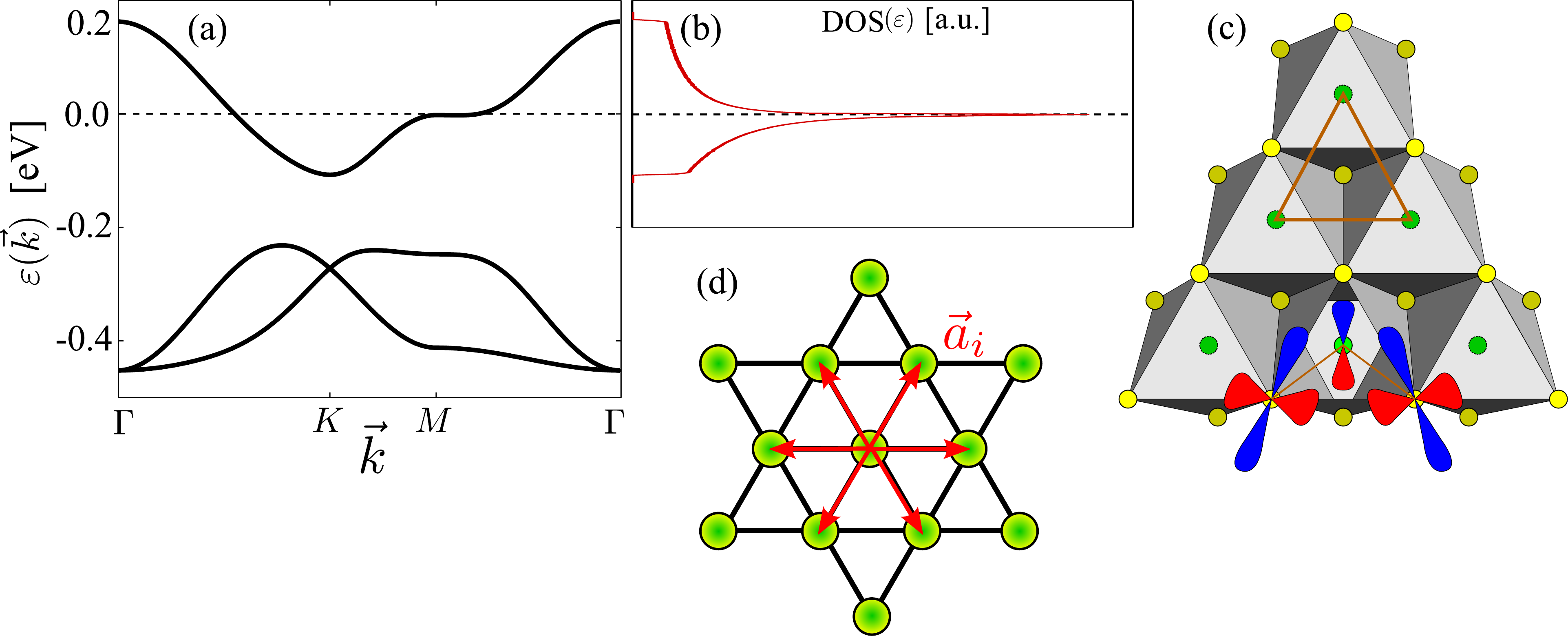}
        \caption{(a) Band structure at van Hove filling ($x=0.09$). Only one band intersects the Fermi energy, resulting in 
        one electron pocket at $\Gamma$. (b) The van Hove singularity (VHS) of this band is clearly visible in the 
        density of states (DOS). (c) The crystal structure of the $\cCo\cO_2$. The yellow circles indicate $\cCo$-sites 
        with $d_{xz}$- and $d_{yz}$-orbitals, while the green circles represent $\cO$-sites with $p_z$-orbitals. (d) 
        Simplified model of the crystal structure: A triangular lattice with three hybridized orbitals at each site. The six 
        nearest-neighbor bonds $\vec{a}_i$ are determined by $(\pm 1,0)$, $(\pm 1/2, \sqrt{3}/2)$ and $(\pm 1/2, -\sqrt{3}/2)$.}
        \label{gr:cobaltates:band_structure}
      \end{figure}
      \begin{equation}
        \begin{split}
          \hat{H} &= \sum \limits_{\langle i,j\rangle} \sum \limits_{\alpha\beta} \sum \limits_{\sigma} \left( \left(t + t'\delta_{\alpha\beta} + D \ \delta_{ij} \right) \hat{c}_{i\alpha\sigma}^{\dagger} \hat{c}_{j\beta\sigma}^{\pdagger} + \HC  \right) \\
                  &+ \mu \sum \limits_{i} \sum \limits_{\alpha} \sum \limits_{\sigma} \hat{n}_{i\alpha\sigma} + U_1 \sum \limits_{i} \sum \limits_{\alpha} \hat{n}_{i\alpha\uparrow} \hat{n}_{i\alpha\downarrow} \\
                  &+ \frac{1}{2} \sum \limits_{i} \sum \limits_{\genfrac{}{}{0pt}{}{\alpha\beta}{\alpha\neq\beta}} \left( U_2 \sum \limits_{\sigma\nu} \hat{n}_{i\alpha\sigma} \hat{n}_{i\beta\sigma\nu} + J_H \sum \limits_{\sigma\nu} \hat{c}_{i\alpha\sigma}^{\dagger} \hat{c}_{i\beta\nu}^{\dagger} \hat{c}_{i\alpha\nu}^{\pdagger} \hat{c}_{i\beta\sigma}^{\pdagger} + J_P \ \hat{c}_{i\alpha\uparrow}^{\dagger} \hat{c}_{i\alpha\downarrow}^{\dagger} \hat{c}_{i\beta\uparrow}^{\pdagger} \hat{c}_{i\beta\downarrow}^{\pdagger} \right) , \label{eq:cobaltates:hamiltonian}
        \end{split}
      \end{equation}
      where $\hat{c}_{i\alpha\sigma}^{\dagger}$ denotes the electron creation operator with spin $\sigma$ in orbital $\alpha$ at site $i$. 
      The occupation number is defined as $\hat{n}_{i\alpha\sigma} = \hat{c}_{i\alpha\sigma}^{\dagger} \hat{c}_{i\alpha\sigma}^{\pdagger}$. 
      In addition, $t$ represents the hopping mediated by $\cO_{p_z}$ and $t'$ corresponds to a direct $\cCo$-$\cCo$-hopping, $D$ is the 
      crystal-field splitting, and $\mu$ is the chemical potential. These parameters are set to $t = 0.1$eV, $t' = -0.02$eV, $D = 0.105$eV~\cite{bourgeois-09prl066402}. 
      The resulting dispersion relation (plotted in Fig.~\ref{gr:cobaltates:band_structure}a) features three bands with only one band 
      intersecting the Fermi level. There is a van Hove singularity (VHS) visible in the DOS, presented in Fig.~\ref{gr:cobaltates:band_structure}b.
      For the weak-coupling FRG calculations, generally only bands which intersect the Fermi level are considered (as discussed in Sec.~\ref{sec:frgimpl}). 
The Fermi surface features one hole pocket around the $\Gamma$-point, i.e. the center of the Brillouin zone. The evolution of 
this pocket is plotted in Figs.~\ref{gr:cobaltates:fermi_surfaces}(b-d) for doping levels $x=\{0.1,0.2,0.3\}$, respectively. 
At $x\approx0.28$, the nesting of the FS is optimal, with three inequivalent nesting wave vectors
      \begin{figure}[t]
        \centering
        \includegraphics[width=0.99\linewidth]{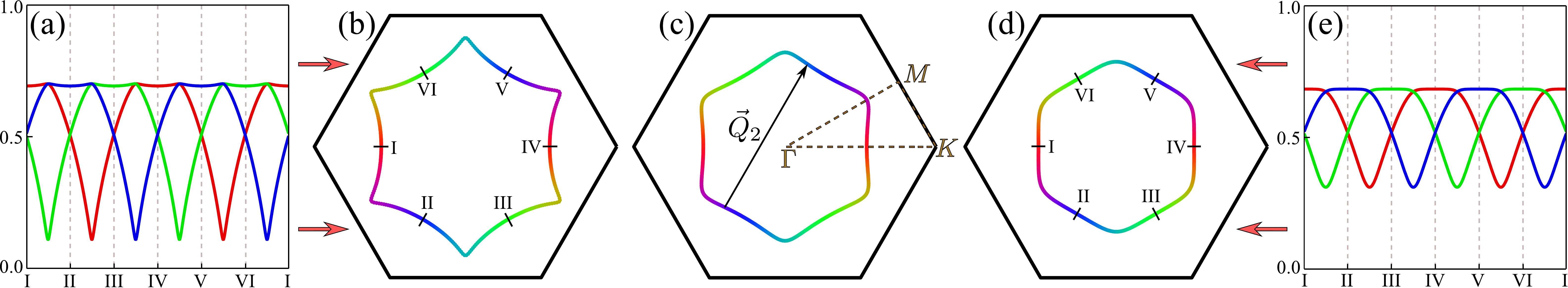}
        \caption{The Fermi surfaces (FS) at (b) $x=0.1$, (c) $x=0.2$ and  (d) $x=0.3$. The 
        different colors indicate the dominant orbital weights at the FS. An explicit plot of the orbital weights at the 
        FS is given in (a) for $x=0.1$ and (e) for $x=0.3$. The marks I to VI are 
        also plotted in the corresponding FSs. The orbital weights change along the FS but never individually decrease to zero.} 
        \label{gr:cobaltates:fermi_surfaces}
      \end{figure}
      \begin{equation}
        \begin{split}
          \vec{Q}_1 = \left( - \sqrt{2}\pi,0 \right) \quad , \quad \vec{Q}_2 = \left( \frac{\pi}{\sqrt{2}},\sqrt{\frac{3}{2}}\pi \right) \quad \text{and} \quad \vec{Q}_3 = \left( \frac{\pi}{\sqrt{2}}, - \sqrt{\frac{3}{2}}\pi \right) \ . \label{eq:cobaltes:nesting_vectors}
        \end{split}
      \end{equation}
Note that it is a generic property of the triangular lattice band structure that the location of van Hove 
filling is not identicial to the filling of perfect nesting when longer range hoppings are taken into consideration. 
As we shall see in the following, this is a recurrent motif to understand the conceptual features of Fermi surface 
instabilities for such models. This is because the unnested van Hove filling usually promotes ferromagnetic fluctuations, 
which are often overshadowed for the nearest-neighbor tight-binding models where the particle-hole fluctuations with 
finite momentum transfer dominate for conincidental perfect nesting and van Hove filling~\cite{honerkamp.2003.PhysRevB.68.104510}.

      All three hybridized orbitals contribute to the FS. The orbital weights, which are the Bogoliubov-transform matrix 
      elements of the transformation from orbital to band representation (see als Eq.~(\ref{eq:bandcoupling}) in Chap.~\ref{chap:fRG}), 
      are presented in Fig.~\ref{gr:cobaltates:fermi_surfaces}a (for $x=0.1$) and Fig.~\ref{gr:cobaltates:fermi_surfaces}e (for $x=0.3$). 
      Here, the individual effective orbitals are represented by a red, green or blue line, respectively. The dominating regions 
      of orbitals are also included in the FSs in Figs.~\ref{gr:cobaltates:fermi_surfaces}(b-d). Each orbital features two antipodal 
      dominant regions, linked by the nesting vectors $\vec{Q}_{1}$, $\vec{Q}_{2}$ or $\vec{Q}_{3}$, respectively.
      The interaction part of Eq.~(\ref{eq:cobaltates:hamiltonian}) is taken to be of general density-density form as 
      for the pnictides in Chap.~\ref{chap:fRG}, and hence features intraorbital Coulomb interaction $U_1$, interorbital 
      Coulomb interaction $U_2$, Hund's rule coupling $J_H$, and pair hopping $J_P$. Note that for the interaction 
      parameters derived from the cluster calculation~\cite{bourgeois-09prl066402}, the screening effects reducing the 
      absolute interaction scales which are kept in the calculation only derive from the electrons in the single cluster, 
      and as are underestimated. Unless stated otherwise, we hence follow~\cite{PhysRevLett.111.097001} and employ the 
      exemplary parameter values $U_1=0.37$eV, $U_2=0.25$eV, and $J_H=J_P=0.07$eV, where the ratio between the parameters 
      are motivated by Ref.~\cite{bourgeois-09prl066402}.

\subsubsection {Phase Diagram} \label{sec:cobaltates:phase_diagram}
      To get a complete picture of the possible phases of the multi-orbital model, the relation $U_1/U_2$, the doping 
      level the and global interaction scale is varied in our FRG analysis. The results are presented in Fig.~\ref{gr:cobaltates:phase_diagram}:
      \begin{figure}[t]
        \centering
        \includegraphics[width=0.7\linewidth]{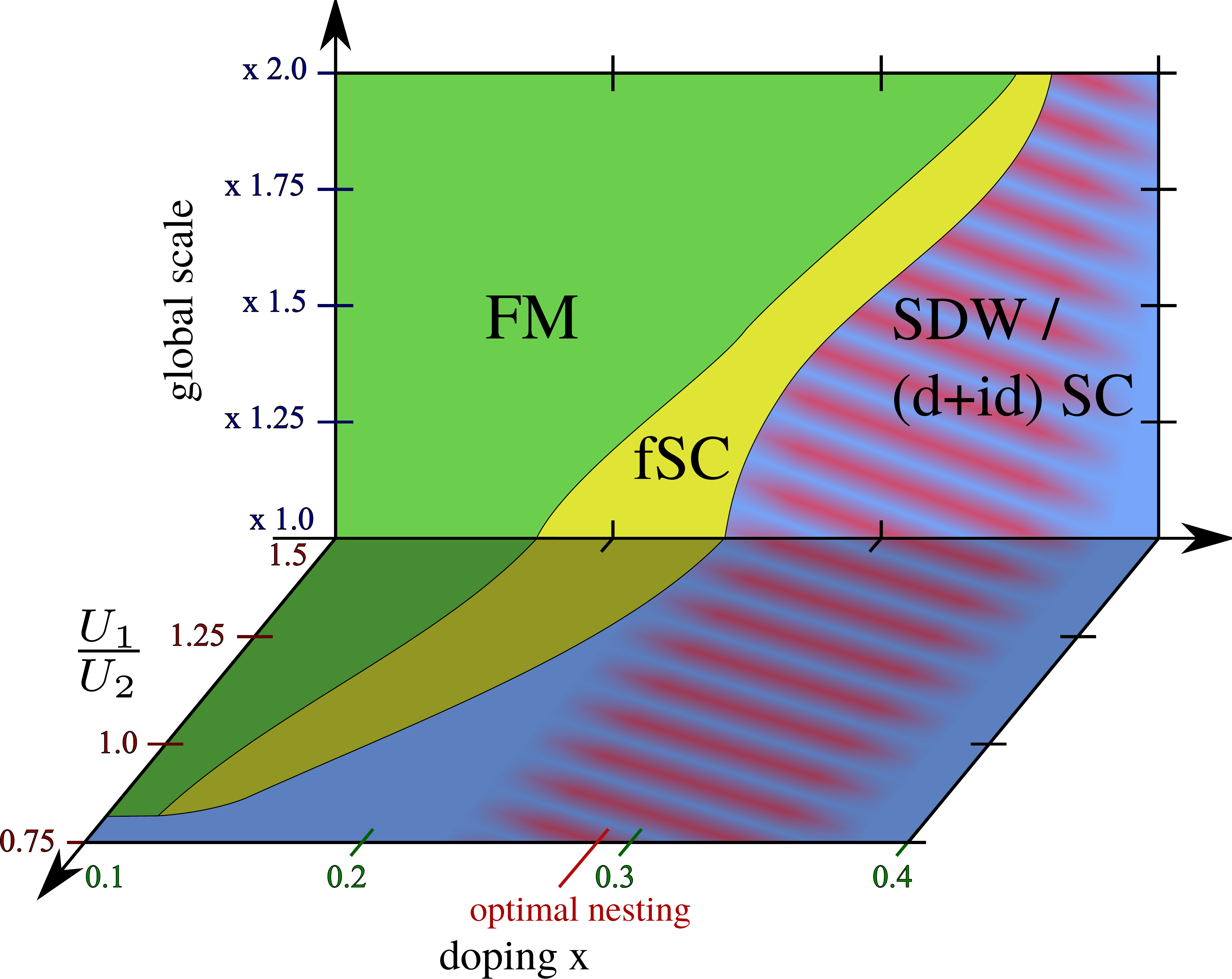}
        \caption{Phase diagram of model~(\ref{eq:cobaltates:hamiltonian}) as a function of doping, interaction ratio, 
        and interaction scale~\cite{PhysRevLett.111.097001}. There are four phases: $d$+$\ii d$-wave superconductivity ($d$+$\ii d$ SC, blue), weak 
        ferromagnetism (weak FM, green), $f$-wave superconductivity ($f$SC, yellow) and a phase with competitive 
        spin-density wave and $d$+$\ii d$-wave superconductivity (SDW / ($d$+$\ii d$) SC, purple and blue shaded). $U_1$ has 
        been varied for given $U_2=0.25$eV and $J_H=J_P=0.07$eV.}
        \label{gr:cobaltates:phase_diagram}
      \end{figure}
      For large interaction strengths, the large DOS at the VHS promotes fluctuations with zero momentum transfer, resulting 
      in weak ferromagnetism (weak FM). We define weak ferromagnetism such that the dominant channel in the FRG flow is 
      ferromagnetic, while the flow does not develop a clearly visible divergency to strong coupling in the FM channel or any other channel.  

With a lowered interaction scale, other fluctuations become more competitive. 
When the nesting of the FS is optimal, strong spin-density wave (SDW) fluctuations 
along with singlet $d$+$\ii d$-wave superconductivity occurs. For low dopings, low interaction scale and $U_1\lessapprox U_2$, 
a clean $d$+$\ii d$ SC is dominating with no competitive SDW background. Between these three phases, in the proximity of weak 
ferromagnetic fluctuations, triplet superconductivity with an $f$-wave form factor ($f$ SC) is dominant.  The form factors of 
the superconducting phases are discussed in the next section.
      We can see that as a consequence of the multiorbital character, the three-band model in Eq.~(\ref{eq:cobaltates:hamiltonian}) 
      exhibits a rich variety of phases in the particle-hole and particle-particle channel, despite its simple single-pocket 
      Fermiology. It stresses once again how fundamentally the general problem set of Fermi surface instabilities is changed 
      due to the multi-orbital character of the Fermiology, to be further analyzed in the following.
    \subsubsection {Form Factors and Superconducting Gap} \label{sec:cobaltates:form_factors}
      The superconducting state is classified by the symmetry of the Cooper-pair wave function and its spectral properties. 
      Taking into account the hexagonal lattice, i.e. the $C_{6v}$ symmetry group, the multiple degeneracies of superconducting 
      solutions at the instability level are due to higher-dimensional irreducible lattice representations (Appendix.~\ref{sec:ppphcond}). 
      This naturally suggests possible chiral superconducting phases, where the preferred superconducting solution to minimize the 
      free energy consists of a complex superposition of the superconducting form factor in order to maximize condensation energy. 
      (Note that this conclusion is generically justified in our case because the single pocket is centered around $\Gamma$, and any nodal 
      superconducting solution should generically intersect the Fermi surface and hence cost condensation energy.) Counting 
      from $l=0$ to $l=3$ relative angular momentum of the condensing Cooper pairs, taking into account the previous elaboration, 
      possible candidates 
are in principle $s$-wave (spin singlet, nodeless), $p$+$\ii p$-wave (spin triplet, nodeless), $d$+$\ii d$-wave (spin singlet, nodeless) 
and $f$-wave (spin triplet, with nodes) symmetry. Note that unless there are multiple pockets between which a sign change can be imposed 
as for the pnictides, $s$-wave can be in principle ruled out from the start because the condensing Cooper pairs pay too much energy penalty 
due to local repulsion between the electrons. This is a generic feature of itinerant superconductivity stemming from repulsive electronic 
interactions.
      
\begin{figure}[t]
        \centering
        \includegraphics[width=0.99\linewidth]{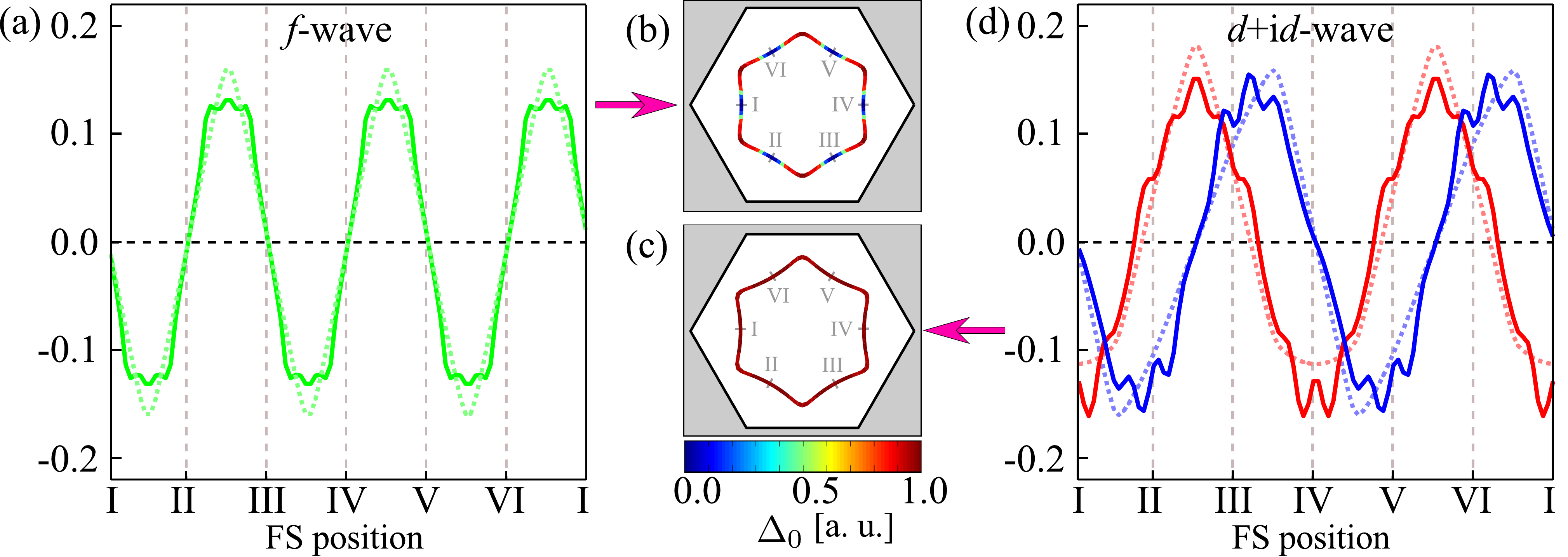}
        \caption{(a) Form factors from Eq.~(\ref{eq:cobaltates:form_factors}) (dotted lines) and the mean-field 
        decoupled pairing channel (solid lines) for $f$-wave superconducting phases with $U_1=0.32$eV, $U_2=0.25$eV, $J_H=J_P=0.07$eV, $x=0.18$. 
        (b) The corresponding gap size along the Fermi surface, where blue regions indicate nodes. The are no nodes in the $d$+$\ii d$-wave superconducting phase, plotted in \textbf{(c)}. 
        The corresponding form factors are presented in (d). (Figure was taken from \cite{PhysRevLett.111.097001} with $U_1=0.13$eV, $U_2=0.25$eV, $J_H=J_P=0.07$eV $x=0.14$.)}
        \label{gr:cobaltates:form_factors}
      \end{figure}

The phase diagram in Fig.~\ref{gr:cobaltates:phase_diagram} includes both a singlet and a triplet dominated SC phase. A mean-field type 
decoupling of the final FRG coupling function $V(\vec{k}_1,\vec{k}_2,\vec{k}_3,\vec{k}_4)$ provides the form factors associated with the 
different instabilities (Sec.~\ref{subsec:decoup}). In our case, the singlet SC instability is doubly degenerate and corresponds to 
$d$-wave symmetry ($E_2$ representation), while the triplet SC instability is non-degenerate and fulfills an $f$-wave symmetry ($B_2$ representation). 
The corresponding form factors for nearest-neighbor pairing are
      \begin{equation}
        \begin{split}
          d\text{-wave:} &\quad \begin{cases} \quad f_{E_2,1} = 2 \cos \left( k_x \right) - \cos \left( \frac{k_x - \sqrt3 k_y}{2} \right) - \cos \left( \frac{k_x + \sqrt3 k_y}{2} \right) \\ \quad f_{E_2,2} = \cos \left( \frac{k_x + \sqrt3 k_y}{2} \right) - \cos \left( \frac{k_x - \sqrt3 k_y}{2} \right) \end{cases} \\
          f\text{-wave:} &\qquad \quad f_{B_2} = \sin \left( k_y \right) - 2 \cos \left( \frac{\sqrt3 k_x}{2} \right) \sin \left( \frac{k_y}{2} \right) . \label{eq:cobaltates:form_factors}
        \end{split}
      \end{equation}
      The calculation of the gap function and the corresponding free energy of the system is explained in Appendix~\ref{sec:mftreat}. 
For the $f$-wave SC, the gap function is associated with a one-dimensional irreducible representation. It cannot take advantage of 
any complex superposition of orthogonal degenerate form factors (as is the case for $d$-wave), and hence necessarily features nodes. 
A comparison of $f$-wave and chiral $d$-wave superconductivity is presented in Fig.~\ref{gr:cobaltates:form_factors}. While there 
are nodes in the absolute gap of the $f$-wave SC (Fig.~\ref{gr:cobaltates:form_factors}b), 
the gap is nodeless and rather homogeneous for the $d$+$\ii d$-wave superconducting state at low sodium doping $x$. (Fig.~\ref{gr:cobaltates:form_factors}c).
The close competition between $d$+$\ii d$ and $f$-wave SC is, as we shall also see in subsequent discussions, a comparably 
generic feature of many Hubbard-type interaction scenarios on hexagonal lattices. $f$-wave is suggestive as it fits with the $60$ 
degree hexagonal symmetry an hence is likely to be driven by electronic interaction between these sectors. Otherwise, $d$-wave, 
due to its doubly degenerate emergence at the instability level, allows to avoid the loss of condensation energy by forming a 
chiral singlet superconducting state. It is then a question of model-dependent microscopics to identify the winner in such a scenario.


    \subsubsection {Gap Anisotropy} \label{sec:cobaltates:gap_ansiotropy}
 It is educating to further analyze the  $d$+$\ii d$ phase  as a function of system parameters. The comparably clean single-harmonic 
 form factor depicted in Fig.~\ref{gr:cobaltates:form_factors}  is located in the \emph{pure} $d$+$\ii d$ superconducting domain 
 without competing SDW background (blue region in Fig.~\ref{gr:cobaltates:phase_diagram}). If e.g. the doping is increased, however, 
 the nesting of the FS improves and as such the SDW instability becomes more competitive. 
      \begin{figure}[t]
        \centering
        \includegraphics[width=0.99\linewidth]{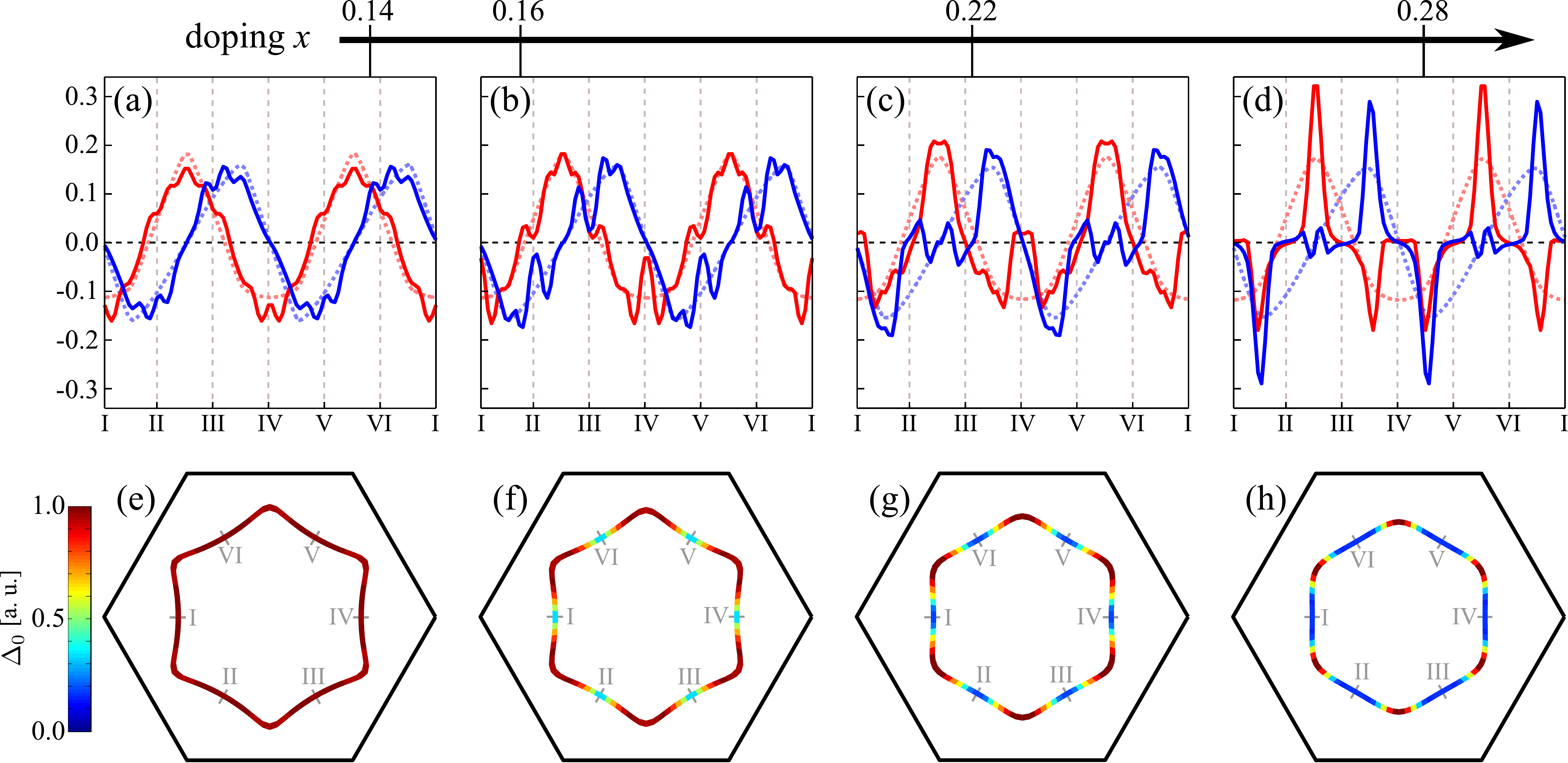}
        \caption{(a-d) Change of the $d$+$\ii d$-wave form factor on doping. The setting is $U_1=0.13$eV, $U_2=0.25$eV, $J_H=J_P=0.07$eV and 
        a doping level of (a,e) $x=0.14$, (b,f) $x=0.16$, (c,g) $x=0.22$ and (d,h) $x=0.28$. While the FRG 
        form factors (solid lines) in the pure $d$+$\ii d$ SC phase (a) match Eq.~(\ref{eq:cobaltates:form_factors}) (dotted lines) 
        accurately for low doping, this decreases with doping. When the SDW state is competitive (d), the form factors differ significantly. 
        (e-h) This influences the superconducting gap in the $d+\ii d$ phase, which develops gap dips upon doping.}
        \label{gr:cobaltates:formfactor_anisotropy}
      \end{figure}
      The doping dependence of the two degenerate $d$-wave form factors and the resulting gap is plotted in Fig.~\ref{gr:cobaltates:formfactor_anisotropy}. 
      It is remarkable that the discrepancy between the calculated FRG form factors and the analytical form factors of Eq.~(\ref{eq:cobaltates:form_factors}) 
      is strongly enhanced as the doping approaches the competitive SDW / $d$+$\ii d$ SC regime. Note, however, that we never find a scenario solely 
      dominated by some further neighbor harmonics. The calculation of the resulting gap function, again via minimization of the free energy, reveals 
      that the linear combination $d_1$+$\ii d_2$ is still energetically favorable, but the structure of the gap changes. Note that for stronger 
      doping where the gap becomes more anisotropic, the corners of the Fermi surface feature the regime with the largest gap, while the edges 
      exhibit the weakest gap. Taken to zero at the edges, this gap structure would be similar to the $f$-wave superconducting gap.    

To better analyze how the gap anisotropy is enhanced with doping, we define the variance of the gap function divided by the mean:
      \begin{figure}[t]
        \centering
        \includegraphics[width=0.7\linewidth]{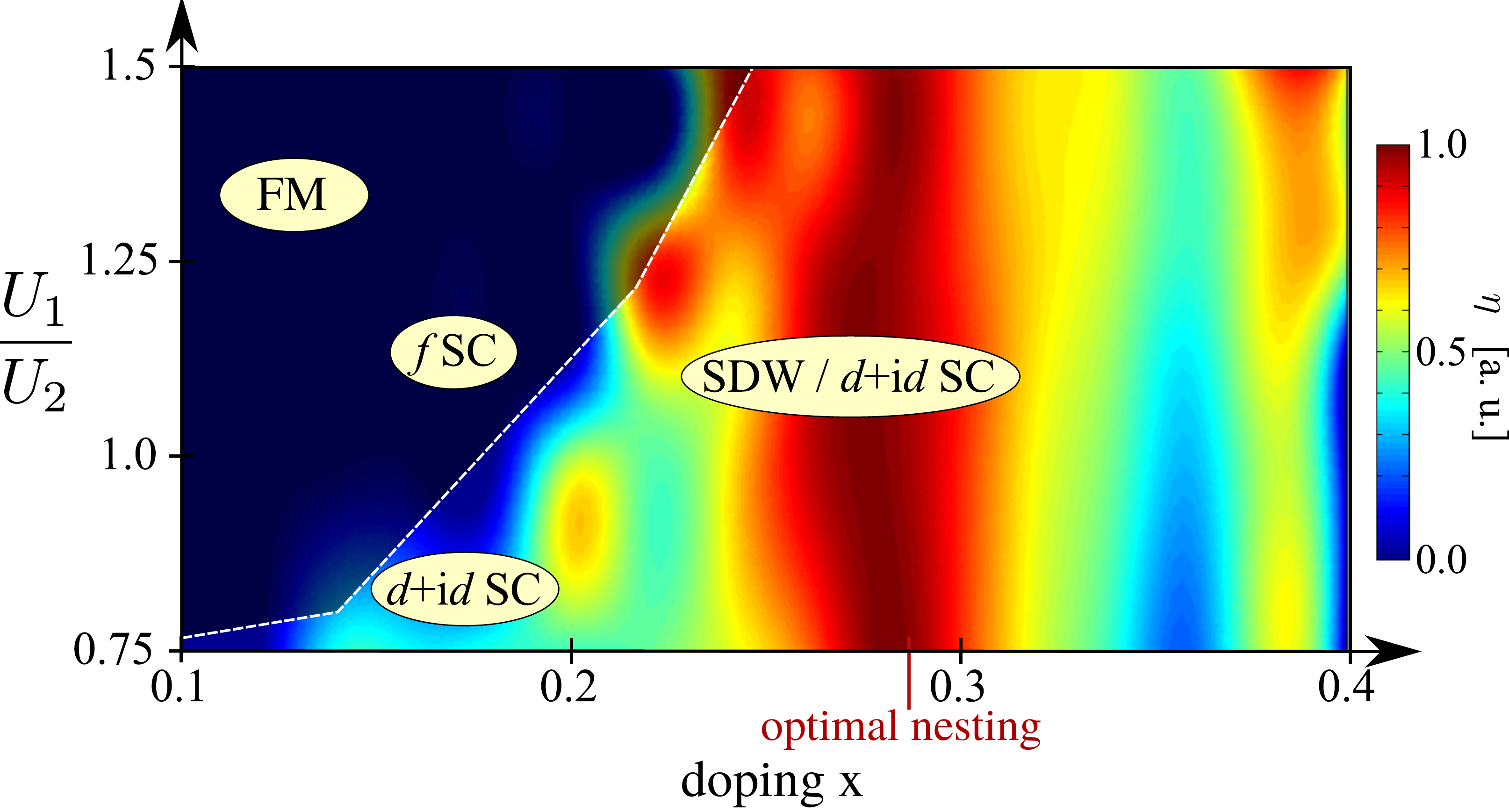}
        \caption{Gap anisotropy $\eta=\frac{\sigma(\Delta_0)}{\overline{\Delta_0}}$ in the $d$+$\ii d$-wave superconductivity phase~\cite{PhysRevLett.111.097001}. 
        Blue regions indicate a homogeneous, red regions a particularly anisotropic gap. At doping according to optimal nesting condition 
        where the competition of different SDW channels is strongest, the gap is most anisotropic.}
        \label{gr:cobaltates:gap_map}
      \end{figure}
      \begin{equation}
        \begin{split}
          \text{gap anisotropy } := \eta = \frac{\sigma(\Delta_0)}{\overline{\Delta_0}}
        \end{split}
      \end{equation}
      In Fig.~\ref{gr:cobaltates:gap_map}, $\eta$ is plotted as a function of doping $x$ and  interaction ratio $\frac{U_1}{U_2}$. The 
      maximum of $\eta$ is at a doping level $x\approx0.3$, insensitive of $\frac{U_1}{U_2}$. This implies that the proximity to the 
      optimal nesting distinctly influences the $d$+$\ii d$-wave, and as such that the Fermiology and not the interaction profile is the crucial parameter for this effect. 
      \begin{figure}[t]
        \centering
        \includegraphics[width=\linewidth]{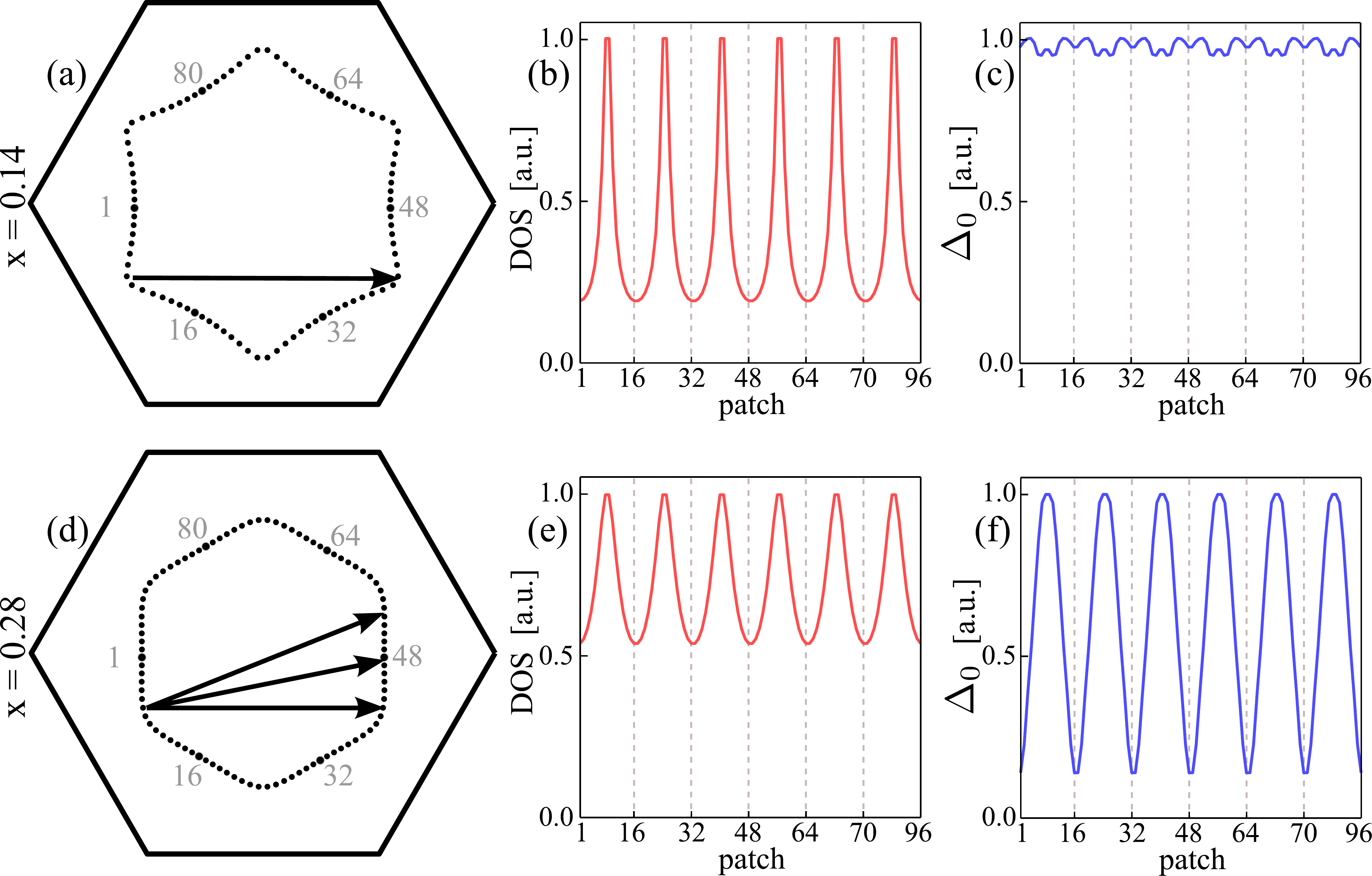}
        \caption{(a-c) $d$+$\ii d$ SC phase with $U_1=0.13$eV, $U_2=0.25$eV, $J_H=J_P=0.07$eV, $x=0.14$. (a) The 
        dominant particle hole channels connect the hot spot corners of the Fermi surface with largest density of states seen in (b). 
        Few SDW channels are promotes which results in an isotropic superconducting chiral $d$-wave gap plotted in (c).  
        (d-f) $d$+$\ii d$ SC phase with $U_1=0.32$eV, $U_2=0.25$eV, $J_H=J_P=0.07$eV, $x=0.28$. The perfect nesting (plotted in (d)) 
        extends over the whole antipodal regions of the Fermi surface, along which the density of states is more homogeneously 
        distributed (plotted in (e)). This drives various competing SDW fluctuations which yield an anisotropic superconducting gap shown in (f).
        Figure was taken from~\cite{PhysRevLett.111.097001}.}
        \label{gr:cobaltates:anisotropy}
      \end{figure}
To resolve the microscopic origin of this change of gap anisotropy, we investigate in detail the change of Fermiology and particle hole 
fluctuation profile from the pure $d$+$\ii d$ superconducting phase domain to the regime with the competitive SDW / $d$+$\ii d$ 
superconducting phase. In Fig.~\ref{gr:cobaltates:anisotropy}, the Fermi surface, the Fermi level density of states, and the 
resulting gap function is depicted for low doping ($x=0.14$) and larger doping ($x=0.28$). 
For $x=0.14$, the density of states along the Fermi surface is strongly peaked at the corners (Fig.~\ref{gr:cobaltates:anisotropy}b). This 
in turn provokes a clearly singled-out particle hole fluctuation channel, connecting the different ``hot spots". As a consequence, 
the resulting d+id gap is rather homogeneous (Fig.~\ref{gr:cobaltates:anisotropy}c). 
For $x=0.28$, i.e. the parameter regime which actually links to the experimentally observed superconducting phase in the cobaltates, 
the situation changes significantly. As the density of states is much more homogeneously distributed along the Fermi 
surface (Fig.~\ref{gr:cobaltates:anisotropy}e), the previous hot spot particle hole channels are not as pronounced as 
for low doping. In turn, along with the fact that the antipodal Fermi surface sections become more nested, several 
particle-hole channels become promoted at the same time. The resulting effect is a formation of $d+id$ where the gap 
become significantly more anisotropic (Fig.~\ref{gr:cobaltates:anisotropy}f). It becomes smallest in the regimes of 
minimal DOS along the Fermi surface, similar to the position of the nodes of the $f$-wave phase which would likewise 
do that in order to minimize the loss of condensation energy due to the nodes. This anisotropic $d+id$ state 
provides an important insight to chiral superconductors in general. Despite the 
fact that the system tends to form a chiral state in order to maximize condensation energy, the joint of effect 
of interactions, Fermiology, and multi-orbital character can often yield a rather anisotropic gap. This has 
fundamental consequences in terms of experimental observation, and, in particular, might be consistent with the 
evidence currently at hand for the cobaltates~\cite{PhysRevLett.111.097001}. From the viewpoint of FRG, it stresses 
that the careful consideration of particle-hole fluctuations as the seed for superconducting order is indispensable 
to reach an adequate description of such effects, which is exactly what the FRG can provide.

\subsection{Honeycomb lattice: Competing Instabilites in Doped Graphene}
\label{chap:graphene}
\begin{figure}[t]
\centering
   {\includegraphics[scale=0.22]{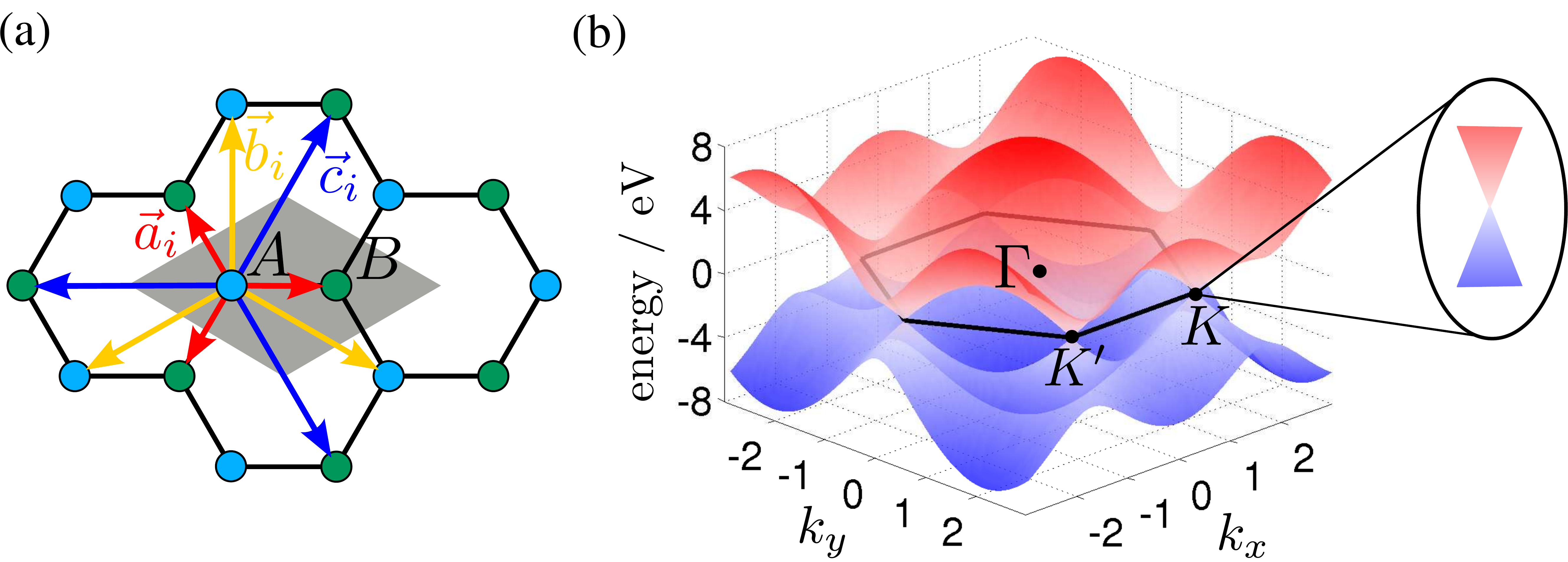}}
\caption{\label{fig:modelhex}(a) Lattice structure of graphene together with the two atomic unit-cell (gray) and the three nearest-neighbor bonds $\vec{a}_i$, $\vec{b}_i$, $\vec{c}_i$. 
Band structure of graphene for $t_1=2.8, t_2=0.7, t_3=0.02$eV within the hexagonal Brillouin zone (black) as well as the zoomed-in Dirac-cone. The first to third 
nearest-neighbor bonds are $\vec{a}_{1,2,3}=(1,0),(-1/2,\pm \sqrt{3}/2)$, 
$\vec{b}_{1,\ldots,6}=(0,\pm \sqrt{3}),(\pm 3/2,\pm \sqrt{3}/2)$ and $\vec{c}_{1,2,3}=(-2,0),(1,\pm 3/2)$.} 
\end{figure}

Besides its remarkable mechanical features, graphene predominantly generated interest due to its unique 
electronic properties~\cite{castronet09rmp109}, such as a room-temperature quantum Hall effect or the realization of Klein tunneling. 
Most of these exceptional properties can, in turn, be ascribed to the close resemblance between the Hamiltonian of graphene around half-filling and that of massless, relativistic particles.
In particular, this similarity is reflected in the respective low-energy spectrum showing linear dispersing bands (see Fig.~\ref{fig:modelhex}) as well as 
a vanishing density of states at the undoped Fermi level. 
The role of electronic interactions in this semi-metallic environment for generic Hubbard interactions on the honeycomb lattice is subject of a vivid debate.
For example, Raghu~\ea~\cite{raghu-08prl156401} pointed out that longer-ranged interactions in graphene can also generate topological Mott 
phases displaying quantum Hall and quantum spin-Hall behavior. Furthermore, Meng~\ea~\cite{meng-10nature847} claimed via quantum Monte Carlo calculations that the honeycomb Hubbard model may host an exotic spin-liquid phase at moderate local interactions $U/t \sim 4.3$ right before antiferromagnetic order  
sets in at higher $U$. Despite the fact that this proposal did not withstand further analysis of Monte Carlo studies of larger system size and more careful extrapolation procedures of the magnetization~\cite{sorella,PhysRevX.3.031010}, it confirms that there is a strong propensity of the honeycomb Hubbard model towards unconventional fluctuation profiles. 

From the viewpoint of Fermi surface instabilties, the honeycomb Hubbard model becomes more interesting as we vary 
the carrier density away from the intrinsic half-filled case ($x=1/2$) where the Fermi level density of states is vanishingly small.  Let the doping $x$ be defined by $x=n_{el}/2$, with $n_{el}$ denoting the number of electrons per site. 
According to Fig.~\ref{fig:graphenedisp}c, the density of states then increases away from half-filling and thus allows for sizable critical scales of electronic instabilities at weak coupling. (Furthermore, note that a large DOS is naturally fulfilled already for half-filling in bilayer graphene, which features quadratic rather than linear band-crossings and which, therefore, displays an interesting arena to investigate Fermi surface instabilities \cite{PhysRevB.86.075467,scherer-arXiv1112.5038,kotov-10arXiv1012.3484}.)
{\it Electron (hole) doping the single-layer graphene further away from half-filling, the Fermi level approaches one of the van-Hove singularity (VHS) at $x=3/8$ ($x=5/8$). Here, the 
diverging density of states as well as the near-nested Fermi-surface (see Fig.~\ref{fig:graphenedisp}a) suggest a variety of competing many-body phases with relatively 
high transition temperatures.}

We study the competing many-body instabilities in the doped honeycomb Hubbard model via FRG and {\it predict a rich phase diagram including topological superconductivity, 
exotic spin-order as well as spin-triplet pairing.} Although there have been
different proposals about superconductivity and magnetic ordering in doped graphene~\cite{gonzalez-08prb205431,nandkishore-12natphys158,PhysRevB.86.020507,PhysRevB.85.035414,li-12epl37001}, 
the competition between these different states as well as the dependence
on the system parameters is a challenging task for theoretical descriptions. The main reasons are that the whole Fermi surface and not just the van Hove hot spots have to be taken into consideration, and that both the particle-particle and the particle-hole channels host Fermi surface instabilities entering the phase diagram. For this reason, we apply the method of functional RG which allows an unbiased investigation of the different many-body phases
and, at the same time, also enables us to include the full band-structure details as well as longer-ranged interactions. 
\subsubsection{Model and Implementation}
Motivated but not constrained to the doped graphene scenario, we assume a single orbital at each lattice site, leaving only the two sublattices of the honeycomb setup as the multi-orbital character.
We allow the corresponding tight-binding Hamiltonian to include up to third nearest-neighbor hopping, which is then given by
\begin{align}\nonumber
H_0 = \Big[ t_1\sum_{\langle i,j\rangle,\sigma} c^{\dagger}_{i,\sigma}c^{\phantom{\dagger}}_{j,\sigma} + 
      t_2\sum_{\langle\langle i,j\rangle\rangle,\sigma}c^{\dagger}_{i,\sigma}c^{\phantom{\dagger}}_{j,\sigma} 
 +  t_3\sum_{\langle\langle\langle i,j\rangle\rangle\rangle,\sigma}c^{\dagger}_{i,\sigma}c^{\phantom{\dagger}}_{j,\sigma} + {\text h.c.} \Big] - \mu n,
\end{align}
where $n=\sum_{i,\sigma} n_{i,\sigma}=\sum_{i,\sigma} c_{i,\sigma}^\dagger c_{i,\sigma}^{\phantom{\dagger}}$ labels the electron density and $c^{\dagger}_{i,\sigma}$, 
$c^{\phantom{\dagger}}_{i,\sigma}$ denote the creation and annihilation operator of an electron with spin $\sigma$ at site~$i$. According to \cite{castronet09rmp109}, the
hopping parameters for doped graphene are determined by $t_1 = 2.8,t_2=0.1,t_3=0.07$ (in units of eV). As the unit-cell of graphene contains two atoms $A,B$ (see Fig.~\ref{fig:modelhex}a), 
the Hamiltonian $H_0$ in momentum-space representation can then be written as 
\begin{align}\label{eq:hammatgraphene}
H_0 = \sum_{\bs{k}\sigma} 
\begin{pmatrix}
c^{\dagger}_{\bs{k}A,\sigma} & c^{\dagger}_{\bs{k}B,\sigma}
\end{pmatrix}
\begin{pmatrix}
A(\bs{k})-\mu & B(\bs{k})+iC(\bs{k}) \\
B(\bs{k})-iC(\bs{k}) & A(\bs{k})-\mu
\end{pmatrix}
\begin{pmatrix}
c^{\phantom{\dagger}}_{\bs{k}A\sigma}\\
c^{\phantom{\dagger}}_{\bs{k}B\sigma}
\end{pmatrix},
\end{align}
with the following abbreviations
\begin{align}\nonumber
A(\bs{k}) &= 2t_2\sum_{j=1,2,3}\cos(\bs{k}\cdot \vec{b}_j)\\\nonumber
B(\bs{k}) &= 2t_1\sum_{j=1,2,3}\cos(\bs{k}\cdot \vec{a}_j) +  2t_3\sum_{j=1,2,3}\cos(\bs{k}\cdot \vec{c}_j)\\\nonumber
C(\bs{k}) &= 2t_1\sum_{j=1,2,3}\sin(\bs{k}\cdot \vec{a}_j) +  2t_3\sum_{j=1,2,3}\sin(\bs{k}\cdot \vec{c}_j)
\end{align}
and bond vectors $\vec{a}_i$, $\vec{b}_i$, $\vec{c}_i$ (see Fig.~\ref{fig:modelhex}a):
\begin{align}\nonumber
&\vec{a}_1 = (1,0),&\quad& \vec{b}_1 = (0,\sqrt{3}),&\quad& \vec{c}_1 = (-2,0)\\\nonumber
&\vec{a}_2  = (-1/2,\sqrt{3}/2),&\quad& \vec{b}_2 = (3/2,-\sqrt{3}/2),&\quad& \vec{c}_2 = (-1,\sqrt{3})\\\nonumber
&\vec{a}_3  = (-1/2,-\sqrt{3}/2),&\quad& \vec{b}_3  = (-3/2,-\sqrt{3}/2),&\quad& \vec{c}_3 = (-1,-\sqrt{3}).
\end{align}
The resulting band structure shown in Fig.~\ref{fig:modelhex}b then consists of two bands by diagonalizing (\ref{eq:hammatgraphene})
\begin{equation}\nonumber
E_{1,2}(\bs{k}) = \pm\sqrt{B(\bs{k})^2+C(\bs{k})^2} + A(\bs{k}) - \mu. \label{graph-band}
\end{equation} 
Most studies of the undoped graphene scenario focus on the linear dispersing part at low energies 
near the two inequivalent  (i.e. not connected through a reciprocal-lattice vector) momenta $\bs{q} = K,K'$. Here, the band structure is given by the massless Dirac particle dispersion.
\begin{figure}[t]
\centering
   {\includegraphics[scale=0.36]{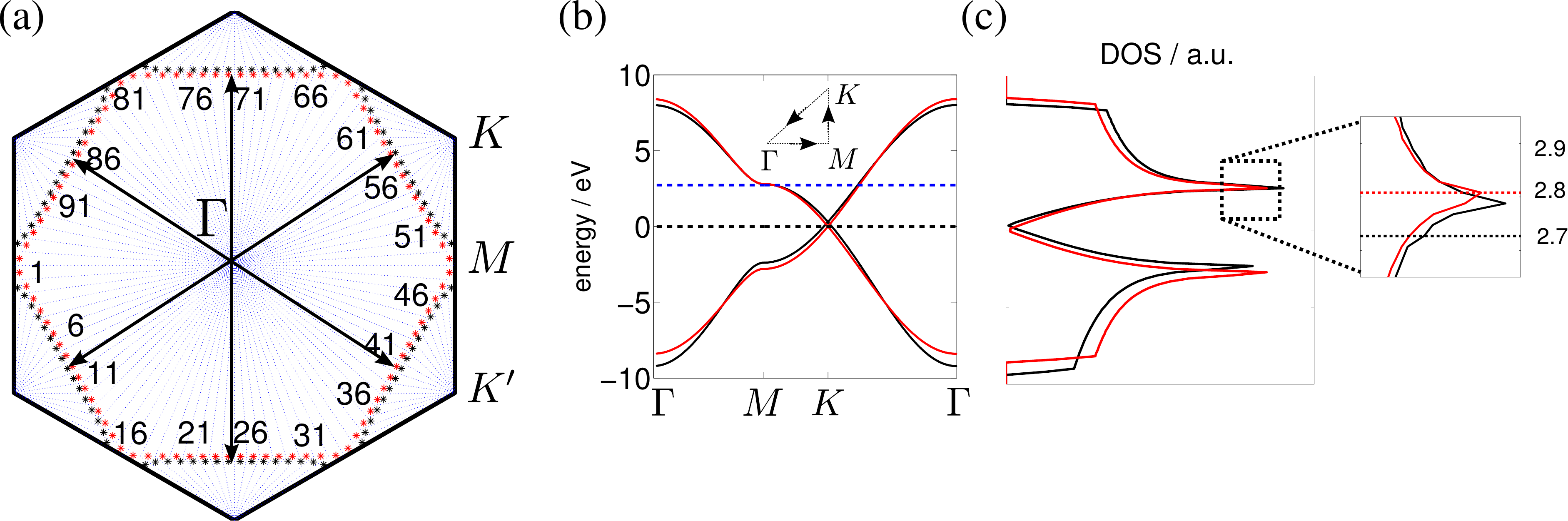}}
\caption{\label{fig:graphenedisp}
(a) Brillouin zone displaying the Fermi surface near the van Hove point (dashed blue level in (b)) with 96 patches used in the implementation of the functional RG 
as well as the (partial) nesting vectors. (b) Band structure of graphene for $t_1=2.8, t_2=t_3=0eV$ (red) and $t_1=2.8, t_2=0.1, t_3=0.07eV$ (black).  (c) Density of states for both band structures in (a). 
The inset shows the position shift of Fermi surface nesting (dashed vertical lines) versus the VHS peak.} 
\end{figure}
Note also that any diagonal part $(\propto I)$ in the Hamiltonian $H_0$ of (\ref{eq:hammatgraphene}) only shifts the Dirac-like cone in energy 
and thereby eliminates particle-hole symmetry. In order to open up a gap $(E \sim \pm v_F\sqrt{\delta\bs{q}^2 + m^2})$ without the inclusion of many-body interaction effects, the diagonal entries of $H_0$ must involve different 
signs. This requirement can for example be achieved by a staggered 
sublattice potential $(m\sigma_z)$ with $\sigma_z=\pm 1$ for the different sublattices or by a spatially varying magnetic field with zero net flux 
$(m\tau_z\sigma_z)$, where $\tau_z = \pm 1$ describes the states at $K$ and $K'$. Here, the former term only breaks inversion symmetry 
and leads to a trivial insulator, whereas the latter term breaks time-reversal symmetry and gives rise to a quantum Hall insulator~\cite{haldane-88prl2015}. Furthermore, another way of opening a gap in (\ref{graph-band}) is to include intrinsic
spin-orbit coupling $(m s_z\tau_z\sigma_z)$ with $s_z = \pm 1$ for different spins and leads to the quantum spin Hall insulator as discovered by Kane and Mele \cite{kane-05prl226801}. As we shall see, many-body interaction are another way to open a gap in graphene, e.g. via the chiral $d$-wave superconducting state as explained in the following.

For the doped case, we find that the band structure features two van-Hove singularities (VHS)  
at $x=3/8$ and $x=5/8$ (see Fig.~\ref{fig:graphenedisp}c). 
Constraining ourselves without loss of generality to the electron-doped case, the $x=5/8$ electron-like Fermi surface 
is shown in Fig.~\ref{fig:graphenedisp}b. As depicted, this is the regime of largely enhanced density of states which we investigate in the following.  
For $t_2=t_3=0$ (red curve in Fig.~\ref{fig:graphenedisp}), the VHS coincides with the partial nesting of different sections of the Fermi 
surface with $Q=(0,2\pi/\sqrt{3}), (\pi, \pi/\sqrt{3})$, and $(\pi, -\pi/\sqrt{3})$. For a realistic band structure estimate of graphene with 
finite $t_2$ and $t_3$~\cite{castronet09rmp109} (black curve in Fig.~\ref{fig:graphenedisp}), this gives an important shift of the perfect nesting position 
versus the VHS and affects the many-body phase found there. 
We assume Coulomb interactions represented by a long range Hubbard Hamiltonian~\cite{wehling-11prl236805}
\begin{equation}\label{eq:grapheneint}
H_{\text{int}}= U_{0} \sum_i n_{i, \uparrow} n_{i, \downarrow} + \frac{1}{2} U_{1} \sum_{\langle i,j\rangle, \sigma, \sigma'} n_{i, \sigma} n_{j, \sigma'}
+ \frac{1}{2} U_{2} \sum_{\langle \langle i,j\rangle \rangle, \sigma, \sigma'} n_{i, \sigma} n_{j, \sigma'},
\end{equation}
where $U_{0,1,2}$ parametrizes the Coulomb repulsion scale from onsite to second nearest-neighbor interactions. 
At the VHS, we assume perfect screening and consider $U_0$ only, while away from the VHS, we investigate the phenomenology of taking $U_1$ and $U_2$ into consideration. 
The typical scale of the effective $U_0$ for graphene has been found to be $10 eV < W$~\cite{wehling-11prl236805}, where $W\sim 17 eV$ is the kinetic bandwidth. We still allow a large range of interaction strengths to obtain an adequate description of the long-range honeycomb Hubbard model motivated, but not constrained by the actual scenario of doped graphene.
Using the above setup, following the procedure explained in Chap.~\ref{chap:fRG} and employed for the pnictides on the square lattice in Chap.~\ref{chap:pnictide}, we then employ the functional RG and study how the renormalized interaction evolves under integrating out high-energy fermionic modes. 
The renormalized interaction at an energy-scale $\Lambda$ then reads as 
$V^{\Lambda}(k_1;k_2,k_3,k_4)\gamma_{k_1s}^{\dagger}\gamma_{k_2s'}^{\dagger}\gamma_{k_3s}^{\phantom{\dagger}}\gamma_{k_4s'}^{\phantom{\dagger}},$
where the flow parameter $\Lambda$ can be interpreted as an effective temperature and $k_1$ to $k_4$ label the incoming and outgoing momenta as well as the associated band-indices. 
Note that we consider the Hamiltonian in band rather than in real space orbital representation, as the interaction is invariant under the sublattice index and hence can still be conveniently formulated in band space. As in the previous illustration of the FRG algorithm, the $k's$ are discretized to take on the values representing the 
different patches of the Brillouin zone as shown in Fig.~\ref{fig:graphenedisp}b for a 96-type patching scheme. We checked for selected representative scenarios 
that our results are converged against supercomputer simulations with 192 patch resolution.
The starting conditions of the RG are then given by the bare
interactions at an energy scale at the order of the bandwidth. Following the flow of the 4-point function $V^{\Lambda}(k_1;k_2,k_3,k_4)$ 
down to low energies, the diverging channels signal the nature of the instability.  
The corresponding $\Lambda_c$ as a function of some given system parameter such as doping gives the same 
qualitative behavior as $T_c$. At an energy scale $\Lambda$ where the leading instability starts to diverge, 
we further decompose the different channels such as the superconducting one SC or the spin-density wave channel SDW into different eigenmode 
contributions and obtain the form factors associated with the different instabilities as discussed in Sec.~\ref{subsec:decoup}. 
\begin{figure}[t]
\centering
   {\includegraphics[scale=0.20]{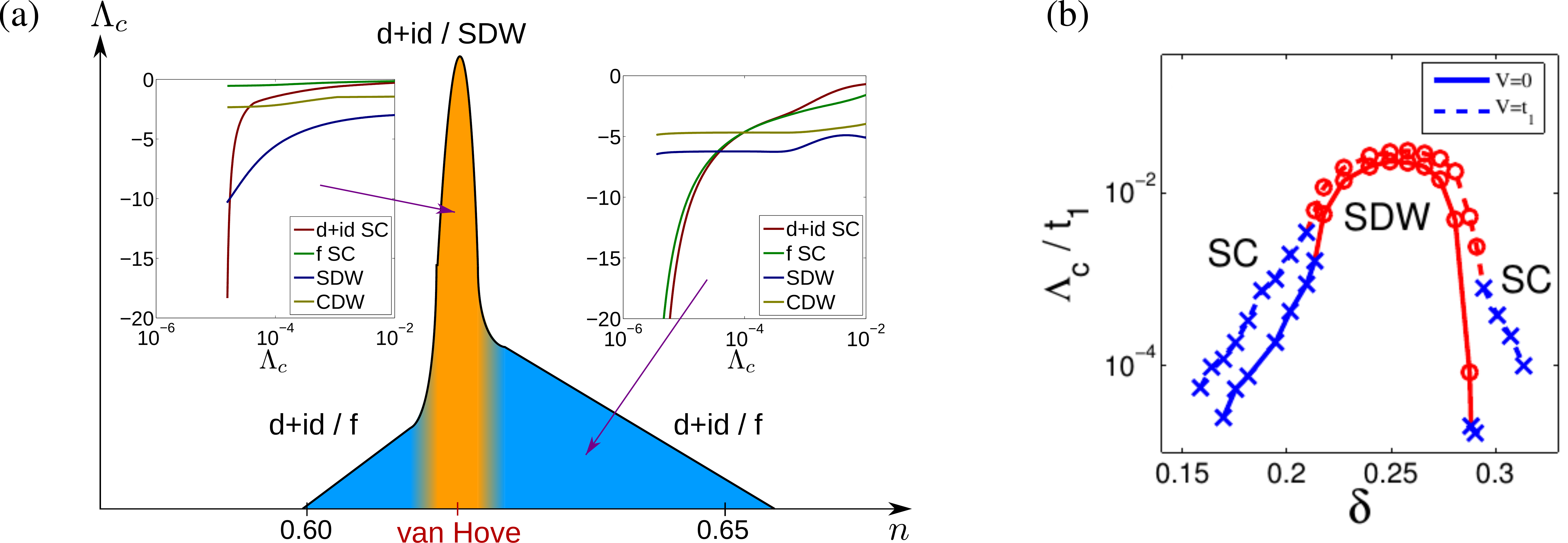}}
\label{graph-diagram}
\caption{\label{fig:graphenephase}(a) Schematic phase diagram displaying the critical instability scale $\Lambda_c \sim T_c$ as a function of doping~\cite{PhysRevB.86.020507}. 
At the van Hove singularity (VHS, light shaded (orange) area), $(d+id)$-pairing competes with the spin density wave (SDW) channel (left flow picture: dominant $(d+id)$-pairing 
instability for $U_0=10eV$ and the band structure in~\cite{mcchesney-10prl136803}). Away from the VHS (dark shaded (blue) area), $\Lambda_c$ drops 
and whether the $d+id$ or $f$-wave pairing instability is preferred depends on the long-rangedness of the interaction 
(right flow picture: $U_1/U_0=0.45$ and $U_2/U_0=0.15$). (b) Phase diagram determined by the singular-mode functional RG calculation 
of Wang~\ea~\cite{PhysRevB.85.035414} for zero and finite nearest-neighbor density-density interactions. The filling scale $n$ in (a)
and the doping scale $\delta$ in (b) translates into $n_e = 2n = 1 + \delta$ electrons per site.} 
\end{figure}
\subsubsection{Phase Diagram}
In Fig.~\ref{fig:graphenephase}, we display the tentative phase diagram of doped graphene for realistic microscopic 
model parameters~\cite{mcchesney-10prl136803,wehling-11prl236805}. Near the van-Hove singularity 
(orange-shaded area in Fig.~\ref{fig:graphenephase}), the density of states is so large that a local Hubbard description 
is appropriate. Here, we find the $(d+id)$-pairing instability to be dominant for $U_0 \sim 10 eV$, whereas a spin-density wave (SDW)
phase becomes leading only for very large scales $U_0>18eV$. If we, on the other hand, change the band structure parameter from red to black in Fig.~\ref{fig:graphenedisp}c such that the shift 
between Fermi-surface nesting and van-Hove singularity gets reduced, the enhanced nesting further promotes particle-hole fluctuations in the SDW channel along the three scattering vector $Q_1$, $Q_2$, and $Q_3$. 
This is in accordance with singular-mode FRG~\cite{PhysRevB.85.035414} (see Fig.~\ref{graph-diagram}b) where a similar change from SC to SDW is observed at van Hove filling as a function of coupling strength. In the optimal $t_1$-only case for SDW where the Fermi-surface nesting coincides with the diverging density of state,
the SDW phase already dominates for $U_0> 8.5eV$. This demonstrates how sensitive the system is towards slight modification of the Fermiology as soon as a van Hove singularity is present. Furthermore, which is beyond the current investigation, it suggests that any degree of disorder which generically spreads and damps the van Hove DOS peak will be a likewise sensitive parameter regarding the preferred Fermi surface instability of the system. (Furthermore, the role of impurities which may spoil the symmetry between 
the two $d$-wave solutions which could yield a nodal single $d$-wave gap beyond sufficient impurity concentration~\cite{florens-05prb094516}.) 
As we move away from the van-Hove singularity (blue-shaded area in Fig.~\ref{fig:graphenephase}), details of the band 
structure become less relevant. In addition, the critical instability scale $\Lambda_c$ drops stronger towards the Dirac point 
than away from it which is mainly due to the smaller density of states (see inset of Fig.~\ref{fig:graphenedisp}b). As the SDW fluctuations are weakened 
away from van-Hove filling, the pairing channels become more dominant. Again assuming rather local Coulomb interactions ($U_1/U_0 < 0.4$), we find that 
the system still favors the $(d+id)$-pairing state. However, allowing for longer-ranged Hubbard interactions, the picture changes and charge-density wave (CDW) 
fluctuations are comparable to the SDW. In this case, the singlet pairing channel which is biased by the SDW fluctuations, 
weakens and the triplet $f$-wave pairing now becomes competitive. 
\subsubsection{Chiral $(d+id)$-Pairing Phase}
\begin{figure}[t]
\centering
   {\includegraphics[scale=0.46]{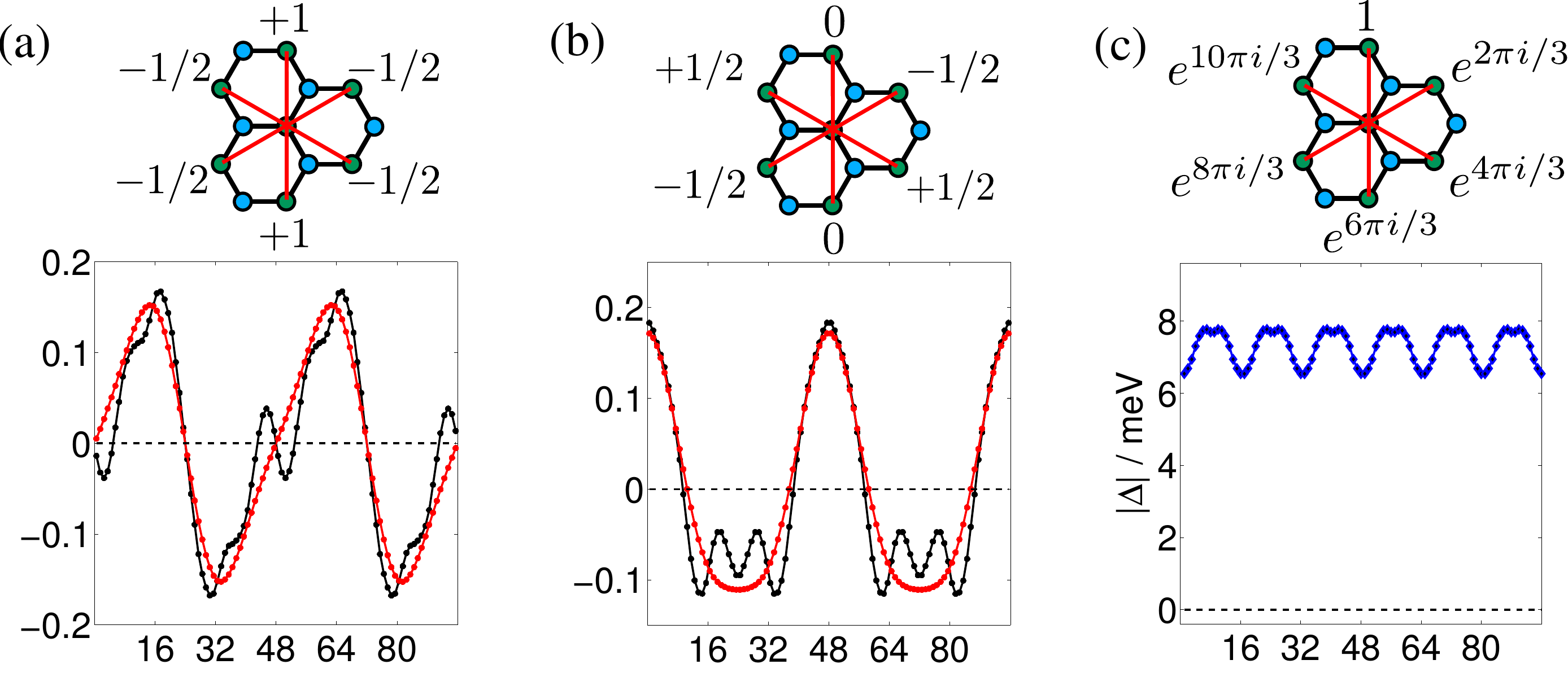}}
\caption{\label{fig:grapheneda}The $d_{x^2-y^2}$-wave (a) and $d_{xy}$-wave (b) form factors (black) for $U_0=10eV$ at van-Hove filling~\cite{PhysRevB.86.020507}, together with
an analytical fit (red) given in the text and a corresponding real-space representation (inset). All form factors are plotted along the Fermi surface
with patch indices defined in Fig.~\ref{fig:graphenedisp}b. (c) The gap profile of $d+id$ along the Fermi 
surface (actual connection to experimental energy scale can still vary by a global factor) and its real-space representation (inset) showing a $4\pi$ phase winding.
The pairing amplitudes $f_{ij}$ in the insets determine the order-parameter in real-space representation 
$O^{SCs} = \sum_{ij} f_{ij}\langle c^{\dagger}_ic^{\dagger}_j\rangle$ as described in Appendix~\ref{sec:realsp}.} 
\end{figure}

We analyze the pairing 
phase at van-Hove filling and moderate interactions. 
Applying the functional RG, we find the leading instability to occur in the pairing channel with two degenerate eigenvalues. 
The corresponding SC form factors can be fit to the following lattice harmonics
\begin{align}\nonumber
d_{x^2-y^2}(k)&=2\cos(\sqrt{3}k_y) - \cos[(\sqrt{3}k_y - 3k_x)/2] -  \cos[(\sqrt{3}k_y + 3k_x)/2] \\\nonumber
d_{xy}(k)&=\cos[(\sqrt{3}k_y - 3k_x)/2] -  \cos[(\sqrt{3}k_y + 3k_x)/2]
\end{align}
and are plotted in Fig.~\ref{fig:grapheneda}(a,b) together with their respective real-space representations. 
As the lattice structure of graphene is characterized by a $C_{6v}$ symmetry, each gap form factor must 
transform in one of the six irreducible representations of this group. Using the character table of $C_{6v}$, as derived in 
Tab.~\ref{fig:c4vc6vcharacter} of Appendix~\ref{sec:ppphcond}, it is now easy to verify that $d_{x^2-y^2}$ and $d_{xy}$ transform in the two-dimensional representation $E_2$. 
This in turn implies that all superpositions $d_{x^2-y^2} + e^{i\theta}d_{xy}$ must have the same 
transition temperatures by symmetry but do not necessarily display the same free energies. In order to figure out the superposition of lowest free energy,
we therefore optimize the corresponding mean-field free energy. The resulting gap function is then depicted 
in Fig.~\ref{fig:grapheneda}c and corresponds to the fully gapped superposition $\hat{\Delta}_{k} = d_{x^2-y^2}(k) + id_{xy}(k)$. 
This is rather generic in a situation of degenerate nodal gap form factors as the complex superposition allows 
the system to avoid gap nodes\cite{cheng-10prb024504}.
The resulting $(d+id)$-pairing state breaks time-reversal symmetry as well as the reflection symmetry with respect to one spatial coordinate (parity), rendering the phase a chiral singlet superconductor which can be characterized by a topological invariant $N$ \cite{volovik-97jetp522}. 
Similar to the equivalence of the integer valued Euler characteristic of a manifold and its integrated curvature,
this Chern number is also integer valued and equals an integrated Berry curvature. A small continuous variation of the Hamiltonian (manifold) 
can therefore never change the Chern number (Euler-characteristic) which is in this sense a topological invariant. The only possible way to change this invariant
is by closing the bulk gap or, in case of the Euler characteristic, to vary the number of holes in the manifold. A transparent way of calculating the Chern number
in case of a chiral superconductor is to determine the phase-winding of the complex gap $\hat{\Delta}_{k} = |\hat{\Delta}_{k}|e^{i\phi(k)}$ 
along the Fermi-surface \cite{volovik-97jetp522, cheng-10prb024504,qi-10prb184516}:   
\begin{equation}\label{eq:chern}
N = \frac{1}{2\pi}\oint_{FS} \nabla_{k}\phi(k)dk.
\end{equation}
From the real-space gap structure pictured in Fig.~\ref{fig:grapheneda}c, it is then apparent that the $(d+id)$-pairing state is characterized by 
Chern number $N=2$, whereas its time-reversed partner $d-id$ reveals $N=-2$. Experimentally, this topological invariant manifests itself
in universal signatures of thermal and spin Hall conductivity due to $N$ low-energy edge modes of the superconducting droplet~\cite{Read-00prb10267,senthil-99prb4245,horovitz-03prb214503}. Note that while an $N=1$ Chern Bogoliubov band suggests the existence of a single Majorana mode in a vortex core at zero energy protected by particle-hole symmetry
\cite{Ivanov01prl268,Read-00prb10267,roy-10prl186401,nayak-08rmp1083}, this vortex core profile of a chiral $d$-wave superconductor is less revealing, as the two Chern modes can recombine and gap out. However, Sato~\ea~\cite{sato-10prb134521} pointed out that the addition of Rashba spin-orbit coupling and Zeeman field in a $(d+id)$-superconductor effectively realizes
the spinless $(p+ip)$-pairing state and therefore could lead to the very same non-Abelian properties sought after.
 
Away from van-Hove filling, the screening is expected to decrease away from van-Hove level and 
we consider longer-range interactions $U_0=10eV$, $U_1=4.5eV$ and $U_2=1.5eV$. 
\begin{figure}[t]
\centering
   {\includegraphics[scale=0.47]{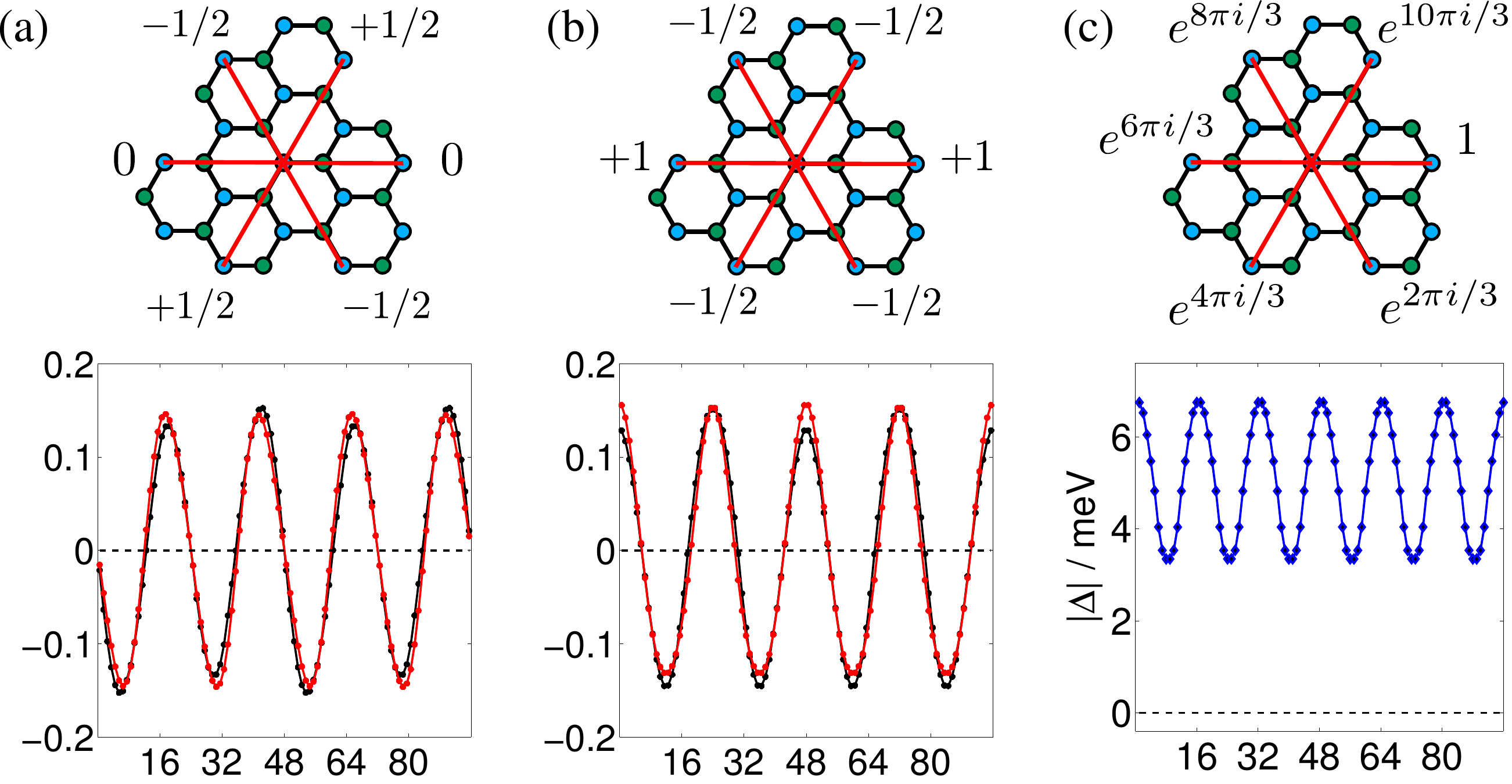}}
\caption{\label{fig:graphenedb}The $d_{x^2-y^2}$-wave (a) and $d_{xy}$-wave (b) form factors (black) plotted as in Fig.~\ref{fig:grapheneda} 
away from van-Hove filling at $x=0.65$~\cite{PhysRevB.86.020507}. Here, we used the longer-range interaction parameters $U_0=10eV$, $U_1=4.5eV$ and $U_2=1.5eV$
to account for a reduced screening. As a consequence, the real-space pairing distances (see Appendix~\ref{sec:realsp}) increase in order to avoid the longer-range Coulomb repulsion.} 
\end{figure} 
In this case, the $(d+id)$-pairing phase
remains energetically preferred but features longer-ranged real-space pairings, as can be seen by comparing Fig.~\ref{fig:grapheneda}
and Fig.~\ref{fig:graphenedb}. This behavior can in turn be understood as a way to avoid the longer-range Coulomb repulsion by means of an increased
Cooper pair distance. The resulting gap form factors can be fit by the following higher harmonic $d$-wave functions
\begin{align}\nonumber
d_{x^2-y^2}(k)&=2\cos(3k_x) - \cos[(3\sqrt{3}k_y - 3k_x)/2] -  \cos[(3\sqrt{3}k_y + 3k_x)/2] \\\nonumber
d_{xy}(k)&=\cos[(3\sqrt{3}k_y - 3k_x)/2] -  \cos[(3\sqrt{3}k_y + 3k_x)/2]
\end{align}
and are depicted in Fig.~\ref{fig:graphenedb}. Note that the additional nodes in the gap form factors also lead to a
shift in the phase winding from $4\pi$ to $8\pi$, indicating a transition from Chern number $2$ to $4$.
\subsubsection{Spin-Triplet Pairing}
\begin{figure}[t]
\centering
   {\includegraphics[scale=0.4]{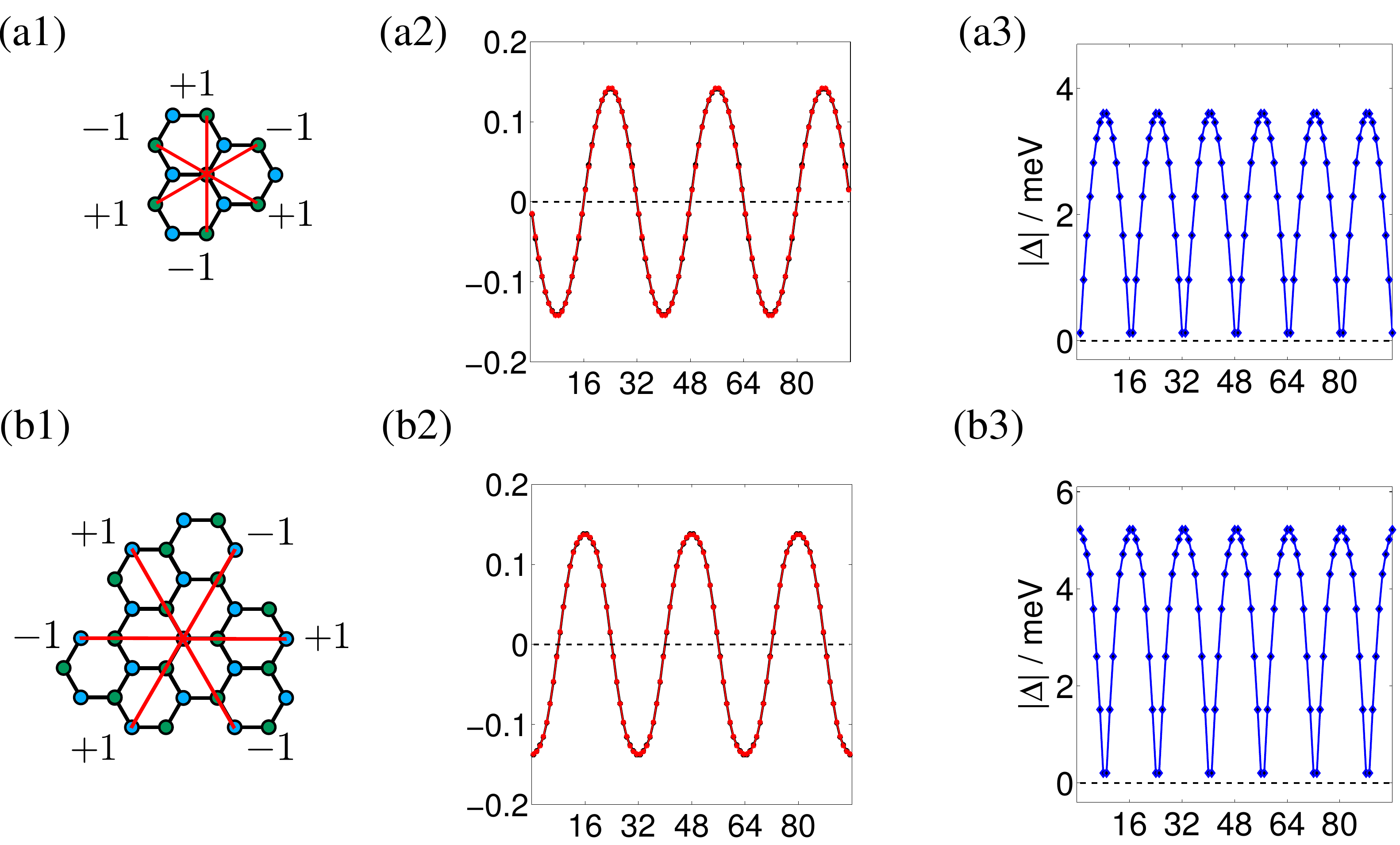}}
\caption{\label{fig:graphenefa}Real-space pairing structure, form factor, and gap profile for the $f$-wave phases~\cite{PhysRevB.86.020507}. We depicted the results for the 
representative filling of $x=0.64$ which is larger than van-Hove filling and chose $U_0=10eV$, $U_2=6eV$ for (a) and $U_0 = 10eV$, $U_1=6eV$, $U_2=2eV$ for (b). 
The gap profiles reveal nodes with positions shifting from (a) to (b).} 
\end{figure}

Besides the $(d+id)$-pairing, it is also interesting to analyze the subleading triplet pairing channel~\cite{raghu-10prb224505} 
which becomes competitive for longer-range Coulomb interaction. We therefore determine the corresponding form factors at the representative filling of 
$x=0.65$ and consider two different interaction parameter settings. The latter dependence is of particular interest as there are different proposals of tuning 
these interaction parameters by means of dielectric substrates, and would certainly be possible for a optical lattice realization of the itinerant honeycomb Hubbard model~\cite{ulf}. For $U_0 = 10eV$, $U_1=5eV$, we obtain the gap form factor shown in Fig.~\ref{fig:graphenefa}(a2), 
which can be fit to the lattice harmonic
\begin{equation}\nonumber
f_{B_1}(k)=\sin(\sqrt{3}k_y) - 2\sin(\sqrt{3}k_y/2)\cos(3k_x/2),
\end{equation}
 transforming in the one-dimensional $B_1$ representation of $C_{6v}$. 
Opposed to the $d+id$ phase, this $f$-wave pairing state has a nodal gap as can be seen from Fig.~\ref{fig:graphenefa}(a3). 
A similar situation occurs for longer-range interactions given by
$U_0 = 10eV$, $U_1=5eV$, $U_2=3eV$. Here, the gap form factor is plotted in Fig.~\ref{fig:graphenefa}(b2) and can be fit to the $B_2$ lattice harmonic  
\begin{equation}\nonumber
f_{B_2}(k)=\sin(3k_x) - 2\sin(3k_x/2)\cos(3\sqrt{3}k_y/2).
\end{equation}
Comparing the corresponding real-space pairing structures of $f_{B_1}$ and $f_{B_2}$ in Fig.~\ref{fig:graphenefa}(a1,b1), one again finds that 
the Cooper pair distance increases in order to avoid the longer-range Coulomb interaction. 
This then leads to a change of the nodal position as apparent from Fig.~\ref{fig:graphenefa}(a2,b2). 
The position of the nodes would hence indicate the Cooper pair distance associated 
with the long-range properties of the Coulomb interaction. It is important to note that in the present formulation of FRG we cannot make any statements
on the spin-structure of the triplet Cooper pair  because we remain in the spin-rotational symmetric state throughout the entire flow. 
For fillings smaller than van-Hove level, the Fermi surface 
becomes disconnected, and the system likely prefers that the nodes of the $f_{B_2}$ gap do not intersect with the Fermi surfaces. In this case, the $f$-wave pairing
phase could even appear nodeless.
\subsubsection{Spin-Density Wave Phase}\label{sec:graphenesdw}
Although the $(d+id)$-pairing is found to be the dominant instability at van-Hove filling for moderate interactions, a slight variation in the band structure parameters can
bias the system towards spin-density wave (SDW) order. Therefore, we also want to study the nature and experimental signatures of those SDW ordered phases.
As the Fermi-surface of graphene near van-Hove filling supports three inequivalent nesting vectors $Q_1,Q_2,Q_3$, there is a variety of possible
SDW states. For example, the condensation could occur at one single nesting vector $Q_k$, i.e. $\langle \vec{S}_{i}\rangle = \vec{M}e^{iQ_k\cdot r_i}$ with 
$\vec{M}$ indicating the direction of magnetization, or at all three nesting vectors $Q_k$
\begin{equation}\label{eq:sdwgraphen}
\langle \vec{S}_{i}\rangle = \vec{M}_1e^{iQ_1\cdot r_i} + \vec{M}_2e^{iQ_2\cdot r_i} + \vec{M}_3e^{iQ_3\cdot r_i}
\end{equation}
with different mutual orientations of $\vec{M}_{1,2,3}$.\par
\begin{figure}[t]
\centering
   {\includegraphics[scale=0.2]{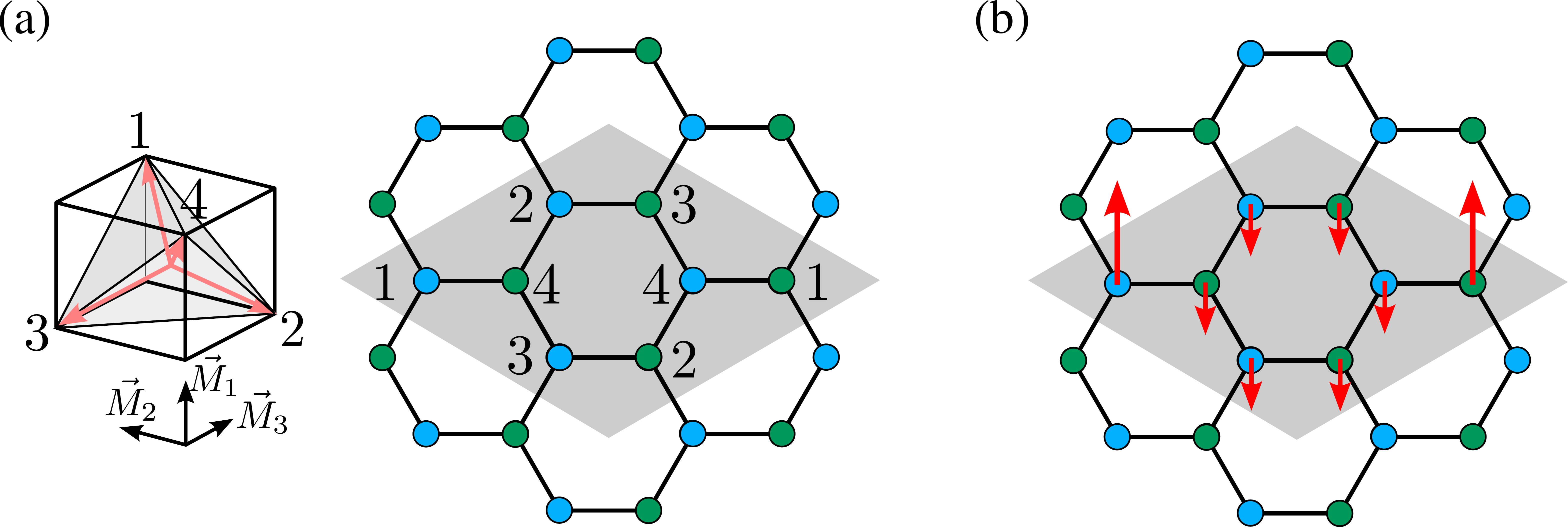}}
\caption{\label{fig:graphene_sdw} (a) Chiral SDW order as proposed in \cite{li-12epl37001,wang-12prb035414} with four neighboring spins forming
         a tetrahedral ordering pattern  (see inset). The spins on positions 1 to 4 are determined by $\vec{M}_1 + \vec{M}_2 + \vec{M}_3$, $-\vec{M}_1 - \vec{M}_2 + \vec{M}_3$,
         $\vec{M}_1 - \vec{M}_2 - \vec{M}_3$, $-\vec{M}_1 + \vec{M}_2 - \vec{M}_3$ with three mutually orthogonal vectors $\vec{M}_1,\vec{M}_2,\vec{M}_3$.
         (b) Uniaxial SDW order as suggested in \cite{nandkishore-12prl227204}. The shaded area in both subfigures indicates the respective magnetic unit cell.}
\end{figure}

In two recent works of Li~\cite{li-12epl37001} and Wang~\ea~\cite{wang-12prb035414}, a chiral SDW state (see Fig.~\ref{fig:graphene_sdw}a) was proposed for doped graphene
near van-Hove filling. Here, the four neighboring spins form a tetrahedron as shown in Fig.~\ref{fig:graphene_sdw}a, which translates into (\ref{eq:sdwgraphen}) 
with three mutually orthogonal vectors $\vec{M}_{1,2,3}$.
The resulting state is fully gapped and breaks time-reversal symmetry as well as parity due to 
the nonzero spin chirality, i.e. $\langle \vec{S}_i\cdot(\vec{S}_j\times\vec{S}_k)\rangle\neq 0$ for neighboring sites $i,j,k$.
As a consequence of this nonzero chirality, the moving electrons feel an effective internal magnetic field and feature a spontaneous Hall effect \cite{taguchi-01science2573}
with $\sigma_{xy} = e^2/h$ \cite{li-12epl37001,martin-08prl156402}.  The insulating state is therefore a Chern-insulator in the sense of a nonzero quantized Hall conductance, 
i.e. $\sigma_{xy} = ne^2/h$ with $n\neq 0$. The study of Nandkishore~\ea~\cite{nandkishore-12prl227204}, on the other hand, suggested an SDW state where all magnetic moments are  
aligned along the same axes (see Fig.~\ref{fig:graphene_sdw}b). Here, the system remains metallic with gapless states all having the same spin-projection~\cite{nandkishore-12prl227204}.\par
From the functional RG implementation described in Sec.~\ref{chap:fRG}, one can not specify which of these SDW orders are favored, as this would require higher-order vertex functions. 
However, as accomplished for the superconducting channel in Sec.~\ref{sec:timereversal}, we can take the functional RG results as an input for a subsequent mean-field analysis. 
For this purpose, we start from the leading spin-density wave correlations at an energy scale $\Lambda$ and consider the effective Hamiltonian 
\begin{equation}\nonumber
H^{\Lambda} = \sum_{ks} \epsilon(k)\gamma^{\dagger}_{ks}\gamma^{\phantom{\dagger}}_{ks} + \sum_{i}\sum_{k,q} W^{\Lambda,Q_i}(k,q)\vec{S}^{\dagger}_{k,Q_i}\vec{S}^{\phantom{\dagger}}_{q,Q_i}.
\end{equation}
Here, the spin operator $\vec{S}^{\phantom{\dagger}}_{q,Q_i} = \sum_{ss'}\gamma^{\dagger}_{qs\phantom{'}}\vec{\sigma}^{\phantom{\dagger}}_{ss'}\gamma^{\phantom{\dagger}}_{q+Q_is'}$
is defined in the basis of Bloch states $\gamma^{\dagger}|0\rangle$, and $Q_{1,2,3}$ labels the three nesting vectors $Q_1$, $Q_2$, $Q_3$.
Neglecting the quadratic fluctuations in the order-parameter field, one obtains the mean-field approximation   
\begin{align}\nonumber
H^{\Lambda}_{MF}  = & \sum_{ks} \epsilon(k)\gamma^{\dagger}_{ks}\gamma^{\phantom{\dagger}}_{ks} + 
                      \sum_i\sum_{k,ss'}\left[(\vec{m}_{ik}\cdot\vec{\sigma})^{\phantom{\dagger}}_{ss'}\gamma^{\dagger}_{ks\phantom{'}}\gamma^{\phantom{\dagger}}_{k+Q_is'} + \text{h.c.}\right]\\\nonumber
                    & \ \ \ \ \ \ \ \ \ \ \ \ \ \ \ \ \ \ \ \ \ \ -\sum_{i}\sum_{k,q} W^{\Lambda,Q_i}(k,q)\langle\vec{S}^{\dagger}_{k,Q_i}\rangle\langle\vec{S}^{\phantom{\dagger}}_{q,Q_i}\rangle\\\label{eq:resum}
                  = &\sideset{}{'}\sum_{k} \Gamma^{\dagger}_{k}A(k)\Gamma^{\phantom{\dagger}}_{k} 
                      -\sum_{i}\sum_{k,q} W^{\Lambda,Q_i}(k,q)\langle\vec{S}^{\dagger}_{k,Q_i}\rangle\langle\vec{S}^{\phantom{\dagger}}_{q,Q_i}\rangle,
\end{align}
with the following shorthand notation
\begin{align}
\vec{m}_{ik} & = \sum_q W^{\Lambda,Q_i}(k,q)\langle\vec{S}^{\phantom{\dagger}}_{q,Q_i}\rangle \\\nonumber
\Gamma^{\dagger}_k & =
\begin{pmatrix}
\gamma^{\dagger}_{k\uparrow} & \gamma^{\dagger}_{k\downarrow} & \gamma^{\dagger}_{k+Q_1\uparrow} & \gamma^{\dagger}_{k+Q_1\downarrow} 
                   & \gamma^{\dagger}_{k+Q_2\uparrow}& \gamma^{\dagger}_{k+Q_2\downarrow}
                   & \gamma^{\dagger}_{k+Q_3\uparrow}& \gamma^{\dagger}_{k+Q_3\downarrow}
\end{pmatrix}\\\label{eq:akmatrix}
A(k) & =
\begin{pmatrix}
\epsilon(k)\sigma^0                        & (\vec{m}_{1k}\cdot\vec{\sigma})                & (\vec{m}_{2k}\cdot\vec{\sigma})                 & (\vec{m}_{3k}\cdot\vec{\sigma}) \\ 
(\vec{m}_{1k}\cdot\vec{\sigma})^{\dagger}  & \epsilon(k+Q_1)\sigma^0                        & (\vec{m}_{3k+Q_1}\cdot\vec{\sigma})             & (\vec{m}_{2k+Q_1}\cdot\vec{\sigma}) \\ 
(\vec{m}_{2k}\cdot\vec{\sigma})^{\dagger}  & (\vec{m}_{3k+Q_1}\cdot\vec{\sigma})^{\dagger}  & \epsilon(k+Q_2)\sigma^0                         & (\vec{m}_{1k+Q_2}\cdot\vec{\sigma}) \\ 
(\vec{m}_{3k}\cdot\vec{\sigma})^{\dagger}  & (\vec{m}_{2k+Q_1}\cdot\vec{\sigma})^{\dagger}  & (\vec{m}_{1k+Q_2}\cdot\vec{\sigma})^{\dagger}   & \epsilon(k+Q_3)\sigma^0 
\end{pmatrix}.
\end{align}
The summation $\sideset{}{'}\sum_{k}$ in (\ref{eq:resum}) restricts to the magnetic Brillouin zone associated with the unit cell of Fig.~\ref{fig:graphene_sdw}, and $\sigma_0$ denotes the  
$(2\times 2)$ unit-matrix. In a first step, we compared the mean-field free-energies of the uniaxial $\langle\vec{S}^{\phantom{\dagger}}_{q,Q_i}\rangle = M\vec{e}_3$
and chiral $\langle\vec{S}^{\phantom{\dagger}}_{q,Q_i}\rangle = M\vec{e}_i$ states through the evaluation of (\ref{eq:resum}). Here, we find the chirally ordered state of Fig.~\ref{fig:graphene_sdw}a
to be energetically favored and content ourselves with this preliminary result. In order to substantiate this finding, further refined studies beyond mean-field level have to be performed to distinguish both scenarios. 

\setcounter{MaxMatrixCols}{12}

 \subsection{Kagome lattice: Exotic phases in the long-range Hubbard model} \label{sec:theory:lattice_kagome}

The kagome lattice (Fig.~\ref{gr:theory:lattice_kagome}), a lattice of cornersharing triangles, is a prototypical scenario for unconventional quantum states of matter at all coupling strengths. This unusual lattice structure is realized in some exotic compounds, e.g. $\cZn\cCu_3(OH)_6\cCl_2$ (herbertsmithite)~\cite{helton.2010.PhysRevLett.104.147201}, $\cSr\cCr_{8-x}\cGa_{4+x}\cO_{19}$~\cite{broholm.1990.PhysRevLett.65.3173}, $\cPr_3\cGa_5\cSi\cO_{14}$ (langasite)~\cite{zhou.2009.PhysRevLett.102.067203} and $\cBa\cNi_3(\cO\cH)_2(\cV\cO_4)_2$ (vesignieite)~\cite{freedman.2012.ChemCommu.48.64}. So far, however, the few kagome layered materials found in nature are located in the Mott-type local spin regime at half-filling and do not suggest a conventient description in terms of itinerant electrons. 
As a consequence, previous research on interacting electrons for the kagome lattice has predominantly focused on the strong coupling regime. It is, however, not unlikely that progress can be made on the material synthesis side of heterostructures and thin films designed such as to provide kagome lattice structures. Furthermore, the herbertsmithites such as $\cZn\cCu_3(OH)_6\cCl_2$ appear as a relevant class of candidates for intermediately coupled materials where the finite bandwidth might still be important~\cite{han.2012.PhysRevLett.108.157202}. In addition, a promising alternative route starts to emerge in optical kagome lattices of ultra-cold fermionic atomic gases such as for the isotopes ${}^6\text{Li}$ and ${}^{40}\text{K}$~\cite{PhysRevLett.108.045305}.

      Multiband descriptions are both implied due to multiple orbitals as seen for the pnictides and cobaltates in Chapter~\ref{chap:pnictide} and Section~\ref{chap:cobaltates}, respectively, but also for multiple sites associated with the unit cell of a given lattice as seen for the doped graphene scenario in Section~\ref{chap:graphene}. The kagome lattice~\cite{elser.1989.PhysRevLett.62.2405} possesses a minimal three-band model due to three sites per unit cell. For the kagome Hubbard model, the three sublattices imply fundamental problems in characterizing its preferred electronic many-body phases. In the strong-coupling limit at half-filling, the kagome spin model exhibits strong quantum-disorder fluctuations and, both in theory and experiment, has become one of the paradigmatic models of frustrated magnetism~\cite{ramirez.1994.AnnuRevMatterSci.24.453,misguich.2004.2dqa,mendels.2010.JPhysScoJpn.79.011001}. While the associated Mott transition at finite coupling might still be described within dynamical mean field theory~\cite{ohashi.2008.ProgTheorPhysSupp.176.97}, the scope of collective electronic phases at intermediate Hubbard strength and general filling is particularly challenging to investigate: {\it In the same way as electronic Bloch states at the Fermi level can involve different orbital admixtures for the multiorbital case, the electronic states in the kagome lattice can be assigned differently among the multiple sublattices.} Furthermore, as the three sublattices spoil particle-hole symmetry, large scale numerical simulations of two-dimensional systems such as quantum Monte Carlo calculations cannot be employed, even at half filling, due to the sign problem.

\subsubsection{Model}

        The kagome lattice features a triangular superlattice where each unit cell contains three sites in a triangular arrangement (Fig.~\ref{gr:theory:lattice_kagome}a). The nearest and next-nearest neighbors are sites from a different sublattice, while the 3rd-nearest neighbors are sites from the same sublattice. In the following we constrain ourselves to a nearest neighbor tight-binding Hamiltonian with long-range Hubbard interactions
        \begin{equation}
          \begin{split}
            \hat{H} = t \sum \limits_{\langle i,j\rangle} \sum \limits_\sigma \left( \hat{c}_{i\sigma}^{\dagger} \hat{c}_{j\sigma}^{\pdagger} + \HC \right) + U_0 \sum \limits_i \hat{n}_{i\uparrow} \hat{n}_{i\downarrow} + \frac{U_x}{2} \sum \limits_{\left[i,j\right]} \sum \limits_{\sigma,\nu} \hat{n}_{i\sigma} \hat{n}_{j\nu} + \mu \sum \limits_i \sum \limits_{\sigma} \hat{n}_{i\sigma} , \label{eq:theory:kagome_hamiltonian}
          \end{split}
        \end{equation}
        where $\hat{n}_{i\sigma}= \hat{c}_{i\sigma}^\dagger \hat{c}_{i\sigma}^{\pdagger}$, and $\hat{c}^\dagger_{i\sigma}$ denotes the electron creation operator of spin $\sigma=\{\uparrow, \downarrow\}$ at site $i$. The hopping is restricted to nearest neighbors, indicated by $\langle i,j\rangle$. The local interaction is $U_0$, while the long-range interaction $U_x$ is implemented for $U_1$, $U_2$ and $U_3$ as an interaction between nearest neighbors, next-nearest neighbors, and 3rd-nearest neighbors. $U_1$ and $U_2$ connect different sublattices, while $U_3$ connects sites of the same sublattice. (We will mostly constrain ourselves to the $U_0$-$U_1$ model in the following.) To obtain the band structure of the non-interacting system, the free part of the Hamiltonian has to be transformed to momentum space. Consequently, the lattice is divided into unit cells, containing three sites each. The electron creation operator in this basis has indices for (super)site, sublattice and spin:
        \begin{figure}[t]
          \centering
          \includegraphics[width=0.99\textwidth]{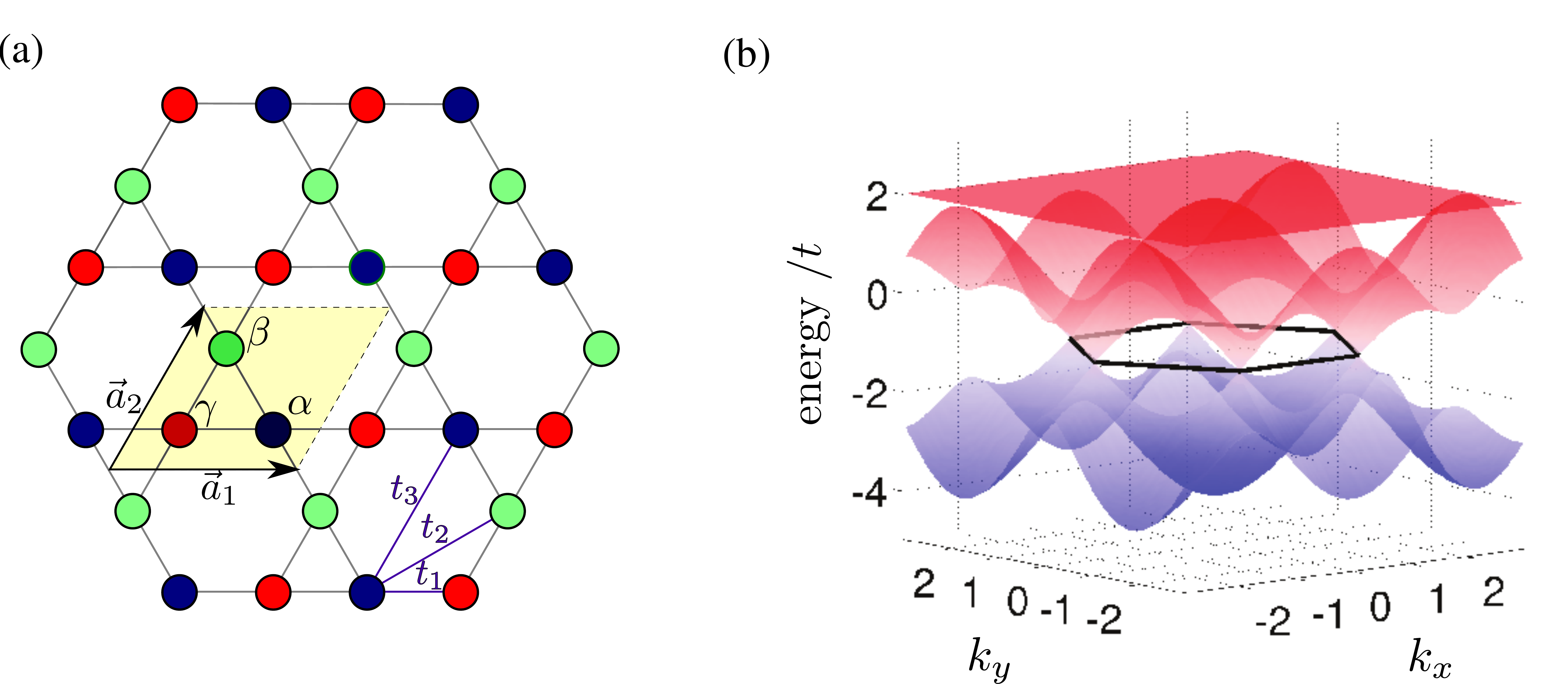}
          \caption{Sublattice structure (a) and tight-banding band structure (b) of the kagome lattice. 
          The different bond vectors $t_i$ are explicitly written out in Appendix~\ref{sec:hexbandint}.}
          \label{gr:theory:lattice_kagome}
        \end{figure}
        \begin{equation}
          \begin{split}
            \hat{c}^{\dagger}_{\underbrace{i}_{\text{site}}\underbrace{m}_{\text{sublattice}}\underbrace{\sigma}_{\text{spin}}} = \frac{1}{\sqrt{N}} \sum \limits_{\vec{k}} \sum \limits_n u^*_{mn}(\vec{k}) \ \hat{c}^{\dagger}_{\underbrace{\vec{k}}_{\text{momentum}}\underbrace{n}_{\text{band}}\underbrace{\sigma}_{\text{spin}}} \ee^{-\ii\vec{k}(\vec{R}_i+\vec{r}_m)} . \label{eq:theory:fourier_transformation}
          \end{split}
        \end{equation}
        After the Fourier transform, the new creation operator has indices for momentum, band and spin. The three bands of the kagome tight-binding model follow the analytic form
        \begin{equation}
          \begin{split}
            \varepsilon_1(\vec{k}) &= - t_1 - t_1 \sqrt{3 + 2 \cos\left( 2 k_x \right) + 2 \cos\left( k_x + \sqrt{3} k_y  \right) + 2 \cos\left( k_x - \sqrt{3} k_y  \right)}\\
            \varepsilon_2(\vec{k}) &= - t_1 + t_1 \sqrt{3 + 2 \cos\left( 2 k_x \right) + 2 \cos\left( k_x + \sqrt{3} k_y  \right) + 2 \cos\left( k_x - \sqrt{3} k_y  \right)}\\
            \varepsilon_3(\vec{k}) &= 2 t_1 \label{eq:theory:kagome_energy_dispersion}
          \end{split}
        \end{equation}
        Note that the band structure of the two dispersive bands is identical to the honeycomb tight-binding model in Sec.~\ref{chap:graphene}, while the third band is completely flat (Fig.~\ref{gr:theory:lattice_kagome}b). 
Due to only nearest-neighbor hopping, the van Hove singularity in the density of states (Fig.~\ref{gr:wcrg:kagome_and_grapheneBS}a) and the perfect nesting condition (Fig.~\ref{gr:wcrg:kagome_and_grapheneFS}a) coincide at the same filling.

  \subsubsection{Sublattice Interference: kagome vs. honeycomb lattice} \label{chap:wcrg}

      \begin{figure}[t]
        \centering
        \includegraphics[width=0.99\linewidth]{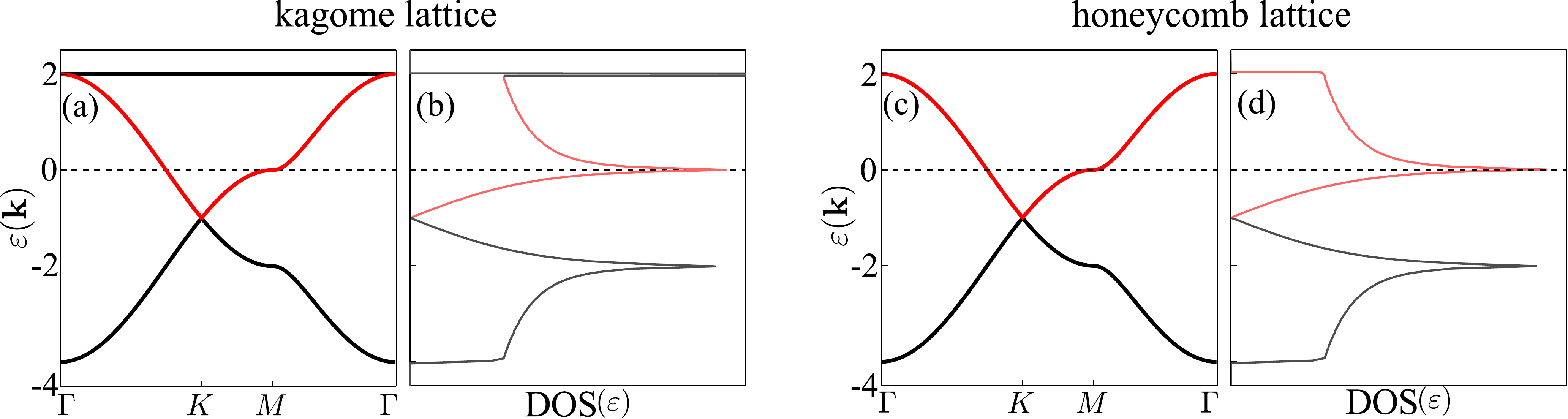}
        \caption{Tight-binding model in the kagome (a,b) and the honeycomb (c,d) lattice. For nearest-neighbor tight-binding, the dispersive bands of both lattices, are identical. The chemical potential is tuned to van Hove filling of the upper dispersive band (red).}
        \label{gr:wcrg:kagome_and_grapheneBS}
      \end{figure}

In order to understand the subtle interplay of multiple sublattices with regard to Fermi surface instabilities on the kagome lattice, it is revealing to compare the kagome scenario with the honeycomb scenario. From a tight-binding perspective (Fig.~\ref{gr:wcrg:kagome_and_grapheneBS}a), in addition to two strongly dispersive bands identical to the honeycomb band structure, the kagome Hubbard model features one flat band which, for appropriate fillings, has been suggested to be particularly susceptible to ferromagnetism along Stoner's criterion~\cite{tanaka.2003.PhysRevLett.90.067204}.  Aside from that, however, the pure view from Fermi surface topology on the dispersive kagome band would suggest an identical profile of Fermi surface instabilities in the kagome Hubbard model as compared to the honeycomb Hubbard model in Sec.~\ref{chap:graphene}, assuming sufficiently weak coupling and appropriate Fermi level such that the additional flat band does not play a role.
The Fermi surface topology is, however, not the complete characterization of the problem. {\it The eigenstates of the Fermi surface matter in the presence of multiple orbitals or sublattices, as they enter as matrix elements in formulating the action of  interactions in momentum space.} The core information is encoded in the transformation coefficients $u_{sn}(\vec{k})$ of the Bogoliubov transformation matrix from the real-space and sublattice picture to the momentum-space and band representation, defined in Eq.~\ref{eq:theory:fourier_transformation}, i.e. 
$          \hat{c}_{is\sigma}^{\dagger} = \sum \limits_{\vec{k}} \sum \limits_n u^*_{sn}(\vec{k}) \ \hat{c}_{\vec{k}n\sigma}^{\dagger} \ee^{-\ii\vec{k}(\vec{R}_i+\vec{r}_s)} . $
      For a given band $n$ and momentum $\vec{k}$ in the BZ, the coefficients obey $\sum_s | u_{sn} (\vec{k}) |^2=1$, where the band index $n$ can be omitted for our analysis because 
we implicitly assume that only the band is considered which intersects the Fermi level. In the following, these coefficients are called sublattice weights.

      \begin{figure}[t]
        \centering
        \includegraphics[width=0.99\linewidth]{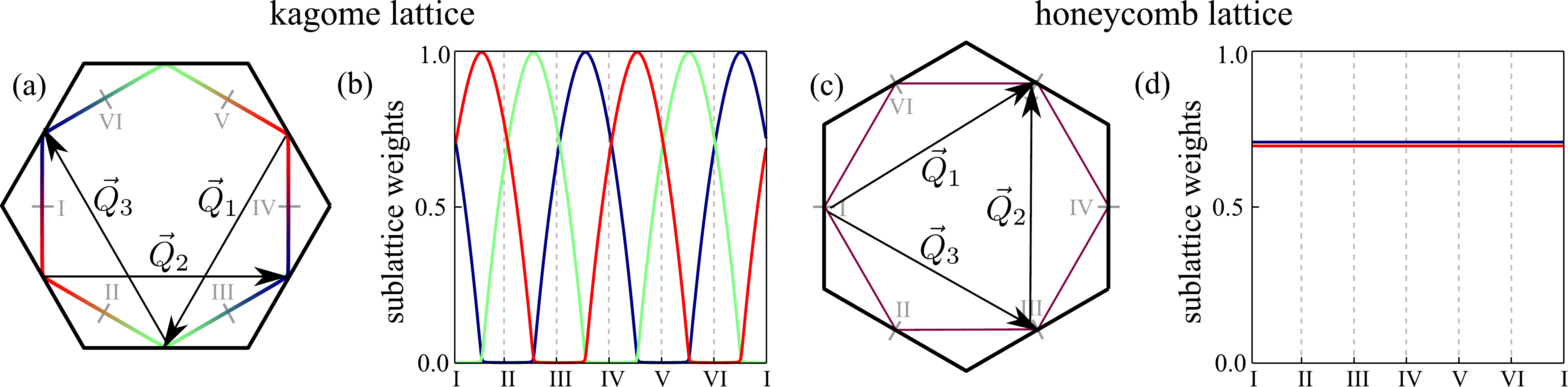}
        \caption{(a) FS of the kagome tight-binding model at $n=5/12$ van Hove filling. It touches the $M$ point of the hexagonal Brillouin zone (BZ), where the DOS is maximal; it exhibits three nesting features denotes $Q_1$, $Q_2$, and $Q_3$. 
 The colors blue, red, and green label the major sublattice occupation of the Fermi-surface band eigenstates. (b) The FS labels I-VI defined in (a) along with the color shift show the change of sublattice occupation weights $|u_s(\vec{k})|$ counterclockwise along the FS. (c) FS of the honeycomb tight-binding model at $n=5/8$ van Hove filling. (d) The sublattice occupation along the FS is homogeneous.}
        \label{gr:wcrg:kagome_and_grapheneFS}
      \end{figure}
To illustrate the {\it sublattice interference mechanism}~\cite{PhysRevB.86.121105}, let us assume only local Hubbard interactions $U_0$. The FS at van Hove filling $n=5/12$ is depicted in Fig.~\ref{gr:wcrg:kagome_and_grapheneFS}a. The interaction vertex takes the simple form
      \begin{equation}
        \begin{split}
          V(\vec{k}_1,\vec{k}_2, \vec{k}_3, \vec{k}_4)= U_0 \sum_s u_{s}^*(\vec{k}_1) u_{s}^*(\vec{k}_2) u_s(\vec{k}_3) u_s(\vec{k}_4). \label{eq:wcrg:vertex}
        \end{split}
      \end{equation}
      From Eq.~\ref{eq:wcrg:vertex}, because of the locality of $U_0$, the only momentum dependence is given by the sublattice weights. Their evolution along the FS is depicted through color coding in Figs.~\ref{gr:wcrg:kagome_and_grapheneFS}a,b. Eq.~\ref{eq:wcrg:vertex} looks very familiar from orbital makeup factors in multiorbital systems such as the pnictides in Chap.~\ref{chap:pnictide}. In the kagome Hubbard model, this similar role is assigned to the sublattice weight distribution. As in the multiorbital case, the sublattice now affects the nesting enhancement of particle-hole fluctuations along the FS. A first guess from Fermi-surface topology without invoking the sublattice distribution would suggest the nesting vectors
      \begin{equation}
        \begin{split}
          \vec{Q}_1 = \pi \left( -\frac{1}{2} , -\frac{\sqrt{3}}{2} \right) \quad , \quad \vec{Q}_2 = \pi \left( 1 , 0 \right) \quad \text{and} \quad \vec{Q}_3 = \pi \left(-\frac{1}{2} , \frac{\sqrt{3}}{2} \right) \ .
        \end{split}
      \end{equation}
      {\it As they connect FS points with mainly different sublattice occupation, however, the interaction vertex~\eqref{eq:wcrg:vertex} will be small as it is diagonal in the sublattice index $s$.} Consequently, this effect is called \tql sublattice interference\tqr.

      It is instructive to reconcile this scenario with the Hubbard model on the honeycomb lattice featuring two lattice sites per unit cell (Sec.~\ref{chap:graphene}). There, the tight-binding band structure matches the dispersive bands of the kagome lattice and allows one to similarly tune the honeycomb model to the equivalent van Hove filling. While the DOS as well as the Fermi-surface topology exactly match with the kagome case (compare Figs.~\ref{gr:wcrg:kagome_and_grapheneFS}a,c), the sublattice weights for the honeycomb model are homogeneous along the FS (Fig.~\ref{gr:wcrg:kagome_and_grapheneFS}d). This in turn suggests that sublattice interference is absent for the honeycomb model.

    \subsubsection{$U_0$-$U_1$ Phase diagram}
      The Hamiltonian of the kagome Hubbard model is written in Eq.~\ref{eq:theory:kagome_hamiltonian}. After the Fourier transform to momentum space, the kinetic part (Eq.~\ref{eq:theory:kagome_energy_dispersion}) and interaction part (Eq.~\ref{eq:theory:kagome_hamiltonian}) are implemented in the fRG formalism.
      \begin{figure}[t]
        \centering
        \includegraphics[width=0.8\textwidth]{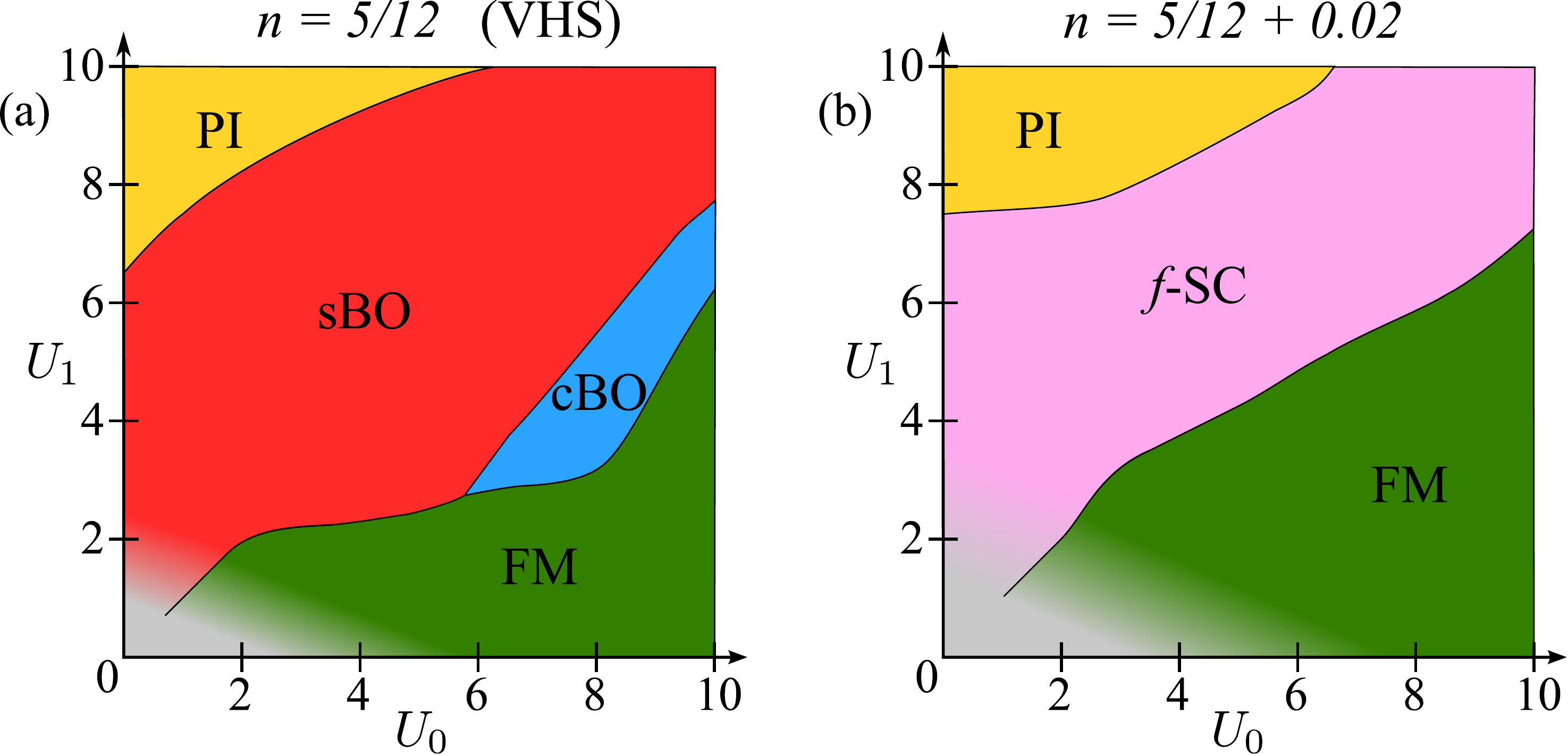}
	\caption{Phase diagram at (a) and in the vicinity (b) of the van Hove singularity (VHS)~\cite{PhysRevLett.110.126405}. In both settings, a ferromagnetic phase (FM, green) with ordering vector $\vec{Q}=(0,0)$ is found for dominant $U_0$, and a Pomeranchuk instability (PI, yellow) for dominant $U_1$. For particularly low interactions $U_0$ and $U_1$, the FRG cannot be employed due to numerical limitations (grey shaded area). (a) At VHS, the nesting is optimal giving rise to a charge bond-ordered phase (cBO, blue area) and a spin bond-ordered phase (sBO, red area). (b) Away from van Hove filling, triplet $f$-wave superconductivity (f-SC) dominates a large part of the phase diagram.}
        \label{gr:kagome:phase_diagram}
      \end{figure}
      The phase diagram obtained by parameter scans is plotted in Fig.~\ref{gr:kagome:phase_diagram}. There, two distinct fillings (at van Hove filling $n=5/12$ and close to van Hove filling $n=5/12+0.02$) are compared against each other. (Note that we allow the interactions to take rather large values, certainly reaching upper bound at which the FRG scheme is applicable.) The first remarkable feature is that only ferromagnetic (FM) fluctuations are present for a strictly local interaction ($U_0>0$, $U_1=0$). These fluctuations with $\vec{Q}=(0,0)$ are driven by the high density of state at van Hove filling, independent of the Fermiology. The order parameter reads
      \begin{equation}
        \begin{split}
          \vec{\mathcal{O}}_\text{FM}= \sum \limits_{\vec{k},l} \sum \limits_{\phantom{\vec{k}}\!\!\!\!\mu,\nu} \left\langle c_{\vec{k}l \mu}^\dagger \vec{\sigma}_{\phantom{\vec{k}}\!\!\!\!\!\!\mu\nu}^{\pdagger} c_{\vec{k}l \nu}^{\pdagger} \right\rangle \ ,
        \end{split}
      \end{equation}
      with the vector of Pauli matrices $\vec{\sigma}$. The dominance of ferrmagnetic fluctuations is a direct consequence of the sublattice interference. In addition, the background of ferromagnetic fluctuations at high energies provides further bias for a spin bond-ordered phase at VHS and the $f$-wave SC phase away from the VHS. Note that due to the discretized patching of the BZ (Fig.~\ref{gr:theory:lattice_kagome}), meaningful results can only be obtained through FRG for a certain amount of finite interaction strength. The infinitesimal limit, in turn, can be described exactly by perturbative RG~\cite{PhysRevB.86.121105}, while both results cannot be continuously interpolated between each other. Especially the $M$-points with the highest local DOS demand careful consideration in the patching scheme of FRG. By altering the patching scheme, so that the resolution in the proximity to the $M$-points is enhanced, the numerically stable parameter space of interaction strengths is extended to smaller coupling. Up to minimal $U_0$ coupling strengths reachable by FRG, FM still dominates. The other phases (and their phase boundaries) elaborated on in the following are only minimally effected by the alternative patching, demonstrating the stability of the FRG technique.
 
      The absence of conventional SDW related to the generic nesting vectors is further discussed in the next section which, judging from the analogous scenario in the honeycomb model (Sec.~\ref{chap:graphene}), explains the absence of a $d$+$\ii d$ superconducting phase. By contrast, FM fluctuations favor spin alignment and enhance the propensity towards triplet SC.  Despite the lack of the type of SDW order known from the honeycomb Hubbard model, under the perfect nesting condition at $n=5/12$, long-range orders are still found with the same expected ordering vectors $\vec{Q}_N$ (Fig.~\ref{gr:kagome:phase_diagram}a) 
       \begin{equation}
         \begin{split}
           \vec{Q}_1 = \pi \left( -\frac{1}{2} , -\frac{\sqrt{3}}{2} \right) \quad , \quad \vec{Q}_2 = \pi \left( 1 , 0 \right) \quad \text{and} \quad \vec{Q}_3 = \pi \left(-\frac{1}{2} , \frac{\sqrt{3}}{2} \right) \ ,
         \end{split}
       \end{equation}
which are pairwise equivalent modulo a reciprocal lattice vector.
Nearest-neighbor interactions $U_1$ are necessary to yield relevant fluctuations in these density wave channels, giving rise to exotic charge bond order (cBO) (i.e. a particle-hole condensate with the quantum numbers relative angular momentum $L=1$ and spin $S=0$), and a spin bond order (sBO) ($L=1$, $S=1$).
There, each nesting vector induces a different independent long-range order, seated on distinct pairs of sublattices. Hence, a superposition of all three orders is possible. For large $U_1$, the kagome Hubbard model also features a Pomeranchuk instability (PI) phase. Here, the ordering vector is $\vec{Q}=(0,0)$ and remains stable away from $n=5/12$, along with an extendend $f$-wave superconducting regime due to the strong presence of FM fluctuations. The PI instability for dominant $U_1$ and the FM instability for dominant $U_0$ are phenomenologically related: due to the onset of this order, the Fermi surface reorganizes itself such as to avoid the high density of state at the Fermi level without breaking the lattice translation symmetry.: While for the FM, the DOS is reduced by a spin-dependent shift of the FS, in the PI phase, discrete lattice symmetries are broken to induce a deformation of the FS to decrease the total DOS. 
It should be noted that the subtle interplay of competing fluctuations makes the kagome Hubbard model a challenging problem for FRG techniques. At present, different formulations of FRG have given largely similar, but partly also conflicting results~\cite{PhysRevLett.110.126405,PhysRevB.87.115135}. (A more detailed discussion is contained in Ref.~\cite{PhysRevLett.110.126405}.) For the purpose of this review, we intend to elaborate and explain what the conceputal mechanisms are that drive unconventional phases in kagome Hubbard model, which is a task that can be unambiguously addressed despite the open questions regarding the differences between different FRG approaches.

    \subsubsection{Suppression of conventional Spin-Density Wave} \label{sec:kagome:no_sdw}
    A phase one would have expected in the kagome Hubbard model at van Hove filling in its analogy to honeycomb scenario is the spin-density wave (SDW) (Sec.~\ref{sec:graphenesdw}). In fact, if the FS features a perfect nesting condition, both the triangular lattice and the honeycomb lattice exhibit strong SDW fluctuations, as depicted in the phase diagrams of Fig.~\ref{gr:cobaltates:phase_diagram} and Fig.~\ref{fig:graphenephase}, respectively. In contrast, the kagome lattice shows only ferromagnetic fluctuations (Fig.~\ref{gr:kagome:phase_diagram}), which is driven by the high DOS at the Fermi level. Even thought there also exists finite DOS, for both the triangular and honeycomb scenario, the FRG analysis suggests that FM fluctuations present due to finite Fermi level DOS are generically overshadowed by finite momentum particle-hole fluctuations for weak coupling, and only at high interaction scales, the FM fluctuations become competitive.
    

As seen before, the source for the missing SDW fluctuations in the kagome Hubbard model stems from the sublattice interference, which we intend to investigate with a bit more rigor. In a numerical experiment, we artificially set the sublattice weights to a constant value of $\frac{1}{\sqrt{3}}$ for all bands and momenta in the kagome scenario, neglecting the sublattice structure. With this simplification, the SDW is the dominating instability, yielding the identical scenario as for the honeycomb lattice.
      Fig.~\ref{gr:wcrg:kagome_and_grapheneFS} depicts the different  sublattice contributions at the Fermi surface for the honeycomb and kagome case. These are the transformation coefficients $u_{sn}(\vec{k})$ of the Bogoliubov transformation matrix from the real-space and sublattice picture to the momentum-space and band representation, defined in Eq.~\ref{eq:theory:fourier_transformation}.
      Only the band intersecting the FS is considered, so the band index $n$ is neglected. The local interaction is diagonal in the sublattice indices $s$:
      \begin{equation}
        \begin{split}
          V(\vec{k}_1,\vec{k}_2, \vec{k}_3, \vec{k}_4)= U_0 \sum_s u_m^*(\vec{k}_1) u_m^*(\vec{k}_2) u_m(\vec{k}_3) u_m(\vec{k}_4). \label{eq:kagome:vertex}
        \end{split}
      \end{equation}
      In the SDW channel, some parts of the FS are connected by a nesting vector $\vec{Q}_n$. Thus, for a strictly local interaction, the nesting condition is modulated by an overlap of the sublattice weights $\sum_{m} u_m(\vec{k}) u_m(\vec{k}+\vec{Q_N})$.
      For the honeycomb scenario, the sublattice weights are homogeneous at the FS, hence the sublattice weight prefactor is unity. 
For the kagome scenario, the nesting vectors connect parts of the FS with mismatching sublattice contributions.
Thus, the nesting condition is strongly weakened and not sufficient to drive the fRG-flow towards an SDW. As a consequence, the ferromagnetic fluctuations are stronger and dominate the FRG flow.

    \subsubsection{f-wave Superconductivity}
      A superconducting instability appears quickly when the filling is chosen away from van Hove filling in the presence of intermediate $U_1$ Coulomb interactions  (Fig.~\ref{gr:kagome:phase_diagram}). The vertex flow diverges in the spin-triplet channel and is non-degenerate, indicating to obey $f$-wave symmetry. The mean-field decoupled instability vertex in the spin-triplet channel (Sec.~\ref{subsec:decoup}), after a transformation to sublattice representation, yields the superconducting form factor. Identifying the leading harmonics in the $f$-wave form factor reveals that the best match is given for pairings between next-nearest neighbors, i.e. sites which are located on different sublattices. As such, inter-sublattice pairing is preferred.
      \begin{figure}[t]
        \centering
        \includegraphics[width=0.75\textwidth]{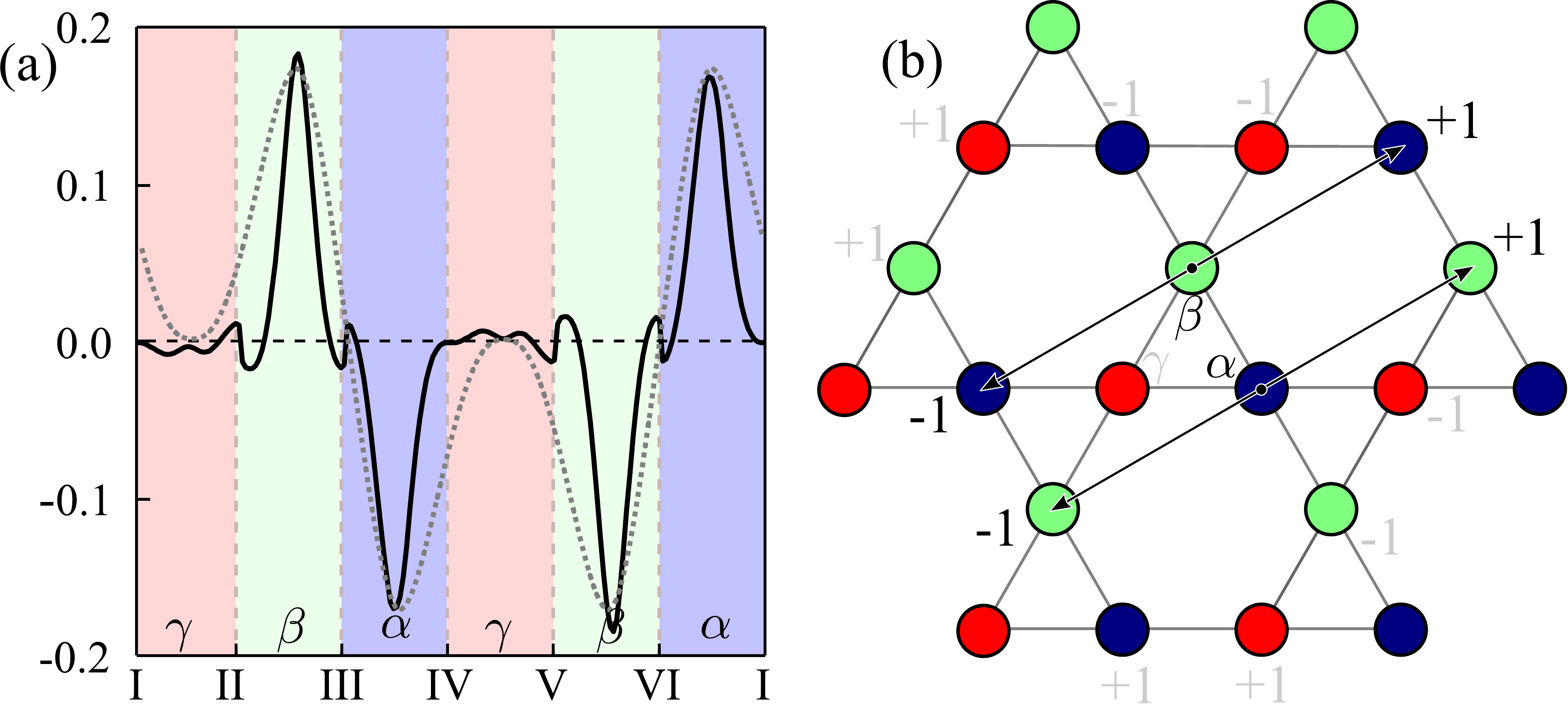}
        \caption{(a) $f$-wave Form factor for hopping between next-nearest neighbors on $\alpha$- and $\beta$-sites at $n=5/12+0.02$ filling. The labels I-VI along the Brillouin zone are defined in Fig.~\ref{gr:wcrg:kagome_and_grapheneFS}e. After the transformation to sublattice space, the fRG form factor (solid line) fits to the analytical form factor for next-nearest-neighbor pairing (Eq.~\ref{eq:kagome:fsc_form_factor}, dashed line). (b) The corresponding real-space pattern of the $\alpha$-$\beta$-pairing. The other two combinations are hinted in light grey. Figure was taken from~\cite{PhysRevLett.110.126405}.}
        \label{gr:kagome:fsc_form_factor}
      \end{figure}
Note that each pair of sublattices is separated from the others as they individually form one-dimensional chains on the kagome lattice. Within these chains, the next-nearest-neighbor hopping is transformed to momentum space, modulated with an $f$-wave formfactor. This gives:
      \begin{equation}
        \begin{split}
          f_{\alpha\beta}(\vec{k}) = \sin\left(\frac{3}{2}k_x + \frac{\sqrt{3}}{2}k_y\right)\\
          f_{\beta\gamma}(\vec{k}) = \sin\left(\frac{3}{2}k_x - \frac{\sqrt{3}}{2}k_y\right)\\
          f_{\alpha\gamma}(\vec{k}) = \sin\left(\sqrt{3}k_y\right)\label{eq:kagome:fsc_form_factor}
        \end{split}
      \end{equation}
      with $f_{m,n} = f_{n,m}$ and $f_{m,m}=0$. In Fig.~\ref{gr:kagome:fsc_form_factor}a, the form factor of the $\alpha$-$\beta$-pairing is compared to the analytical leading harmonic prediction. It fits well with the mean-field results. The pair forms between next-nearest neighbors because, due to an increased pairing distance, the (non-local) nearest-neighbor Coulomb repulsion is avoided~\cite{PhysRevB.86.020507}. The final superposition of all three sublattice pairings restore the full $f$-wave symmetry ($B_2$ element of the $C_{6v}$ group). Together, they form the order parameter
      \begin{equation}
        \begin{split}
          \mathcal{O}_\text{fSC}= \sum \limits_{\vec{k},m,n} \left\langle c_{\vec{k}m\uparrow}^\dagger c_{\vec{k}n\downarrow}^\dagger + c_{\vec{k}m\downarrow}^\dagger c_{\vec{k}n\uparrow}^\dagger \right\rangle f_{m,n}\left(\vec{k}\right) \ ,
        \end{split}
      \end{equation}
      which can be rotated within the spin-triplet sector to the $S_z=\pm1$ states.

    \subsubsection{Charge and Spin Bond Order} \label{sec:kagome:bond}
At van Hove filling $n=5/12$, the phase diagram in Fig.~\ref{gr:kagome:phase_diagram} features two unconventional density-wave phases, the spin bond order (sBO) and the charge bond order (cBO). (The labels of these phases relate to the real-space patterns discussed later.) Since the fRG is executed in momentum space, a Fourier transform has to be performed to determine the correct real-space picture (Eq.~\ref{eq:theory:fourier_transformation}). Again, the sublattice structure of the kagome lattice has to be carefully taken into consideration: The general final interaction vertex $V(\vec{k}_1,\vec{k}_2,\vec{k}_3,\vec{k}_4)$ consists of $4$ creation/annihilation operators, which have to be transformed separately. Each combination of sublattice indices $m_1,m_2,m_3,m_4$ with $m_x\in\{\alpha,\beta,\gamma\}$ constitutes an individual sector. Additionally,  three ordering 
vectors $\vec{Q}_{1}$, $\vec{Q}_{2}$ and $\vec{Q}_{3}$ have to be considered. For all $3^5$ combinations, the mean-field decoupling in the SDW or CDW channel (i.e. indirect or direct particle-hole channel)  is performed. As before, we are interested in the leading instability, i.e. the dominant eigenvalue of the decomposition. From the analysis of the final FRG vertex, we obtain a sixfold degeneracy of the dominant eigenvalue for $V(\vec{r}_1,m_1,\vec{r}_2,m_2,\vec{r}_3,m_3,\vec{r}_4,m_4)$ at both the SDW and CDW channel:
      \begin{center}
        \begin{tabular}{c|c|c|c|c}
          $m_1$    &    $m_2$ &    $m_3$ &    $m_4$ & $\vec{Q}_x$ \\ \hline
          $\alpha$ & $\alpha$ & $\gamma$ & $\gamma$ & $2$ \\
          $\gamma$ & $\gamma$ & $\alpha$ & $\alpha$ & $2$ \\
          $\beta$  & $\beta$  & $\gamma$ & $\gamma$ & $1$ \\
          $\gamma$ & $\gamma$ & $\beta$  & $\beta$  & $1$ \\
          $\alpha$ & $\alpha$ & $\beta$  & $\beta$  & $3$ \\
          $\beta$  & $\beta$  & $\alpha$ & $\alpha$ & $3$
        \end{tabular}
      \end{center}
      These combinations are identical for both density wave phases, i.e. once found in the direct particle-hole channel and once in the indirect particle-hole channel. In terms of band structure effects from the mean-field dcoupling of this order, it yields
      \begin{equation}
        \begin{split}
          V(\vec{r}_1,\alpha,\vec{r}_2,\alpha,\vec{r}_3,\gamma,\vec{r}_4,\gamma) \sim \hat{c}^{\dagger}_{\alpha} \hat{c}^{\dagger}_{\alpha} \hat{c}^{\pdagger}_{\gamma} \hat{c}^{\pdagger}_{\gamma} \approx \bra \hat{c}^{\dagger}_{\alpha} \hat{c}^{\pdagger}_{\gamma} \ket \hat{c}^{\dagger}_{\alpha} \hat{c}^{\pdagger}_{\gamma}
        \end{split}
      \end{equation}
Indeed, there is a directed hopping expectation value $\bra \hat{c}^{\dagger}_{\alpha} \hat{c}^{\pdagger}_{\gamma} \ket$, resulting in an additional hopping on bonds between sublattices $\alpha$ and $\gamma$. The form factor is fitted best with a $\sin(k_x)$, which corresponds to a correlation between nearest neighbors. This form factor can be rewritten as $\bra \hat{c}^{\dagger}_{\vec{k}\gamma} \hat{c}^{\pdagger}_{\vec{k}+\vec{Q}\alpha} \ket = \sin(\vec{T}\vec{k}) \cdot \Phi$  with translation vector $\vec{T}=(1,0)=\frac{\vec{Q}_2}{\pi}$, so the corresponding ordering vector is parallel to the direction of these bonds, forming quasi one-dimensional chains. The expectation value reveals the real-space order~\cite{nayak-00prb4880}:
       \begin{equation}
\tilde{\Phi}= \sin^2(\vec{T}\vec{k}) \cdot \Phi =     \sum \limits_{\vec{R}_j} \left( \bra \hat{c}^{\dagger}_{\vec{R}_j+\vec{r}_\alpha-\vec{T}} \hat{c}^{\pdagger}_{\vec{R}_j+\vec{r}_\alpha} \ket - \bra \hat{c}^{\dagger}_{\vec{R}_j+\vec{r}_\alpha+\vec{T}} \hat{c}^{\pdagger}_{\vec{R}_j+\vec{r}_\alpha} \ket \right) \frac{1}{2\ii} \ee^{-\ii \vec{Q} (\vec{R}_j+\vec{r}_\alpha)}
\label{eq:kagome:bond1}
       \end{equation}
      Here, the position vector of a site is decomposed in the vector of the unit cell $\vec{R}_i$ and the vector within the unit cell $\vec{r}_x$. The square of the orbital weights is invariant under reflection, so it analogously gives
      \begin{figure}[t]
        \centering
        \includegraphics[width=0.6\textwidth]{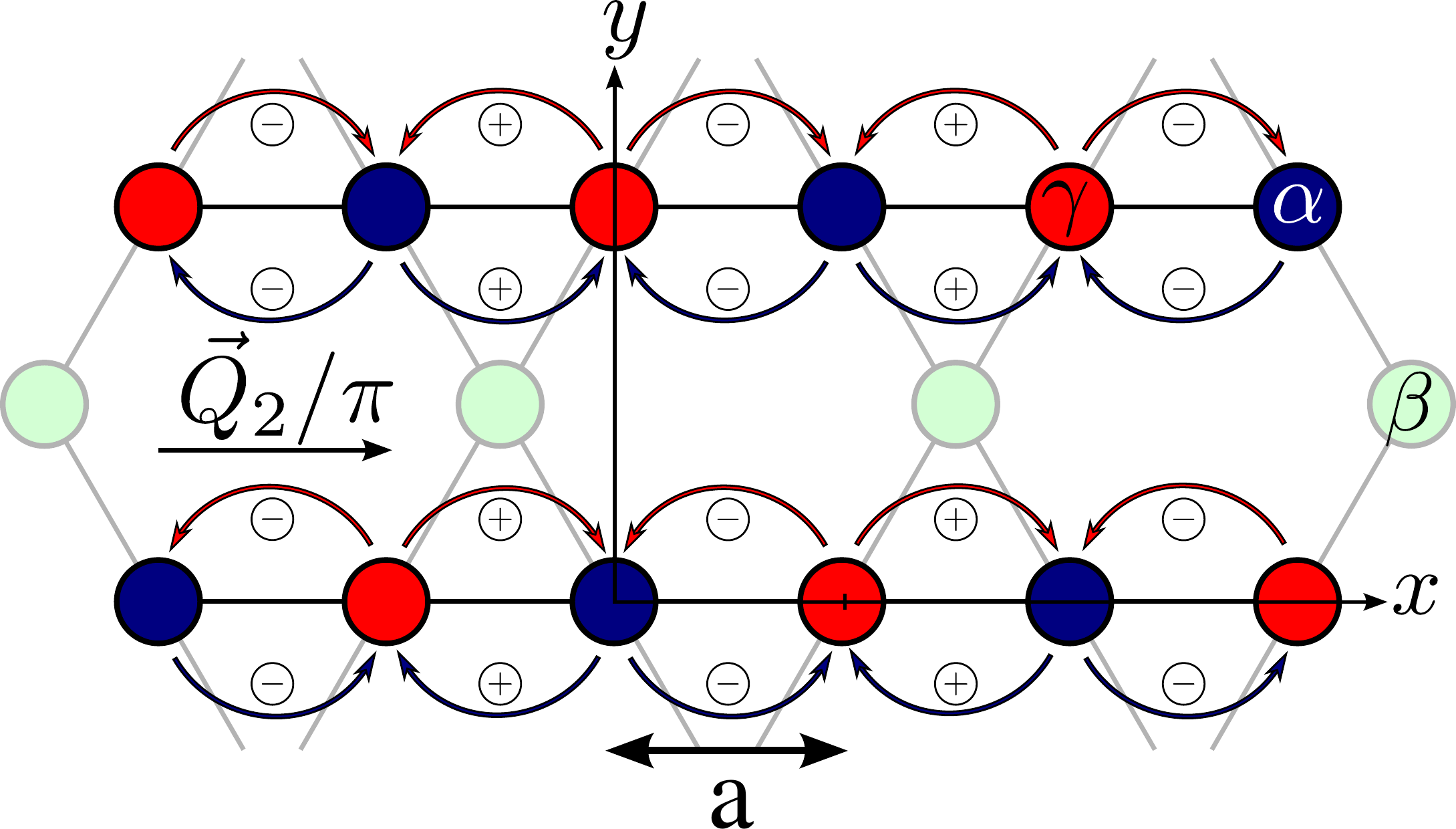}
        \caption{Formation of the cBO associated with $Q_2$: The hopping from $\alpha$- to $\gamma$-sites is modulated by Eq.~\ref{eq:kagome:bond1} (red arrows), the hopping from $\gamma$- to $\alpha$-sites by Eq.~\ref{eq:kagome:bond2} (blue arrows). The ordering vector is parallel to the $\alpha$-$\gamma$-chains, so $\beta$-sites are unaffected. The ordering vector causes a minus sign between neighboring chains. As consequence, the hopping amplitude between the bonds is alternatingly reduced and enhanced.}
        \label{gr:kagome:bond_construction}
      \end{figure}
      \begin{equation}
        \begin{split}
          \bra \hat{c}^{\dagger}_{\vec{k}\alpha} \hat{c}^{\pdagger}_{\vec{k}+\vec{Q}\gamma} \ket = \sin(\vec{T}\vec{k}) \cdot &\Phi\\
     \tilde{\Phi} =    \sum \limits_{\vec{R}_j} \left( \bra \hat{c}^{\dagger}_{\vec{R}_j+\vec{r}_\gamma-\vec{T}} \hat{c}^{\pdagger}_{\vec{R}_j+\vec{r}_\gamma} \ket - \bra \hat{c}^{\dagger}_{\vec{R}_j+\vec{r}_\gamma+\vec{T}} \hat{c}^{\pdagger}_{\vec{R}_j+\vec{r}_\gamma} \ket \right) \frac{1}{2\ii} &\ee^{-\ii \vec{Q} (\vec{R}_j+\vec{r}_\gamma)} \label{eq:kagome:bond2}
        \end{split}
      \end{equation}
      Together, the mean field effect of this order forms an alternating modulation of the hopping between nearest neighbors in the chains specified by the given pair of sublattices. The construction of the real-space pattern is presented in Fig.~\ref{gr:kagome:bond_construction}, while the final pattern is plotted in Fig.~\ref{gr:kagome:bond_real_space_pattern}a, featuring a bond-ordered structure. Note that aside from the scenario discussed here,  such bond orders promise experimental realization in dipolar fermion models~\cite{bhongale.2012.PhysRevLett.108.145301}.

      For a mean-field analysis of the resulting band structure of e.g. the cBO phase related to a given nesting vector $Q_2$, the non-interacting Hamiltonian in Eq.~\ref{eq:theory:kagome_energy_dispersion} has to be adapted and expanded with a Weiss field $\Delta_{\text{cBO}}$. Due to the broken translational symmetry, the new unit cell includes six sites (Fig.~\ref{gr:kagome:bond_real_space_pattern}) and, hence, a six-band calculation is needed, which is illustrated in Appendix~\ref{sec:frg:app_mean_field_hamiltonian}. As a consequence of the cBO, the FS is partially gaped. The same derivation is possible for the sBO (with a spin dependence of the additional hopping term), with the resulting real-space pattern plotted in Fig.~\ref{gr:kagome:bond_real_space_pattern}b. Furthermore, each pattern is threefold degenerate, where the other patterns are obtained by rotations by $2\pi/3$, corresponding to the ordering vectors $\vec{Q}_1$ and $\vec{Q}_3$.
      \begin{figure}[t]
        \centering
        \includegraphics[width=0.7\textwidth]{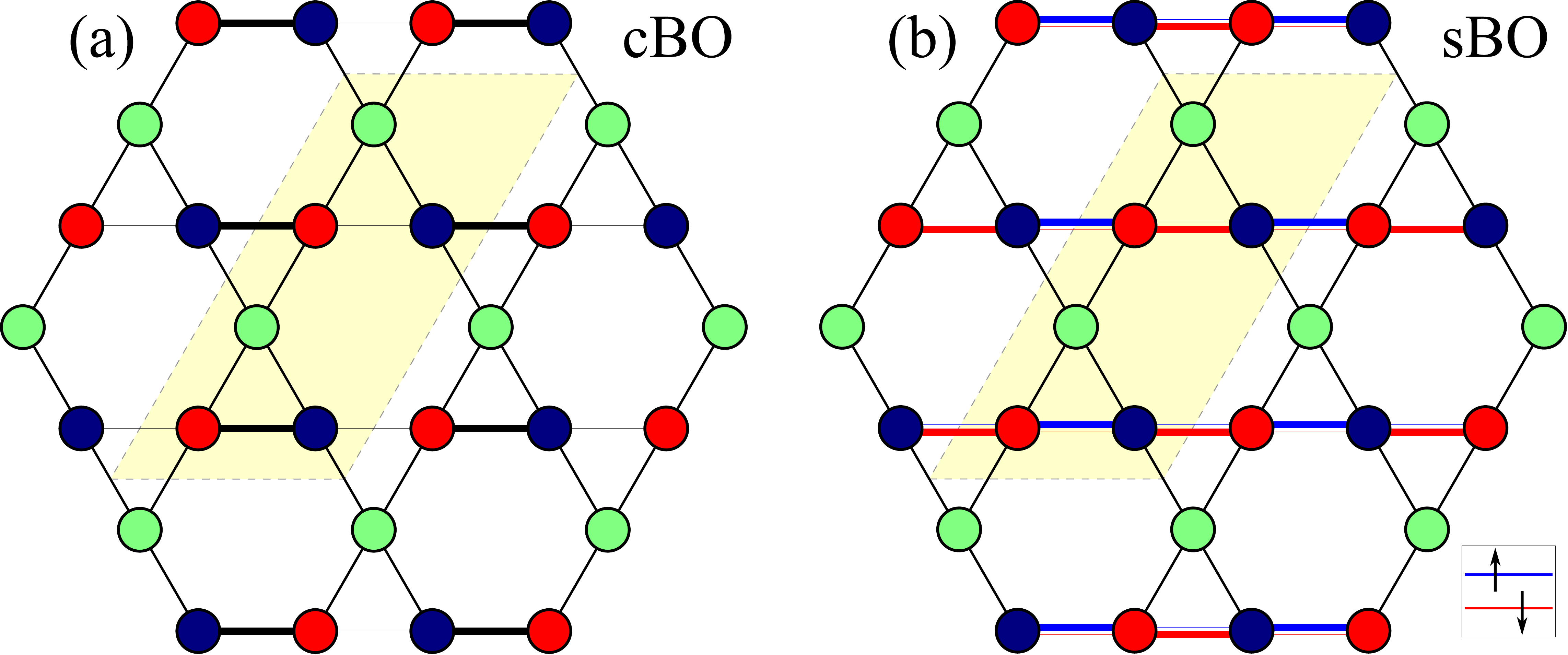}
        \caption{Real space pattern of the bond-ordered phases with ordering vector $\vec{Q}_2$~\cite{PhysRevLett.110.126405}: (a) cBO with the parameter setting $U_1=6.0$ and $n=5/12$ (VHS). There is an alternating hopping amplitude between $\alpha$- and $\gamma$-sites. The unit cell contains $6$ sites. (b) sBO with the parameter setting $U_0=8.0$, $U_1=6.0$ and $n=5/12$ (VHS). The real space pattern is similar to (a) but the hopping modulation depends on the spin polarization of the electrons.}
        \label{gr:kagome:bond_real_space_pattern}
      \end{figure}
      It is important to note that there is a direct correspondence between the cBO (sBO) related to one of the three nesting vectors $Q_n$ and the pair of sublattices where a charge (spin) modulation is imposed: Both real-space patterns of the order for $\vec{Q}_2$ (Fig.~\ref{gr:kagome:bond_real_space_pattern}) affect only bonds between $\alpha$-sites (blue) and $\gamma$-sites (red). The other bonds are included only in the orders for $\vec{Q}_1$ or $\vec{Q}_3$, respectively. Consequently, the three different orders do not interfere with each other and a simultaneous formation is possible.
      \begin{figure}[t]
        \centering
        \includegraphics[width=0.7\textwidth]{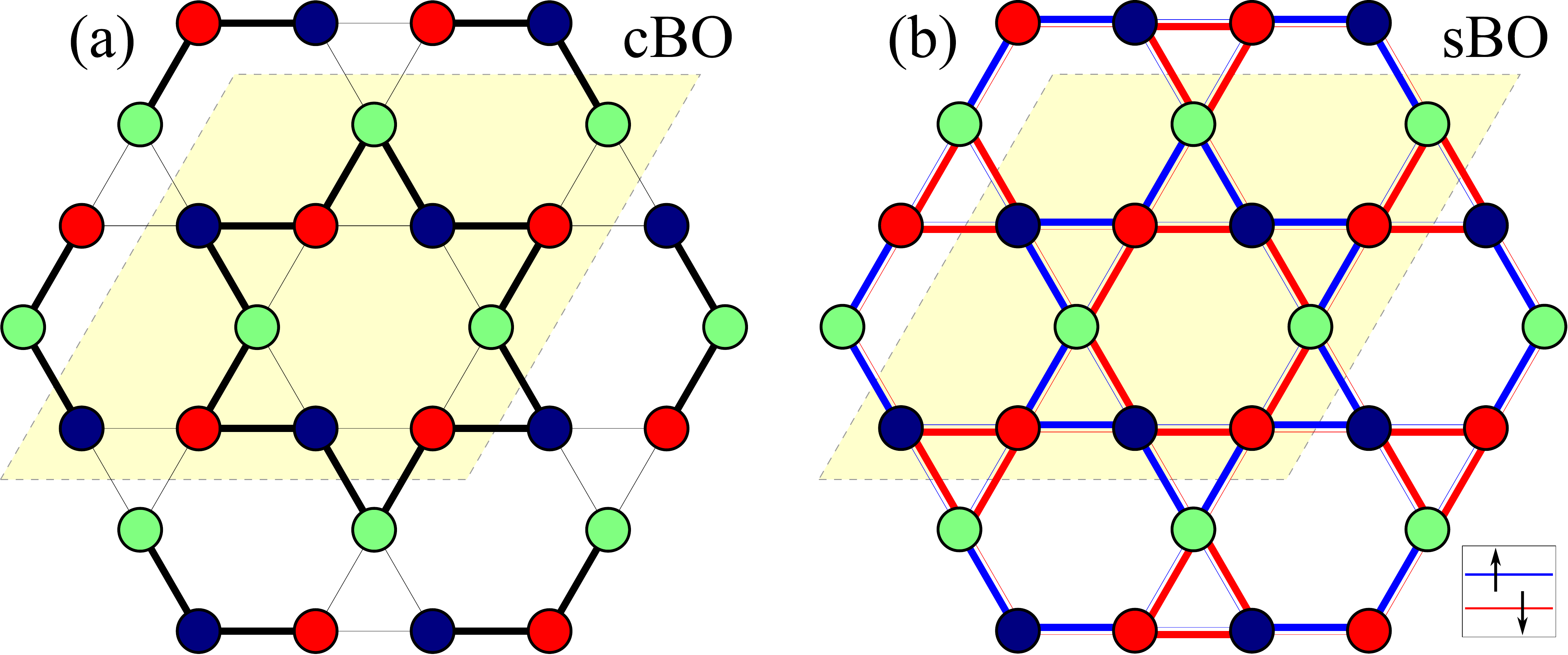}
        \caption{Complete real space pattern of bond-ordered phases. In analogy to a specified ordering vector in Fig.~\ref{gr:kagome:bond_real_space_pattern}, all bond order fields associated $\vec{Q}_{1}$, $\vec{Q}_{2}$ and $\vec{Q}_{3}$ are simultaneously applied. The unit cell contains $12$ sites.}
        \label{gr:kagome:bond_real_space_pattern3}
      \end{figure}
Following up on this observation, i.e. assuming that a cBO order for one $\vec{Q}_n$ reduces the free energy of the system and the orders for each $\vec{Q}_i$ are independent, the simultaneous formation of all three orders should further optimize the free energy. To verify this statement, a mean-field analysis is done in Appendix~\ref{sec:frg:app_mean_field_hamiltonian}, now containing $12$ bands because all order parameter fields are considered at a time. We find that the system linearly gains energy from forming the $3$ individual mean fields, so the ordering formation along the individual bond directions is indeed independent. Therefore, the new BZ is reduced to a quarter of the size and retains its hexagonal structure. This illustrates that the $C_{6v}$ symmetry, which is broken if only one order with $\vec{Q}_m$ is applied, is restored. In the reduced BZ, the spectral weight at the FS between the $\Gamma$- and $M$-points is shifted away from the Fermi energy when the Weiss fields $\Delta_{\text{
cBO}}$ are switched on.

      The real-space patterns of the superpositions are presented in Fig.~\ref{gr:kagome:bond_real_space_pattern3}, including the enlarged unit cell with $12$ sites. Their order parameters are given by
      \begin{equation}
        \begin{split}
          \vec{\mathcal{O}}_\text{sBO}=& \sum \limits_{\vec{k},\mu,\nu\atop l,m,n} \left\langle c_{\vec{k}l \mu}^\dagger \vec{\sigma}_{\phantom{\vec{k}}\!\!\!\!\!\!\mu\nu}^{\pdagger} c_{\vec{k}+\vec{Q}_m n \nu}^{\pdagger} \right\rangle  \sin\left( \frac{\vec{Q}_m\vec{k}}{\pi} \right) \left| \epsilon_{lmn} \right|\\
          \mathcal{O}_\text{cBO}=& \sum \limits_{\vec{k},\mu,\nu\atop l,m,n} \left\langle c_{\vec{k}l \mu}^\dagger \hat{1}_{\phantom{\vec{k}}\!\!\!\!\!\!\mu\nu}^{\pdagger} c_{\vec{k}+\vec{Q}_m n \nu}^{\pdagger} \right\rangle  \sin\left( \frac{\vec{Q}_m\vec{k}}{\pi} \right) \left| \epsilon_{lmn} \right|
        \end{split}
      \end{equation}
      with the Levi-Civita tensor $\epsilon_{lmn}$.

    \subsubsection{Pomeranchuk Instability} \label{sec:kagome:pi}
\begin{figure}[t]
        \centering
        \includegraphics[width=0.99\textwidth]{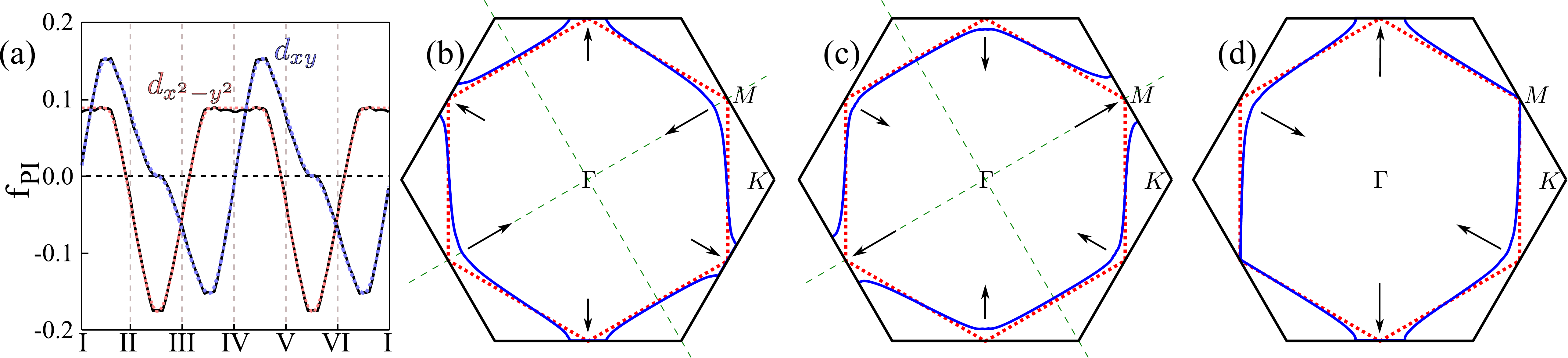}
        \caption{(a) Form factors of the Pomeranchuk instability~\cite{PhysRevLett.110.126405}. The calculated form (solid black lines) fits very well to the analytical ones (dotted lines, Eq.~\ref{eq:kagome:pi_formfactors}). (b) Deformation of the FS with the PI field $\Delta_0=0.1$ and $\rho=-\frac{\pi}{3}$ (Eq.~\ref{eq:kagome:pi_hamiltonian}). At two opposite $M$-points, which are the points of the FS with the highest DOS, the FS is shifted towards the center of the BZ, while it is shifted away at the remaining four $M$-points. The symmetry group is reduced to $C_{2v}$. (c) Same as (b) but $\rho=\frac{2\pi}{3}$. The distortion of the FS is opposed to (b). (d) Same as (b) but $\rho=\frac{\pi}{6}$. Here, the FS includes no reflection symmetries, so the symmetry group is $C_2$.}
        \label{gr:kagome:pi_fermisurface}
      \end{figure}

      For strong nearest-neighbor interaction $U_1$, the phase diagram (Fig.~\ref{gr:kagome:phase_diagram}) exhibits a Pomeranchuk instability (PI). In the mean-field decoupling, it is identical to the CDW channel with zero ordering momentum $\vec{Q}=(0,0)$, i.e. it does not break translational symmetry. In the comparable scenario for the cuprates (one-band model in the quadratic lattice), the PI results in a deformation of the FS, breaking its tetragonal symmetry~\cite{halboth.2000.PhysRevLett.85.5162,davis}. Accordingly, a similar effect can be expected for the kagome lattice. However, due to the two-dimensional irreducible representation of the $d$-wave channel, the mean-field decoupling of the PI channel in the kagome scenario yields a doubly degenerate instability~\cite{PhysRevLett.110.126405,maharaj}. After transformation from band representation to sublattice representation, we find that the correlations between the \textit{same} sublattice are dominating, which is implicitly demanded in order not to break the lattice translation symmetry. The fRG form factors are presented in Fig.~\ref{gr:kagome:pi_fermisurface}a. The analytical form factors for an $E_2$ representation on the 3rd-nearest neighbors read:
      \begin{equation}
        \begin{split}
          f_{d_{x^2-y^2}}\left(\vec{k}\right) &= \cos\left(2k_x\right) - \cos\left(k_x\right) \cos\left(\sqrt{3} k_y\right)\\
          f_{d_{xy}}\left(\vec{k}\right) &= \sqrt{3} \sin\left(k_x\right) \sin\left(\sqrt{3} k_y\right) \label{eq:kagome:pi_formfactors}
        \end{split}
      \end{equation}
      The accordance between analytical and calculated form factors is complete within numerical accuracy. After mean-field decoupling, the effective Hamiltonian reads:
       \begin{equation}
         \begin{split}
           \hat{H}_\text{PI} &= \sum \limits_{\sigma} \sum \limits_{\vec{k}} \varepsilon (\vec{k}) \hat{c}^{\dagger}_{\vec{k}\sigma} \hat{c}^{\pdagger}_{\vec{k}\sigma} + \frac{1}{N} \sum \limits_{\sigma} \sum \limits_{\vec{k}\vec{q}} V_{\text{PI}}(\vec{k},\vec{q}) \ \left( \hat{c}^{\dagger}_{\vec{k}\sigma} \hat{c}^{\dagger}_{\vec{q}\sigma} \hat{c}^{\pdagger}_{\vec{k}\sigma} \hat{c}^{\pdagger}_{\vec{q}\sigma} \right)\\
                             &= \sum \limits_{\sigma} \left( \sum \limits_{\vec{k}} \varepsilon (\vec{k}) \hat{c}^{\dagger}_{\vec{k}\sigma} \hat{c}^{\pdagger}_{\vec{k}\sigma} - \frac{1}{N} \sum \limits_{\vec{k}\vec{q}} V_{\text{PI}}(\vec{k},\vec{q}) \ \left( \bra \hat{c}^{\dagger}_{\vec{k}\sigma} \hat{c}^{\pdagger}_{\vec{k}\sigma} \ket \hat{c}^{\dagger}_{\vec{q}\sigma} \hat{c}^{\pdagger}_{\vec{q}\sigma} + \hat{c}^{\dagger}_{\vec{k}\sigma} \hat{c}^{\pdagger}_{\vec{k}\sigma} \bra \hat{c}^{\dagger}_{\vec{q}\sigma} \hat{c}^{\pdagger}_{\vec{q}\sigma} \ket - \bra \hat{c}^{\dagger}_{\vec{k}\sigma} \hat{c}^{\pdagger}_{\vec{k}\sigma} \ket \bra \hat{c}^{\dagger}_{\vec{q}\sigma} \hat{c}^{\pdagger}_{\vec{q}\sigma} \ket \right) \right)\\
                             &= \sum \limits_{\sigma} \left( \sum \limits_{\vec{k}} \varepsilon (\vec{k}) \hat{c}^{\dagger}_{\vec{k}\sigma} \hat{c}^{\pdagger}_{\vec{k}\sigma} - \sum \limits_{\vec{q}} \Delta_{\vec{q}} \hat{c}^{\dagger}_{\vec{q}\sigma} \hat{c}^{\pdagger}_{\vec{q}\sigma} - \sum \limits_{\vec{k}} \Delta_{\vec{k}} \hat{c}^{\dagger}_{\vec{k}\sigma} \hat{c}^{\pdagger}_{\vec{k}\sigma} + \frac{1}{N} \sum \limits_{\vec{k}\vec{q}} V_{\text{PI}}(\vec{k},\vec{q}) \bra \hat{c}^{\dagger}_{\vec{k}\sigma} \hat{c}^{\pdagger}_{\vec{k}\sigma} \ket \bra \hat{c}^{\dagger}_{\vec{q}\sigma} \hat{c}^{\pdagger}_{\vec{q}\sigma} \ket \right)\\
                             &= \sum \limits_{\sigma} \left( \sum \limits_{\vec{k}} \left( \underbrace{\varepsilon (\vec{k}) - 2 \Delta_{\vec{k}}}_{E(\vec{k})} \right) \hat{c}^{\dagger}_{\vec{k}\sigma} \hat{c}^{\pdagger}_{\vec{k}\sigma} + \sum \limits_{\vec{k}} \Delta_{\vec{k}} \bra \hat{c}^{\dagger}_{\vec{k}\sigma} \hat{c}^{\pdagger}_{\vec{k}\sigma} \ket \right) \ ,
         \end{split}
       \end{equation}
      where the definition of the gap function
      \begin{equation}
        \begin{split}
          \Delta_{\vec{k}} := \frac{1}{N} \sum \limits_{\vec{q}} V_{\text{PI}}(\vec{k},\vec{q}) \bra \hat{c}^{\dagger}_{\vec{q}\sigma} \hat{c}^{\pdagger}_{\vec{q}\sigma} \ket \in \mathbb{R}
        \end{split}
      \end{equation}
      is used. The reality constraint essentially derives from the hermiticity of $\bra \hat{c}^{\dagger}_{\vec{q}\sigma} \hat{c}^{\pdagger}_{\vec{q}\sigma} \ket$. (As we have seen before, this is not generic for finite $q$ particle-hole condensates $\bra \hat{c}^{\dagger}_{\vec{q}+Q\sigma} \hat{c}^{\pdagger}_{\vec{q}\sigma} \ket$ where a complex  $\Delta_{\vec{k},Q}$ is allowed.)  
The PI adds a 3rd-nearest-neighbor hopping to the Hamiltonian, which corresponds to hoppings within the same sublattice. For the $E_2$ Pomeranchuk instability, there is still a degree of freedom in the choice of the $d$-wave form factor because any linear (real) combination of $d_{x^2-y^2}$ and $d_{xy}$ is a solution. Rewriting the individual via a $d$+$\ii d$-superposition with an additional phase $\rho$, the 
imaginary part vanishes because of its Hermitian counterpart, so $\rho$ enables a continuous choice within the space of solutions composed by $d_{x^2-y^2}$ and $d_{xy}$. After Fourier transform, depending on $\rho$, the 3rd-nearest-neighbor hopping is not equivalent for all three hopping axes, and  the dispersion relation (Eq.~\ref{eq:theory:kagome_energy_dispersion}) is extended to
      \begin{equation}
        \begin{split}
          \tilde{\varepsilon}_n(\vec{k}) &= \varepsilon_n(\vec{k}) + \Delta_0 \left[ \ee^{\ii\rho} \left( d_{x^2-y^2} + \ii d_{xy} \right) \right] + \HC\\
                                         &= \varepsilon_n(\vec{k}) + 2 \Delta_0 \re \left[ \ee^{\ii\rho} \left( \cos\left(2k_x\right) - \cos\left(k_x\right) \cos\left(\sqrt{3} k_y\right) + \ii \sqrt{3} \sin\left(k_x\right) \sin\left(\sqrt{3} k_y\right) \right) \right] \label{eq:kagome:pi_hamiltonian}
        \end{split}
      \end{equation}
      This results in a deformation of the FS, as presented in Figs.~\ref{gr:kagome:pi_fermisurface}b-d, where the FS is shifted away from the six $M$ points, which are the momenta of highest DOS. The order parameter is also adapted to this superposition:
      \begin{equation}
        \begin{split}
          \mathcal{O}_\text{PI}=& \sum \limits_{\vec{k},l,\mu} \left\langle c_{\vec{k}l \mu}^\dagger c_{\vec{k}l \mu}^{\phantom{\dagger}} \right\rangle \re \left[ \ee^{\ii\phi} \left( f_{d_{x^2-y^2}}(\vec{k}) + \ii f_{d_{xy}} (\vec{k}) \right) \right]
        \end{split}
      \end{equation}
In the two-dimensional PI phase, the symmetry of the resulting FS depends on the choice of superposition, i.e. $\rho$,
      \begin{equation}
        \text{point group }\begin{cases}
          C_{2v} & \text{if } \rho = \frac{\pi}{3} z \quad \text{with } z \in \mathbb{Z}\\
          C_2 & \text{else,}
        \end{cases}
      \end{equation}
      while the symmetry of the lattice remains unbroken. This is a generic feature of PI on hexagonal lattice systems~\cite{maharaj}. 

 \subsection{Summary and Outlook} \label{hexa-summary}

As we have seen in this chapter, {\it the lattice symmetry plays a profound role in characerizing Fermi surface instabilities.} The core insight is that for an interaction profile which is invariant under discrete lattice symmetry transformations, both particle-particle and particle-hole condensates are associated with an irreducible lattice representation unless it is the Fermi surface instability itself (such as for Pomeranchuk instabilities) which breaks the discrete lattice symmetry. {\it For hexagonal systems, this promotes a natural two-fold degeneracy of the $d$-wave channel, which we have seen to yield chiral singlet d-wave superconductivity in the triangular and honeycomb lattice as well as a doubly degenerate $d$-wave Pomeranchuk instability on the kagome lattice.} In case of nested Fermi surface features, the hexagonal symmetry gives a natural propensity for the Fermiology to yield three nesting vectors, as seen for the triangular, honeycomb, and kagome lattice tuned to nested fillings or nested band structure parameters. (Note that for the pnictides on the square lattice, for example, one finds two nesting vectors according to the 90 degree rotation symmetry. For the cuprates, this even reduces to a single nesting channel along the main diagonal in the square Brillouin zone.)  These features alone already have a relevant impact on the Fermi surface instabilitiy profile of hexagonal models, as particularly illustrated by the spin-density wave phase in the honeycomb Hubbard model near van Hove filling as well as in the independent three-directional bond order in the kagome Hubbard model. 

{\it The full understanding of interacting itinerant electrons on hexagonal lattices is, however, only reached when we also consider multi-orbital effects stemming from different physical orbitals on a given lattice site as seen for the cobaltate triangular model in Sec.~\ref{chap:cobaltates} or the multiple sublattices for the honeycomb and kagome models in Sec.~\ref{chap:graphene} and Sec.~\ref{sec:theory:lattice_kagome}.} The kagome Hubbard model, in particular, allows for a remarkable interplay between sublattice and interaction effects, which suppresses the natural spin-density wave instability associated with a particle-hole condensate of zero relative angular momentum and, instead, together with a ferromagnetic fluctuations background, gives rise to highly unconventional Fermi surface instabilities such as particle-hole condensates with finite relative angular momentum (i.e. charge and spin bond order). Given the promising current developments in experiments, ranging from hexagonal arsenides over cobaltate-type layered triangular systems to ultra-cold atomic gas scenarios, it is likely that major discoveries can be hoped for along this direction, and that the FRG will establish a valuable analysis tool for microscopic descriptions of such systems.

\section{Conclusion}
\label{chap:conclusion}


{\it The central aim of this review has been two-fold: (i), to convey the multi-orbital functional renormalization group (FRG) as a 
powerful tool to address Fermi surface instabilities in multi-orbital systems of interacting electrons, and (ii), to use this 
multi-orbital scheme in combination with experiment} to help characterizing universal and more material-dependent aspects of the 
electronic problem.

With the precise and pedagogical description of the FRG in Chap.~\ref{chap:fRG}, the frame has been given to develop an intuitive 
understanding for ordered electronic quantum phases stemming from repulsive Coulomb-type interactions. The FRG approach at work has 
been illustrated for the iron pnictides in Chap.~\ref{chap:pnictide} and hexagonal lattice Hubbard scenarios in Chap.~\ref{chap:hexa}. 
It has allowed us to investigating models exhibiting a plethora of unconventional phases, ranging from ferromagnetism, collinear and 
chiral spin density waves over extended $s$-wave, $f$-wave and topological chiral $d+id$-wave superconductivity to two-fold $d$-wave 
Pomeranchuk instabilities as well as spin and charge bond order. 

From all detailed investigations and comparisons to experiment elaborated on in the review, {\it various universal features emerge 
that provide "overarching'' principles of Fermi surface instabilities.} This includes the natural propensity of hexagonal systems 
to exhibit time-reversal symmetry breaking in the $d$-wave superconducting channel, or the gap anisotropy generically triggered by 
competing magnetic fluctuation as well as multi-orbital effects which we have seen for the extended $s$-wave superconducting state 
in the pnictides (i.e. $s \pm$-frustration) and the chiral $d$-wave superconductor on the triangular lattice model for sodium doped 
cobaltates. It stresses, in particular, that the physical picture of spin fluctuations which seed the superconducting phases allows 
for a diverse and rather accurate perspective on a broad range of material classes, as accomplished by FRG.

{\it These universal features also come along with several not less important material-dependent, or lattice-dependent, properties 
that have a significant impact on the nature of the electronic state.} Examples mentioned in the review are the absence of contingent 
electron pockets for hole-doped KFe${}_2$As${}_2$, the possible irrelevance of pnictide Fermi pockets because of their $d_{z^2}$ 
orbital content, or the sublattice interference mechanism for the kagome Hubbard model. This kind of features can be taken into 
consideration by the precise knowledge of material-dependent Fermiology entering the FRG approach.

As an overbranching insight, only the combination of universal features and material-dependent properties provide a sufficient 
accuracy for modeling correlated electron systems in order to ensure a fruitful interplay between theoretical prediction and 
experimental endeavor. To our conviction, the functional renormalization group establishes a valuable tool along this direction.

\section*{Acknowledgments}
We particularly wish to thank Maximilian L. Kiesel for joint
collaborations many of which relate to topics addressed in this
review. Furthermore, we thank D. A. Abanin, B. A. Bernevig,
J. Checkelsky, W. Cho, C. Fang, M. Fischer, M. Z. Hasan, C. Honerkamp,
J. Hu, S. A. Kivelson, A. Maharaj, T. Neupert, P. Ong, S. Raghu,
M. Rice, M. Sigrist, Q.-H. Wang, L. Wray, Y. Wu, F.-C. Zhang, and
S. C. Zhang for collaborations on related topics as well as
G. Baskaran, U. Bissbort, S. Blundell, S. Borisenko, J. van den Brink,
P. Brydon, B. B\"uchner, S. Chakravarty, A. V. Chubukov, A. Coldea,
M. Daghofer, S. Davis, H: Ding, T. Devereaux, I. Eremin, R. Fernandez,
I. R. Fisher, R. Hackl, P. Hirschfeld, D. Johrendt, D.-H. Lee,
G. Lonzarich, A. MacKenzie, Y. Matsuda, I. Mazin, W. Metzner,
R. Prozorov, A. Rost, M. Salmhofer, D. J. Scalapino, J. Schmalian,
H. Takagi, L. Taillefer, Z. Tesanovic, R. Valenti, F. Wang, and
A. Yazdani for fruitful discussions. C.P., W.H., and R.T. are
supported by DFG-SPP 1458/2. R.T. further acknowledges support by the
ERC project TOPOLECTRICS, ERC-StG-2013-336012.

 \section{Appendix}
 \label{chap:appendix}
  \subsection{Ward-Identities}\label{sec:ward}
 In Chap.~\ref{chap:fRG}, we discussed the effect of symmetry on vertex functions and utilized these insights
 to reduce the complexity of the corresponding flow equations. Instead of looking at the implications of symmetry 
 on a given vertex function $\gamma^{2m}$, we now want to derive exact relations between the different
 $m$-point vertex functions, also known as Ward identities. While not
 immediately important for the present formulation of FRG employed in
 the review, it will certainly prove relevant for various possible
 future step in FRG method development. 

Consider a typical action such as
  \begin{equation}
 S(\overline{\psi},\psi)  = -(\overline{\psi},Q\psi) + V_{int}(\overline{\psi},\psi),
 \end{equation}
 to which we apply an infinitesimal field transformation 
 \begin{equation}\label{eq:inffieldmap}
 \psi' = \psi-T\cdot\psi,\quad
 \overline{\psi}' = \overline{\psi}-\overline{\psi}\cdot\overline{T}
 \end{equation}
 characterized by some generators $T,\overline{T}$. The only restriction we require here is that (\ref{eq:inffieldmap}) does neither change $V_{int}$ 
 nor the functional integration measure $D(\overline{\psi},\psi)$. Besides these two conditions, the field transformation 
 in (\ref{eq:inffieldmap}) can be chosen arbitrarily, and the following arguments do not depend on its precise form.
 Note that, for the sake of clarity, we used the same shortened matrix-vector like notation as in Sec.~\ref{sec:floweq}. If we then
 perform (\ref{eq:inffieldmap}), the action changes only in the quadratic part and hence gives rise to
  \begin{equation}\label{eq:actionchange}
 S(\overline{\psi}',\psi') = S(\overline{\psi},\psi) + \left(\overline{\psi},\Delta Q\psi\right),
 \end{equation}
 with an additional quadratic term $\Delta Q$ defined by 
 \begin{equation}
 \Delta Q = QT+\overline{T}Q.
 \end{equation}
 Due to the invariance of the functional integration measure, we further obtain
 \begin{align}\nonumber
    \int  D(\overline{\psi},\psi) e^{-S(\overline{\psi},\psi) + (\overline{\eta},\psi) + (\overline{\psi},\eta)}
 &= \int D(\overline{\psi}',\psi') e^{-S(\overline{\psi}',\psi') + (\overline{\eta},\psi') + (\overline{\psi}',\eta)}\\\label{eq:fieldrename}
  = \int D(\overline{\psi},\psi) e^{-S(\overline{\psi},\psi) + (\overline{\eta},\psi) + (\overline{\psi},\eta)}
   &\left[1-\left(\overline{\psi},\Delta Q\psi\right) + \left(\overline{\eta},T\cdot\psi\right) + \left(\overline{\psi},\overline{T}\cdot\eta\right)\right],
 \end{align}
 where we relabeled the fields $(\overline{\psi},\psi)\rightarrow (\overline{\psi}',\psi')$ and applied (\ref{eq:actionchange}).
 Subtracting the left hand side of (\ref{eq:fieldrename}) and employing the definition of the generating functional $W[\eta,\overline{\eta}]$ in (\ref{eq:W}), 
 we derive the following identity
 \begin{align}\nonumber
 0 &= \int D(\overline{\psi},\psi) e^{-S(\overline{\psi},\psi) + (\overline{\eta},\psi) + (\overline{\psi},\eta)}
      \left[\left(\overline{\psi},\Delta Q\psi\right) + \left(\overline{\eta},T\cdot\psi\right) - \left(\overline{T}\cdot\eta,\overline{\psi}\right)\right]\\\label{eq:wardW}
   &= \left\{\left(\partial_{\eta},\Delta Q\partial_{\overline{\eta}}\right) 
    + \left(\overline{\eta},T\cdot\partial_{\overline{\eta}}\right) + \left(\overline{T}\cdot\eta,\partial_{\eta}\right)\right\}W[\eta,\overline{\eta}].
 \end{align}
 Now, starting from this result, we can generate exact relations between the $m$-point and $(m+2)$-point Green functions, just by 
 taking the appropriate functional derivatives $\partial/\partial\overline{\eta},\partial/\partial\eta$ and by setting $\eta=\overline{\eta}=0$ afterwards.
 However, as we are mainly dealing with flow equations for the 1PI vertex functions, we now want derive Ward identities 
 between the different $m$-point 1PI vertex functions. We therefore insert $W = e^{-\mathcal{G}}$ into (\ref{eq:wardW}),
 which equals the definition of the generating functional for the connected Green functions $\mathcal{G}$ in (\ref{eq:connectG}), and multiply
 with $e^{\mathcal{G}}$ from the right:
 \begin{align}\label{eq:wardconnect}
 0 = e^{\mathcal{G}[\eta,\overline{\eta}]}\left(\partial_{\eta},\Delta Q\partial_{\overline{\eta}}\right)e^{-\mathcal{G}[\eta,\overline{\eta}]} 
     -\left(\overline{\eta},T\cdot\partial_{\overline{\eta}}\right)\mathcal{G}[\eta,\overline{\eta}] - \left(\overline{T}\cdot\eta,\partial_{\eta}\right)\mathcal{G}[\eta,\overline{\eta}]. 
 \end{align}
 What is striking here is that the first term in
 (\ref{eq:wardconnect}) has exactly the same form as the flow equation (\ref{eq:mathcalGflowfirst}),
 only with $\dot{Q}^{\Lambda}$ replaced by $\Delta Q$. Due to this similarity, we can simply apply the calculation steps leading 
 from the flow equation for $\mathcal{G}^{\Lambda}$ in (\ref{eq:mathcalGflowfirst}) to the one for the effective action $\Gamma^{\Lambda}$ in (\ref{eq:partialres}).
 The resulting expression then reads as
 \begin{align}\label{eq:wardgamma}
 0 &= \left(\overline{\zeta},\Delta Q\zeta\right) + \text{tr}\left(\Delta Q\left(\left(\bs{\partial}^2\Gamma[\zeta,\overline{\zeta}]\right)^{-1}\right)_{11}\right)\\\nonumber
   &- \left(T\cdot\zeta,\partial_{\zeta}\right)\Gamma[\zeta,\overline{\zeta}] - \left(\overline{\zeta},\overline{T}\cdot\partial_{\overline{\zeta}}\right)\Gamma[\zeta,\overline{\zeta}]
 \end{align}
 and hence generates all Ward-identities for the 1PI vertex functions.\par
\begin{figure}[t]
 \begin{center}
 {\includegraphics[scale=0.65]{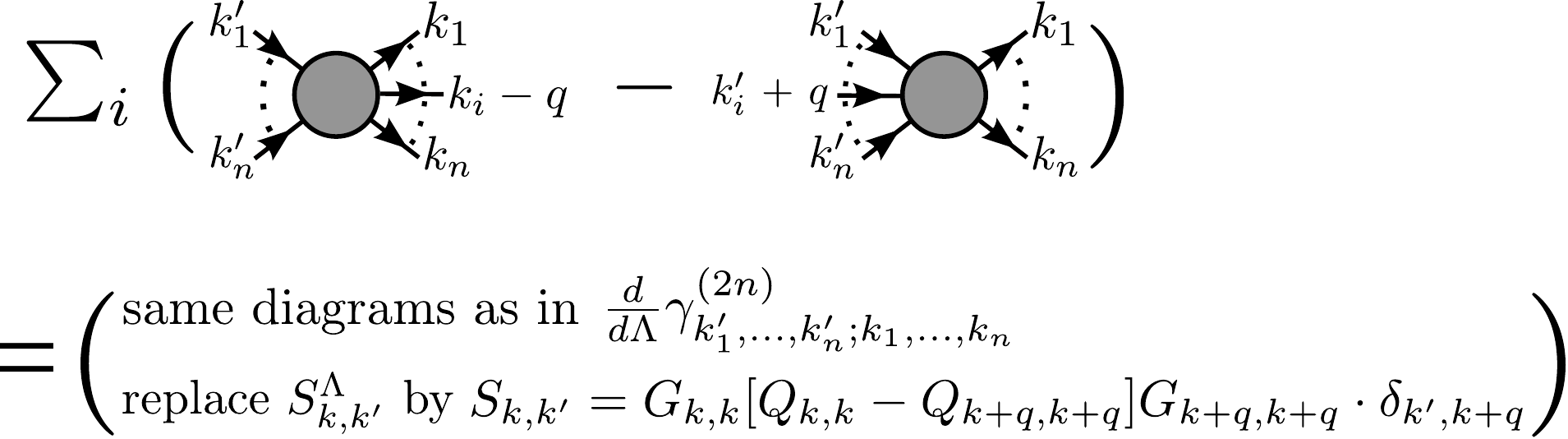}}
 \end{center}
 \caption{\label{pic:ward}\label{wardpic} $U(1)$-Ward identities for the 1PI vertex functions. The right-hand side
         equals the flow equation for $\gamma^{(2n)}$ with a replaced single-scale propagator $S^{\Lambda}$, and the left hand side shows
         the sum of differences of $\gamma^{(2n)}$ functions with $q$ subtracted (added) in the $i$-th outgoing (ingoing) leg.}
 \end{figure}

 Now, after this general discussion, we want to apply (\ref{eq:wardgamma}) and derive the Ward-identities associated with a $U(1)$-phase transformation 
 in the one-band Hubbard model
 \begin{equation}\label{eq:onebandaction}
 S(\overline{\psi},\psi)  = -\int_k\overline{\psi}_kQ_{k,k}\psi_k + U\int_{k_1,k_2,k'_1,k'_2}
                             \overline{\psi}_{k_1}\overline{\psi}_{k_2}\psi_{k'_1}\psi_{k'_2}. 
 \end{equation}
 Here, the quadratic part is given by $Q_{k,k} = ik_0 - \xi(\bs{k})$ with a dispersion relation $\xi(\bs{k})$ and the condensed notation of
 $k=(k_0,\bs{k})$ introduced in (\ref{eq:bareaction}). We now consider the space-time dependent field-transformations
 $\psi'_{r} = e^{i\alpha(r)}\psi'_{r}$, $\overline{\psi}'_{r} = e^{-i\alpha(r)}\overline{\psi}_{r}$
 which, in its infinitesimal form, read as
 \begin{equation}\label{eq:u1fieldtransform}
 \psi'_{r} = \psi_{r}+i\alpha(r)\psi_{r},\quad
 \overline{\psi}'_{r} = \overline{\psi}_{r}-i\alpha(r)\overline{\psi}_{r}.
 \end{equation}
 Note that we switched here from frequency- and momentum-dependent fields to its space-time representation with $r=(\tau,\bs{r})$ including 
 imaginary-time $\tau$ and spatial coordinates $\bs{r}$. The transformation in (\ref{eq:u1fieldtransform})
 then describes a local $U(1)$ phase transformation and, by switching back to   
 frequency and momentum space, the same transformation is given by the convolution
 \begin{equation}
 \psi'_{k} = \psi_{k}+i\int_q\alpha(-q)\psi_{k+q},\quad
 \overline{\psi}'_{k} = \overline{\psi}_{k}-i\int_q\alpha(-q)\overline{\psi}_{k-q}.
 \end{equation}
 The corresponding generators therefore read as 
 \begin{equation}
 T_{kk'} = i\int_q\alpha(-q)\delta_{k',k+q}, \quad \overline{T}_{kk'} = -i\int_q\alpha(-q)\delta_{k,k'-q}.
 \end{equation}
 Using (\ref{eq:wardgamma}) and taking the functional derivative with
 respect to $\alpha(-q)$, we obtain the following identity
 \begin{align}\label{eq:u1wardgamma}
 \int_k\frac{\partial\Gamma}{\partial\zeta_k}\zeta_{k+q} - \frac{\partial\Gamma}{\partial\overline{\zeta}_{k+q}}\overline{\zeta}_{k}&
 =\int_k\left[Q_{k,k} - Q_{k+q,k+q}\right]
        \left[\left(\left(\bs{\partial}^2\Gamma\right)^{-1}_{k+q,k}\right)_{11} + \overline{\zeta}_k \zeta_{k+q}\right].
 \end{align}
 Starting from this latter expression, we can now generate all orders of Ward identities just by
 expanding in powers $\overline{\zeta},\zeta$ and by comparing coefficients. In the $n$-th order Ward-identity,
 i.e. in the identity for the coefficients of order $\mathcal{O}((\overline{\zeta}\zeta)^{n})$,
 the right-hand side of (\ref{eq:u1wardgamma}) generates all diagrams of the flow equation
 in Fig.~\ref{pic:floweq}
 but with a modified single-scale propagator
 \begin{equation}
 S_{kk'} = G_{k,k}\cdot\left[Q_{k,k} - Q_{k+q,k+q}\right]\cdot G_{k+q,k+q}\cdot\delta_{k',k+q}.
 \end{equation}
 Here, $G$ denotes the full propagator and $Q$ is the quadratic part of the action (\ref{eq:onebandaction}). 
 The left-hand side of the $n$-th order Ward-identity then simply consists 
 of a sum over $n$ differences where $\gamma^{(2n)}$ with $q$ added to the $i$-th ingoing leg is subtracted from 
 $\gamma^{(2n)}$ with $-q$ added to the $i$-th outgoing leg. The pictorial representation of all expansion orders
 of (\ref{eq:u1wardgamma}) is shown in Fig.~\ref{pic:ward}.
 For the first order Ward identity, we therefore obtain the following exact 
 relation between the 2-point 1PI-vertex function or self-energy $\Sigma$ and the 4-point 1PI-vertex function $\gamma^{(4)}$: 
 \begin{align}\label{eq:2ptward}
 \Sigma_{w,w} - \Sigma_{w+q,w+q} = \int_k\left[Q_{k,k} - Q_{k+q,k+q}\right]
                                   \gamma^{(4)}_{w,k+q;w+q,k}G_{k+q,k+q}G_{k,k}.
 \end{align}
 Note that this statement is based on the very general assumption that both the functional integration measure and the non-quadratic part
 of the action remain invariant under the infinitesimal field-transformation (\ref{eq:inffieldmap}). Within the functional RG approach,
 we equipped the quadratic part of the action with some parameter
 dependence $Q_{k,k}\rightarrow Q^{\Lambda}_{k,k}$ in order to extrapolate from a solvable model at large values of $\Lambda$
 to the original full model at $\Lambda \rightarrow 0$. As this introduced $\Lambda$-dependence only changes the quadratic part
 of the action, the requirements for (\ref{eq:2ptward}) remain valid and equation (\ref{eq:2ptward}) holds on all scales $\Lambda$. 
 Yet, if one truncates the infinite hierarchy of flow equations, for example by setting $\gamma^{(2n)} = 0$ for all $n>2$, the second order
 Ward identity is no longer fulfilled as this would also include finite contributions from $\gamma^{(6)}$.
 Even the first order Ward-identity (\ref{eq:2ptward}) is then
 violated by terms of the order of $\mathcal{O}((\gamma^{(4)})^3)$
 as shown by Katanin \cite{katanin04prb115109}. There, it was shown that the case of $q=(q_0,\bs{0},\sigma)$ in (\ref{eq:2ptward}) leads to
 \begin{align}\label{eq:partialward}
 \partial_{w_0}\Sigma^{\Lambda}_{w,w}= -\int_k\gamma^{(4)\Lambda}_{w,k;k,w}\left(G^{\Lambda}_{k}\right)^2
                           = \int_{\widetilde{k}}
                               \left\{V^{(4)\Lambda}_{\widetilde{w},\widetilde{k};\widetilde{k},\widetilde{w}}
                                      -2V^{(4)\Lambda}_{\widetilde{w},\widetilde{k};\widetilde{w},\widetilde{k}}
                               \right\}\left(G^{\Lambda}_{\widetilde{k}}\right)^2,
 \end{align}
 where, assuming spin-rotational invariance, the spin sum was already performed with the notation of (\ref{eq:v4define}) and
 the $\Lambda$ dependence was introduced. Now, if (\ref{eq:partialward}) is fulfilled, the $\Lambda$-derivative of the right-side should
 equal the $w_0$-derivative of the flow equation $\partial_{\Lambda}\Sigma^{\Lambda}_{w,w}$ due to
 $\partial_{\Lambda}\partial_{w_0}\Sigma^{\Lambda}_{w,w} = \partial_{w_0}\partial_{\Lambda}\Sigma^{\Lambda}_{w,w}$.
 This latter equation and therefore the Ward identity (\ref{eq:partialward}) does not hold if the hierarchy 
 of flow-equations is truncated with $\gamma^{(2n)} = 0$ for all $n>2$ as shown in \cite{katanin04prb115109}.
 Including certain terms from the neglected $\gamma^{(6)}$ contribution to $\gamma^{(4)}$, Katanin showed that
 this improves the fulfillment of (\ref{eq:partialward}) to correction terms in the order of $\mathcal{O}((\gamma^{(4)})^4)$.
 A simplified version of this scheme, known as Katanin truncation, which replaces the single-scale propagator $S^{\Lambda}$ 
 by the total derivative $dG^{\Lambda}/d\Lambda$, still leads to an essential improvement \cite{salmhofer-04ptp943,thesisreuther,cluster,reuther-10prb144410,PhysRevB.83.024402} 
 and, for example, recovers dressed RPA. 
  \subsection{Renormalized Mean-Field Analysis}\label{sec:mftreat}
 In this section, we describe the mean-field treatment of an effective low-energy theory obtained from the functional RG. Similar to the 
 implementation presented by Reiss $et\ al.$ \cite{reiss-07prb075110}, the flow is stopped at an energy scale $\Lambda_{MF}\gtrsim\Lambda_{c}$, and the renormalized 
 interaction part is decoupled into its leading mean-field channels. For the moment, we 
 assume that the most leading low-energy correlations appear exclusively within the pairing channel, i.e.  
 \begin{equation}\label{decomp}
\gamma^{(4)\Lambda_{MF}}(k_1,k_2;k_3,k_4)\approx V_{s_1,\ldots,s_4}^{pair}(\bs{k}_1,\bs{k}_3)\delta_{\bs{k}_2,-\bs{k}_1}\delta_{\bs{k}_4,-\bs{k}_3},
\end{equation}
with $V_{s_1,\ldots,s_4}^{pair}(\bs{k}_1,\bs{k}_3)=\gamma^{(4)\Lambda_{MF}}(\bs{k}_1s_1,-\bs{k}_1s_2;\bs{k}_3s_3,-\bs{k}_3s_4)$. 
Note that this latter assumption can of course be easily extended to additional correlation channels, which then merely
increases the number of mean-fields in the calculation. The effective low-energy theory taken from the functional RG is then given 
by the following reduced Hamiltonian
\begin{equation}\label{eq:red}
H^{\Lambda} = \sum_{\bs{k}s}\xi(\bs{k})c_{\bs{k}s}^{\dagger}c_{\bs{k}s}^{\phantom{\dagger}} + 
              \frac{1}{2N}\sum_{s_1,\ldots,s_4}\sum_{\bs{k},\bs{q}}
               V_{s_1,\ldots,s_4}^{pair}(\bs{k},\bs{q})
               c_{-\bs{k}s_1}^{\dagger}c_{\bs{k}s_2}^{\dagger}c_{\bs{q}s_3}^{\phantom{\dagger}}c_{-\bs{q}s_4}^{\phantom{\dagger}}.
\end{equation}
Mean field theory provides an exact solution of these reduced types of Hamiltonians in the thermodynamic limit \cite{muehlschlaegel-62jmp522}.
We replace the quartic interaction part by 
\begin{align}\nonumber
c_{-\bs{k}s_1}^{\dagger}c_{\bs{k}s_2}^{\dagger}c_{\bs{q}s_3}^{\phantom{\dagger}}c_{-\bs{q}s_4}^{\phantom{\dagger}}
&= \langle c_{-\bs{k}s_1}^{\dagger}c_{\bs{k}s_2}^{\dagger}\rangle c_{\bs{q}s_3}^{\phantom{\dagger}}c_{-\bs{q}s_4}^{\phantom{\dagger}}
+ c_{-\bs{k}s_1}^{\dagger}c_{\bs{k}s_2}^{\dagger}\langle c_{\bs{q}s_3}^{\phantom{\dagger}}c_{-\bs{q}s_4}^{\phantom{\dagger}}\rangle \\\nonumber
&-\langle c_{-\bs{k}s_1}^{\dagger}c_{\bs{k}s_2}^{\dagger}\rangle\langle c_{\bs{q}s_3}^{\phantom{\dagger}}c_{-\bs{q}s_4}^{\phantom{\dagger}}\rangle\\\label{eq:meanfielddec}
&+\left(c_{-\bs{k}s_1}^{\dagger}c_{\bs{k}s_2}^{\dagger}  - \langle c_{-\bs{k}s_1}^{\dagger}c_{\bs{k}s_2}^{\dagger}\rangle \right)
    \left(c_{\bs{q}s_3}^{\phantom{\dagger}}c_{-\bs{q}s_4}^{\phantom{\dagger}} - \langle c_{\bs{q}s_3}^{\phantom{\dagger}}c_{-\bs{q}s_4}^{\phantom{\dagger}}\rangle \right)
\end{align}
and perform the usual mean-field approximation of neglecting order-parameter fluctuations in the last line of (\ref{eq:meanfielddec}). 
Introducing the parameters
\begin{align}\nonumber
\Delta_{\bs{k},ss'} & = -\frac{1}{N}\sum_{\bs{k}',\sigma\sigma'} V_{s',s,\sigma,\sigma'}^{pair}(\bs{k},\bs{k}') 
                       \langle c_{\bs{k}'\sigma}^{\phantom{\dagger}}c_{-\bs{k}'\sigma'}^{\phantom{\dagger}}\rangle\\\nonumber
\Delta^{*}_{\bs{k},ss'} & = -\frac{1}{N}\sum_{\bs{k}',\sigma\sigma'} V_{\sigma',\sigma,s,s'}^{pair}(\bs{k}',\bs{k}) 
                       \langle c_{-\bs{k}'\sigma'}^{\dagger}c_{\bs{k}'\sigma}^{\dagger}\rangle\\\nonumber
                        & = \frac{1}{N}\sum_{\bs{k}',\sigma\sigma'} V_{\sigma',\sigma,s',s}^{pair}(\bs{k}',-\bs{k}) 
                       \langle c_{-\bs{k}'\sigma'}^{\dagger}c_{\bs{k}'\sigma}^{\dagger}\rangle,
\end{align}
the Hamiltonian in (\ref{eq:red}) then reads as
\begin{align}\nonumber
H^{\Lambda} &= \sum_{\bs{k}s}\xi(\bs{k})c_{\bs{k}s}^{\dagger}c_{\bs{k}s}^{\phantom{\dagger}}
              + \frac{1}{2}\sum_{\bs{k},s_1s_2} \Delta_{\bs{k},s_1s_2}
               c_{\bs{k}s_1}^{\dagger}c_{-\bs{k}s_2}^{\dagger} 
               - \frac{1}{2}\sum_{\bs{q},s_3s_4}\Delta^{*}_{-\bs{q},s_4s_3}
                c_{-\bs{q}s_4}^{\phantom{\dagger}}c_{\bs{q}s_3}^{\phantom{\dagger}} + \mathcal{K}\\\label{eq:meanfieldham}
             &= \frac{1}{2}\sum_{\bs{k}} 
 C_{\bs{k}}^{\dagger}
  \begin{pmatrix}
 \xi(\bs{k})\sigma_0 & \hat{\Delta}^{\phantom{\dagger}}_{\bs{k}} \\
 \hat{\Delta}^{\dagger}_{\bs{k}} & -\xi(-\bs{k})\sigma_0
 \end{pmatrix}
 C^{\phantom{\dagger}}_{\bs{k}} + \sum_{\bs{k}}\xi(\bs{k}) + \mathcal{K},
\end{align}
with $\sigma_0$ denoting the two-dimensional unit matrix and $\hat{\Delta}^{\phantom{\dagger}}_{\bs{k}}$, $C_{\bs{k}}^{\dagger}$ given by
 \begin{equation}
 \hat{\Delta}^{\phantom{\dagger}}_{\bs{k}} = 
 \begin{pmatrix}
 \Delta^{\phantom{\dagger}}_{\bs{k},\uparrow\uparrow} & \Delta^{\phantom{\dagger}}_{\bs{k},\uparrow\downarrow} \\
 \Delta^{\phantom{\dagger}}_{\bs{k},\downarrow\uparrow} & \Delta^{\phantom{\dagger}}_{\bs{k},\downarrow\downarrow}
 \end{pmatrix},
 \quad
 C_{\bs{k}}^{\dagger} =
 \begin{pmatrix}
 c_{\bs{k}\uparrow}^{\dagger} & c_{\bs{k}\downarrow}^{\dagger} & c_{-\bs{k}\uparrow}^{\phantom{\dagger}} & c_{-\bs{k}\downarrow}^{\phantom{\dagger}}
 \end{pmatrix}.
 \end{equation}
 Note that in (\ref{eq:meanfieldham}), we also made use of 
 $\hat{\Delta}^{\dagger}_{\bs{k}} = (\hat{\Delta}^{T}_{\bs{k}})^* = -(\hat{\Delta}_{-\bs{k}})^{*}$ and
 introduced the $c$-number term $\mathcal{K}$ determined by
 \begin{equation}
\mathcal{K} = - \frac{1}{2N}\sum_{s_1,\ldots,s_4}\sum_{\bs{k},\bs{q}}
                V_{s_1,\ldots,s_4}^{pair}(\bs{k},\bs{q})\langle c_{-\bs{k}s_1}^{\dagger}c_{\bs{k}s_2}^{\dagger}\rangle
                 \langle c_{\bs{q}s_3}^{\phantom{\dagger}}c_{-\bs{q}s_4}^{\phantom{\dagger}}\rangle.
\end{equation}
Assuming a unitary pairing state, which means that 
$\hat{\Delta}^{\phantom{\dagger}}_{\bs{k}}\hat{\Delta}^{\dagger}_{\bs{k}}$ is proportional to the unit matrix 
(see Sigrist and Ueda~\cite{sigrist-91rmp239} for details), we can diagonalize 
the quadratic Hamiltonian (\ref{eq:meanfieldham}):
\begin{align}
H^{MF} &= \frac{1}{2}\sum_{\bs{k}} 
C_{\bs{k}}^{\dagger}
\begin{pmatrix}
\xi(\bs{k})\sigma_0 & \hat{\Delta}^{\phantom{\dagger}}_{\bs{k}} \\
\hat{\Delta}^{\dagger}_{\bs{k}} & -\xi(-\bs{k})\sigma_0
\end{pmatrix}
C^{\phantom{\dagger}}_{\bs{k}} + \sum_{\bs{k}}\xi(\bs{k}) + \mathcal{K}\\\nonumber
&= \frac{1}{2}\sum_{\bs{k}} C_{\bs{k}}^{\dagger}U_{\bs{k}}^{\phantom{\dagger}} 
\begin{pmatrix}
E_{\bs{k}}\sigma_0 & 0 \\
0 & -E_{\bs{k}}\sigma_0 \\
\end{pmatrix}
U_{\bs{k}}^{\dagger}C^{\phantom{\dagger}}_{\bs{k}} + \sum_{\bs{k}}\xi(\bs{k}) + \mathcal{K}\\\label{eq:meanhamdiag}
       &= \sum_{\bs{k}s} E_{\bs{k}}\alpha^{\dagger}_{\bs{k}s}\alpha^{\phantom{\dagger}}_{\bs{k}s} - \sum_{\bs{k}} E_{\bs{k}} + \sum_{\bs{k}}\xi(\bs{k}) + \mathcal{K}. 
\end{align}
Here, the two twofold degenerate eigenvalues $\lambda_{1,2}=\pm E_{\bs{k}}$ were obtained
by using $\text{det}\begin{pmatrix}A & B \\C & D\end{pmatrix} = \text{det}(AD-BC)$ for commuting two-dimensional matrices $C,D$, which then yields the associated energy spectrum 
\begin{equation}\label{eq:excspec}
 E_{\bs{k}}=\sqrt{\xi_{\bs{k}}^2 + \text{tr}(\hat{\Delta}^{\phantom{\dagger}}_{\bs{k}}\hat{\Delta}^{\dagger}_{\bs{k}})/2}.
\end{equation}
The unitary matrix $U_{\bs{k}}$ in (\ref{eq:meanhamdiag}) can likewise be identified as
\begin{equation}\label{eq:Umatrix}
U_{\bs{k}} = 
\begin{pmatrix}
\hat{u}_{\bs{k}} & \hat{v}_{\bs{k}} \\
\hat{v}^*_{-\bs{k}} & \hat{u}^*_{-\bs{k}}
\end{pmatrix}
=
\frac{1}{\sqrt{2E_{\bs{k}}(E_{\bs{k}} + \xi_{\bs{k}})}}
\begin{pmatrix}
\sigma_0(E_{\bs{k}} + \xi_{\bs{k}}) & -\hat{\Delta}_{\bs{k}}^{\phantom{\dagger}} \\
\hat{\Delta}_{\bs{k}}^{\dagger} &  \sigma_0(E_{\bs{k}} + \xi_{\bs{k}})
\end{pmatrix}
\end{equation}
and determines the quasi-particle excitations $\alpha^{\dagger}_{\bs{k}s}, \alpha^{\phantom{\dagger}}_{\bs{k}s}$
mixing particle and hole states:
\begin{equation}\label{eq:bogol}
\alpha_{\bs{k}s} = \sum_{s'}u^*_{\bs{k},ss'}c^{\phantom{\dagger}}_{\bs{k}s'} + v_{-\bs{k}ss'}c^{\dagger}_{-\bs{k}s'}.
\end{equation}
In matrix-vector notation, (\ref{eq:bogol}) can also be written as
\begin{equation}\label{eq:bogomat}
 \begin{pmatrix}
 \alpha_{\bs{k}\uparrow}^{\dagger} & \alpha_{\bs{k}\downarrow}^{\dagger} & \alpha_{-\bs{k}\uparrow}^{\phantom{\dagger}} & \alpha_{-\bs{k}\downarrow}^{\phantom{\dagger}}
 \end{pmatrix}
 =
 \begin{pmatrix}
 c_{\bs{k}\uparrow}^{\dagger} & c_{\bs{k}\downarrow}^{\dagger} & c_{-\bs{k}\uparrow}^{\phantom{\dagger}} & c_{-\bs{k}\downarrow}^{\phantom{\dagger}}
 \end{pmatrix}
 U_{\bs{k}}.
\end{equation}
The grand canonical potential of (\ref{eq:meanhamdiag}) can be easily expressed as   
\begin{equation}\label{eq:grandpot}
\Omega = - T \sum_{\bs{k}s}\ln \left(1 + e^{-\beta E_{\bs{k}}}\right) - \sum_{\bs{k}} E_{\bs{k}} + \sum_{\bs{k}}\xi(\bs{k}) + \mathcal{K},
\end{equation}
and the calculation of a stationary point in $\Omega$ turns out to be equivalent to the self-consistent gap equation: 
\begin{equation}\label{eq:gapeq}
0 = \frac{\partial\Omega}{\partial\langle c_{-\bs{k}s_1}^{\dagger}c_{\bs{k}s_2}^{\dagger}\rangle} \Longleftrightarrow\Delta_{\bs{k},s_2s_1}=
-\frac{1}{N}\sum_{\bs{q},s's} V^{pair}_{s_1s_2s's}(\bs{k},\bs{q}) \frac{\Delta_{\bs{q},s's}}{2E(\bs{q})}\tanh\left(\frac{E(\bs{q})}{2T}\right).
\end{equation}
In order to evaluate $\Omega$ in a stationary point and to compare the free energies of different gap solutions, we apply (\ref{eq:gapeq}) and 
obtain the following expression for the free energy in a stationary solution
\begin{align}\nonumber
\Omega^{stat} =& - T \sum_{\bs{k}s}\ln \left(1 + e^{-\beta E(\bs{k})}\right) 
                 + \frac{1}{2}\sum_{\bs{k},s_1s_2} \frac{\Delta_{\bs{k},s_2s_1}^{\dagger}
                   \Delta_{\bs{k},s_2s_1}}{2E(\bs{k})}\tanh\left(\frac{E(\bs{k})}{2T}\right)\\\label{eq:omegaeval}
               & + \sum_{\bs{k}}\xi(\bs{k}) -\sum_{\bs{k}}E(\bs{k}).
\end{align}
For the subsequent analysis, we then take $V^{pair}$ from the functional RG at a scale $\Lambda_{MF}$ and 
minimize the grand-canonical potential (\ref{eq:grandpot}). This in turn can be achieved by locating the stationary points
of $\Omega$ using (\ref{eq:gapeq}) and by comparing the associated free energies via (\ref{eq:omegaeval}). Note that this 
combined functional RG and mean-field approach goes far beyond conventional mean-field studies as the functional RG takes into account all 
fluctuations down to an energy scale $\Lambda_{MF}\gtrsim \Lambda_c$. For a reasonable estimate, we can also consider the linear gap-equation as a 
first-order expansion of (\ref{eq:gapeq}) in $\hat{\Delta}_{\bs{k}}$, which becomes exact in the vicinity of the superconducting 
transition. The self-consistent gap equation then reads as
\begin{align}\nonumber
\Delta_{\bs{k},s_2s_1}&=
-\frac{1}{N}\sum_{\bs{q},s's} V^{pair}_{s_1s_2s's}(\bs{k},\bs{q}) \frac{\Delta_{\bs{q},s's}}{2\xi(\bs{q})}\tanh\left(\frac{\xi(\bs{q})}{2T}\right)
 + \mathcal{O}(\Delta^2)\\\label{eq:lingapeqpre}
&=\frac{1}{\lambda}\sum_{s's}\langle V^{pair}_{s_1s_2s's}(\bs{k},\bs{q})\Delta_{\bs{q},s's}\rangle_{\bs{q}\in FS} + \mathcal{O}(\Delta^2) 
\end{align}
with a constant $\lambda$ being equivalent to~\cite{sigrist-91rmp239}:
\begin{equation}
\frac{1}{\lambda} = -N_0 \int_0^{\Lambda_c}d\xi\frac{\tanh\left(\frac{\xi(\bs{q})}{2T}\right)}{\xi(\bs{q})} = -N_0\ln (1.13\Lambda_c/T). 
\end{equation}
 In linear order, the gap-equation (\ref{eq:lingapeqpre}) therefore reduces to an eigenvalue problem in the form of
\begin{equation}\label{eq:lingapeq}
\lambda\Delta_{\bs{k},s_2s_1} = \sum_{s's}\langle V^{pair}_{s_1s_2s's}(\bs{k},\bs{q})\Delta_{\bs{q},s's}\rangle_{\bs{q}\in FS},
\end{equation}
and the gap $\Delta$ of largest negative eigenvalue $\lambda$ determines the transition temperature $T_c$ through
\begin{equation}\label{eq:tctest}
T_c = 1.13\Lambda_ce^{1/(N_0\lambda)}\le 1.13\Lambda_c.
\end{equation}
We will usually compare the flow of eigenvalues in order to identify the favored type of order.
From (\ref{eq:tctest}) it is also apparent that the critical energy scale $\Lambda_c$ already gives an upper bound for $T_c$.   
At $T=0$, we can further apply (\ref{eq:omegaeval}) to determine the ground-state energy 
\cite{mineev-book}: 
\begin{align}\nonumber
\Omega_{T=0} &= \frac{1}{2}\sum_{\bs{k},s}(\xi(\bs{k}) - E(\bs{k})) + 
\frac{1}{4}\sum_{\bs{k}}\text{tr}\left(\Delta^{\dagger}_{\bs{k}}\Delta^{\phantom{\dagger}}_{\bs{k}}\right)E(\bs{k})\\\label{eq:groundstateenergy}
&\approx - N_0\text{tr}\langle(\Delta^{\dagger}_{\bs{k}}\Delta^{\phantom{\dagger}}_{\bs{k}})\rangle_{\bs{k}\in FS}.
\end{align}
Here, (\ref{eq:groundstateenergy}) implies that gap zeros tend to be unfavorable due to a lower condensation energy. 

\subsection{Symmetry Classification of Particle-Particle and Particle-Hole Condensates}\label{sec:ppphcond}
 
\subsubsection{General scope}
\begin{figure}[t]
  \centering
    {\includegraphics[scale=0.6]{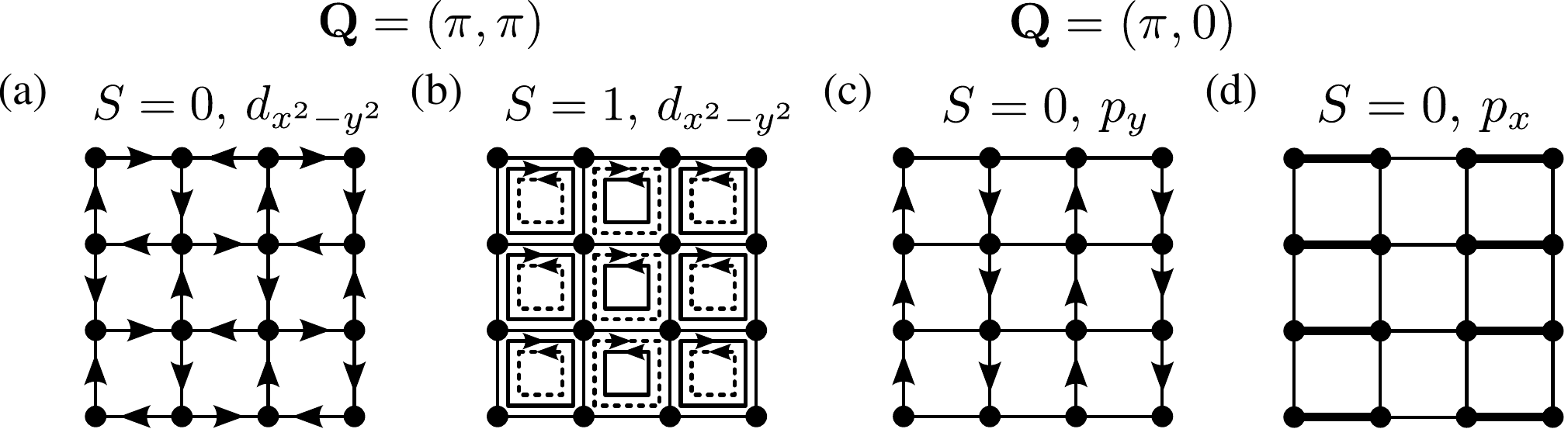}}
  \caption{\label{fig:phorder} Different examples of particle-hole condensates with finite angular momentum on the square lattice. (a) Singlet $d$-density wave phase with alternating plaquette currents,
   (b) triplet $d$-density wave phase with alternating currents of spin-up (full-lined arrows) and spin-down (dashed-line arrows).  (c) Alternating currents along the vertical bonds and 
   (d) Peierls state with enhanced (thick lines) and reduced (thin lines) hopping amplitudes. (a) and (c) break time-reversal symmetry, (a)-(d) break the $C_{4v}$ point-group symmetry 
   as well as translational symmetry, (b) breaks also spin-rotational symmetry.} 
  \end{figure}

According to Landau's theory of phase transitions, different states of matter can be characterized from the perspective
 of spontaneous symmetry breaking. Here, the symmetry of the ground-state is spontaneously reduced below a certain critical temperature, 
 and the system develops some kind of additional order. As a consequence, one can find a field variable which
 acquires a finite expectation value, known as order parameter. In the vicinity of such a phase transition, 
 the free energy can be expanded in polynomials of the order-parameter field $\phi$, i.e.
 \begin{equation}\nonumber
 F[\phi] \propto (T-T_c)\phi^2 + u\phi^4 + \ldots,
 \end{equation}
 and it is directly apparent that the fields $\phi$ of generically
 identical $T_c$ form an irreducible representation of the underlying symmetry group. Using the Pauli-matrix notation with
 $\widetilde{\bs{\sigma}}=(\sigma^+,\sigma^3,\sigma^-)$ as well as the antisymmetric tensor $\epsilon_{\alpha\beta}$, the particle-particle like
 order-parameter fields are characterized by
 \begin{align}\nonumber
 \langle c^{\dagger}_{-\bs{k}\alpha}c^{\dagger}_{\bs{k}+\bs{Q}\beta}\rangle &= \Phi_{\bs{Q}}(\bs{k})\epsilon_{\alpha\beta}\\\label{eq:ppgeneral}
 \langle c^{\dagger}_{-\bs{k}\alpha}c^{\dagger}_{\bs{k}+\bs{Q}\beta}\rangle &= \left(\bs{\vec{\Phi}}_{\bs{Q}}(\bs{k})\cdot\widetilde{\bs{\sigma}}\right)_{\alpha\gamma}\epsilon_{\gamma\beta}.
 \end{align}
 Here, we also in principle allow for a nonzero center-of-mass momentum $\bs{Q}$ which leads to an additional breaking of translational symmetry, known
 as Fulde-Ferrell-Larkin-Ovchinnikov (FFLO) state~\cite{fulde-64pr550}. 
 Analogous to the spin-singlet and triplet particle-particle fields in (\ref{eq:ppgeneral}),
 one can also decompose the particle-hole fields into irreducible representations of the spin-rotation group, which gives rise to analogous spin-singlet and triplet 
 order-parameter fields
 \begin{align}\nonumber
 \langle c^{\dagger}_{\bs{k}\alpha}c^{\phantom{\dagger}}_{\bs{k}+\bs{Q}\beta}\rangle &= \Phi_{\bs{Q}}(\bs{k})\delta_{\alpha\beta}\\\label{eq:phgeneral}
 \langle c^{\dagger}_{\bs{k}\alpha}c^{\phantom{\dagger}}_{\bs{k}+\bs{Q}\beta}\rangle &= \left(\bs{\vec{\Phi}}_{\bs{Q}}(\bs{k})\cdot\bs{\sigma}\right)_{\alpha\beta}.
 \end{align}
 Here, $\delta_{\alpha\beta}$ indicates the Kronecker delta and $\bs{\sigma}$ is given by $\bs{\sigma} = (\sigma^1,\sigma^2,\sigma^3)$ with
 the usual Pauli-matrices $\sigma^{1,2,3}$. 
 Note that the singlet and triplet contributions in (\ref{eq:ppgeneral})
 and (\ref{eq:phgeneral}) are quite different due to the distinct transformation 
 behavior of particle- and hole-operators under spin-rotation. On the other hand, the $\bs{k}$-dependent parts $\Phi_{\bs{Q}}(\bs{k})$ and $\bs{\vec{\Phi}}_{\bs{Q}}(\bs{k})$
 transform in both particle-particle and particle-hole cases as irreducible representations of the space group which leaves $\bs{Q}$ invariant modulo reciprocal lattice vectors.
 In Fig.~\ref{fig:phorder}, we illustrate four examples of unconventional (non $s$-wave) particle-hole condensates on the square lattice. The corresponding
 order-parameter fields for (a)-(d) read as follows:
 \begin{align}\nonumber
 \bs{Q}=(\pi,\pi)&,\ S=0,\ L\sim d_{x^2-y^2}:\ \langle c^{\dagger}_{\bs{k}\alpha}c^{\phantom{\dagger}}_{\bs{k}+(\pi,\pi)\beta}\rangle\propto  (\cos(k_x) - \cos(k_y))\delta_{\alpha\beta}\\\nonumber
 \bs{Q}=(\pi,\pi)&,\ S=1,\ L\sim d_{x^2-y^2}:\ \langle c^{\dagger}_{\bs{k}\alpha}c^{\phantom{\dagger}}_{\bs{k}+(\pi,\pi)\beta}\rangle\propto  (\cos(k_x) - \cos(k_y))(\bs{n}\cdot\bs{\sigma})_{\alpha\beta}\\\nonumber
 \bs{Q}=(\pi,0)&,\ S=0,\ L\sim p_{y}:\hspace{0.82cm}\langle c^{\dagger}_{\bs{k}\alpha}c^{\phantom{\dagger}}_{\bs{k}+(\pi,0)\beta}\rangle\propto  \sin(k_y)\delta_{\alpha\beta}\\\nonumber
 \bs{Q}=(\pi,0)&,\ S=0,\ L\sim p_{x}:\hspace{0.82cm}\langle c^{\dagger}_{\bs{k}\alpha}c^{\phantom{\dagger}}_{\bs{k}+(\pi,0)\beta}\rangle\propto  \sin(k_x)\delta_{\alpha\beta}.
 \end{align}
 Note that the usual charge- and spin-density waves of ordering vector $\bs{Q}$ are similarly described by an $s$-wave particle-hole condensate in the spin-singlet and triplet channel, respectively.
 For a detailed description of the experimental signatures and the excitation spectrum in unconventional particle-hole condensates we refer to the articles of Nayak~\ea~\cite{nayak-00prb4880}
 and Garcia-Aldea~\cite{garcia-10prb184526}.\par
 During the next section, we present a construction recipe for calculating all basis functions $\Phi_{\bs{Q}}(\bs{k})$ and $\bs{\vec{\Phi}}_{\bs{Q}}(\bs{k})$ in a 
 given irreducible point-group representation. This in turn provides a complete characterization
 of two-particle type order parameter fields on arbitrary lattice geometries. In addition, the presented scheme 
 can be easily extended to the classification of order-parameters containing more than two operators.
 \begin{table}[h!]
 \begin{tabular}{|l|c|}
 \hline
 \cellcolor{hellgrau}$E$: &\cellcolor{hellblau2} identity operation\\
 \hline
 \cellcolor{hellgrau}$C_{n}$: &\cellcolor{hellrot} rotation through $2\pi/n$ with $n=1,2,3,4,6$ \\
 \cellcolor{hellgrau}         &\cellcolor{hellrot} the axis of highest $n$ is called principle  \\
 \hline
 \cellcolor{hellgrau}$\sigma_v$: &\cellcolor{hellblau2} vertical reflection plane - passing through\\
 \cellcolor{hellgrau}            &\cellcolor{hellblau2} the origin and the principle axis\\ 
 \hline
 \cellcolor{hellgrau}$\sigma_d$: &\cellcolor{hellrot} special case of $\sigma_v$ but also bisecting the angle between\\
 \cellcolor{hellgrau}            &\cellcolor{hellrot} two two-fold rotational axes perpendicular to the principal axis\\
 \hline
 \cellcolor{hellgrau}$\sigma_h$: &\cellcolor{hellblau2} horizontal reflection plane - passing through\\
 \cellcolor{hellgrau}            &\cellcolor{hellblau2} the origin and perpenticular to the principle axis.\\ 
 \hline
 \cellcolor{hellgrau}$S_n$: &\cellcolor{hellrot} rotation through $2\pi/n$ followed by a reflection in the plane\\
 \cellcolor{hellgrau}       &\cellcolor{hellrot} perpendicular to the axis of rotation\\
 \hline
 \end{tabular}
 \caption{\label{tab:schoenflies}Symmetry operations of point groups in Schoenflies notation.}
 \end{table}        
 \subsubsection{Elements of Representations Theory}\label{sec:elemreptheory}
\begin{figure}[h!]
  \begin{center}
  {\includegraphics[scale=1.4]{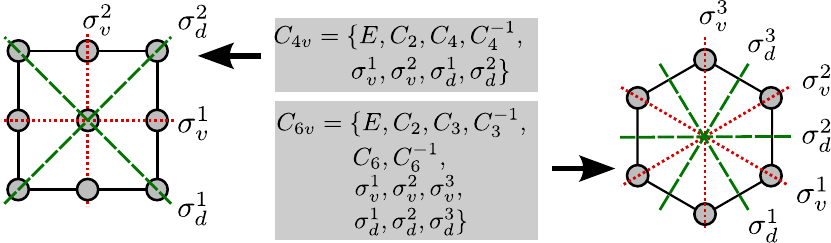}}
  \end{center}
  \caption{\label{fig:c4vc6v}Point group elements of the square-lattice $C_{4v}$ (left) and hexagonal-lattice structure $C_{6v}$ (right).}
  \end{figure}

In the following section, we want to recap the basic notions of representation theory in order to calculate the
above basis functions $\Phi_{\bs{Q}}(\bs{k})$ and $\bs{\vec{\Phi}}_{\bs{Q}}(\bs{k})$ for different lattice systems. 
However, to begin with, we start with the definition of an abstract group consisting of a
set of elements $\mathcal{G}$ together with some operation $\circ$ acting on that set.
\begin{itemize}\itemsep8pt
\item[]\underline{\textit{Definition (Group)}}
\item[] A group $\mathcal{G}=(G,\circ)$ consists of a set $G$ and an operation $\circ$ such that:
\item[]
  \begin{itemize}\itemsep8pt
  \item[1)] For all $a,b\in G$, the result $a\circ b$ is also in $G$
  \item[2)] The operation $\circ$ is associative, i.e. $a\circ (b \circ c) = (a\circ b)\circ c$ holds for all elements $a,b,c\in G$
  \item[3)] There exists a unit element such that $a\circ e = e\circ a = a$ for $a\in G$
  \item[4)] For each $a\in G$, there exists an inverse element $a^{-1}\in G$ such that $a\circ a^{-1} = a^{-1}\circ a = e$.
  \end{itemize}
\end{itemize}
As an example, we consider the set of transformations which map a given lattice into itself by leaving
one point fixed. This point does not necessarily has to be a lattice point, and the corresponding transformations are commonly termed as point 
groups. The elements in these point groups typically consist of lattice transformations like the one shown in Tab.~\ref{tab:schoenflies}.
Note that the last two operations of Tab.~\ref{tab:schoenflies} are contained in the first four 
in case of a two-dimensional lattice structure. As an example, we
depicted the point-groups for the square- and hexagonal-lattice structures in Fig.~\ref{fig:c4vc6v}, which play a pivotal role in
the discussion of Chap.~\ref{chap:pnictide} and Chap.~\ref{chap:hexa}. Here, $C^{-1}_{3,4,6}$ denote the inverse elements of $C_{3,4,6}$, and all other elements equal their own inverse.
Another definition which is essential when discussing representation theory is the notion of conjugate elements and classes given in the following.
 \begin{itemize}\itemsep8pt
 \item[]\underline{\textit{Conjugate Elements and Classes in a Group}}
 \item[]
  \begin{itemize}\itemsep8pt
  \item[1)] Two elements $a,b$ of the group $\mathcal{G}=(G,\circ)$ are conjugate, written $a\sim b$, if and only if there is another element $x\in G$ such that
             $b=x\circ a\circ x^{-1}$.
  \item[2)] A class is the entirety of conjugate elements. 
       \end{itemize}
 \end{itemize}
 In order to calculate the classes of a given group, 
 one can simply compute all conjugate elements for each group element. At this, it turns
out that the group $C_{4v}$ and $C_{6v}$ consist of 5 and, respectively, 6 different classes written down below
\begin{align}\nonumber
C_{4v} &: \{E\},\ \{C_2\},\ 2C_{4}=\{C_4,C^{-1}_4\},\ 2\sigma_v = \{\sigma^1_v,\sigma^2_v\},\ 2\sigma_d := \{\sigma^1_d,\sigma^2_d\}\\\nonumber
C_{6v} &: \{E\},\ \{C_2\},\ 2C_{3}=\{C_3,C^{-1}_3\},\ 2C_{6}=\{C_6,C^{-1}_6\},\\\label{eq:classes}
       &\hspace{2.5cm} 3\sigma_v = \{\sigma^1_v,\sigma^2_v,\sigma^3_v\},\ 3\sigma_d := \{\sigma^1_d,\sigma^2_d,\sigma^3_d\}.
\end{align}
Later, it will turn out that the number of classes in a group equals the number of irreducible representations, but, before that,
we want to review the basic concepts of representation theory. The idea here is to avoid the calculation with abstract group elements and to represent 
each element by objects that are more convenient to deal with, for example linear operators in the form of square matrices.  
Besides that, representation theory often allows to make statements about degeneracies in energy-spectra
or enables to determine matrix elements without explicit integration. For this reason, we summarize the following notions
required for later discussions.
\begin{itemize}\itemsep8pt
\item[]\underline{\textit{Representation of a Group}}
\item[]A group $\mathcal{R} = (R,\cdot)$ is a representation of another group $\mathcal{G}=(G,\circ)$ if there is a one-to-one mapping $M:G\mapsto R$ such that
       $M(a\circ b) = M(a)\cdot M(b)$ for all $a,b$ in $G$.
       In all cases considered here, $R$ is a set of $n$-dimensional square matrices, and ``$\cdot$'' denotes the usual matrix multiplication. In the following,
       we use the abbreviation $\Gamma$ to denote a representation of a group $\mathcal{G}$.
\item[]
      \begin{itemize}\itemsep8pt
      \item[1)] If the representation  $\Gamma$ consists of $(n\times n)$-matrices, it is termed $n$-dimensional.
      \item[2)] Two $n$-dimensional representations $\Gamma_1, \Gamma_2$ of a group $\mathcal{G}$ are equivalent, 
            if there is a regular $(n\times n)$-matrix $U$ such that
            $N(a) = U\cdot M(a)\cdot U^{-1}$ for all elements $a$ in $\mathcal{G}$. Here, $N(a)$ and $M(a)$ denote the matrices assigned to 
            $a$ in the representations $\Gamma_1$ and $\Gamma_2$, respectively.
      \item[3)] A representation  $\Gamma$ is denoted as reducible if it is equivalent in terms of (2) to a representation where all matrices have a common block structure
              $M(a)=\bigl(\begin{smallmatrix}
                         M_1(a) & 0 \\
                            0   & M_2(a)
                     \end{smallmatrix}\bigr)$. If this is not possible, the representation is denoted as irreducible.
      \end{itemize}
\end{itemize}
Two immediate consequences follow from this definition. First, the representation matrix of
the unit element $e$ is always given by a unit matrix, and, secondly, each group 
has a trivial representation provided by assigning $M(g)=1$ to all group elements $g\in\mathcal{G}$.
In order to determine all irreducible representations of a given group or to figure out whether a given representation is reducible,
we need the following concept of characters.
\begin{itemize}\itemsep8pt
\item[]\underline{\textit{The Usage of Characters}}
\item[]The character of a group element $g$ in a representation $\Gamma_i$ is determined by the trace $\chi^{i}(g)=\text{tr}(D(g))$
       of its representing matrix $D(g)$.
\item[]
       \begin{itemize}\itemsep8pt
       \item[1)] The dimension $n_i$ of a representation $\Gamma_i$ is given by the character 
             of the identity map $e$, i.e. $n_i = \chi_i(e)$.
       \item[2)] The number of classes $n_c$ in a given group equals the number of (inequivalent) irreducible representations.
       \item[3)] The characters $\chi_i$ of a representation $\Gamma_i$ are equivalent for all elements in the same class, and,
             since there are as many classes as (inequivalent) irreducible representations, we can arrange all characters
             in a $(n_c\times n_c)$-character table:
             \vspace{0.1cm}
             \item[]
              \begin{center}
              \begin{tabular}[t]{|c|c|c|c|}
              \hline
              \cellcolor{hellgelb}$\mathcal{G}$ & \cellcolor{hellgrau}$\mathcal{C}_1$ & \cellcolor{hellgrau}$\cdots$ & \cellcolor{hellgrau}$\mathcal{C}_{n_c}$\\
              \hline
              \cellcolor{dunkelgrau}$\Gamma_1$ & $\chi_1(\mathcal{C}_1)$ & $\cdots$ & $\chi_1(\mathcal{C}_{n_c})$ \\
              \hline
              \cellcolor{dunkelgrau}$\vdots$ & $\vdots$ &   & $\vdots$\\
              \hline
              \cellcolor{dunkelgrau}$\Gamma_{n_c}$ & $\chi_{n_c}(\mathcal{C}_1)$ & $\cdots$ & $\chi_{n_c}(\mathcal{C}_{n_c})$\\
              \hline
              \end{tabular}
              \end{center}
             \vspace{0.1cm}
       \item[4)] There are two very elegant orthogonality relations which allow to compute such character tables. The first one 
             states that the scalar product of two columns gives
             \begin{equation}\label{eq:firstortho}
             \sum_{i=1}^{n_c}\chi_i(\mathcal{C}_q)\chi^*_i(\mathcal{C}_{q'}) = \delta_{qq'}N/h_q,
             \end{equation}
             with $N$ denoting the number of elements in the entire group, and $h_q$ labeling the number of elements in class $\mathcal{C}_q$.
             The second relation provides a similar relation for the weighted scalar product of two rows in the character table:    
             \begin{equation}\label{eq:secondortho}
             \sum_{q=1}^{n_c}h_q\chi_i(\mathcal{C}_q)\chi^*_j(\mathcal{C}_q) = \delta_{ij}N.
             \end{equation}
       \end{itemize}
     \end{itemize}
It is now an easy task to determine the character tables for the two point groups $C_{4v}$ and $C_{6v}$. 
Applying (\ref{eq:firstortho}) to the classes $\mathcal{C}_q = \mathcal{C}_q' = \{e\}$ and using $\chi_i(e) = n_i$, we obtain 
\begin{align}\nonumber
C_{4v}&:\quad n_1^2 + n_2^2 + n_3^2 + n_4^2 + n_5^2 = 8\\\nonumber
C_{6v}&:\quad n_1^2 + n_2^2 + n_3^2 + n_4^2 + n_5^2 + n_6^2= 12
\end{align}
as the group $C_{4v}$ ($C_{6v}$) consists of 8 (12) group elements and 5 (6) classes.
This in turn implies that $C_{4v}$ has one two-dimensional and four one-dimensional irreducible representations, whereas 
$C_{6v}$ reveals two two-dimensional and four one-dimensional representations. Therefore, the first columns of the character 
tables in Fig.~\ref{fig:c4vc6vcharacter} are already determined, and, as each group has
  \begin{figure}[h!]
 \centering
 \begin{minipage}[t]{0.4\textwidth}
 \begin{tabular}[t]{|r|r|r|r|r|r|}
 \hline
 \cellcolor{hellgelb}$C_{4v}$ & \cellcolor{hellgrau}$E$ & \cellcolor{hellgrau}$C_2$ & \cellcolor{hellgrau}$2C_4$ & \cellcolor{hellgrau}$2\sigma_v$ & \cellcolor{hellgrau}$2\sigma_d$\\
 \hline
 \cellcolor{dunkelgrau}$A_{1}$ & 1 & 1 & 1 & 1 & 1 \\
 \hline
 \cellcolor{dunkelgrau}$A_{2}$ & 1 & 1 & 1 & -1 & -1 \\
 \hline
 \cellcolor{dunkelgrau}$B_{1}$ & 1 & 1 & -1 & 1 & -1 \\
 \hline
 \cellcolor{dunkelgrau}$B_{2}$ & 1 & 1 & -1 & -1 & 1 \\
 \hline
 \cellcolor{dunkelgrau}$E_{1}$ & 2 & -2 & 0 & 0 & 0 \\
 \hline
 \end{tabular}
 \end{minipage}
 \hspace{0.5cm}
 \begin{minipage}[t]{0.52\textwidth}
 \begin{tabular}[t]{|r|r|r|r|r|r|r|}
  \hline
  \cellcolor{hellgelb}$C_{6v}$ & \cellcolor{hellgrau}$E$ & \cellcolor{hellgrau}$C_2$ & \cellcolor{hellgrau}$2C_3$ & \cellcolor{hellgrau}$2C_6$ & \cellcolor{hellgrau}$3\sigma_v$ & \cellcolor{hellgrau}$3\sigma_d$\\
  \hline
  \cellcolor{dunkelgrau}$A_{1}$ & 1 & 1 & 1 & 1 & 1 & 1\\
  \hline
  \cellcolor{dunkelgrau}$A_{2}$ & 1 & 1 & 1 & 1 & -1 & -1\\
  \hline
  \cellcolor{dunkelgrau}$B_{1}$ & 1 & -1 & 1 & -1 & 1 & -1\\
  \hline
  \cellcolor{dunkelgrau}$B_{2}$ & 1 & -1 & 1 & -1 & -1 & 1\\
  \hline
  \cellcolor{dunkelgrau}$E_{1}$ & 2 & -2 & -1 & 1 & 0 & 0\\
  \hline
  \cellcolor{dunkelgrau}$E_{2}$ & 2 & 2 & -1 & -1 & 0 & 0\\
  \hline
  \end{tabular}
 \end{minipage}
 \caption{\label{fig:c4vc6vcharacter}Character tables for the two point groups $C_{4v}$ (left) and $C_{6v}$ (right). The different rows
          denote the different irreducible representations, the columns label the various classes in each 
          group (see Eq.~(\ref{eq:classes})).}
 \end{figure}
 a trivial representation with $M(g) = 1$ for all $g\in\mathcal{G}$, one has in addition one trivial row in each character table (see Fig.~\ref{fig:c4vc6vcharacter}).
 Using the orthogonality of columns and rows according to Eq.~(\ref{eq:firstortho}) and (\ref{eq:secondortho}), we can also determine the remaining entries of the character tables
 illustrated in Fig.~\ref{fig:c4vc6vcharacter}.\par
 In order to decide whether a given representation of a group $\mathcal{G}$ is irreducible, one
 simply has to compute its characters and see whether these coincide with a row in the associated character table. 
 If this is not the case, the representation is reducible
 and there is an equivalent representation where all matrices $M_{red}(g)$
 are of the same block structure. Each of these blocks then forms an irreducible representation
 of the group $\mathcal{G}$, and we can write $M_{red}(g)$ as a direct sum
 \begin{equation}\label{eq:reducing}
 M_{red}(g) = c_1M_1(g) \oplus c_2M_2(g) \oplus \cdots c_{n_c}M_{n_c}(g) 
 \end{equation}
 for all $g$ in $\mathcal{G}$ with
 \begin{equation}\nonumber
 c_i = \frac{1}{N}\sum_{q = 1}^{n_c} h_q\chi_{red}(g)\chi_q^*(g).
 \end{equation}
 The character table therefore allows an efficient way of fully reducing a given representation,
 and, for example, gives information on crystal field splittings as 
 the irreducible representations of the full rotation group
 operating on the atomic states is reducible in the lower symmetry subgroup of
 a crystal structure. This then leads to a splitting of the former degenerate states according to (\ref{eq:reducing}) with
 $c_i$ non-degenerate levels which transform with $M_i$.\par 
   \begin{figure}[b]
 \centering
 \begin{minipage}[c]{0.4\textwidth}
 {\includegraphics[scale=1.0]{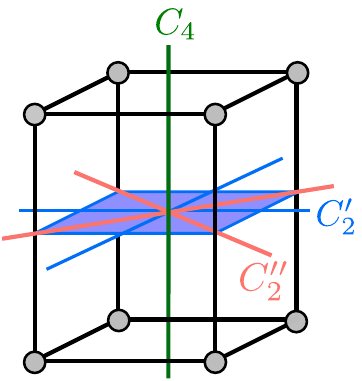}}
 \end{minipage}
 \hspace{1cm}
  \begin{minipage}[c]{0.4\textwidth}
 \begin{tabular}[c]{|r|r|r|r|r|r|}
 \hline
 \cellcolor{hellgelb}$C_{4v}$ & \cellcolor{hellgrau}$E$ & \cellcolor{hellgrau}$C_2$ & \cellcolor{hellgrau}$2C_4$ & \cellcolor{hellgrau}$2\sigma_v$ & \cellcolor{hellgrau}$2\sigma_d$\\
 \hline
 \cellcolor{dunkelgrau}$A_{1}$ & 1 & 1 & 1 & 1 & 1 \\
 \hline
 \cellcolor{dunkelgrau}$A_{2}$ & 1 & 1 & 1 & -1 & -1 \\
 \hline
 \cellcolor{dunkelgrau}$B_{1}$ & 1 & 1 & -1 & 1 & -1 \\
 \hline
 \cellcolor{dunkelgrau}$B_{2}$ & 1 & 1 & -1 & -1 & 1 \\
 \hline
 \cellcolor{dunkelgrau}$E_{1}$ & 2 & -2 & 0 & 0 & 0 \\
 \hline
 \end{tabular}
 \end{minipage}
 \hspace{3.0cm}
 \begin{minipage}[c]{0.3\textwidth}
 \vspace{-3cm}
 \begin{tabular}[c]{|r|r|r|}
  \hline
  \cellcolor{hellgelb}$C_{I}$ & \cellcolor{hellgrau}$E$ & \cellcolor{hellgrau}$I$\\
  \hline
  \cellcolor{dunkelgrau}$A_{1}$ & 1 & 1\\
  \hline
  \cellcolor{dunkelgrau}$A_{2}$ & 1 & -1\\
  \hline
  \end{tabular}
 \end{minipage}
 \caption{\label{fig:c4vcIcharacter}Tetragonal symmetry (left) with one four-fold and four two-fold rotational axes as
          well as the character tables for the two point groups $C_{I}$ (middle) and $C_{4v}$ (right).}
 \end{figure}
 Moreover, if we have the character tables for two different groups $\mathcal{G}$ and $\mathcal{K}$, we can
 easily write down the character table for the direct product $\mathcal{G}\otimes\mathcal{K}$:
 \begin{equation}\label{eq:charmult}
 \chi^{\mathcal{G}\otimes\mathcal{K}}_{ij}(gk) = \chi^{\mathcal{G}_1}_i(g)\chi^{\mathcal{G}_2}_j(k).
 \end{equation}
  \begin{table}[h!]
  \centering
  \begin{tabular}[t]{|r|r|r|r|r|r||r|r|r|r|r|}
  \hline
  \cellcolor{hellgelb}$D_{4h}$ & \cellcolor{hellgrau}$E$ & \cellcolor{hellgrau}$C_2$ & \cellcolor{hellgrau}$2C_4$ & \cellcolor{hellgrau}$2\sigma_v$ & \cellcolor{hellgrau}$2\sigma_d$
                               & \cellcolor{hellgrau}$I$ & \cellcolor{hellgrau}$C_2I$ & \cellcolor{hellgrau}$2C_4I$ & \cellcolor{hellgrau}$2\sigma_vI$ & \cellcolor{hellgrau}$2\sigma_dI$\\
  \hline
  \cellcolor{dunkelgrau}$A_{1g}$ & 1 & 1 & 1 & 1 & 1 & 1 & 1 & 1 & 1 & 1\\
  \hline
  \cellcolor{dunkelgrau}$A_{2g}$ & 1 & 1 & 1 & -1 & -1 & 1 & 1 & 1 & -1 & -1\\
  \hline
  \cellcolor{dunkelgrau}$B_{1g}$ & 1 & 1 & -1 & 1 & -1 & 1 & 1 & -1 & 1 & -1\\
  \hline
  \cellcolor{dunkelgrau}$B_{2g}$ & 1 & 1 & -1 & -1 & 1 & 1 & 1 & -1 & -1 & 1\\
  \hline
  \cellcolor{dunkelgrau}$E_{g}$ & 2 & -2 & 0 & 0 & 0 & 2 & -2 & 0 & 0 & 0\\
  \hline
  \hline
  \cellcolor{dunkelgrau}$A_{1u}$ & 1 & 1 & 1 & 1 & 1 & -1 & -1 & -1 & -1 & -1\\
  \hline
  \cellcolor{dunkelgrau}$A_{2u}$ & 1 & 1 & 1 & -1 & -1 & -1 & -1 & -1 & 1 & 1\\
  \hline
  \cellcolor{dunkelgrau}$B_{1u}$ & 1 & 1 & -1 & 1 & -1 & -1 & -1 & 1 & -1 & 1\\
  \hline
  \cellcolor{dunkelgrau}$B_{2u}$ & 1 & 1 & -1 & -1 & 1 & -1 & -1 & 1 & 1 & -1\\
  \hline
  \cellcolor{dunkelgrau}$E_{u}$ & 2 & -2 & 0 & 0 & 0 & -2 & 2 & 0 & 0 & 0\\
  \hline
  \end{tabular}
  \caption{\label{tab:tetra}Character table for the tetragonal symmetry group $D_{4h} = C_{4v}\otimes C_I$ calculated by multiplying the character
           tables in Fig.~\ref{fig:c4vcIcharacter} according to Eq.~(\ref{eq:charmult}).
           In the literature one often uses the following equivalent notations $\sigma_h = C_{2}I$, $2S_4 = 2C_4I$, $C_2'=2\sigma_v I$ and $C_2''=2\sigma_dI$.
           Note that the additional rotational axes $C_2'$ and $C_2''$ 
           are depicted in Fig.~\ref{fig:c4vcIcharacter}.}
  \end{table}
  From this, we can then derive the character tables for 
  $D_{4h} = C_{4v}\otimes C_I$ or $C_{nh} = C_{n}\otimes C_I$,
  where $C_I$ is the group consisting only of the identity map 
  and of the space inversion $(x,y,z)\rightarrow (-x,-y,-z)$. 
  As an example, we determine the character table of the 
  tetragonal symmetry group $D_{4h}$ in Tab.~\ref{tab:tetra}.
 \subsubsection{Basis Functions for Irreducible Representations}
\begin{figure}[t]
 \centering
   {\includegraphics[scale=0.3]{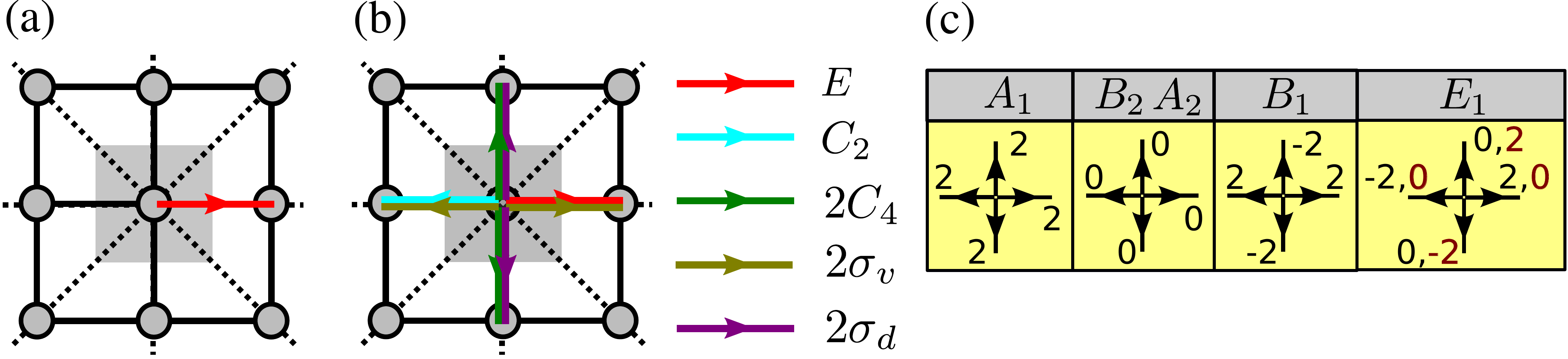}}
 \caption{\label{fig:c4vbasis}Projection method for the nearest-neighbor basis functions $\phi(r_i,r_j)$ on the square lattice with $C_{4v}$ symmetry.
          Starting with a nearest-neighbor bond (a), we obtain the bond structure in (b) with indicated color coding, and, if we assign the complex-conjugated
          characters according to Table \ref{fig:c4vc6vcharacter}, we end up with the real-space functions $\phi(r_i,r_j)$ in (c). The black and red numbers in the representation $E_1$ of (c)
          correspond to the two required basis functions.} 
 \end{figure}

 As discussed in Sec.~\ref{sec:ppphcond}, the particle-particle $\phi=\langle c^{\dagger}c^{\dagger}\rangle$ and 
 particle-hole $\phi=\langle c^{\dagger}c\rangle$ like order-parameter fields can be classified in terms of irreducible representations
 of the underlying symmetry group. Separating off the spin part from $\phi$, we obtained the following contributions of total spin $S=1$ and total spin $S=0$:  
  \begin{align}\nonumber
 \langle c^{\dagger}_{-\bs{k}\alpha}c^{\dagger}_{\bs{k}+\bs{Q}\beta}\rangle &= \Phi_{\bs{Q}}(\bs{k})\epsilon_{\alpha\beta},\hspace{1.8cm} 
 \langle c^{\dagger}_{\bs{k}\alpha}c^{\phantom{\dagger}}_{\bs{k}+\bs{Q}\beta}\rangle = \Phi_{\bs{Q}}(\bs{k})\delta_{\alpha\beta}\\\nonumber
 \langle c^{\dagger}_{-\bs{k}\alpha}c^{\dagger}_{\bs{k}+\bs{Q}\beta}\rangle &= \left(\bs{\vec{\Phi}}_{\bs{Q}}(\bs{k})\cdot\widetilde{\bs{\sigma}}\right)_{\alpha\gamma}\epsilon_{\gamma\beta},\hspace{0.5cm}
 \langle c^{\dagger}_{\bs{k}\alpha}c^{\phantom{\dagger}}_{\bs{k}+\bs{Q}\beta}\rangle = \left(\bs{\vec{\Phi}}_{\bs{Q}}(\bs{k})\cdot\bs{\sigma}\right)_{\alpha\beta}
 \end{align}
 with $\Phi_{\bs{Q}}$ and the components of $\bs{\vec{\Phi}}_{\bs{Q}}$ transforming in the irreducible representation of the space-group that
 leaves $\bs{Q}$ invariant (modulo reciprocal lattice vectors). In the following section, we calculate all possible 
 basis functions $\Phi_{\bs{Q}}$ and $\bs{\vec{\Phi}}_{\bs{Q}}$ in a systematic way.\\
 First of all, we note that the transformation behavior under lattice translation $T$ is already determined by the center-of-mass momentum $\bs{Q}$, i.e.
 \begin{equation}\nonumber
 T\phi_{\bs{Q}}(\bs{k}) = e^{i\bs{Q}\bs{r}}\phi_{\bs{Q}}(\bs{k}),
 \end{equation}
 and it is therefore sufficient to derive the basis functions only for the lattice point-group that leaves $\bs{Q}$ invariant modulo reciprocal lattice vectors.
 For this purpose, we consider the real-space representation 
 \begin{equation}\nonumber
 \phi(r_1,r_2) = \sum_{\bs{k}} \phi_{\bs{Q}}(\bs{k})\exp(i\bs{k}(r_1-r_2)),
 \end{equation}
 with $\phi_{\bs{Q}}(\bs{k})$ used in place of $\Phi_{\bs{Q}}$ and in place of the components of $\bs{\vec{\Phi}}_{\bs{Q}}$. 
 In order to construct wave functions $\phi(r_1,r_2)$ that transform
 as an irreducible representation of the point-group, we make use of a fundamental projection theorem stating that the operator
 \begin{equation}\label{eq:projop}
 \mathcal{P}(\Gamma_i)=\sum_{g}\chi^*_i(g)g
 \end{equation}
 projects out the contribution which transforms in the $i$-th irreducible representation $\Gamma_i$.
 Here, the sum runs over all point-group operations $g$ with the corresponding complex-conjugate characters $\chi^*_i(g)$. Let us now apply this 
 result to the simplest case of a square lattice with point-group $C_{4v}$ and project out the contributions in $\Gamma_i$ from the trial 
 wave function given by
 \begin{equation}\label{eq:formfacstart}
 \phi^{start}(r_i,r_j) = 1\delta_{i,i+x}.
 \end{equation}
 The trial wave function $\phi^{start}$ is apparently nonzero only for nearest-neighbor bonds in $x$-direction as indicated in Fig.~\ref{fig:c4vbasis}a. 
 Using the characters of Tab.~\ref{fig:c4vc6vcharacter}, we then apply the projection operator (\ref{eq:projop}) to $\phi^{start}$ and obtain
 the following nearest-neighbor basis function for the trivial representation $A_1$: 
 \begin{align}\nonumber
 \mathcal{P}(A_1)\phi^{start} &= \chi^*_i(E)\delta_{i,i+x} + \chi^*_i(C_2)\delta_{i,i-x} + \chi^*_i(C_4)\delta_{i,i+y} + \chi^*_i(C^{-1}_4)\delta_{i,i-y}\\\nonumber
                              &+ \chi^*_i(\sigma_v^1)\delta_{i,i+x} + \chi^*_i(\sigma_v^2)\delta_{i,i-x} + \chi^*_i(\sigma_d^2)\delta_{i,i+y} + \chi^*_i(\sigma_d^1)\delta_{i,i-y}\\\label{eq:formfacen}
                              &= 2\delta_{i,i+x} + 2\delta_{i,i-x} + 2\delta_{i,i+y} + 2\delta_{i,i-y}.
 \end{align}
  Repeating this procedure with the characters of the 4 other representations $A_2$, $B_1$, $B_2$ and $E_2$
 (see Tab.~\ref{fig:c4vc6vcharacter}), we obtain the remaining basis functions as depicted in Fig.~\ref{fig:c4vbasis}c. Here, it turns out that the basis function in
 $B_2$ and $A_2$ vanish for nearest-neighbor bonds.
 In the case of the two dimensional representation $E_1$, a single basis function is not sufficient, and we have to apply the projection to another linear independent trial wave functions, for example
 $\phi^{start}(r_i,r_j) = 1\delta_{i,i+y}$, which then provides the two black and red 
 colored wave functions of Fig.~\ref{fig:c4vbasis}:
 \begin{align}\nonumber
 \phi^{E_1}_1(r_i,r_j) = \delta_{i,i+x} - \delta_{i,i-x}, \quad\phi^{E_1}_2(r_i,r_j) = \delta_{i,i+y} - \delta_{i,i-y}.
 \end{align}
 Iterating this process for longer ranged bonds and transforming back to $\bs{k}$-space, 
 we obtain a complete set of basis functions shown below for up to fourth nearest-neighbors:  
 \begin{align}\nonumber
 \phi^{A_1}(\bs{k}) :&\ \cos(k_x) + \cos(k_y), \cos(k_x)\cos(k_y), \cos(2k_x) + \cos(2k_y), \\\nonumber
                     &\ \cos(k_x)\cos(2k_y) + \cos(2k_x)\cos(k_y),\ldots \\\nonumber
 \phi^{A_2}(\bs{k}) :&\ 0,0,0,\sin(k_x)\sin(2k_y) - \sin(2k_x)\sin(k_y),\ldots\\\nonumber
 \phi^{B_1}(\bs{k}) :&\ \cos(k_x) - \cos(k_y), 0, \cos(2k_x) - \cos(2k_y),\\\nonumber
                     &\ \cos(2k_x)\cos(k_y) - \cos(k_x)\cos(2k_y),\ldots \\\nonumber
 \phi^{B_2}(\bs{k}) :&\ 0, \sin(k_x)\sin(k_y), 0, \sin(2k_x)\sin(k_y) + \sin(k_x)\sin(2k_y),\ldots\\\nonumber
 \phi_1^{E_1}(\bs{k}):&\ \sin(k_x),\sin(k_x + k_y), \sin(2k_x), \sin(2k_x+k_y),\ldots\\\nonumber
 \phi_2^{E_1}(\bs{k}):&\ \sin(k_y), \sin(-k_x+k_y),\sin(2k_y), \sin(-k_x+2k_y),\ldots.
 \end{align}
 The above basis functions hold for all particle-particle and particle-hole like order-parameter fields 
 on the square lattice where the center-of-mass momentum $\bs{Q}$ is also invariant under $C_{4v}$, 
 as it is the case for $Q=(0,0)$ and $Q=(\pi,\pi)$. On the other hand, for $\bs{Q}=(0,\pi)$, we have to use the characters 
 of the $C_{2v}$ group since the rotated momentum $\bs{Q}=(\pi,0)$ cannot be connected through reciprocal lattice vectors.
 It is further important to note that in the particle-particle case, the exchange symmetry requires the spin part of the $E_1$ representation 
 to be of spin-triplet type, whereas all other representations have to be of spin-singlet structure.\par 
 The described projection scheme of course works out for all kind of lattice geometries, and for the triangular lattice we then obtain the following nearest-neighbor basis functions 
 (see Fig.~\ref{fig:c6vbasis}):
 \begin{figure}[h!]
 \centering
   {\includegraphics[scale=0.3]{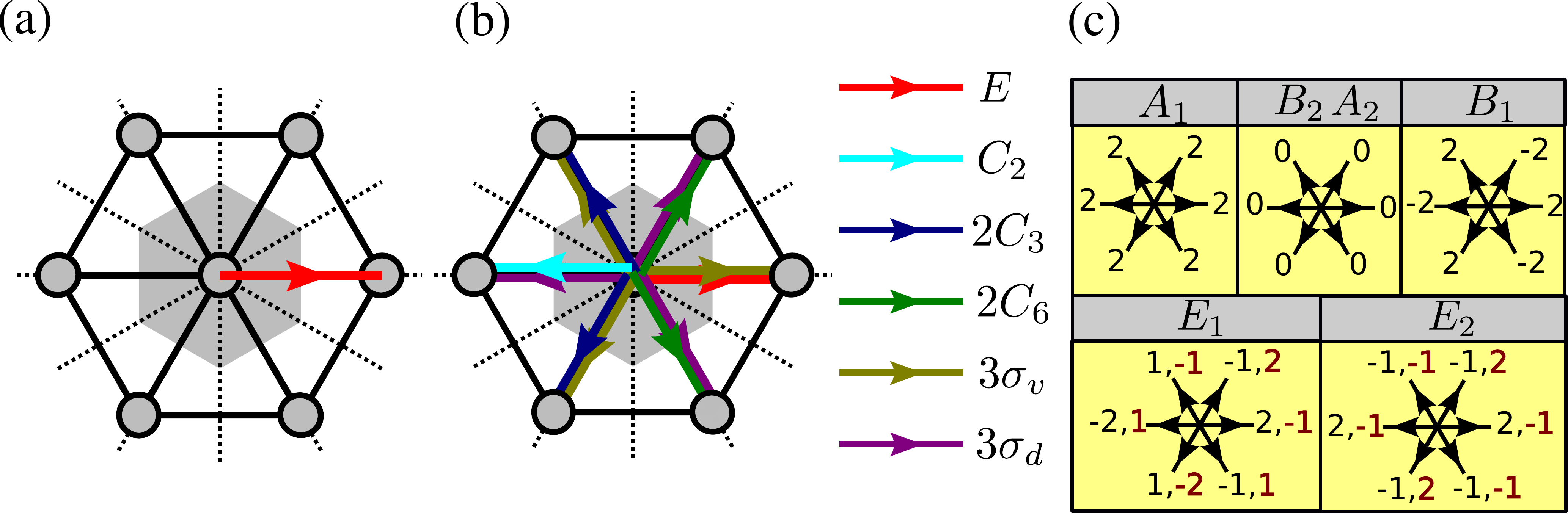}}
 \caption{\label{fig:c6vbasis} Nearest-neighbor basis functions as in Fig.~\ref{fig:c4vbasis} for the triangular lattice with $C_{6v}$ symmetry.} 
 \end{figure}
 \begin{align}\nonumber
 \phi^{A_1}(\bs{k}) :&\ \cos((k_x + \sqrt{3}k_y)/2) + \cos((k_x - \sqrt{3}k_y)/2) + \cos(k_x/2),\ldots\\\nonumber
 \phi^{A_2}(\bs{k}) :&\ 0,\ldots\\\nonumber
 \phi^{B_1}(\bs{k}) :&\ \sin(k_x/2)(1-2\cos(\sqrt{3}k_y/2)),\ldots\\\nonumber
 \phi^{E_1}_1(\bs{k}) :&\ -2\cos(k_x/2)\cos(\sqrt{3}k_y/2) + 2\sin(k_x),\ldots\\\nonumber
 \phi^{E_1}_2(\bs{k}) :&\ 3\cos(k_x/2)\cos(\sqrt{3}k_y/2) + \sin(k_x/2)\sin(\sqrt{3}k_y/2)-\sin(k_x),\ldots\\\nonumber
 \phi^{E_2}_1(\bs{k}) :&\ 2\cos(k_x)-\cos((k_x + \sqrt{3}k_y)/2)-\cos((k_x - \sqrt{3}k_y)/2),\ldots\\\nonumber
 \phi^{E_2}_2(\bs{k}) :&\ -\cos(k_x)+2\cos((k_x + \sqrt{3}k_y)/2)-\cos((k_x - \sqrt{3}k_y)/2)\ldots\quad .
 \end{align}
 In multi-orbital or multi-sublattice systems, the above symmetry classification is more intricate since here the orbital or lattice degrees of freedom itself
 transform under point-group operations. For instance, in the honeycomb lattice with a $C_{6v}$ point-group symmetry, sublattice $A$ is mapped to $B$ under $\pi/3$-rotation,
 and the characterizing symmetry of the order-parameter field apparently depends on its internal orbital structure. For the intraorbital case
 $\langle c^{\dagger}_{-\bs{k}A\alpha}c^{\dagger}_{\bs{k}A\beta}\rangle$ and $\langle c^{\dagger}_{-\bs{k}B\alpha}c^{\dagger}_{\bs{k}B\beta}\rangle$,
 we depicted the nearest-neighbor pairing amplitudes in Fig.~\ref{fig:honeybasis}c. Here, it is important to note that
 the $A_1$, $B_1$ as well as $A_2$, $B_2$ and $E_1$, $E_2$ representations only differ by a relative sign between the two different sublattices.
 In addition, it also turns out that the spin structure in $E_1$, $E_2$ is not determined and therefore contains both a spin-singlet and triplet 
 representation as indicated by the upper and lower signs in Fig.~\ref{fig:honeybasis}c.    
 also changes the sublattice. Therefore, each representation has both, singlet and triplet realizations which are indicated by the upper and lower signs of Fig.~\ref{fig:honeybasis}c.
 Also striking here is that the $A_1$ and $B_1$ as well as the $E_1$ and $E_2$ representations only differe in a relative sign between the two sublattice. So one actually has
 only three different irreducible representations which is related to the fact that the honeycomb lattice essentially has the lower symmetry of $C_{3v}$ with three irreducible representations
 if the fixed reference point is chosen as a lattice site. In all multi-dimensional representations, it sometimes proves useful to orthogonalize the basis functions with regard to the scalar product 
 $\int_{BZ} \phi^*_i(\bs{k})\phi_j(\bs{k}) = \delta_{ij}$ which was not implemented here.
 Another feature of the derived basis functions is that $\int_{BZ} \phi_i(\bs{k}) = 0$ if $\phi_i$ does not transform as the trivial representation $A_1$.
 \begin{figure}[h!]
 \centering
   {\includegraphics[scale=0.2]{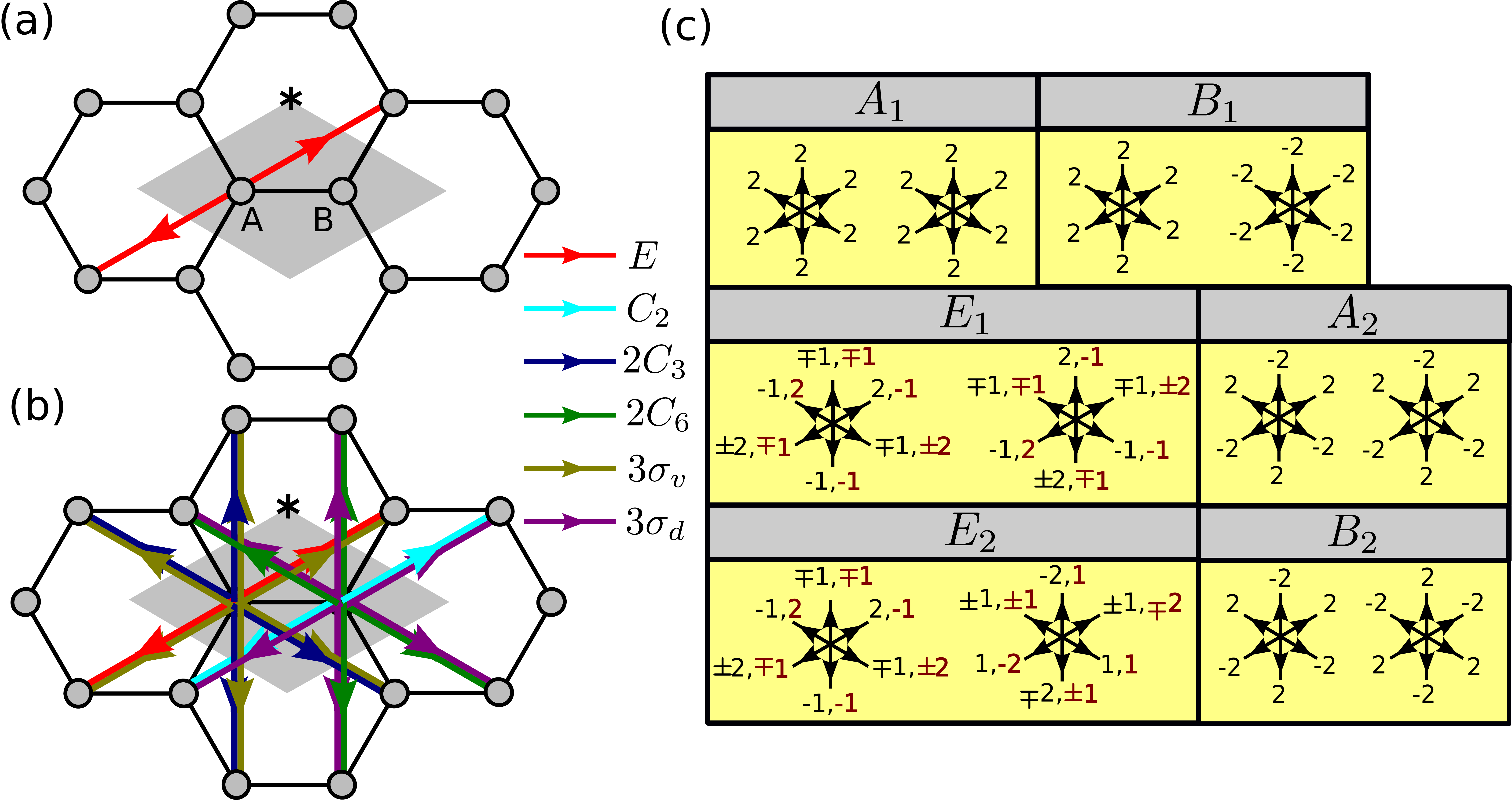}}
 \caption{\label{fig:honeybasis}Projection method to obtain the nearest-neighbor pairing wave functions $\phi_{AA}(r_i,r_j)$ and $\phi_{BB}(r_i,r_j)$ on the honeycomb 
          lattice. Upper and lower signs in $E_1$, $E_2$ denote the corresponding spin-singlet and spin-triplet realizations. The representations $A_1$, $B_1$ as well as 
          $A_2$, $B_2$ and $E_1$, $E_2$ in (c) only differ by a relative sign
          between the two sublattices $A$ (left) and $B$ (right).} 
 \end{figure}
 
 \subsection{Real Space Ordering}\label{sec:realsp}
In the following section, the real-space and orbital structure of the
different ordering tendencies as e.g. found in FeSCs (see Fig.~\ref{fig:formfacs}) is analyzed.
For this purpose, we first invert the transformation from orbital- to band-basis by using
\begin{equation}\label{eq:backtransform}
\gamma_{\bs{k}ms}^{\phantom{\dagger}} = \sum_{a = 1}^5 u_{am}^*(\bs{k})c_{\bs{k}as}^{\phantom{\dagger}},
\end{equation}
and we then rewrite the order parameters (\ref{eq:orderparameter}) in terms of $c_{\bs{k}as}^{\dagger}$, $c_{\bs{k}as}^{\phantom{\dagger}}$. 
For the spin-singlet pairing case, this gives rise to 
\begin{align}\nonumber
O^{SCs} &= \sum_{k} f^{SCs}(k)\langle \gamma_{k\uparrow}^{\phantom{\dagger}}\gamma_{-k\downarrow}^{\phantom{\dagger}}
                                                     -\gamma_{k\downarrow}^{\phantom{\dagger}}\gamma_{-k\uparrow}^{\phantom{\dagger}}\rangle\\\nonumber
&= \sum_{k,ab} f^{SCs}(k)u^*_{am}(\bs{k})u^*_{bm}(-\bs{k})\langle c_{\bs{k}a\uparrow}^{\phantom{\dagger}}c_{-\bs{k}b\downarrow}^{\phantom{\dagger}}
           -c_{\bs{k}a\downarrow}^{\phantom{\dagger}}c_{-\bs{k}b\uparrow}^{\phantom{\dagger}}\rangle\\\nonumber
&= \sum_{ij,ab}\tilde{f}^{SCs}_{ab}(\bs{r}_i-\bs{r}_j)\langle c_{ia\uparrow}^{\phantom{\dagger}}c_{jb\downarrow}^{\phantom{\dagger}}
           -c_{ia\downarrow}^{\phantom{\dagger}}c_{jb\uparrow}^{\phantom{\dagger}}\rangle,
\end{align}
where we defined the real-space and orbital-based form factor
\begin{equation}\nonumber
\tilde{f}^{SCs}_{ab}(\bs{r}_i-\bs{r}_j) = \sum_{k}e^{i\bs{k}(\bs{r}_i-\bs{r}_j)}f^{SCs}(k)u^*_{am}(\bs{k})u^*_{bm}(-\bs{k}).
\end{equation}
Note here that the $k$-sum in the last two expressions includes the summation over momenta $\bs{k}$ and associated band-indices $m$. 
$\tilde{f}^{SCs}_{ab}$ can be interpreted as a wave function of two paired electrons 
in orbital $a$ and $b$. 
\begin{figure}[t]
\centering
   {\includegraphics[scale=0.18]{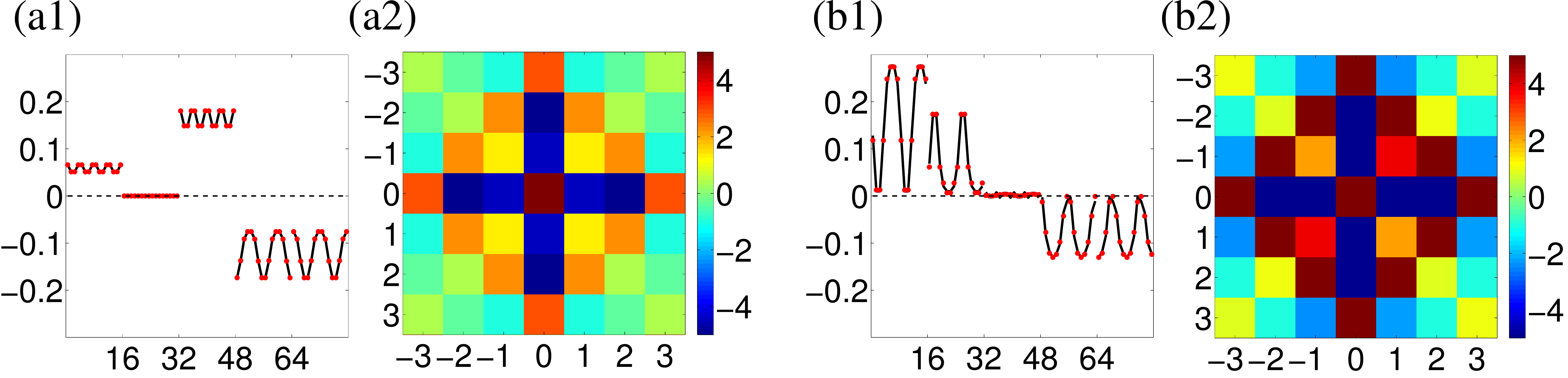}}
\caption{\label{fig:realspacesc} 
    Form factors $\tilde{f}^{SCs}_{ab}(k)$ (red points) and harmonic-fit (black lines) for the intra-orbital pairing in the $d_{X^2-Y^2}$-orbital (a1) and $d_{XZ}$-orbital (b1).
    (a2,b2) Corresponding pair wave functions $\tilde{f}^{SCs}_{aa}(\bs{d})$ in dependence of the relative distance $\bs{d} = \bs{r}_i - \bs{r}_j$ plotted in units of the lattice constant. 
    The pair wave function for the electrons in orbital $d_{YZ}$ is obtained by 90 degrees rotation of the one for $d_{XZ}$. Note
    that the $d_{XZ}$-orbital here points to the next-nearest neighbor site.} 
\end{figure} 
In Fig.~\ref{fig:realspacesc}, we then depicted $\tilde{f}^{SCs}_{ab}(k) = f^{SCs}(k)u^*_{am}(\bs{k})u^*_{bm}(-\bs{k})$ (red points) for the leading 
intra-orbital pairing $a=b$ between electrons in the $d_{X^2-Y^2}$-orbital (a1) and $d_{XZ}$-orbital (b1) along the Fermi-surface. The case of $d_{YZ}$ intra-orbital pairing
simply follows from the one of $d_{XZ}$ by $\pi/2$-rotation, and the pairing between orbitals other than $d_{XZ}$, $d_{YZ}$ and $d_{X^2-Y^2}$ is strongly suppressed 
as these three contribute by far the largest values in
$u_{am}(\bs{k})$. In addition, the pairing between different orbitals
is disfavored because $\bs{k}$ and $-\bs{k}$
share the same orbital weights by symmetry. After that, we further apply a harmonics fit (black lines), which leads to the real-space pair wave functions $f^{SCs}_{ab}(\bs{r}_i-\bs{r}_j)$. 
As the total Cooper-pair momentum is zero, the real-space pairing function features the same translational invariance as the underlying lattice, 
and we therefore depicted $f^{SCs}_{ab}(\bs{d})$ in Fig.~\ref{fig:formfacs}(a2,b2) as a function of the relative distance $\bs{d} = \bs{r}_i - \bs{r}_j$.
Here, it becomes apparent that the matrix elements $u_{am}(\bs{k})$ of (\ref{eq:backtransform})
obscure the relation between the form factors in Fig.~\ref{fig:formfacs} and its corresponding real-space ordering pattern. This is, by the way, also 
the reason why the $(\pi,0)/(0,\pi)$-SDW gap shows a nodal behavior although its real-space ordering is of conventional structure.\par
Unlike the spin-density phase, the leading charge-density wave channel shows a $p_x/p_y$-wave real-space structure for the ordering momenta $(\pi,0)$, $(0,\pi)$ 
which cannot be ascribed to charge modulations but characterizes a Peierls ordering as shown in the (CDW) inset of Fig.~\ref{fig:formfacs}. Here, certain
hopping ampliudes are weakened (thin lines) or enhanced (thick lines), and the corresponding order parameter turns out to be imaginary due to
\begin{align}\nonumber
 O^{CDW*} &= \sum_{i,a,s}(-1)^{i_x}\langle c^{\dagger}_{ias}c^{\phantom{\dagger}}_{i+xas} - c^{\dagger}_{ias}c^{\phantom{\dagger}}_{i-xas}\rangle^*\\\nonumber 
          &= \sum_{i,a,s}(-1)^{i_x}\langle c^{\dagger}_{i+xas}c^{\phantom{\dagger}}_{ias} - c^{\dagger}_{i-xas}c^{\phantom{\dagger}}_{ias}\rangle = - O^{CDW}.
\end{align}
Note that the real-space patterns displayed in the insets of Fig.~\ref{fig:formfacs} only represent the nearest-neighbor representation compatible
with the symmetry in each channel. Similar to the pair wave function in Fig.~\ref{fig:realspacesc}, there are in general also sizeable contributions from higher harmonics.\par  
The real-space form factors of the leading ferromagnetic and Pomeranchuk channels both reveal an $s$-wave structure
in band- and orbital space and its nearest-neighbor representations are shown in the respective insets. 
In fact, the $s$-wave Pomeranchuk phase does not break any symmetries and just corresponds to a uniform shift in the hopping-parameters. Other so-called nematic
ordering channels with $Q=(0,0)$ and non $s$-wave form factors, which
only break point-group symmetries, are subleading to its $s$-wave counterparts.\par 
\setcounter{MaxMatrixCols}{12} 
    \subsection{Mean-Field Hamiltonian for Long-Range Orders in the Kagome Lattice} \label{sec:frg:app_mean_field_hamiltonian}
      For long-range orders with finite ordering vector $\vec{Q}\neq(0,0)$, the translational symmetry of the system is broken. Hence, the unit cell has to be expanded. This results in a reduction of the Brillouin zone (BZ) and a back folding of the band dispersion, i.e. the number of bands is increased.\par
      As an example, the charge bond order (cBO) in the kagome lattice
      is considered (see Sec.~\ref{sec:kagome:bond}). There are three independent ordering vectors
      \begin{equation}
        \begin{split}
          \vec{Q}_1 = \pi \left( -\frac{1}{2} , -\frac{\sqrt{3}}{2} \right) \quad , \quad \vec{Q}_2 = \pi \left( 1 , 0 \right) \quad \text{and} \quad \vec{Q}_3 = \pi \left(-\frac{1}{2} , \frac{\sqrt{3}}{2} \right) \ ,
        \end{split}
      \end{equation}
      each corresponding to another real-space order. The expectation value $\bra \hat{c}^{\dagger}_{\vec{k}\gamma} \hat{c}^{\pdagger}_{\vec{k}+\vec{Q}\alpha} \ket$ (Eq.~\ref{eq:kagome:bond1}) features creation and annihilation operators at distinct sublattices and momenta. Hence, the Bogoliubov transform is increased in dimension.\par
      First, only $\vec{Q}_2$ is regarded, so the vector of creation operators reads
      \begin{equation}
        \begin{split}
          \hat{C}_2 = \begin{pmatrix} \hat{c}_{\vec{k}\alpha}^{\dagger} & \hat{c}_{\vec{k}\beta}^{\dagger} & \hat{c}_{\vec{k}\gamma}^{\dagger} & \hat{c}_{\vec{k}+\vec{Q}_2\alpha}^{\dagger} & \hat{c}_{\vec{k}+\vec{Q}_2\beta}^{\dagger} & \hat{c}_{\vec{k}+\vec{Q}_2\gamma}^{\dagger} \end{pmatrix} \ .
        \end{split}
      \end{equation}
      Consequently, the non-interacting Hamiltonian reads
      \begin{equation}\label{eq:bondmfham}
        \begin{split}
          \hat{H_0} &= \sum \limits_{\vec{k}}\hat{C}_2 \begin{pmatrix} -\mu & A_0 & B_0 & 0 & 0 & Y_0 \\ A_0 & -\mu & C_0 & 0 & 0 & 0 \\ B_0 & C_0 & -\mu & Y_0 & 0 & 0 \\ 0 & 0 & Y^*_0 & -\mu & A_{\vec{Q}_2} & B_{\vec{Q}_2} \\ 0 & 0 & 0 & A_{\vec{Q}_2} & -\mu & C_{\vec{Q}_2} \\ Y^*_0 & 0 & 0 & B_{\vec{Q}_2} & C_{\vec{Q}_2} & -\mu \end{pmatrix} \hat{C}^{\dagger}_2
        \end{split}
      \end{equation}
      with the hopping amplitudes
      \begin{equation}
        \begin{split}
          A_{\vec{p}} &= - 2 t_1 \cos \left( \frac{k_x-p_x}{2}-\sqrt{3}\frac{k_y-p_y}{2} \right)\\
          B_{\vec{p}} &= - 2 t_1 \cos \left( k_x-p_x \right)\\
          C_{\vec{p}} &= - 2 t_1 \cos \left( \frac{k_x-p_x}{2}+\sqrt{3}\frac{k_y-p_y}{2} \right)\\
        \end{split}
      \end{equation}
      and the Weiss field
      \begin{equation}
        \begin{split}
          Y_{\vec{p}} &= \ii \Delta \sin \left( k_x-p_x \right) \ .
        \end{split}
      \end{equation}
      The one-particle spectrum of the mean-field Hamiltonian (\ref{eq:bondmfham}) then reveals the opening of a partial gap at the FS.\par
      As discussed in Sec.~\ref{sec:kagome:bond}, it is possible to apply $\vec{Q}_{1}$, $\vec{Q}_{2}$ and $\vec{Q}_{3}$ simultaneously. Hence, the unit cell is expanded to $12$ sites. Analogously to the above derivation, a $12$-vector is defined for the creation operators:
      \begin{equation}
        \begin{split}
          \hat{C}_{12} &= \left( \hat{c}_{\vec{k}\alpha}^{\dagger} \quad \hat{c}_{\vec{k}\beta}^{\dagger} \quad \hat{c}_{\vec{k}\gamma}^{\dagger} \quad \hat{c}_{\vec{k}+\vec{Q}_1\alpha}^{\dagger} \quad \hat{c}_{\vec{k}+\vec{Q}_1\beta}^{\dagger} \quad \hat{c}_{\vec{k}+\vec{Q}_1\gamma}^{\dagger}\dots\right.\\
                       &\qquad\dots\left.\hat{c}_{\vec{k}+\vec{Q}_2\alpha}^{\dagger} \quad \hat{c}_{\vec{k}+\vec{Q}_2\beta}^{\dagger} \quad \hat{c}_{\vec{k}+\vec{Q}_2\gamma}^{\dagger} \quad \hat{c}_{\vec{k}+\vec{Q}_3\alpha}^{\dagger} \quad \hat{c}_{\vec{k}+\vec{Q}_3\beta}^{\dagger} \quad \hat{c}_{\vec{k}+\vec{Q}_3\gamma}^{\dagger} \right) \ .
        \end{split}
      \end{equation}
      The Weiss fields corresponding to $\vec{Q}_1$ and $\vec{Q}_3$ are introduced as
      \begin{equation}
        \begin{split}
          X_{\vec{p}} &= \ii \Delta \sin \left( -\frac{k_x-p_x}{2}-\sqrt{3}\frac{k_y-p_y}{2} \right)\\
          Z_{\vec{p}} &= \ii \Delta \sin \left(-\frac{k_x-p_x}{2}+\sqrt{3}\frac{k_y-p_y}{2} \right) \ .
        \end{split}
      \end{equation}
      Additionally, one can show that $\vec{Q}_1-\vec{Q}_2\hat{=}\vec{Q}_3$. So, the complete $12\times12$ matrix can be built up:
       \begin{equation}
         \begin{split}
           \hat{H_0} &= \sum \limits_{\vec{k}}\hat{C}_{12} 
            \begin{pmatrix} 
            -\mu & A_0 & B_0 & 0 & 0 & 0  & 0 & 0 & Y_0 & 0 & Z_0 & 0 \\
             A_0 & -\mu  & C_0   & 0    & 0 & X_0 & 0 & 0 & 0 & Z_0 & 0 & 0 \\ 
             B_0 & C_0   & -\mu  & 0    & X_0 & 0 & Y_0 & 0 & 0 & 0 & 0 & 0 \\ 
             0   & 0     & 0     & -\mu        & A_{\vec{Q}_1} & B_{\vec{Q}_1} & 0 & Z_{\vec{Q}_1} & 0 & 0 & 0 & Y_{\vec{Q}_1} \\ 
             0   & 0     & X^*_0 & A_{\vec{Q}_1} & -\mu & C_{\vec{Q}_1} & Z_{\vec{Q}_1} & 0 & 0 & 0 & 0 & 0 \\ 
             0   & X^*_0 & 0     & B_{\vec{Q}_1} & C_{\vec{Q}_1} & -\mu & 0 & 0 & 0 & Y_{\vec{Q}_1} & 0 & 0 \\ 
             0   & 0     & Y^*_0 & 0 & Z^*_{\vec{Q}_1} & 0 & -\mu & A_{\vec{Q}_2} & B_{\vec{Q}_2} & 0 & 0 & 0 \\ 
             0   & 0     & 0     & Z^*_{\vec{Q}_1} & 0 & 0 & A_{\vec{Q}_2} & -\mu & C_{\vec{Q}_2} & 0 & 0 & X_{\vec{Q}_2} \\ 
           Y^*_0 & 0     & 0     & 0 & 0 & 0 & B_{\vec{Q}_2} & C_{\vec{Q}_2} & -\mu & 0 & X_{\vec{Q}_2} & 0 \\ 
             0   & Z^*_0 & 0     & 0 & 0 & Y^*_{\vec{Q}_1} & 0 & 0 & 0 & -\mu & A_{\vec{Q}_3} & B_{\vec{Q}_3} \\ 
           Z^*_0 & 0     & 0     & 0 & 0 & 0 & 0 & 0 & X^*_{\vec{Q}_2} & A_{\vec{Q}_3} & -\mu & C_{\vec{Q}_3} \\ 
             0   & 0     & 0     & Y^*_{\vec{Q}_1} & 0 & 0 & 0 & X^*_{\vec{Q}_2} & 0 & B_{\vec{Q}_3} & C_{\vec{Q}_3} & -\mu 
             \end{pmatrix} 
            \hat{C}^{\dagger}_{12} \\
           &= \sum \limits_{\vec{k}} \sum \limits_{n=1}^{12} \mathcal{E}_n(\vec{k}) \ \hat{c}^{\dagger}_{\vec{k}n} \hat{c}^{\pdagger}_{\vec{k}n} \label{eq:kagome:app_12band_ham}
         \end{split}
       \end{equation}
The result is that the generic FS is completely gapped. Only the $M$- and $\Gamma$-points remain gapless.\par
      For the calculation of the free energy, the full Hamiltonian can be rewritten with a mean-field decoupling:
      \begin{equation}
        \begin{split}
          \hat{H}_{\text{cCP}} &= \sum \limits_{\vec{k}} \sum \limits_{n=1}^{3} \varepsilon_n(\vec{k}) \ \hat{c}^{\dagger}_{\vec{k}n} \hat{c}^{\pdagger}_{\vec{k}n} + \frac{1}{N} \sum \limits_{\vec{k}\vec{p}} \sum \limits_{i=1}^{3} V_{\text{cCP}} \ \hat{c}^{\dagger}_{\vec{k}} \hat{c}^{\dagger}_{\vec{p}} \hat{c}^{\pdagger}_{\vec{p}-\vec{Q}_i} \hat{c}^{\pdagger}_{\vec{k}+\vec{Q}_i}\\
          &= \sum \limits_{\vec{k}} \sum \limits_{n=1}^{3} \varepsilon_n(\vec{k}) \ \hat{c}^{\dagger}_{\vec{k}n} \hat{c}^{\pdagger}_{\vec{k}n} +\frac{1}{N} \sum \limits_{\vec{k}\vec{p}} \sum \limits_{i=1}^{3} V_{\text{cCP}} \left( - \bra \hat{c}^{\dagger}_{\vec{k}} \hat{c}^{\pdagger}_{\vec{k}+\vec{Q}_i} \ket \bra \hat{c}^{\dagger}_{\vec{p}} \hat{c}^{\pdagger}_{\vec{p}-\vec{Q}_i} \ket \right.\\
          &\qquad+ \left. \bra \hat{c}^{\dagger}_{\vec{k}} \hat{c}^{\pdagger}_{\vec{k}+\vec{Q}_i} \ket \hat{c}^{\dagger}_{\vec{p}} \hat{c}^{\pdagger}_{\vec{p}-\vec{Q}_i} + \hat{c}^{\dagger}_{\vec{k}} \hat{c}^{\pdagger}_{\vec{k}+\vec{Q}_i} \bra \hat{c}^{\dagger}_{\vec{p}} \hat{c}^{\pdagger}_{\vec{p}-\vec{Q}_i} \ket \right)\\
          &= \sum \limits_{\vec{k}} ( \sum \limits_{n=1}^{3} \varepsilon_n(\vec{k}) \ \hat{c}^{\dagger}_{\vec{k}n} \hat{c}^{\pdagger}_{\vec{k}n} + \sum \limits_{i=1}^{3} K_i\\
          &\qquad+ \sum \limits_{i=1}^{3} \underbrace{\frac{1}{N} \sum \limits_{\vec{p}} V_{\text{cCP}} \ \bra \hat{c}^{\dagger}_{\vec{k}} \hat{c}^{\pdagger}_{\vec{k}+\vec{Q}_i} \ket}_{\vec{p}\leftrightarrow\vec{k}, \ \vec{Q}\leftrightarrow-\vec{Q}, \ \Rightarrow \Delta^{\text{cCP}}_{\vec{k}i}} \hat{c}^{\dagger}_{\vec{p}} \hat{c}^{\pdagger}_{\vec{p}-\vec{Q}_i} + \sum \limits_{i=1}^{3} \underbrace{\frac{1}{N} \sum \limits_{\vec{p}} V_{\text{cCP}} \ \bra \hat{c}^{\dagger}_{\vec{p}} \hat{c}^{\pdagger}_{\vec{p}-\vec{Q}_i} \ket}_{\Delta^{\text{cCP}}_{\vec{k}i}} \hat{c}^{\dagger}_{\vec{k}} \hat{c}^{\pdagger}_{\vec{k}+\vec{Q}_i} )\\
          &= \sum \limits_{\vec{k}} ( \sum \limits_{n=1}^{3} \varepsilon_n(\vec{k}) \ \hat{c}^{\dagger}_{\vec{k}n} \hat{c}^{\pdagger}_{\vec{k}n} + \sum \limits_{i=1}^{3} ( 2\Delta^{\text{cCP}}_{\vec{k}i} \hat{c}^{\dagger}_{\vec{k}} \hat{c}^{\pdagger}_{\vec{k}+\vec{Q}_i} +  K_i))
        \end{split}
      \end{equation}
      Here, an analytic Bogoliubov transformation is not feasible due
      to the effective $12$-band structure. Within mean-field
      approximation, the one-particle excitation energies can still be
      calculated numerically ($\mathcal{E}(\vec{k})$ in
      Eq.~(\ref{eq:kagome:app_12band_ham})), so that the total free energy reads
      \begin{equation}
        \begin{split}
          \Omega = - T \underbrace{\sum \limits_{\vec{k}} \sum \limits_{n=1}^{12} \ln \left( 1 + \ee^{-\frac{\mathcal{E}_n(\vec{k})}{T}} \right)}_A - 2 \underbrace{ \sum \limits_{i=1}^{3} \sum \limits_{\vec{k}} \bra \hat{c}^{\dagger}_{\vec{k}} \hat{c}^{\pdagger}_{\vec{k}+\vec{Q}_i} \ket \Delta^{\text{cCP}}_{\vec{k}i}}_B \ .\label{eq:frg:app_grand_canonical_potential_ccp}
        \end{split}
      \end{equation}
      The second addend $B$ is a sum over three mean fields, each
      corresponding to a quasi one-dimensional bond order
      (c.f. Fig.~\ref{gr:kagome:bond_real_space_pattern}). Under a
      rotation by $\frac{\pi}{3}$, these real space patters can be
      transformed into each other, hence the energy gain for the
      creation of each mean field is identical. In other words, it is
      linear in the application of the individual mean fields. For
      addend $A$, this statement is less direct to be identified, but a straightforward numerical evaluation yields the linear relation. Consequently, the ordering formation along the individual bond directions is indeed independent.\newpage
\subsection{Longer-Ranged Interactions in Hexagonal Lattices}\label{sec:hexbandint}
The following section summarizes the coupling terms occuring in the honeycomb and kagome lattices with longer-ranged Coulomb interactions.
\begin{figure}[h!]
	\centering
    \includegraphics[width=0.99\textwidth]{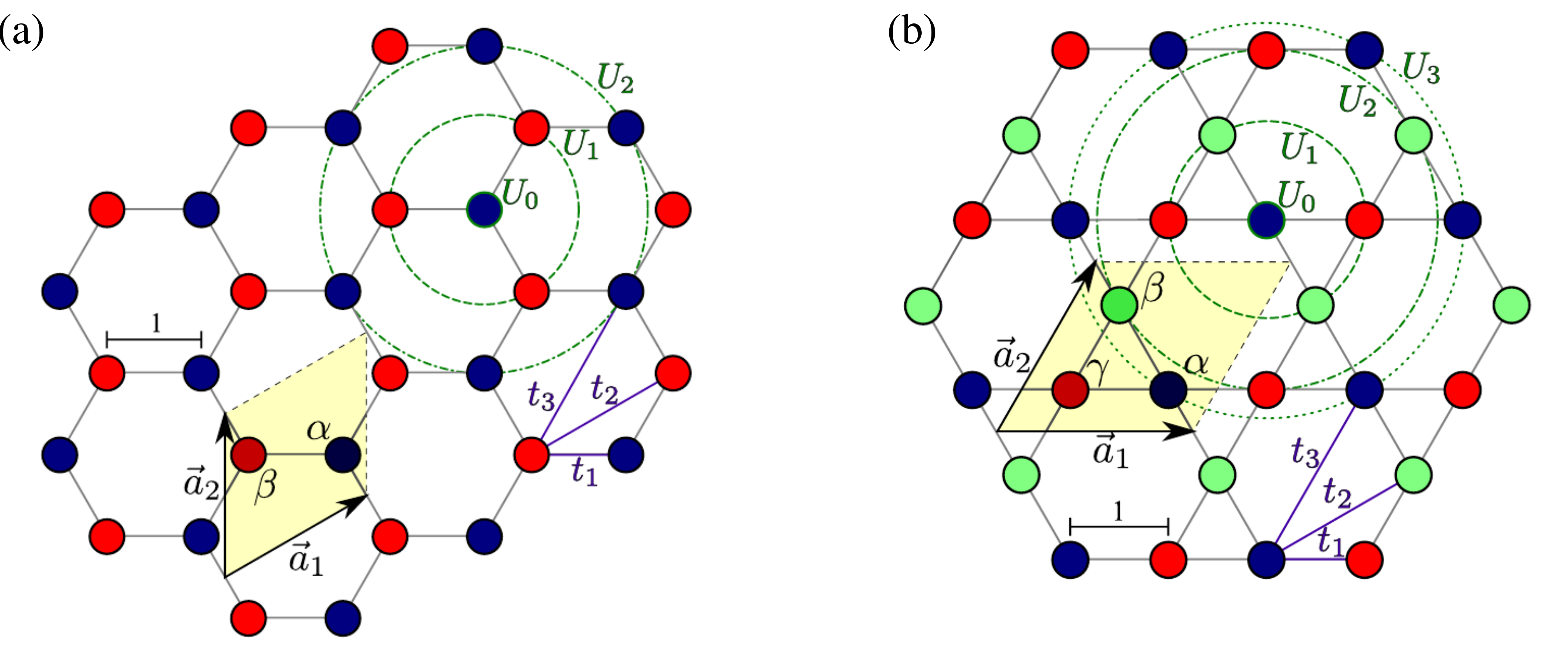}
    \caption{(a) Honeycomb and (b) Kagome lattice structure with two $(\alpha,\beta)$ and three $(\alpha,\beta,\gamma)$ sublattices, respectively.}
    \label{fig:hexlattices}
\end{figure}
Employing the notation of Sect.~\ref{subsec:bandorb}, the tight-binding Hamiltonian $H_{tot} = H_0 + H_{int}$ 
for the honeycomb and kagome lattice in band basis is given by
\begin{align}
H_{total} & = \sum_{\bs{k},s}\sum_{m}E^{\phantom{\dagger}}_{m}(\bs{k})\gamma_{\bs{k}ms}^{\dagger}\gamma_{\bs{k}ms}^{\phantom{\dagger}} \\\nonumber
+ &\sum_{\bs{k}_1,\ldots\bs{k}_4\atop s,s'}\sum_{m_1,\ldots,m_4}V_0(\bs{k}_1m_1,\bs{k}_2m_2,\bs{k}_3m_3,\bs{k}_4m_4)
\gamma_{\bs{k}_1m_1s}^{\dagger}\gamma_{\bs{k}_2m_2s'}^{\dagger}\gamma_{\bs{k}_3m_3s}^{\phantom{\dagger}}\gamma_{\bs{k}_4m_4s'}^{\phantom{\dagger}}.
\end{align}
The coupling function for the longer-ranged interactions in Fig.~\ref{fig:hexlattices} then reads as
\begin{align}\nonumber
V_0(\bs{k}_1m_1,\bs{k}_2m_2,\bs{k}_3m_3,\bs{k}_4m_4) 
  = & \\
U_i\sum_{ab}WW_{ab}(\bs{k}_3-\bs{k}_1)&u^*_{am_1}(\bs{k}_1)u^*_{bm_2}(\bs{k}_2)u_{am_3}(\bs{k}_3)u_{bm_4}(\bs{k}_4),
\end{align}
where the sum runs over all sublattice combinations in the tables below, and $WW(\bs{k}_3-\bs{k}_1)$ is determined by the corresponding entry in 
the rightmost column. 

\begin{center}
   \begin{tabular}{c|c|c|c}
    a & b & $\bs{r}$ (1st neighbor honeycomb)  & $WW_{ab}(\bs{q})$  \\\hline
    $\alpha$ & $\beta$ & $\left\{ \begin{pmatrix} -1 \\ 0 \end{pmatrix}, \begin{pmatrix} 1/2 \\ \sqrt{3}/2 \end{pmatrix}, \begin{pmatrix} 1/2 \\ -\sqrt{3}/2 \end{pmatrix} \right\}$ &  
    $e^{-q_x}+2e^{q_x/2}\cos(\sqrt{3}q_y/2)$ \\ 
    $\beta$  & $\alpha$& $\left\{ \begin{pmatrix} 1 \\ 0 \end{pmatrix}, \begin{pmatrix} -1/2 \\ \sqrt{3}/2 \end{pmatrix}, \begin{pmatrix} -1/2 \\ -\sqrt{3}/2 \end{pmatrix} \right\}$ &  
    $e^{q_x}+2e^{-q_x/2}\cos(\sqrt{3}q_y/2)$ \\\hline\hline 
    a & b & $\bs{r}$ (2nd neighbor honeycomb)                   & $WW_{ab}(\bs{q})$  \\\hline
    $\alpha$ & $\alpha$ & $\left\{ \begin{pmatrix} 0 \\ \pm\sqrt{3} \end{pmatrix}, \begin{pmatrix} \pm 3/2 \\ \pm\sqrt{3}/2 \end{pmatrix}, \begin{pmatrix} \pm 3/2 \\ \mp\sqrt{3}/2 \end{pmatrix} \right\}$ &  
    $2\cos(\sqrt{3}q_y) + 4\cos(3q_x/2)\cos(\sqrt{3}q_y/2)$ \\ 
    $\beta$  & $\beta$& $\left\{ \begin{pmatrix} 0 \\ \pm\sqrt{3} \end{pmatrix}, \begin{pmatrix} \pm 3/2 \\ \pm\sqrt{3}/2 \end{pmatrix}, \begin{pmatrix} \pm 3/2 \\ \mp\sqrt{3}/2 \end{pmatrix} \right\}$ &  
    $2\cos(\sqrt{3}q_y) + 4\cos(3q_x/2)\cos(\sqrt{3}q_y/2)$
    \end{tabular} 
\end{center}

\begin{center}
   \begin{tabular}{c|c|c|c}
    a & b & $\bs{r}$ (1st neighbor kagome)  & $WW_{ab}(\bs{q})$  \\\hline
    $\alpha$ & $\beta$ & $\left\{ \begin{pmatrix} \pm 1/2 \\ \mp\sqrt{3}/2 \end{pmatrix} \right\}$ &  $2\cos((q_x - \sqrt{3}q_y)/2)$ \\
    $\alpha$ & $\gamma$& $\left\{ \begin{pmatrix} \pm 1 \\ 0 \end{pmatrix} \right\}$ &  $2\cos(q_x)$ \\
    $\beta$  & $\alpha$& $\left\{ \begin{pmatrix} \pm 1/2 \\ \mp\sqrt{3}/2 \end{pmatrix} \right\}$ &  $2\cos((q_x - \sqrt{3}q_y)/2)$ \\
    $\beta$ & $\gamma$ & $\left\{ \begin{pmatrix} \pm 1/2 \\ \pm\sqrt{3}/2 \end{pmatrix} \right\}$ &  $2\cos((q_x + \sqrt{3}q_y)/2)$ \\
    $\gamma$ & $\alpha$ & $\left\{ \begin{pmatrix} \pm 1 \\ 0 \end{pmatrix} \right\}$ &  $2\cos(q_x)$ \\
    $\gamma$ & $\beta$ & $\left\{ \begin{pmatrix} \pm 1/2 \\ \pm\sqrt{3}/2 \end{pmatrix} \right\}$ &  $2\cos((q_x + \sqrt{3}q_y)/2)$ \\\hline\hline
    a & b & $\bs{r}$ (2nd neighbor kagome)  & $WW_{ab}(\bs{q})$  \\\hline
    $\alpha$ & $\beta$ & $\left\{ \begin{pmatrix} \pm 3/2 \\ \pm\sqrt{3}/2 \end{pmatrix} \right\}$ &  $2\cos((3q_x + \sqrt{3}q_y)/2)$ \\
    $\alpha$ & $\gamma$& $\left\{ \begin{pmatrix} 0 \\ \sqrt{3} \end{pmatrix} \right\}$ &  $2\cos(\sqrt{3}q_y)$ \\
    $\beta$  & $\alpha$& $\left\{ \begin{pmatrix} \pm 3/2 \\ \pm\sqrt{3}/2 \end{pmatrix} \right\}$ &  $2\cos((3q_x + \sqrt{3}q_y)/2)$ \\
    $\beta$ & $\gamma$ & $\left\{ \begin{pmatrix} \pm 3/2 \\ \mp\sqrt{3}/2 \end{pmatrix} \right\}$ &  $2\cos((3q_x - \sqrt{3}q_y)/2)$ \\
    $\gamma$ & $\alpha$ & $\left\{ \begin{pmatrix} 0 \\ \pm\sqrt{3} \end{pmatrix} \right\}$ &  $2\cos(\sqrt{3}q_y)$ \\
    $\gamma$ & $\beta$ & $\left\{ \begin{pmatrix} \pm 3/2 \\ \mp\sqrt{3}/2 \end{pmatrix} \right\}$ &  $2\cos((3q_x - \sqrt{3}q_y)/2)$ \\\hline\hline
    a & b & $\bs{r}$ (3rd neighbor kagome)  & $WW_{ab}(\bs{q})$  \\\hline
    $\alpha$ & $\alpha$ & $\left\{ \begin{pmatrix} \pm 2 \\ 0 \end{pmatrix},\begin{pmatrix} \pm 1 \\ \pm\sqrt{3} \end{pmatrix},\begin{pmatrix} \pm 1 \\ \mp\sqrt{3} \end{pmatrix} \right\}$ 
     &  $2\cos(2q_x) + 4\cos(q_x)\cos(\sqrt{3}q_y)$ \\
    $\beta$ & $\beta$ & $\left\{ \begin{pmatrix} \pm 2 \\ 0 \end{pmatrix},\begin{pmatrix} \pm 1 \\ \pm\sqrt{3} \end{pmatrix},\begin{pmatrix} \pm 1 \\ \mp\sqrt{3} \end{pmatrix} \right\}$ 
     &  $2\cos(2q_x) + 4\cos(q_x)\cos(\sqrt{3}q_y)$ \\
     $\gamma$ & $\gamma$ & $\left\{ \begin{pmatrix} \pm 2 \\ 0 \end{pmatrix},\begin{pmatrix} \pm 1 \\ \pm\sqrt{3} \end{pmatrix},\begin{pmatrix} \pm 1 \\ \mp\sqrt{3} \end{pmatrix} \right\}$ 
     &  $2\cos(2q_x) + 4\cos(q_x)\cos(\sqrt{3}q_y)$ \\
    \end{tabular}
\end{center}


 \vspace{36pt}

\section{Obtaining the tADP2e Class file}\label{FTP}

\subsection{Via the Taylor \& Francis website}

\noindent This Guide for Authors and the tADP2e.cls Class file may be obtained via the Instructions for Authors on the Taylor \& Francis homepage for the journal ({\tt{http://www.tandf.co.uk/journals/titles/00018732.asp}}).

Please note that the Class file calls up the following open-source LaTeX packages, which will, for convenience, unpack with the downloaded Guide for Authors and Class file: amsbsy.sty, amsfonts.sty, amsmath.sty, amssymb.sty, enumerate.sty, epsfig.sty, graphicx.sty, mathbbol.sty, natbib.sty, rotating.sty, subfigure.sty.

\subsection{Via e-mail}

This Guide for Authors, the Class file and the associated open-source LaTeX packages are also available by e-mail. Requests should be addressed to {\tt latex.helpdesk@tandf.co.uk} clearly stating for which journal you require the Guide for Authors and/or Class file.


\label{lastpage}

\end{document}